\documentclass{aa}  
\usepackage{graphicx}
\usepackage{txfonts}
\usepackage[T1]{fontenc}
\usepackage{graphicx}
\usepackage{multirow}
\usepackage{blindtext}
\usepackage{amsmath}
\usepackage[usenames, dvipsnames]{color}
\usepackage{booktabs}
\usepackage{scrextend}
\usepackage{mathtools}
\usepackage{amsmath}
\usepackage[shortlabels]{enumitem}
\usepackage{lipsum}
\usepackage[normalem]{ulem}
\usepackage{diagbox}
\newcommand{\Hb}{H$\beta$\xspace}
\newcommand{\Ha}{H$\alpha$\xspace}

\newcommand{\LHa}{$ L_{\rm H\alpha}$\xspace}

\newcommand{\LOII}{$L_{\rm{\left[\mathrm{O\,\textrm{\sc{ii}}}\right]}}$\xspace}

\newcommand{\OI}{$\left[\mathrm{O\,\textrm{\sc{i}}}\right]$\xspace}
\newcommand{\SII}{$\left[\mathrm{S\,\textrm{\sc{ii}}}\right]$\xspace}
\newcommand{\OIII}{$\left[\mathrm{O\,\textrm{\sc{iii}}}\right]$\xspace}
\newcommand{\OII}{$\left[\mathrm{O\,\textrm{\sc{ii}}}\right]$\xspace}
\newcommand{\NII}{$\left[\mathrm{N\,\textrm{\sc{ii}}}\right]$\xspace}

\newcommand{\parent}{{\em parent-ELG}\xspace}
\newcommand{\mainsel}{{\em main-ELG}\xspace}
\definecolor{mygray}{gray}{0.5}

\begin{document} 
   \title{Characterizing the ELG luminosity functions in the nearby Universe}
   \titlerunning{\em ELG LFs at $z\sim0.1$}
   \author{G. Favole,\thanks{E-mail: gfavole@iac.es}
          \inst{1,2}
          V. Gonzalez-Perez,\inst{3,4} Y. Ascasibar,\inst{3,4}
 P. Corcho-Caballero,\inst{3,5} A. D. Montero-Dorta,\inst{6} A. J. Benson,\inst{7} J. Comparat,\inst{8} S. A. Cora,\inst{9,10} D. Croton\inst{11} H. Guo,\inst{12} D. Izquierdo-Villalba,\inst{13,14} A. Knebe,\inst{3,4,15} \'A. Orsi,\inst{16} \\
 D. Stoppacher,\inst{3,17,18} C. A. Vega-Mart\'inez\,\inst{19,20}
          }  
          \authorrunning{\em Favole et al. 2023}

\institute{Instituto de Astrof\'{\i}sica de Canarias, s/n, E-38205, La Laguna, Tenerife, Spain
\and
Departamento de Astrof\'{\i}sica, Universidad de La Laguna, E-38206, La Laguna, Tenerife, Spain
\and
Departamento de F\'isica Te\'orica, Facultad de Ciencias, Universidad Aut\'onoma de Madrid, E-28049, Spain
\and
Centro de Investigaci\'on Avanzada en F\'isica Fundamental, Facultad de Ciencias, Universidad Aut\'onoma de Madrid, E-28049 Madrid, Spain
\and
Australian Astronomical Optics, Macquarie University, 105 Delhi Rd, North Ryde, NSW 2113, Australia
\and
Departamento de F\'isica, Universidad T\'ecnica Federico Santa Mar\'ia, Casilla 110-V, Avda. Espa\~na 1680, Valpara\'iso, Chile
\and
Carnegie Observatories, 813 Santa Barbara Street, Pasadena, CA 91101, USA
\and
Max-Planck-Institut f\"{u}r extraterrestrische Physik (MPE), Giessenbachstrasse 1, D-85748 Garching bei M\"{u}nchen, Germany
\and
Instituto de Astrof\'isica de La Plata (CCT La Plata, CONICET, UNLP), Paseo del Bosque s/n, B1900FWA, La Plata, Argentina
\and
Facultad de Ciencias Astron\'omicas y Geof\'isicas, UNLP, Paseo del Bosque s/n, B1900FWA, La Plata, Argentina
\and
Centre for Astrophysics \& Supercomputing, Swinburne University of Technology, P.O.B. 218, Hawthorn, Victoria 3122, Australia
\and
Key Laboratory for Research in Galaxies and Cosmology, Shanghai Astronomical Observatory, Shanghai 200030, China
\and
Dipartimento di Fisica ``G. Occhialini'', Universit\`{a} degli Studi di Milano-Bicocca, Piazza della Scienza 3, I-20126 Milano, Italy
\and
INFN, Sezione di Milano-Bicocca, Piazza della Scienza 3, 20126 Milano, Italy
\and
International Centre for Radio Astronomy Research, The University of Western Australia, Crawley, WA 6009, Australia
\and
PlantTech Research Institute Limited. South British House, 4th Floor, 35 Grey Street, Tauranga 3110, New Zealand
\and
Instituto de Astrof\'isica, Pontificia Universidad Cat\'olica de Chile, Campus San Joaqu\'in, Avda. Vicu\~na Mackenna 4860, Santiago, Chile
\and
Facultad de F\'isicas, Universidad de Sevilla, Avda.\ Reina Mercedes s/n, Campus de Reina Mercedes, 41012 Sevilla, Spain
\and
Instituto de Investigaci\'on Multidisciplinar en Ciencia y Tecnolog\'ia, Universidad de La Serena, Ra\'ul Bitr\'an 1305, La Serena, Chile
\and
Departamento de Astronom\'ia, Universidad de La Serena, Av. Juan Cisternas 1200 Norte, La Serena, Chile
             }

   \date{Received xxx, 2023; accepted xxx}

  \abstract
   {Nebular emission lines are powerful diagnostics for the physical processes at play in galaxy formation and evolution. Moreover, emission-line galaxies (ELGs) are one of the main targets of current and forthcoming spectroscopic cosmological surveys.}
   {We investigate the contributions to the line luminosity functions (LFs) of different galaxy populations in the local Universe, providing a benchmark for future surveys of earlier cosmic epochs.}
   {The large statistics of the observations from the SDSS DR7 main galaxy sample and the MPA-JHU spectral catalog enabled us to precisely measure the \Ha, \Hb, \OII, \OIII, and, for the first time, the \NII, and \SII emission-line LFs over $\sim2.4$ Gyrs in the low-$z$ Universe, $0.02<z<0.22$. We present a generalized $1/V_{\rm max}$ LF estimator capable of simultaneously correcting for spectroscopic, $r-$band magnitude, and emission-line incompleteness. We studied the contribution to the LF of different types of ELGs classified using two methods: (i) the value of the specific star formation rate (sSFR), and (ii) the line ratios on the Baldwin--Phillips--Terlevich (BPT) and the WHAN (i.e., \Ha equivalent width, EW$_{\rm H\alpha}$, versus the \NII/\Ha line ratio) diagrams.}
   {
   The ELGs in our sample are mostly star forming, with $84$ percent having ${\rm sSFR}>10^{-11}{\rm yr^{-1}}$.
   When classifying ELGs using the BPT+WHAN diagrams, we find that $63.3$ percent are star forming, only $0.03$ are passively evolving, and $1.3$ have nuclear activity (Seyfert). The rest are low-ionization narrow emission-line regions (LINERs) and composite ELGs.
   We found that a Saunders function is the most appropriate to describe all of the emission-line LFs, both observed and dust-extinction-corrected (i.e., intrinsic).
   They are dominated by star-forming regions, except for the bright end of the \OIII and \NII LFs (i.e., $L_{[\rm NII]}>10^{42}{\rm erg\,s^{-1}}$, $L_{[\rm OIII]}>10^{43}{\rm erg\,s^{-1}}$), where the contribution of Seyfert galaxies is not negligible.
   In addition to the star-forming population, composite galaxies and LINERs are the ones that contribute the most to the ELG numbers at $L < 10^{41}\,{\rm erg\,s^{-1}}$. We do not observe significant evolution with redshift of our ELGs at $0.02<z<0.22$.  
   All of our results, including data points and analytical fits, are publicly available.
   }
   {Local ELGs are dominated by star-forming galaxies, except for the brightest $[\rm NII]$ and $[\rm OIII]$ emitters, which have a large contribution of Seyfert galaxies. The local line luminosity functions are best described by Saunders functions. We expect these two conclusions to hold up at higher redshifts for the ELG targeted by current cosmological surveys, such as DESI and Euclid.
   }
               
   \maketitle
   
%

\section{Introduction}
Current and upcoming spectroscopic cosmological surveys, such as the Dark Energy Spectroscopic Instrument\,\citep[DESI;][]{desi} and Euclid\,\citep{euclid}, rely on galaxies with strong spectral emission, or emission-line galaxies (ELGs), to build accurate and deep 3D cosmic maps and infer the cosmological composition and evolution of the Universe. 
According to the origin of their spectral lines, different types of ELGs might trace different regions of the cosmic web, or might be the result of a different evolution: for instance, quasars (QSOs) are more strongly clustered than star-forming (SF) galaxies\,\citep{Zhao2021}.
Until now, the statistical errors of cosmological surveys have been larger than the uncertainties due to our lack of understanding of the galaxy formation and evolution processes~\citep{Avila2020,Raichoor2021}. However, this might change with the new generation of Stage-IV cosmological surveys, such as DESI \citep{desi}, Euclid \citep{euclid}, the 4-metre Multi-Object Spectroscopic Telescope \citep[4MOST;][]{dejong2012}, Subaru Prime Focus Spectrograph \,\citep[PSF;][]{psf}, or SphereX \citep{spherex}.

ELGs are also interesting as they have enabled us to reconstruct the cosmic star formation history (SFH) out to $z\sim 2$ \citep[e.g.,][]{1998ARA&A..36..189K, 1998ApJ...498..106M, 2002A&A...387..396A, 2002AJ....124.3135K, Kewley2004, 2003ApJ...599..971H, 2007ApJ...666..870C, 2010ApJ...714.1256C, 2006ApJ...642..775M,2007ApJS..173..267S, 2007ApJ...671..333K, 2009ApJ...703.1672K, 2009ApJ...692..556R, 2010ApJ...719.1191T}. 
The study of star formation from emission lines has been possible thanks to an immense observational effort over the course of the last decades.
In the past, high-sensitivity infra-red (IR) space telescopes, such as Spitzer\footnote{\url{http://irsa.ipac.caltech.edu/data/SPITZER/docs/}} or Herschel\footnote{\url{http://sci.esa.int/herschel/}}, enabled the calibration of monochromatic star formation rate (SFR) indicators in nearby galaxies \citep[]{2013seg..book.....F, 2013seg..book..419C}, complementing the efforts in the UV and optical channels to map the SFR evolution of galaxies out to $z\sim9$ \citep[]{2004ApJ...600L.103G, 2009ApJ...705..936B, 2010ApJ...709L.133B}.

While SF regions constitute the main origin of the spectral emission lines for ELGs \citep[e.g.,][]{Kennicut1992, Sobral2013, Pirzkal2018, Xiao2018, Kewley2019}, other origins are also possible, such as active galactic nuclei\,\citep[AGN; e.g.,][]{Marziani2017,Lin2022}, shocks~\citep[e.g.,][]{hirschmann2022} and old stellar populations\,\citep[e.g.,][]{Kennicut1992,Sansom2015,Byler2019,Nersesian2019,Clarke2021}.
Emission-line diagnostic ratios, such as the BPT diagram \citep{Baldwin1981} or the $\rm D_n(4000)$ break index \citep{Bruzual1983, Balogh1999}, have been used to separate SF ELGs from AGN, as well as older and younger stellar population contributions \citep[e.g.,][]{Kewley2001, Kewley2006, Kauffmann2003, Kauffman2003D4000, Gallazzi2005, Belfiore2016, Wu2018, Angthopo2020}.
The WHAN diagram, relating the equivalent width of the \Ha line and the \NII/\Ha ratio \citep[e.g.,][]{Stasi2006, Cid2011} provides additional information to discriminate between SF and active galaxies, and the relation between EW$_{\rm H\alpha}$ and the D4000 index (the so-called aging diagram) has been proposed to identify sudden changes in the recent star formation activity \citep{Casado15, CorchoCaballero20, CorchoCaballero21b, CorchoCaballero23}.

Over the years, several studies have combined ELG observations both from spectroscopic and imaging surveys in order to constrain the emission-line luminosity functions.
The \Ha \citep{1995ApJ...455L...1G, Tresse2002} and the \OII \citep{Gallego2002} LFs were among the first ones to be characterized in the local Universe. 
\citet[]{Fujita2003} at $z=0.24$ and \citet{Ly2007} at $0.07<z<1.47$ used broad-band galaxy colors to discriminate \Ha from other lines, finding that the \Ha LF evolution is stronger in the faint end than in the bright one.
\citet[]{Gilbank2010} explored  the \OII, \Ha, and $u$-band luminosities as SFR indicators at $z<0.2$, finding that, in the high-mass end (i.e., $\rm M_\star>10^{10}\,M_\odot$), \OII needs a larger correction to compensate for the effects of metallicity dependence and dust extinction. 
\citet[]{2013MNRAS.433.2764G} studied the \Ha LF and SFR density at $z<0.35$, observing an increasing number of SF galaxies in the faint end.
\citet[]{Sobral2013} studied the SFH and \Ha LF evolution at $0.40<z<2.2$, finding that the \Ha line traces the bulk of star formation over the last 11\,Gyr.
In this period, the SF activity has produced $\sim95$ percent of the total stellar mass density observed locally, half of which was assembled within 2\,Gyr between $1.2< z< 2.2$.
\citet{Mehta2015} studied the bivariate \Ha-\OIII LF at $z\sim1$ using galaxies from the WFC3 Infrared Spectroscopic Parallel \citep[WISP;][]{Atek2010} survey. They showed that the \Ha LF can be determined by exclusively fitting \OIII data.
\citet{Zhu2009} and \citet{Comparat2015} studied the \OII LF evolution at $0.75<z<1.45$ and $0.1 <z< 1.65$, respectively. \citet{Comparat2016} measured the \OII, \OIII, and, for the first time, the \Hb LFs over the last nine billion years. They found that both the characteristic luminosity and the density of all LFs increase with redshift.
\cite{Saito2020} used photometric data to model galaxy spectral energy distributions (SEDs) and emission-line fluxes and used them to derive accurate predictions for the \Ha and \OII LF up to $z=2.5$. 

All the studies above show that, so far, the focus has been mainly on \Ha, \OII and \OIII lines. Here we propose a novel analysis aimed at exploring also other lines, namely \Hb, \NII and \SII. We want to split the different galaxy contributions to the ELG production to understand the impact of each one on the line LF. This work will be directly relevant to future high-redshift studies \citep[see e.g.,][]{Violeta2020, Zhai2019}. 
In particular, the aim of our work is twofold: (i) to measure the \Ha, \Hb, \OII, \OIII, \NII, and \SII luminosity functions in the nearby Universe with high accuracy, using a uniform procedure to select our galaxy sample and account for statistical incompleteness; (ii) to establish the contribution of different ELG types to the total LF.

For this study we use a subsample of the SDSS DR7 Main galaxy sample \citep{Strauss2002} at $0.02<z<0.22$, with spectral properties from the MPA-JHU\footnote{\url{https://www.sdss.org/dr12/spectro/galaxy_mpajhu/}\label{mpanote}} release, where the SFR were computed from the \Ha line luminosity as described in \citet{Brinchmann2004}. We classify the selected ELGs based on their specific star formation rate (sSFR, star formation rate divided by the stellar mass), and their position in the BPT and WHAN diagrams.
These diagnostics allow us to classify galaxies beyond the star-forming and passive split, to distinguish composite galaxies from those with spectral emission lines produced in jets or shocks, which, in many cases, host active galactic nuclei, that is, Seyfert galaxies.

The paper is organized as follows. In Sec.\,\ref{sec:data} we describe the SDSS Main galaxy sample, its MPA-JHU spectral properties, the sample selections performed, and their incompleteness effects. In Sec.\,\ref{sec:LFcalculation} we present a generalized $1/V_{\rm max}$ LF estimator capable of simultaneously correcting from spectroscopic, $r-$band magnitude, and emission-line incompleteness.
In Sec.\,\ref{sec:classification} we explain the methods adopted to classify 
ELGs. In Sec.\,\ref{sec:LFresults}, we present the measured LFs, both observed (i.e., dust attenuated) and intrinsic ones (i.e., corrected from dust extinction). Our findings are summarized in Sec.\,\ref{sec:disc}.

Throughout the paper we adopt the MultiDark Planck 2 cosmology consistent with \cite{Planck2016}. Our parameters are: $\Omega_{\rm m} = 0.3071$,  $\Omega_{\rm b} = 0.0482$, $\Omega_\Lambda = 0.6928$, $h=0.6777$, $\sigma_8 = 0.8228$ and $n_s = 0.96$.

\begin{figure}
\centering 
   \includegraphics[width=0.95\linewidth]{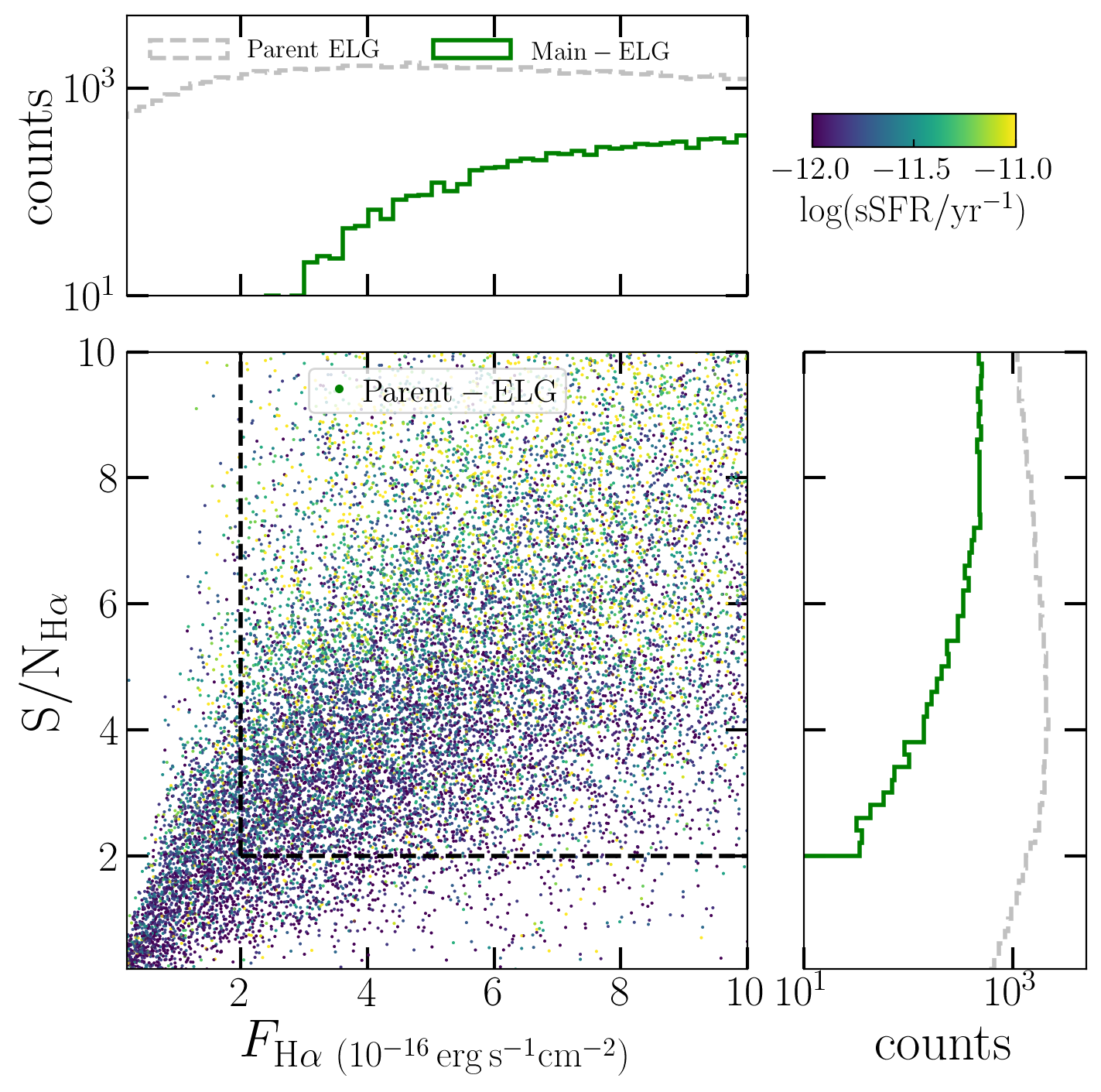}\hfill
\caption{\parent signal-to-noise as a function of the \Ha line flux, color-coded by sSFR. Here we are representing a random subset of the total population, 30 percent of it, to avoid crowding. The black-dashed lines in the main panel represent the flux and S/N cuts we impose on the \parent sample to obtain our \mainsel sample: $F>2\times 10^{-16}{\rm erg\,s^{-1}\,cm^{-2}}$ and ${\rm S/N>2}$ (see Sec.\,\ref{sec:selections}). The top and right panels show the flux and S/N histograms of the \parent (gray-dashed lines) and the \mainsel (green-solid lines) samples. The marginal distributions displayed both here and in Fig.\,\ref{fig:megacortealllines} for the other lines motivate the flux and S/N cuts chosen to select a complete ELG sample.}
  \label{fig:megacorte}
\end{figure}

\section{Observational data}
\label{sec:data}

In this work we aim at characterizing the luminosity functions for a range of spectral emission lines in the local Universe. In particular, we study the following lines: \Ha${\lambda\,6563}$\,\AA,  \Hb${\lambda\,4861}$\,\AA, \OII${\lambda\,3727,3729}$\,\AA, \OIII${\lambda\,5007}$\,\AA, \NII${\lambda\,6584}$\,\AA, \SII${\lambda\,6717,6731}$\,\AA. Here we describe how we generate a sample of ELGs with adequate fluxes and signal-to-noise ratios (S/N) to then study their completeness and measure their LFs.

\subsection{The \parent sample}
\label{sec:parentELGsample}

We select galaxies with good spectra, (i.e., with \texttt{ZWARNING=0}) from the SDSS DR7 Main sample \citep[]{Strauss2002} using the NYU-Value Added Galaxy Catalog\footnote{\url{http://cosmo.nyu.edu/blanton/vagc/}} \citep[]{2005AJ....129.2562B}. We spectroscopically match these galaxies to the MPA-JHU DR7\footref{mpanote} spectral release to obtain further properties, such as star formation rates, stellar masses, spectral emission-line fluxes and equivalent widths \citep{Brinchmann2004, Tremonti2004}. 

The SDSS Main galaxy sample covers an effective area of 7300\,deg$^2$ and is limited in $r-$band petrosian magnitude at $r_{ \rm p}<17.77$. The SDSS spectra span wavelengths of 3800--9200\,\AA, with a resolution that varies from $R=1500$ at $\lambda=3800\,$\AA, to $R=2500$ at $\lambda=9000\,$\AA\, \citep{Stoughton2002}.
We limit our sample to the redshift range $0.02<z<0.22$. The lower redshift cut ensures that we are studying galaxies beyond the local group, reducing the cosmic variance in our sample. The upper limit is chosen to mimic the SDSS Main selection in \cite{Favole2017} and \cite{Guo2015}, minimizing the effect of k-corrections and cosmic evolution.
This matched sample, hereafter ``\parent," is composed of 426625 galaxies.

We calculate the observed (i.e., dust attenuated) luminosities of the \parent sample from the observed fluxes $F$ provided in the MPA-JHU catalog as \citep[e.g.,][]{2003ApJ...599..971H, Favole2017}:
\begin{equation}
L[{\rm erg\,s^{-1}}]=4\pi D_{\rm L}^2(z) F,
\label{eq:lum}
\end{equation}
where $D_{\rm L}(z)$ is the luminosity distance as a function of redshift and cosmology, and the fluxes are given in units of erg~s$^{-1}$~cm$^{-2}$.

The SDSS fluxes were measured by fitting the spectra using \cite{Bruzual2003} stellar population synthesis models, accounting for stellar absorption. We note that, in the case of the \OII${\lambda\,3727,3729}$\,\AA\,, and \SII${\lambda\,6717,6731}$\,\AA\, doublets, the flux is the sum of the individual line fluxes.

\subsection{Fiber aperture correction}
\label{sec:aperture}

The observed fluxes in Eq.\,\ref{eq:lum} need to be corrected for fiber aperture to take into account that only the portion of the flux within each SDSS fiber ($\sim 3^{\prime\prime}$ diameter) was detected by the spectrograph \citep{Strauss2002}. 
Following \citet{2003ApJ...599..971H} and \citet{Gilbank2010}, we estimate the aperture-correction factor for each \parent that is not classified as a candidate active galactic nuclei (AGN; see Sec.\,\ref{sec:classification}) from its total and fiber magnitudes.
The aperture-corrected line luminosity $L^{\rm ap-corr}$ is related to the measured luminosity $L$, given in Eq.\,\ref{eq:lum}, as follows:
\begin{equation}
L^{\rm ap-corr}[{\rm erg\,s^{-1}}]=10^{-0.4(m_{\rm p}-m_{\rm{fib}})}\,L\,,
\label{eq:lumapcorr}
\end{equation}
where the exponent $(m_{\rm p}-m_{\rm{fib}})$ represents the aperture correction as a function of the SDSS petrosian magnitude $m_{\rm p}$, used as a proxy for the total magnitude of the galaxy \citep{Blanton2001}, and the fiber magnitude $m_{\rm fib}$ that accounts for the light enclosed within the diameter of the fiber.

To implement the above correction, we use the magnitudes measured with the SDSS broadband filters \citep{Gunn1998, 1996AJ....111.1748F}\footnote{From \citet{1996AJ....111.1748F}, we see that the $u$ filter peaks at about 3500 \AA, with a full width at half maximum (FWHM) of 600 \AA, and covers the range 3000-4000 \AA; $g$ peaks at $\sim4800$ \AA, with a FWHM of 1400 \AA, and covers the range 4000-5500 \AA; $r$ peaks at about $\sim6250$ \AA, with a FWHM of 1400 \AA, and covers the range 5500-7000 \AA; $i$ peaks at about 7700 \AA, with a FWHM of 1500 \AA, and covers the range 7000-8500 \AA; $z$ peaks at about 9100 \AA, with a FWHM of 1200 \AA, and covers the range 8500-10000 \AA.}.
Table\,\ref{tab:magsap} summarizes the wavelength $\lambda_0$ of our emission lines of interest, emitted in the rest frame of the galaxy, as well as the value $\lambda_z = \lambda_0(1+z)$ observed at the Earth at the minimum, mean, and maximum redshifts of the sample, together with the corresponding SDSS filter. 
For each galaxy, we select the appropriate band for each emission line based on the observed redshift and then use Eq.\,\ref{eq:lumapcorr} to derive the aperture-corrected luminosity. Note that the \SII line at $z=0.02$ falls at the gap between the $r$ and $i$ filters; we choose the latter since it has higher transmission.

\begin{table*}
    \centering
    \setlength{\tabcolsep}{3pt}
    \begin{tabular}{|c|c|c|c|c|c|c|}
    \hline
       $\lambda_z$& \Ha & \Hb &\OII&\OIII&\NII&\SII\\
       \hline
    $z=0$ &6563 ($r$) &4861 ($g$)&3727-3729 ($u$)&5007 ($g$) &6584 ($r$) &6717-6731 ($r$)\\
    $z=0.02$&6694 ($r$) &4958 ($g$)&3801-3803 ($u$)&5107 ($g$) &6716 ($r$) &6851-6865 {($i$)}\\
    $z=0.12$&7350 ($i$)& 5444 ($g$)&4174-4176 ($g$)&5608  ($r$)&7374 ($i$)&7523-7539 ($i$)\\
    $z=0.22$&8006 ($i$)&5930 ($r$)&4546-4549 ($g$)&6108  ($r$)&8032 ($i$)&8195-8211($i$)\\
    \hline
    \end{tabular}
    \caption{ 
    Wavelengths of our six emission lines of interest, together with the SDSS filter (in brackets) in which they fall for a selection of redshifts, for illustration. This information is used for the aperture corrections performed on emission-line luminosities, as described in Sec.~\ref{sec:aperture}. From top to bottom, we tabulate emission-line wavelenghts at: rest-frame ($\lambda_0$, first row), $z=0.02$ (second), $z=0.12$ (third), $z=0.22$ (fourth). The relation between them is: $\lambda_z=\lambda_0(1+z)$. 
    }
    \label{tab:magsap}
\end{table*}

\begin{figure}
\centering 
   \includegraphics[width=\linewidth]{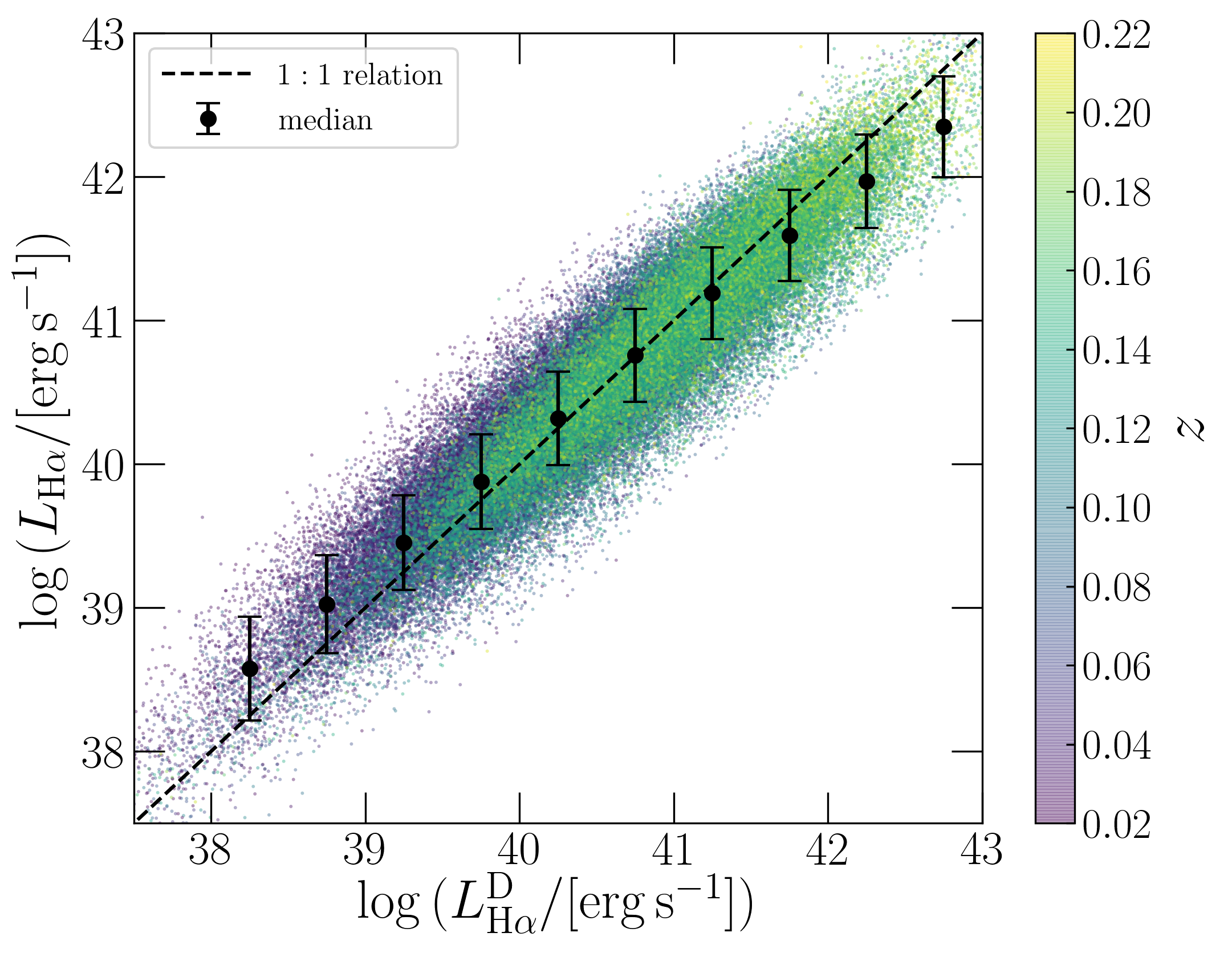}\hfill
    \caption{{\em Parent-ELG} \Ha luminosity computed using our our default aperture correction based on \citet{2003ApJ...599..971H} ($y-$axis) versus the same quantity computed using an approximation to  the \citet{Duarte2017} correction ($z-$axis), color-coded by redshift. We randomly show only 30 percent of the \parent sample to avoid saturation. We overplot the median and $1\,\sigma$ dispersion of our \LHa in bins of $L_{\rm H\alpha}^{\rm D}$, as well as the 1:1 relation for comparison.}
  \label{fig:apcorrvsduarte}
\end{figure}

The \citet{2003ApJ...599..971H} prescription implicitly assumes that the emission measured through the fiber is characteristic of the whole galaxy, that is, the line equivalent width (EW) remains constant across its surface.
To quantify the uncertainty associated this simplification, we compare our approach with the method proposed by \citet{IP2016} and \citet{Duarte2017} to take into account variations of EW across a galaxy.
They fit the growth curves (i.e., integrated flux inside an aperture as a function of radius) of the emission-lines as a function of the petrosian half-light radius, $R_{50}$, enclosing half the petrosian flux.
We have approximated the aperture correction based on the work from \citet{Duarte2017} by using their fifth-order polynomial fit as a function of $R_{\rm 50}$ (i.e., ${\rm X}(\alpha_{\rm 50})$ in their Eq.\,4).
Fig.\,\ref{fig:apcorrvsduarte} shows, as a function of redshift, the difference in the \Ha luminosity of the \parent sample between applying our default aperture correction ($y-$axis) and that of \citet{Duarte2017} (superscript ``D", $x-$axis).  We overplot the median and $1\,\sigma$ dispersion of our \LHa in bins of $L_{\rm H\alpha}^{\rm D}$, as well as the 1:1 relation to help the comparison.

This result shows that the two corrections are consistent in the luminosity range $10^{40}-10^{41.5}\rm erg\,s^{-1}$, while the largest discrepancies arise in both the faint and bright ends, where the lower- and higher-$z$ emitters respectively concentrate. 
Our aperture correction factor has typical values in the range 2-10, and below $10^{39}\,\rm erg\,s^{-1}$ (above $10^{42}\,\rm erg\,s^{-1}$) it returns \Ha luminosities up to 0.5\,dex higher (lower) than those from \citet{Duarte2017}. 
The scatter of \LHa and $L_{\rm H\alpha}^{\rm D}$ in Figure\,\ref{fig:apcorrvsduarte} are comparable, suggesting that the aperture-correction has an associated uncertainty on the order of a factor $\sim 3$.
We thus conclude that our default aperture-correction, assuming EW is constant across a galaxy, is adequate for the purposes of the present study, within this level of uncertainty.

\subsection{The \mainsel selection}
\label{sec:selections}

We aim at selecting a complete population of bright ELGs with well measured fluxes in all of the following six emission lines: \Ha${\lambda\,6563}$\,\AA, \Hb${\lambda\,4861}$\,\AA, \OII${\lambda\,3727,3729}$\,\AA, \OIII${\lambda\,5007}$\,\AA, \NII${\lambda\,6584}$\,\AA, and \SII${\lambda\,6717,6731}$\,\AA. To achieve this, we extract a subsample of the \parent sample above, and then we impose a combination of cuts in emission-line flux and signal-to-noise (S/N) in all the six lines of interest. 
We define the signal-to-noise as the ratio between the observed flux and its error, $\sigma_F$, as given by the MPA-JHU DR7 catalogs: ${\rm S/N}=F/\sigma_F$.

\begin{figure*}
\centering 
   \includegraphics[width=0.45\linewidth]{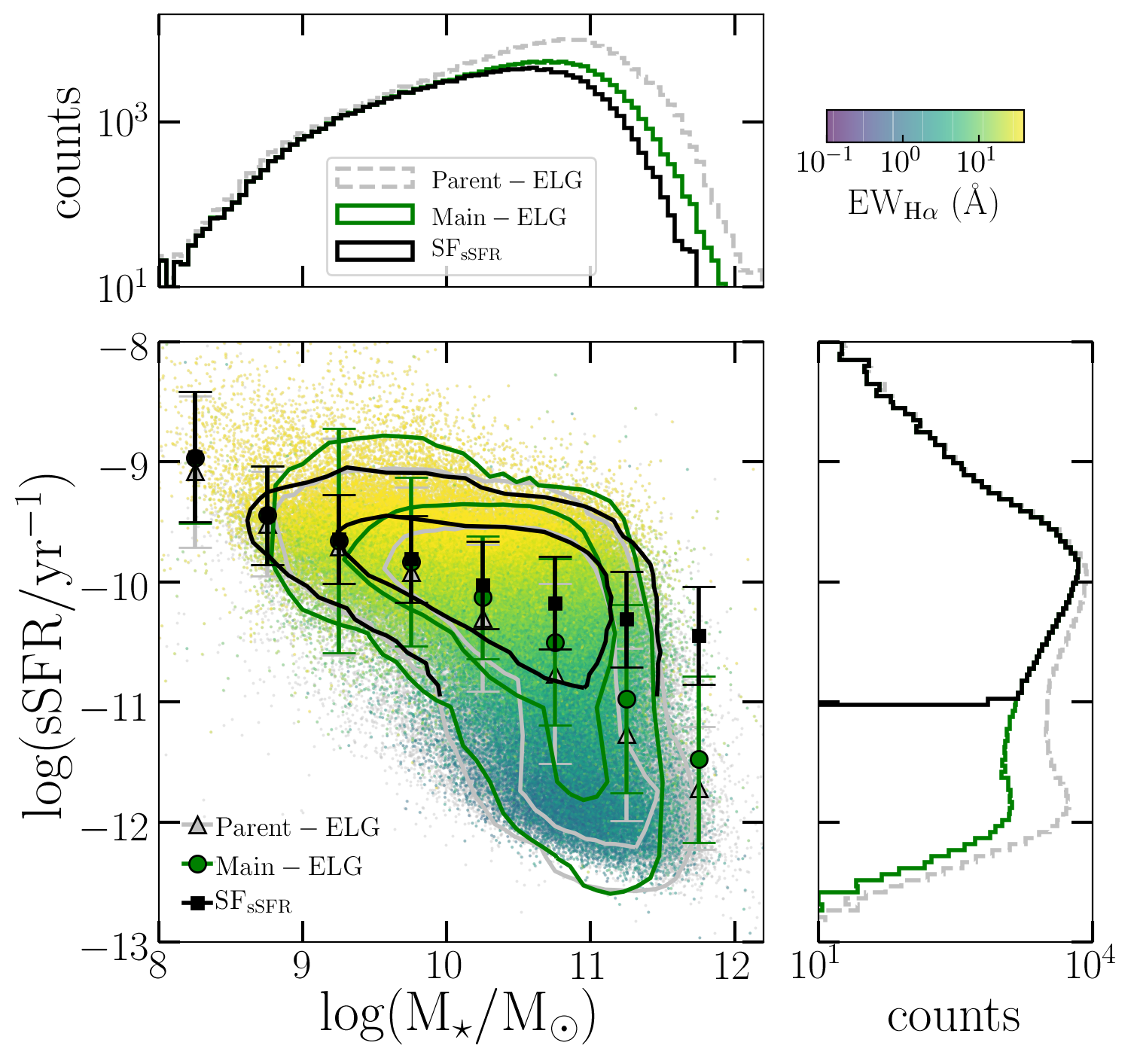}\hfill
   \includegraphics[width=0.45\linewidth]{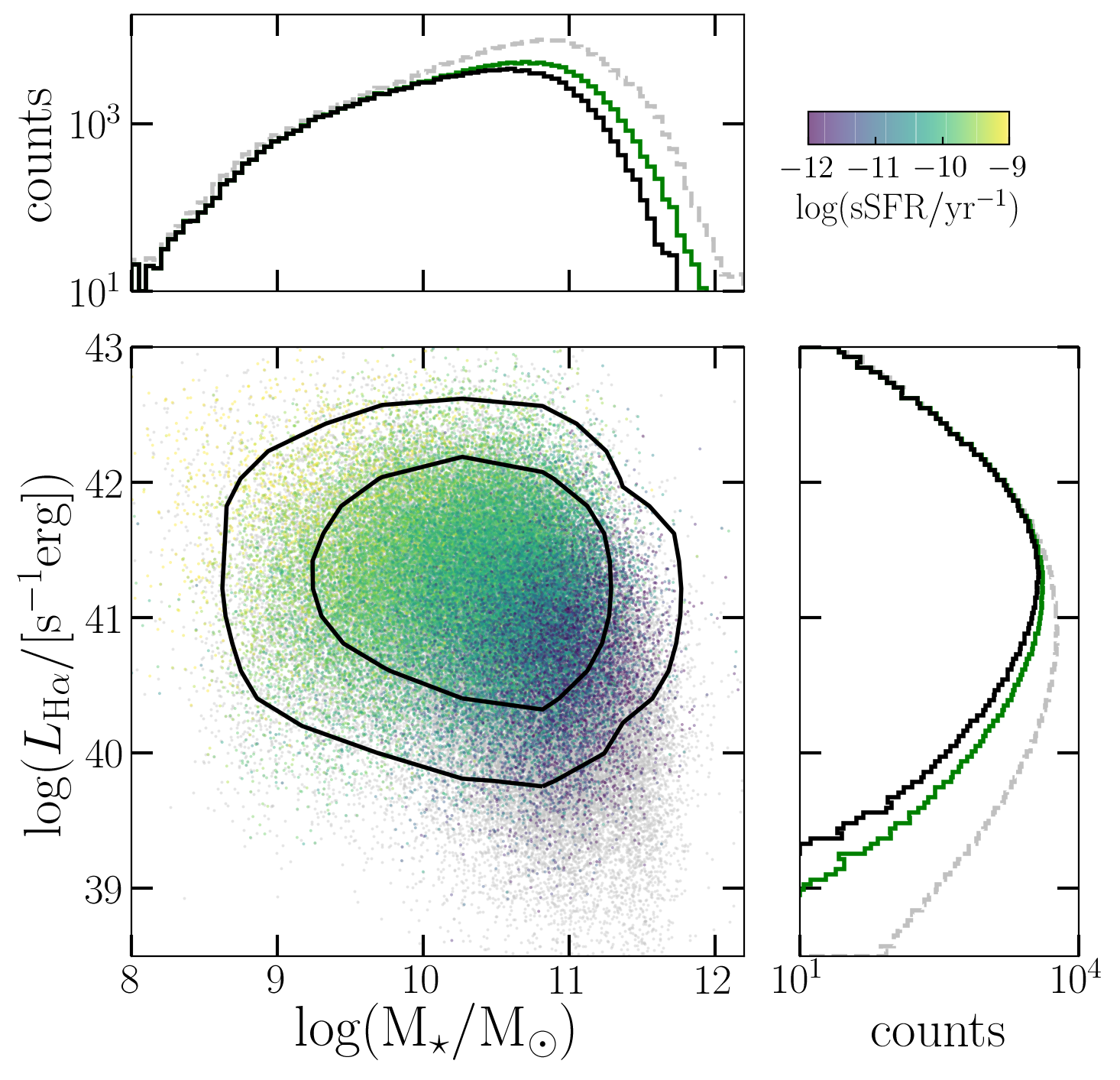}
    \caption{The left panel shows the \mainsel sSFR as a function of stellar mass, color-coded by \Ha EW. On the background we also show the \parent distribution (silver triangles). Here we are representing random subsets of both populations, 30 percent of them, to avoid crowding. We overplot the corresponding 68 and 95 (inner and outer lines, respectively) percent contours as green and silver lines. In addition, we show in black the contours of the star-forming (SF) population selected at ${\rm sSFR}>10^{-11}{\rm yr^{-1}}$. The large markers with error bars display the corresponding sSFR means and 1\,$\sigma$ deviations in bins of stellar. The side histograms show the sSFR and M$_\star$ marginal distributions of the \mainsel (green) and SF populations (black lines), and compare them to the \parent sample (gray), which has no cuts. The right panel shows the \Ha luminosity as a function of stellar mass, color-coded by sSFR, and corresponding marginal distributions, with the same colors as in the left panel.}
  \label{fig:charplot}
\end{figure*}

We cut the \parent sample at $F>2\times10^{-16}{\rm erg\,s^{-1}\,cm^{-2}}$ and ${\rm S/N}>2$ in all the six lines above. Furthermore, we remove any spurious object with nonphysical flux uncertainty by limiting our selection at $\sigma_{F}<10^{-12}\,\rm erg\,s^{-1}\,cm^{-2}$, and $\rm EW\geq0\,$\AA\, in all the six lines under study. 
The resulting ELG sample, hereafter ``\mainsel", is composed of 162733 emitters (about 38 percent of the parent sample).
The characteristics of this sample are discussed in Sec.\,\ref{sec:characteristics}.

Fig.\,\ref{fig:megacorte} shows the effect of the \mainsel selection on the signal-to-noise -- $F$ plane for the \Ha line, color-coded by specific star formation rate (sSFR, star formation rate divided by stellar mass); the effect on the other emission lines, color-coded by both sSFR and EW, is shown in Fig.\,\ref{fig:megacortealllines}.
In all cases, the marginal probability distributions of the measured flux and SN are observed to decay below our adopted thresholds, suggesting that completeness would be very difficult to guarantee beyond that point.


\subsection{{\em Main-ELG} properties}
\label{sec:characteristics}

Here we analyze the impact of the S/N and emission-line flux cuts performed in Sec.\,\ref{sec:selections} on the sSFR, stellar mass, and EW distributions of the \mainsel sample.
Fig.\,\ref{fig:charplot}, left panel, shows the \mainsel sSFR as a function of stellar mass (green lines and colorful dots), compared to the distribution of the \parent sample (gray contours and dots). Individual galaxies are shown as dots and the contours correspond to the 
$1\,\sigma$ and $2\,\sigma$ density distribution. The green contours correspond to the \mainsel sample. The average sSFR and standard deviations in bins of stellar mass are shown by markers. Both the contours and the average values show that the \mainsel population is a fair sample of the \parent one. 

In Fig.\,\ref{fig:charplot}, galaxies from the \mainsel sample are color-coded by the \Ha EW. Here we can see that ELGs with a high sSFR are also those with higher $\rm EW$. As expected, the three selections are consistent with each other up to stellar masses $\sim 10^{11}{\rm M}_{\odot}$.

Similar trends are found for the other spectral lines under study. The corresponding plots are shown in Fig.\,\ref{fig:charplotallines}.

On each side of the figure we display the marginal sSFR and $\rm M_\star$ distributions for the SF and \mainsel samples, and we compare them with the \parent sample (silver).
The \mainsel sample includes galaxies with relatively low sSFR values, that will not be considered as star-forming, neither in terms of their sSFR nor in relation with the so-called star formation main sequence.
We quantify the numbers of these populations below.

The right panel in Fig.\,\ref{fig:charplot} displays the \Ha luminosity as a function of stellar mass, color-coded by sSFR. Fig.\,\ref{fig:lummassallines} shows similar plots for the rest of lines under study. Here we notice that \Ha ELGs with lower star-formation activity (i.e., ${\rm sSFR}\lesssim10^{-11}{\rm yr^{-1}}$) are also the most massive and least luminous ones, whereas SF ELGs with ${\rm sSFR}\gtrsim10^{-11}{\rm yr^{-1}}$ tend to concentrate toward the low-mass and high-luminosity end of the distribution.

These results highlight that ELGs selected with a combination of cuts in signal-to-noise and line flux, that is, the \mainsel sample, are not equivalent to ELGs selected by using a sharp cut in sSFR or, similarly, in EW. This agrees with theoretical studies that have shown that the small-scale clustering is different for samples selected either based on SFR or emission line fluxes \citep{Violeta2020}.
In fact, the selection based on flux and S/N returns a heterogeneous population of galaxies, covering a similar range in both sSFR and stellar mass as the \parent sample.
This guarantees that the number density of galaxies, in particular the fainter ones, is preserved, maximizing the completeness of the luminosity function.


\subsection{Incompleteness effects and redshift evolution}
\label{sec:incompleteness}
\begin{figure*}
\centering 
   \includegraphics[width=0.8\linewidth]{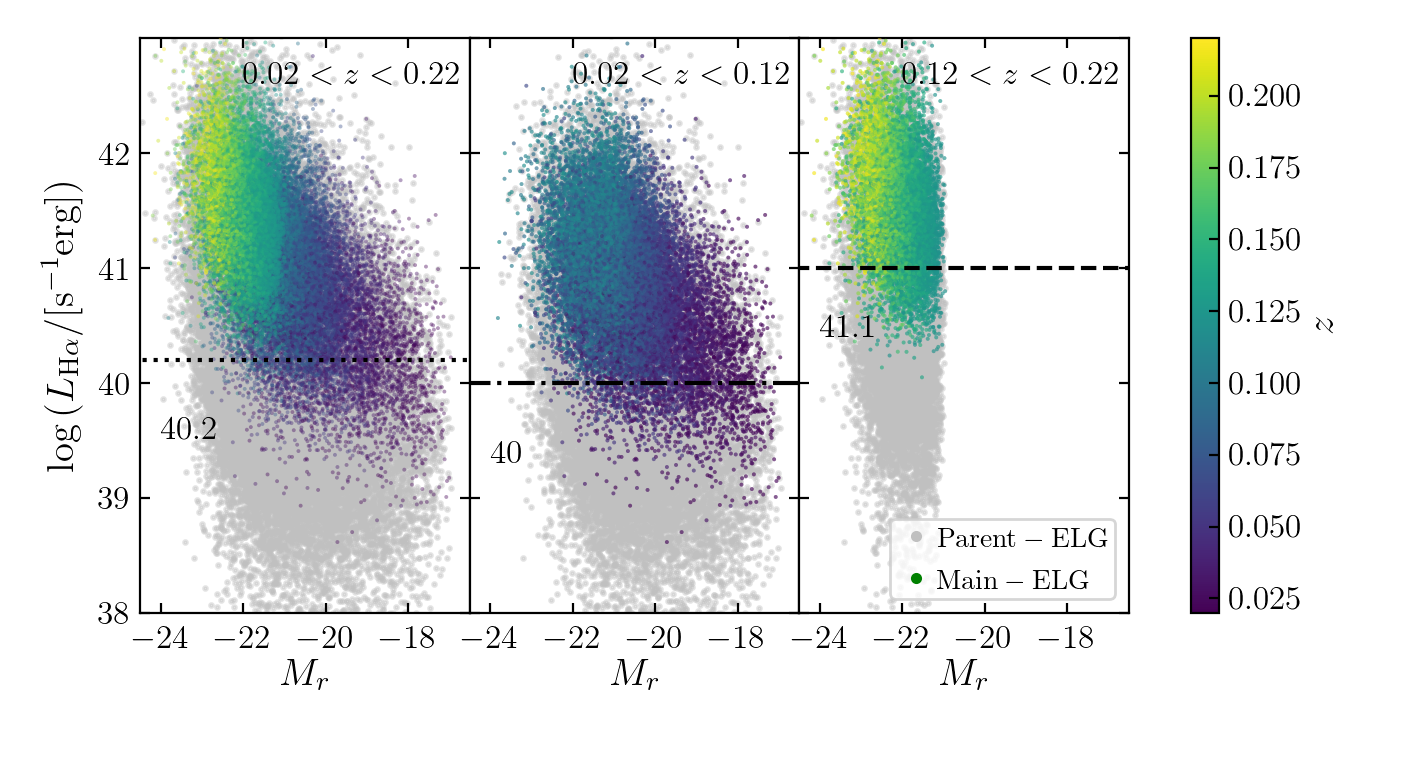}
  \caption{{\em Main-ELG} \Ha luminosity as a function of the $r-$band absolute magnitude, color-coded by redshift. On the background we show in gray the \parent sample distribution, which has no cuts in flux nor S/N. For both populations we display random subsets of 30 percent of the total to avoid saturation. From left to right, we show the full sample ($0.02<z<0.22$), the lower-$z$ bin ($0.02<z<0.12$), and the higher-$z$ ($0.12<z<0.22$) one. The horizontal lines represent our lower completeness limits in luminosity (see the text for details).}
  \label{fig:LHaMr}
  \end{figure*}
Fig.\,\ref{fig:LHaMr} displays the \Ha luminosity of the\,\mainsel sample as a function of the $r-$band absolute magnitude, $M_r$, color-coded by redshift. We compare this distribution to that of the \parent sample, selected at $r_{\rm p}>17.77$. To better understand its evolution, we analyze the result in three redshift bins: the full sample at $0.02<z<0.22$, the lower-$z$ bin at $0.02<z<0.12$, and the higher-$z$ one at $0.12<z<0.22$. Fig.\,\ref{fig:LMralllines} shows similar plots for the other lines under study. 

We find that the \Ha flux cut is not independent of $M_r$ and hence from the limit $r_{\rm p}<17.77$ intrinsic to the \parent sample. A similar result is found for the other lines. The impact of such dependency is stronger as the redshift increases. In other words, when we cut in flux or S/N, we are also removing a fraction of galaxies below a certain line luminosity that varies in a nontrivial way with redshift.

Our modified $1/V_{\rm max}$ method for ELGs (see Sec.\,\ref{sec:LFcalculation}) is capable of individually correcting from flux-limited selection effects, but not from statistical correlations between the line luminosities and broadband magnitudes.
We therefore set a lower completeness limit for all emission-line luminosities in order to ensure that these correlations do not significantly affect the LF measurement. 
For the \Ha line, we set this threshold to $L=\{10^{40.2},\,10^{40},\,10^{41.1}\}\,\rm erg\,s^{-1}$ in the full sample, low-$z$, and high-$z$ bin, respectively. These limits for the other emission lines are provided in Sec.\,\ref{sec:appendixevol}.
All these values are chosen by eye, based on the completeness that the \mainsel sample shows in Figs.\,\ref{fig:LHaMr} and \ref{fig:LMralllines}.
Specifically, we set as threshold the luminosity value where the density of ELGs in these figures starts to degrade, indicated as dotted, dot-dashed and dashed lines in Figs.\,\ref{fig:LHaMr} and \ref{fig:LMralllines}.


\section{Volume correction}
\label{sec:LFcalculation}
\begin{figure}
\centering 
   \includegraphics[width=0.88\linewidth]{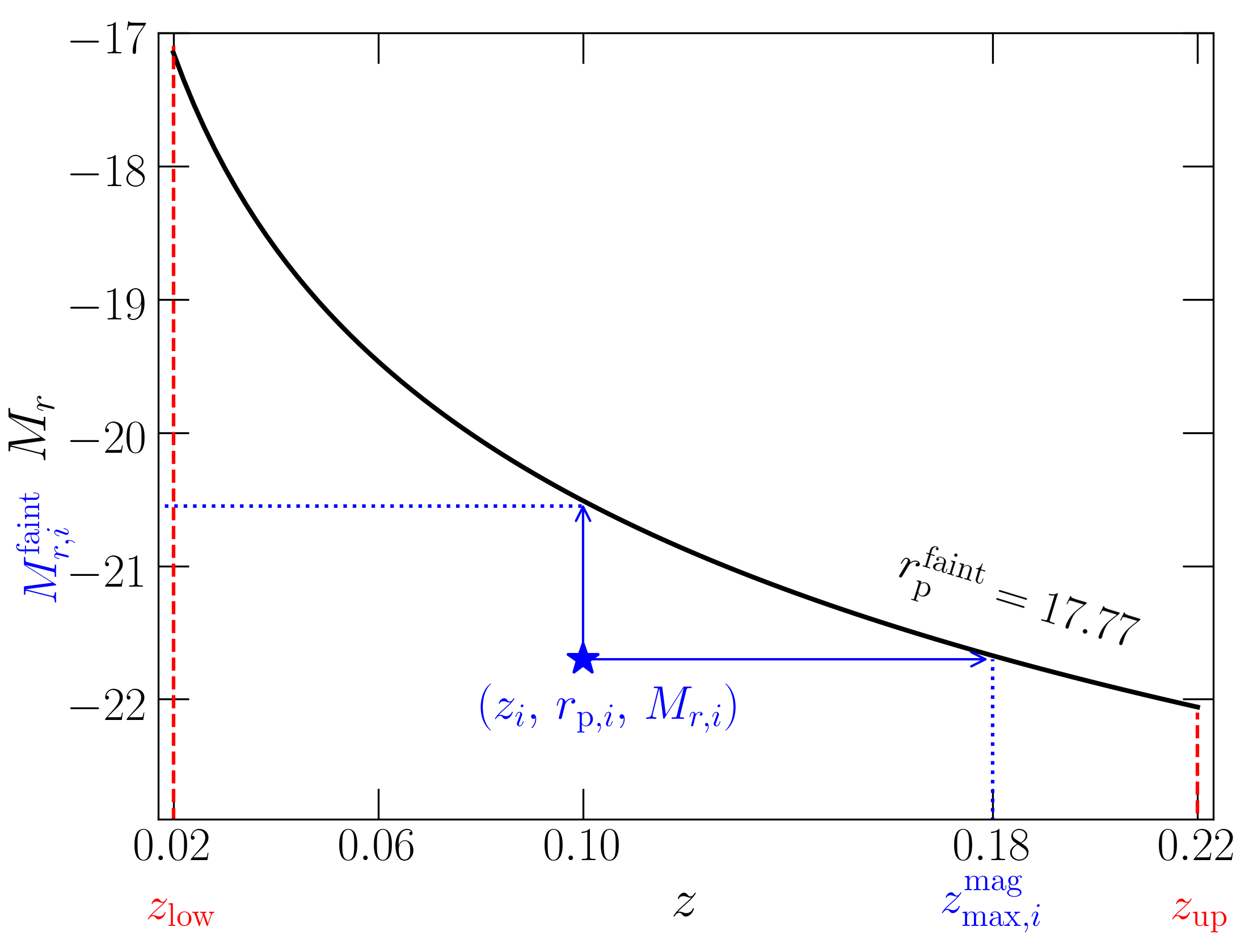}\quad
   \includegraphics[width=0.88\linewidth]{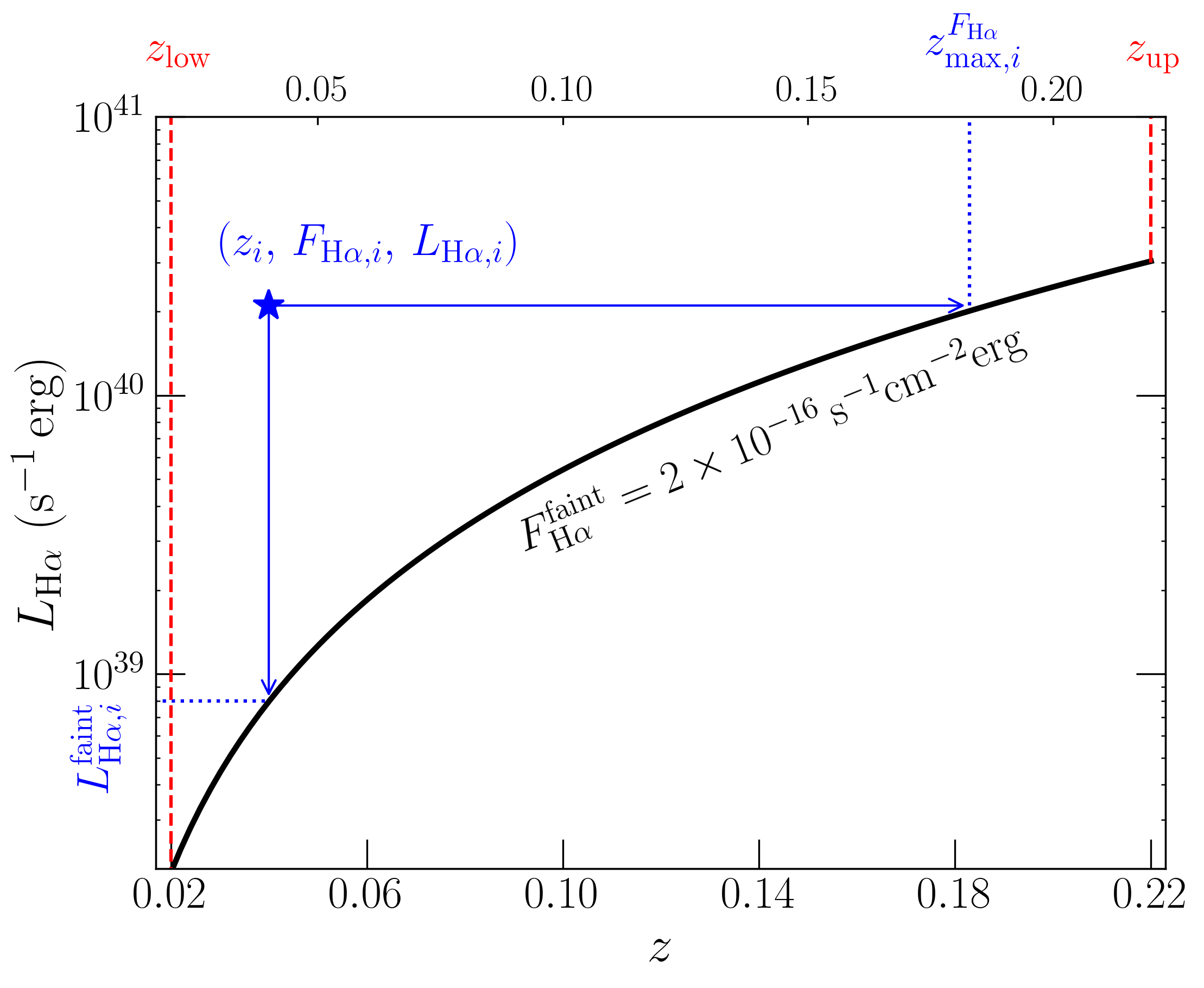}
  \caption{$V_{\rm max}$ computation scheme for a galaxy from the \mainsel sample (star symbol). {\em Top panel:} we take into account the survey $r-$band magnitude limit $r_{\rm p}^{\rm faint}=17.77$; note that we are omitting the $K-$corrections ($K(z_i)=0$ in Eq.~\ref{eq:maglim}) for this representation. {\em Bottom panel:} similar plot as the top one for a survey with a limit in the \Ha line flux of $F_{\rm H\alpha}^{\rm faint}=2\times 10^{-16}\,{\rm erg\,s^{-1}}$. For the \OII, \OIII, \Hb, \NII, and \SII lines the methodology is identical (see Sec.\,\ref{sec:selections}). The lower and upper redshift limits of the survey, $z_{\rm low}=0.02$ and $z_{\rm up}=0.22$, are highlighted by dashed vertical red lines in both panels. The maximum redshifts that the galaxy can have and still be included in the sample, considering its magnitude and \Ha flux limits, as well as the faintest luminosity, are shown by dotted vertical blue lines.}
  \label{fig:Vmaxscheme}
\end{figure}

\begin{figure}
\centering 
   \includegraphics[width=0.84\linewidth]{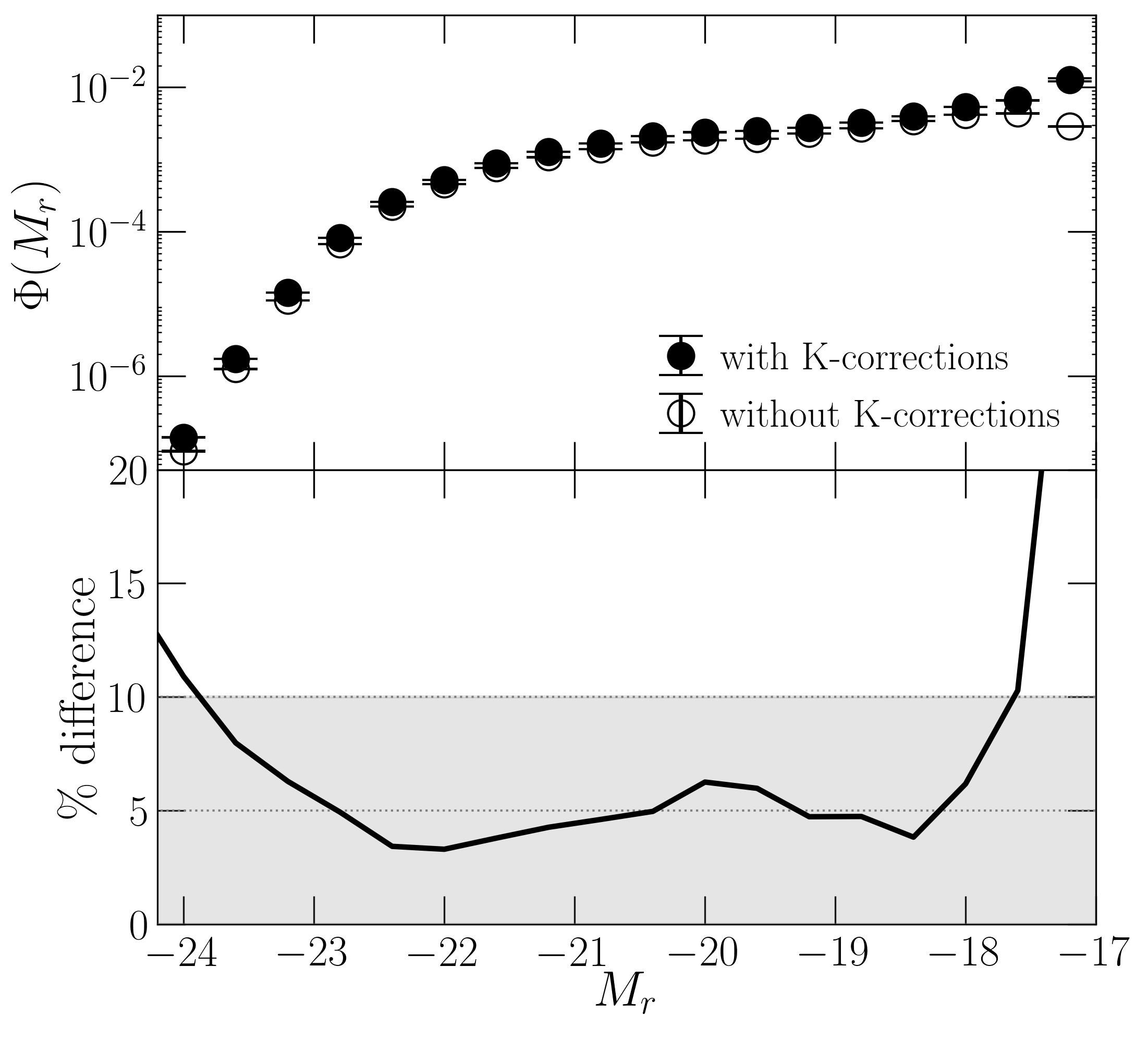}
  \caption{SDSS $M_r$ luminosity functions including (full markers) and omitting (empty markers) $K-$corrections. The effect is less than 7 percent at $-22.5<M_r<-18$, while it grows up to $\sim30$ percent in the faint and luminous tails of the distribution. The gray band highlights the 10 percent difference range.}
  \label{fig:Kcorr}
\end{figure}
    
The differential luminosity function is defined as the number, $N$, of galaxies per unit luminosity interval and comoving volume, $V$,  as:
\begin{equation}
    \Phi(\log L,z)=\frac{dN}{d\log L\,dV(z)},
    \label{eq:phi}
\end{equation}
where $V$ is a function of redshift.
The $1/V_{\rm max}$ estimator \citep[]{Schmidt1968, Felten1976} allows us to correct the LF from the Malmquist bias, that is, the fact that faint objects tend to be detected only in a small volume, while bright ones are observed in the entire sample volume \citep[see e.g.,][]{Weigel2016}. Other methods to estimate the galaxy LF are the C$^-$ method by \citet{Linden1971},  the parametric maximum-likelihood STY method proposed by \citet{Sandage1978}, or the Stepwise Maximum Likelihood Method \citep[SWML;][]{Efstathiou1988, Norberg2002} that does assume any functional form.

Here we focus on emission-line LFs. The galaxy counts need to include their observational incompletness, usually given as a weight. In the \parent sample we have different sources of incompleteness to take into account. In fact, the \mainsel sample is a $r-$band magnitude limited sample, on top of which we have imposed a combination of cuts in flux and signal-to-noise for the six spectral lines under study.
In this section we describe the methodology used to estimate the line LFs taking into account the incompleteness induced by the thresholds we have imposed.

In practice, Eq.\,\ref{eq:phi} is evaluated by counting the number of galaxies in each $\Delta\log{L}$ bin, $N_{k}$, and weighting it by the maximum volume $V_{\rm max}$ in which each galaxy can be observed, given the survey limits and its luminosity. In the $k-$th bin of luminosity and for a sample of $i=1,...,N_k$ galaxies we have:
\begin{equation}
    \Phi_{1/V_{\rm max}}^{k}=\frac{1}{\Delta\log{L}^k}\sum_{i=1}^{\rm N_k}\frac{1}{V_{{\rm max},\,i}}\,.
    \label{eq:Vmaxstandard}
\end{equation}
To estimate $V_{{\rm max},\,i}$ we need to determine the maximum redshift, $z_{{\rm max},\,i}$ at which a galaxy could still be observed as part of the \mainsel sample, given its observational limits. Explicitly this is:
\begin{equation}
    V_{{\rm max},\,i}=\frac{A}{3}\left(\frac{\pi}{180}\right)^2\left(D_{\rm c}^3(z_{{\rm max},\,i})-D_{\rm c}^3(z_{\rm low})\right)\,,
    \label{eq:Vmaxvolume}
\end{equation}
where $A=7300\,\rm deg^2$ is the survey area, $D_{\rm c}(z)$ is the galaxy comoving distance depending on redshift and cosmology, and $z_{\rm low}=0.02$ is the lower redshift limit of the \mainsel sample.  

We modify the standard $1/V_{\rm max}$ formulation in Eq.\,\ref{eq:Vmaxstandard} to correct the \mainsel sample from the spectroscopic, $r-$band magnitude, and luminosity selection effects. 
To correct from spectroscopic incompleteness in the SDSS sample (i.e., the fact that SDSS did not obtain the spectra of all the targets above its magnitude limit), we weight Eq.\,\ref{eq:Vmaxstandard} by $w_{{\rm c},\,i}=c_i^{-1}$, that is, the inverse of the SDSS spectroscopic completeness. Explicitly we have:
\begin{equation}
    \Phi_{1/V_{\rm max}}^{k}=\frac{1}{\Delta\log{L}^k}\sum_{i=1}^{\rm N}\frac{w_{{\rm c},\,i}}{V_{{\rm max},\,i}}\,.
    \label{eq:Vmaxspec}
\end{equation}
This is a small correction, as the \mainsel sample is more than 80 percent complete in spectroscopy \citep{Blanton2003_1}. 

To correct from the limits in $r-$band, line flux and S/N, we define the maximum redshift, $z_{{\rm max},\,i}$, of a galaxy in our sample as a function of the observational cuts imposed (see Sec.\,\ref{sec:selections}):
\begin{equation}
    z_{{\rm max},\,i}={\rm min}\left( z_{{\rm max},\,i}^{\rm mag}, \,z_{{\rm max},\,i}^F, \,z_{{\rm max},\,i}^{\rm S/N},\,z_{\rm up}\right)\,,
    \label{eq:zlim}
\end{equation}
where the superscripts indicate the contributions based on magnitude (mag), flux (F), and signal-to-noise (S/N) limits, while $z_{\rm up}=0.22$ is the upper limit of the \mainsel sample. The flux and S/N are grouped vectors, $F=(F_{\rm H\alpha},\,F_{\rm [O_{II}]},\,F_{\rm [O_{III}]},\,F_{\rm H\beta},\,F_{\rm [N_{II}]},\,F_{\rm [S_{II}]})$ and $\rm S/N=(S/N_{H\alpha},\,S/N_{[O_{II}]},\,S/N_{[O_{III}]},\,S/N_{H\beta},\,S/N_{[N_{II}]},\,S/N_{[S_{II}]})$. 

\begin{figure}
\centering
    \includegraphics[width=0.88\linewidth]{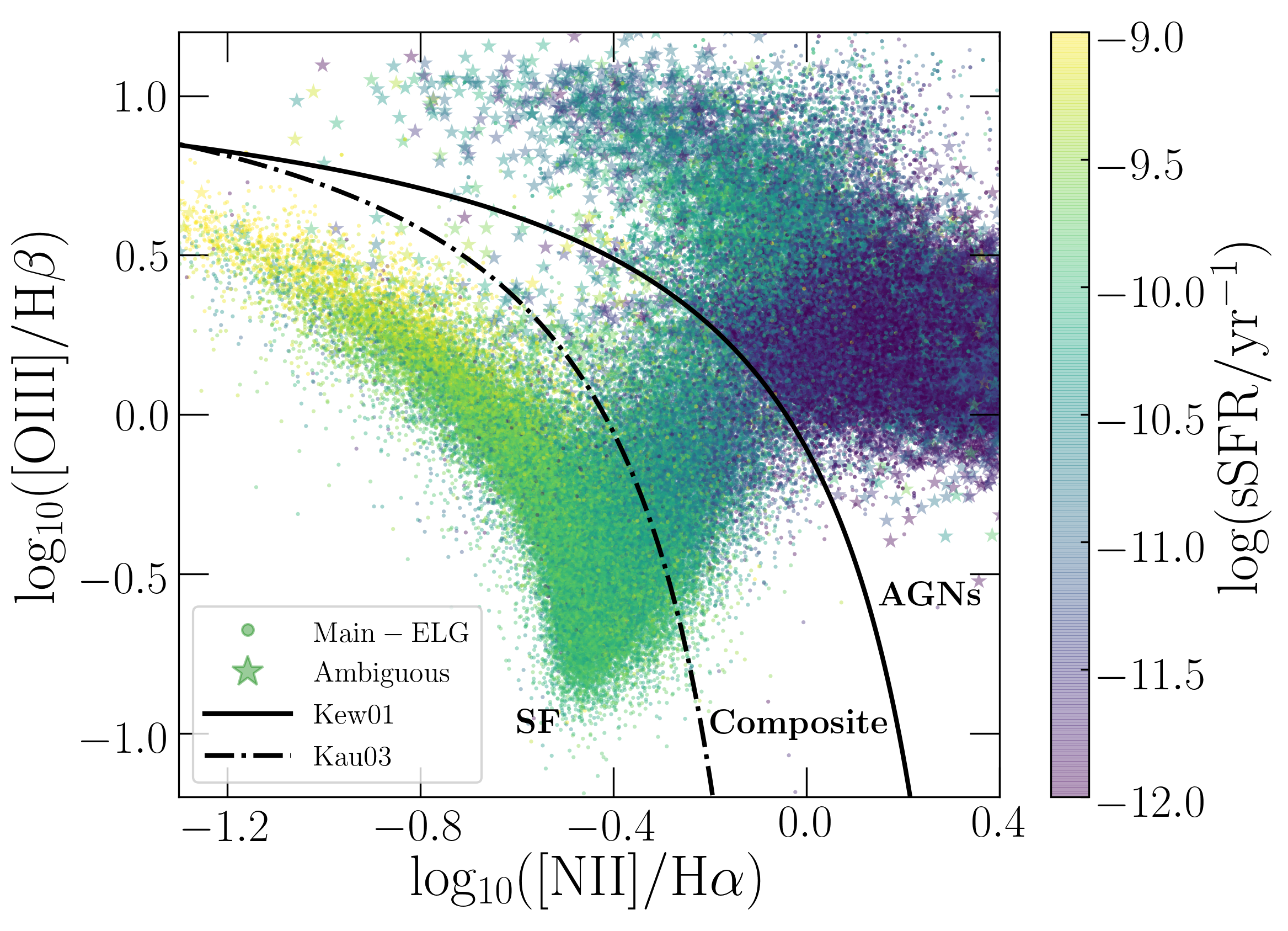}\quad
    \includegraphics[width=0.88\linewidth]{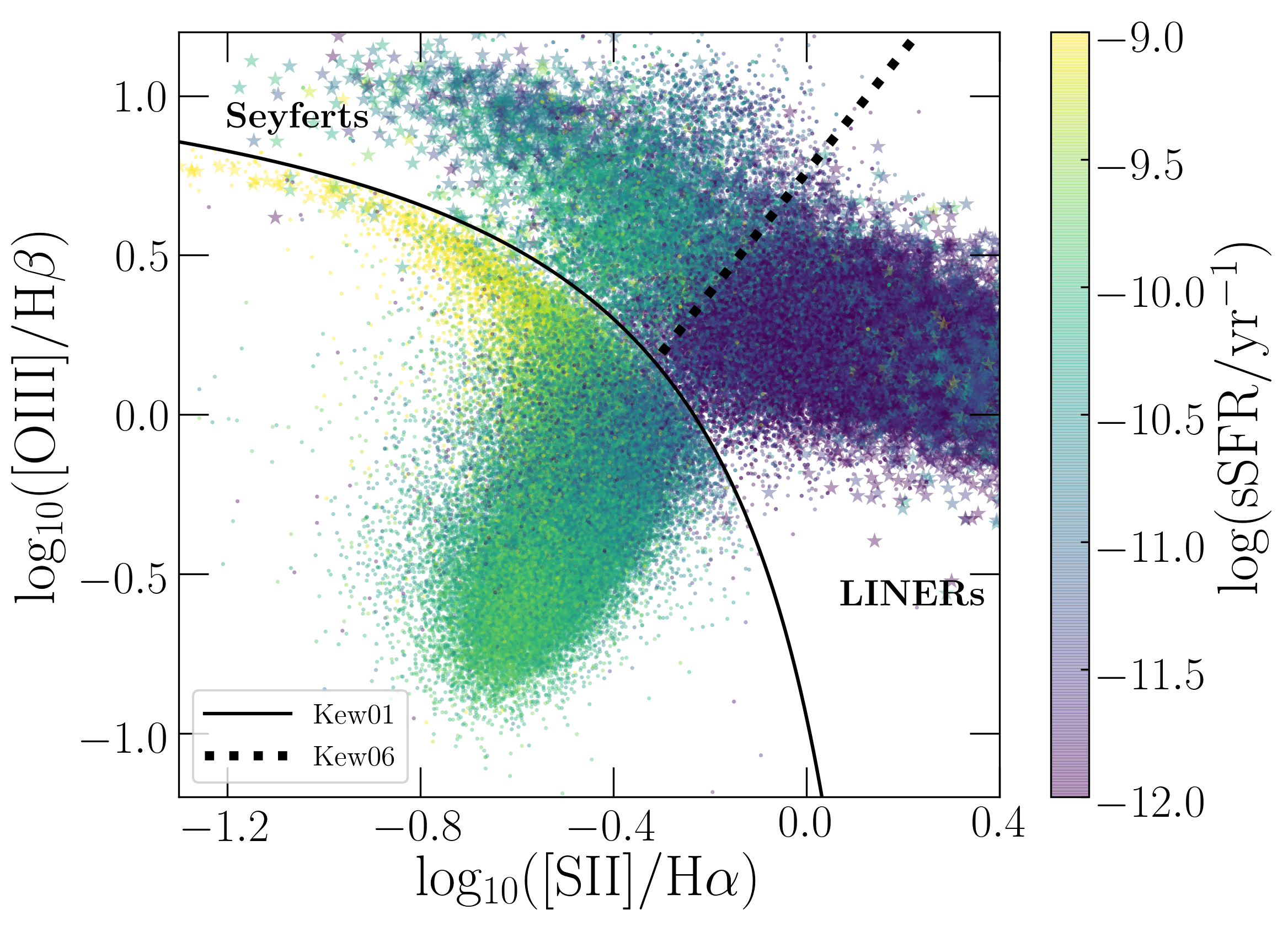}\quad
    \includegraphics[width=0.65\linewidth]{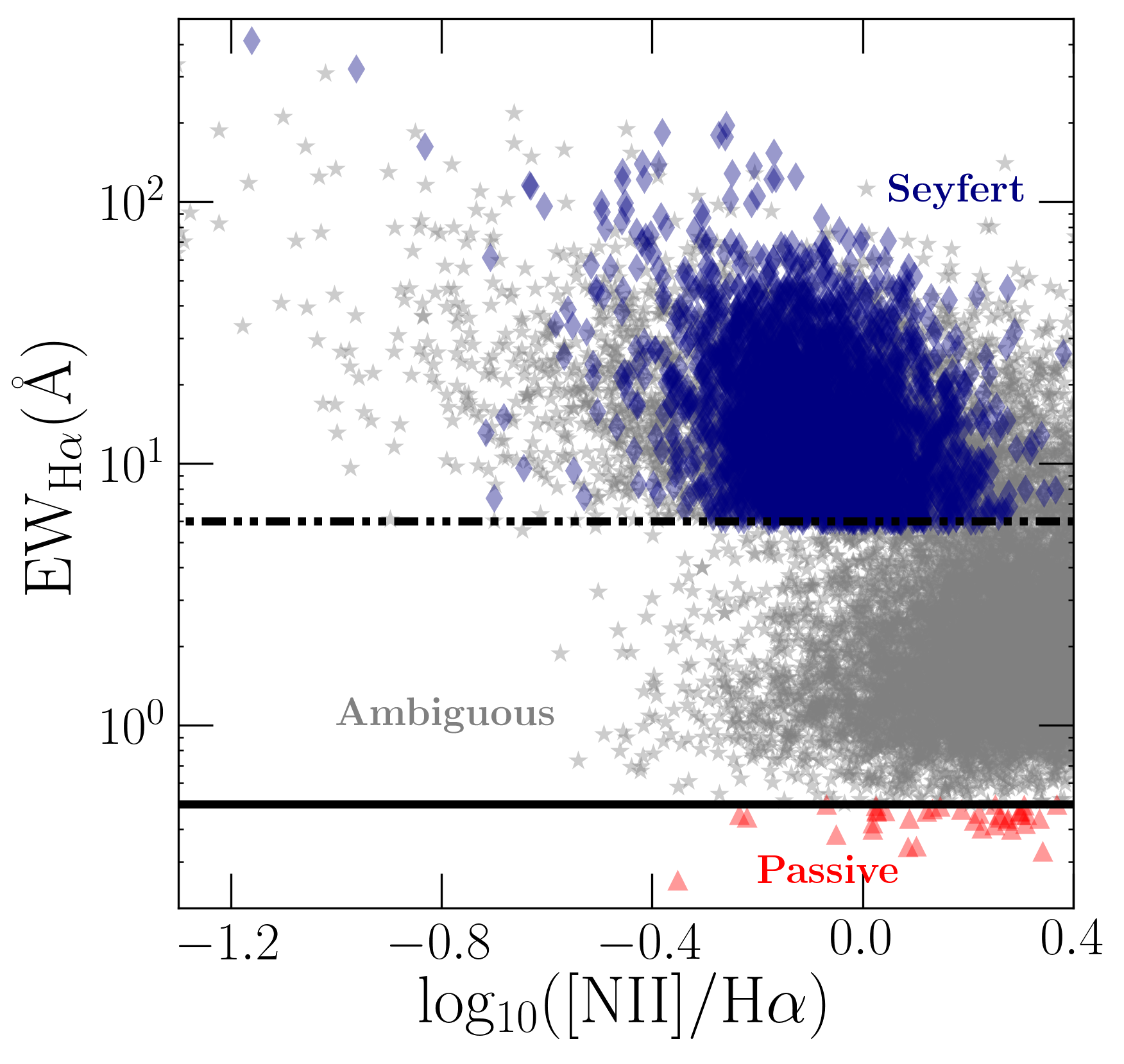}
\caption{Diagnostic diagrams used to provide our {\em BPT classification} (see \S~\ref{sec:BPT}). {\em Top and middle panels:} \mainsel \NII and \SII BPT diagrams. Galaxies are color-coded by their sSFR; here we show only 60 percent, randomly sampled, of each population to avoid saturation. We overplot the \citet{Kewley2001}, \citet{Kauffmann2003} and \citet{Kewley2006} demarcation lines (black solid, dot-dashed, and thick dotted lines, respectively) separating the SF, AGN, LINER and Seyfert contributions (see Sec.\,\ref{sec:BPT}). Ambiguous objects are represented as stars. \textit{Bottom:} Seyfert, passive and Ambiguous components of the \mainsel sample represented in the WHAN diagram, together with the EW cuts we use to select them: the ${\rm EW=6}\,$\AA\, limit (dot-dashed line) separating Seyferts from LINERs, and the ${\rm EW=0.5}\,$\AA\, cut \citep{Cid2011} (solid) to isolate passive ELGs from the rest.}
  \label{fig:BPT}
\end{figure}

As shown in the top panel of Fig.\,\ref{fig:Vmaxscheme}, the faintest $r-$band absolute magnitude that a \mainsel can have while being part of a sample limited at $r_{\rm p}^{\rm faint}=17.77$ is \citep[][]{Blanton2003_1}:
\begin{equation}
M_{r,\,i}^{\rm faint}=r_{\rm p}^{\rm faint}-{\rm DM}(z_i)-K(z_i)\,,
\label{eq:maglim}
\end{equation}
where ${\rm DM}(z)$ is the distance modulus estimated at redshift $z$ in our fiducial cosmology, and $K(z)$ is the $K-$correction. To calculate it we use {\textsc{Kcorrect v4$\_$3}}\footnote{\url{http://kcorrect.org}} \citep{Blanton2007}. Fig.\,\ref{fig:Kcorr} compares the SDSS $M_r$ luminosity functions computed with and without $K-$corrections. In the redshift range under study, the effect of $K-$corrections is less than 7 percent at $-22.5<M_r<-18$, while it grows up to 30 percent in the faintest galaxies in our sample. Note that $K-$corrections are not needed when dealing with emission-line luminosities for which the redshift is known. We choose not to apply any evolution correction, as this is negligible at $0.02<z<0.22$ \citep{Blanton2001}, and would require optimizing the model template to our ELG selection. 

The  maximum redshift, $z_{{\rm max},\,i}$, of a galaxy in our magnitude-limited sample, is found as the root of the following equation:
\begin{equation}
M_{r,\,i}^{\rm faint}-r_{\rm p}^{\rm faint}+{\rm DM}\left(z_{{\rm max},\,i}^{\rm mag}\right)+K(z_{{\rm max},\,i}^{\rm mag})=0\,,
\label{eq:zmax}
\end{equation}
which is solved iteratively by interpolating the $M_{r,\, i}(z)$ -- redshift relation.

The faintest \Ha ELG luminosity that a galaxy can have and still be in the sample, when this is limited in line flux, is obtained in a similar manner, as shown in Fig.\,\ref{fig:Vmaxscheme}. For a flux limit $F_{\rm H\alpha}^{\rm faint}$ (Sec.\,\ref{sec:selections}) we derive the corresponding faintest luminosity in that line as:
\begin{equation}
L_{{\rm H\alpha},\,i}^{\rm faint}\,{\rm [erg\,s^{-1}]}=4\pi\,D_{\rm L}^2(z_i)\,F_{\rm H\alpha}^{\rm faint}\,, 
\end{equation}
where the luminosity distance $D_{\rm L}(z_i)=(1+z_i)\,D_{\rm c}(z_i)$ is measured in [Mpc], and the line flux in [$\rm erg\,s^{-1}\,Mpc^{-2}$]. The maximum redshift, $z_{{\rm max},\,i}$, the galaxy can have in the \Ha flux-limited sample is the root of the following equation:
\begin{equation}
\left(1+z_{{\rm max},\,i}^{F_{\rm H\alpha}}\right)\,D_{\rm c}\left(z_{{\rm max},\,i}^{F_{\rm H\alpha}}\right)-\sqrt{\frac{L_{{\rm H\alpha},\,i}}{4\pi\,F_{\rm H\alpha}^{\rm faint}}}=0\,.
\label{eq:zmaxF}
\end{equation}
This is solved by interpolating and inverting the $D_{\rm c}(z)$ -- redshift relation.
For the \OII, \OIII, \Hb, \NII, and \SII lines we adopt the same procedure with the corresponding flux limit chosen for each line. In our case we choose the same cut for all the lines: $F>2\times10^{-16}\rm{erg\,s^{-1}\,cm^{-2}}$ (see Sec.\,\ref{sec:selections}).

Finally, the faintest \Ha flux a galaxy can reach in the \mainsel sample, when this is limited in ${\rm S/N}_{\rm H\alpha}^{\rm lim}$ (Sec.\,\ref{sec:selections}), is:
\begin{equation}
F_{{\rm H\alpha},\,i}^{\rm faint}={\rm S/N}_{\rm H\alpha}^{\rm lim}\times F_{{\rm err},\,i}\,,
\end{equation}
where $F_{{\rm err},\,i}$ is the line flux uncertainty. By substituting the above expression in Eq.\,\ref{eq:zmaxF}, we obtain $z_{{\rm max},\,i}^{\rm S/N_{H\alpha}}$. Again, for the rest of the lines the procedure is identical, using fixed signal-to-noise limit in our sample: $\rm S/N>2$ (see Sec.\,\ref{sec:selections}).

\section{ELG classification}
\label{sec:classification}
Strong spectral emission lines can have different origins, the most common being the gas heated by newly forming stars. Galaxies hosting super massive black holes actively accreting mass, AGN and QSOs, also present strong emission lines produced in jets and shock regions. The number density of AGN and QSOs is lower than SF galaxies, and their line ratios are different \citep[see e.g.,][]{Kewley2019}. Old stellar populations can also produce strong emission lines \citep[see e.g.,][]{Kennicut1992, holmes, Cid2011, Sansom2015,Byler2019,Nersesian2019,Clarke2021}.

One of the goals of this work is to understand the contribution to the LF of local ELGs classified according to the most likely origin of their emission lines. We split the \mainsel sample using two selection criteria: (i) a sharp cut in sSFR to separate star-forming (SF) from passively evolving galaxies (Sec.\,\ref{sec:sSFRselection}), and (ii) the line ratios in the BPT and WHAN diagrams (Sec\,\ref{sec:BPT}).
In Sec.~\ref{sec:LFresults} we study the luminosity functions for each of these ELG types.


\subsection{Classification using the sSFR}
\label{sec:sSFRselection}

We select star-forming galaxies as those with ${\rm sSFR}>10^{-11}{\rm yr^{-1}}$ in the \mainsel sample. These galaxies constitute $84$ percent of the sample, including the volume correction.
The value chosen for this cut corresponds to the classical threshold adopted to separate SF from passive galaxies \citep[e.g.,][]{Ilbert2015, Donnari19, CorchoCaballero21}.

\subsection{Classification with the BPT and WHAN diagrams}
\label{sec:BPT}

As illustrated in Fig.\,\ref{fig:BPT}, we classify the origin of the \mainsel spectral lines using the emission-line ratios in the Baldwin-Phillips-Terlevich (BPT) and the EW$_{\rm{H\alpha}}$ versus \NII/\Ha (WHAN) diagnostic diagrams \citep[e.g.,][]{Stasi2006, Cid2011}.

We build the BPT diagrams for the \mainsel \NII and \SII lines and adopt the demarcation criteria from \citet{Kewley2001} and \citet{Kauffmann2003} (``Kew01" and ``Kau03", hereafter) to separate ELGs into SF, Composite galaxies and AGN. 
The Kew01 line marks the upper envelope of the H\,{\sc ii} region in \citet{Kewley2001} photoionization models. Above this threshold, the origin of emission lines is expected to be different from young O and B stars \citep[see also][]{Belfiore2016}. The Kau03 demarcation line is derived from an empirical relation to separate SF galaxies. Between this line and that from Kew01, the regions where emission lines originate may be due to star formation and/or other ionization sources.

For those galaxies above the Kew01 line in the BPT \SII diagram, we further split the possible origin of their emission lines using the \citet{Kewley2006} criterion (``Kew06", hereafter) coupled with the $\rm EW\geq6$\,\AA\, condition from \citet{Cid2011} in the WHAN diagram, that is, the plane defined by the \Ha EW values as a function of ${\rm log([N_{II}]/H\alpha)}$. This separation allows us to better distinguish AGN candidates into Seyfert galaxies and low-ionization narrow emission-line regions \citep[LINERs;][]{Heckman1980}. 

LINERs are characterized by lower luminosities compared to Seyfert galaxies and QSOs.
It is well known that most nearby AGN with \OII, \SII or \OI emission are dominated by LINERs \citep[e.g.,][]{Ho1995, Ho1997, Kauffmann2003, Kewley2006, Singh2013, Belfiore2016}. Considering the intensity of their emissions, Seyfert sources and LINERs are often referred to as ``strong" and ``weak" AGN, respectively \citep[see e.g.,][]{Cid2011}.
On the other hand, these line ratios have also been observed in the outskirts of galaxies \citep[e.g.,][]{califa}, and therefore it is unclear whether they may actually be produced by other mechanisms.

By adopting the above criteria, we finally classify the galaxies in our \mainsel sample into the following {\em BPT classification}: (i) \textit{Star-forming} (SF): below Kau03 in \NII~BPT and Kew01 in \SII~BPT; (ii) \textit{Passive}: $\rm EW_{H\alpha}<0.5\,$\AA\, as in \citet{Cid2011}; (iii) \textit{Seyfert} (Sy): above Kew01 in both BPT diagrams, above Kew06 in \SII~BPT, and $\rm EW_{H\alpha}\ge 6$\,\AA; (iv) \textit{LINERs}: above Kew01 in both BPT diagrams, below Kew06 in \SII~BPT; (v) \textit{Composite}: between Kau03 and Kew01 in \NII~BPT; (vi) \textit{Ambiguous}: galaxies that either do not fall within any of the previous classifications (mostly Seyfert galaxies with $\rm EW_{H\alpha}< 6$\,\AA), or that belong to more than one class at the same time.

The above classification is widely used in the literature, and it provides an ideal benchmark to characterize the contributions to emission line LFs from different physical origins.

\subsection{Comparison of the classifications}
\begin{table}
    \centering
    \setlength{\tabcolsep}{3pt}
    \begin{tabular}{|c|c|c|c|}
    \hline
    BPT+WHAN & Total & \multicolumn{2}{c|}{Intersection} \\ 
    type & fraction&${\rm sSFR}>10^{-11}/{\rm yr}$ & ${\rm sSFR}\leq 10^{-11}/{\rm yr}$\\
        &          &(84\% of total) &  (16\% of total)\\
    \hline
    SF& 63.3 & 100.0&0.0 \\
    Passive& 0.03 & 1.1&98.9 \\
    Seyferts& 1.3 & 79.8&20.2 \\
    LINERs& 3.4 & 10.6&89.4 \\
    Composite& 18.0 & 83.8&16.2 \\
    Ambiguous& 13.97 & 33.5&66.5\\ \hline
    
    \hline
    \end{tabular}
    \caption{{\em Second column:} volume-corrected percentages of the different types of ELGs as classified using the BPT+WHAN diagrams. {\em Last two columns:} percent split of the total fraction for each type based on ${\rm sSFR}=10^{-11}{\rm yr^{-1}}$.}
    \label{tab:types}
\end{table}

Table\,\ref{tab:types} compares the volume-corrected percentages of galaxies, classified using the BPT+WHAN diagrams, with those that have a sSFR either above or below ${\rm sSFR}=10^{-11}{\rm yr^{-1}}$. It is clear from this table that, according to the BPT+WHAN classification, spectral emission lines originate from star-forming regions only for 63.3 percent of the \mainsel sample. Spectral emission lines are not originated in SF regions for an important fraction of ELGs with ${\rm sSFR}>10^{-11}{\rm yr^{-1}}$. The origin of these lines is likely to be shocks, as the combined total fraction of SF Seyfert, LINERs and composite ELGs is 22.7 percent. 

In Fig.~\ref{fig:BPT} we show how the six \mainsel types distribute as a function of sSFR in the \NII (top panel) and \SII (middle) BPT planes. Composite and passive ELGs constitute 18 and 0.03 percent of the total, respectively, and show lower sSFR values compared to the SF population (i.e., ${\rm sSFR}\lesssim10^{-9.8}{\rm yr^{-1}}$).
LINERs ($3.4$ percent of the ELG) exhibit even smaller sSFR values, that is, ${\rm sSFR}\lesssim10^{-11.4}{\rm yr^{-1}}$. Ambiguous galaxies make up 13.97 percent of the \mainsel sample. They also feature very small sSFR values, and they tend to preferentially occupy the AGN region of the BPT diagram.
Finally, Seyfert ELGs are a mixed population in terms of sSFR. While most of them will be classified as star-forming, a nonnegligible fraction (i.e., 20.2 percent) of them display sSFR below our adopted threshold of $10^{-11}{\rm yr^{-1}}$.

\begin{figure}
\centering
    \includegraphics[width=0.88\linewidth]{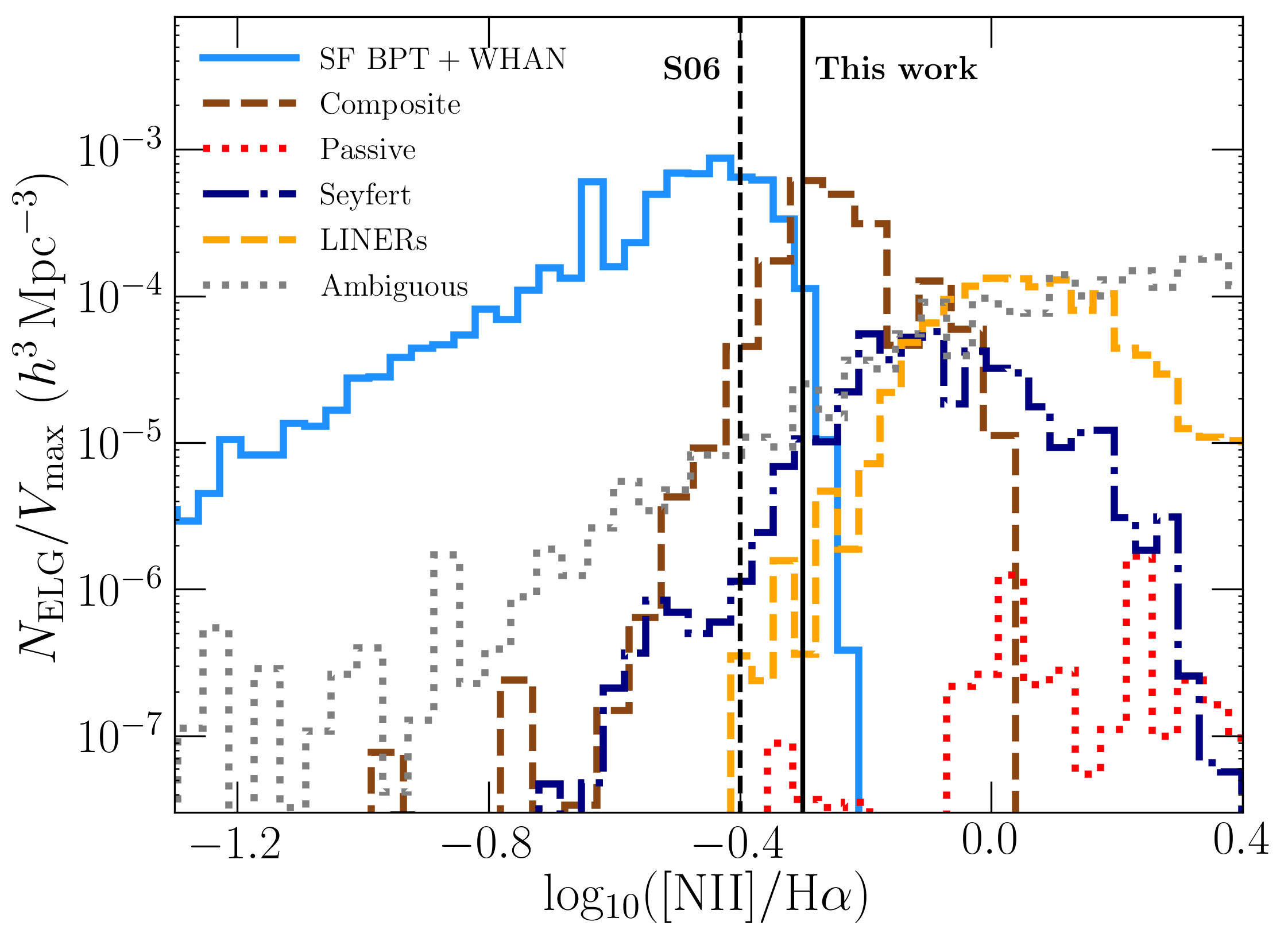}
\caption{Volume-weighted distributions of the \NII/\Ha line ratios resulting from our ELG {\em BPT classification}. The criteria to separate SF ELGs from the rest, solely based on the ratio \NII/\Ha proposed by \citet[][S06]{Stasi2006} is shown by a vertical dashed line, $\rm log([N_{II}]/H\alpha)= -0.4$. Our sample is better separated by a slightly different value, $\rm log([N_{II}]/H\alpha)= -0.3$, also indicated by a vertical solid line. Note that we do not apply any of these two cuts in our analysis. }
  \label{fig:WHAN}
\end{figure}

In the lower panel of Fig.~\ref{fig:BPT}, we display how Seyfert, passive, and Ambiguous ELGs are located in the WHAN diagram. We overplot as horizontal lines the $\rm EW=0.5\,$\AA\, threshold \citep{Cid2011} used to separate passive ELGs from the rest, as well as the $\rm EW=6\,$\AA\, criterion used to separate Seyfert ELGs from LINERs.

Fig.~\ref{fig:WHAN} shows the volume-weighted \NII/\Ha distributions of the ELG components resulting from our {\em BPT classification}. We overplot, as vertical dashed line, the \citet[][]{Stasi2006} $\rm log([N_{II}]/H\alpha)= -0.4$ criterion (``S06" hereafter) separating SF galaxies from the rest \citep[see also][]{Cid2011}. This condition is exclusively based on the \Ha/\NII line ratios and ignores the \OIII/\Hb ones. By looking at the distribution of SF ELGs, we propose $\rm log([N_{II}]/H\alpha)= -0.3$ as an alternative criterion to S06 to better separate SF \mainsel from the rest.  Note, however, that we do not apply any of these two cuts in our analysis, as we select SF galaxies exclusively based on Kew01, Kew06 and Kau03 demarcation lines in the BPT diagrams. This result shows that a significant fraction (18.8 percent) of SF ELGs selected from BPTs spills into the non-SF region of the WHAN plane, as defined by S06, while only 2 percent of composite ELGs spills into the SF plane. If instead of S06 we applied our proposed criterion, the fraction of SF ELGs in the non-SF region would decrease to 0.8 percent, while that of composite in the SF plane would go up to 26 percent.

\begin{figure}
\centering
    \includegraphics[width=0.88\linewidth]{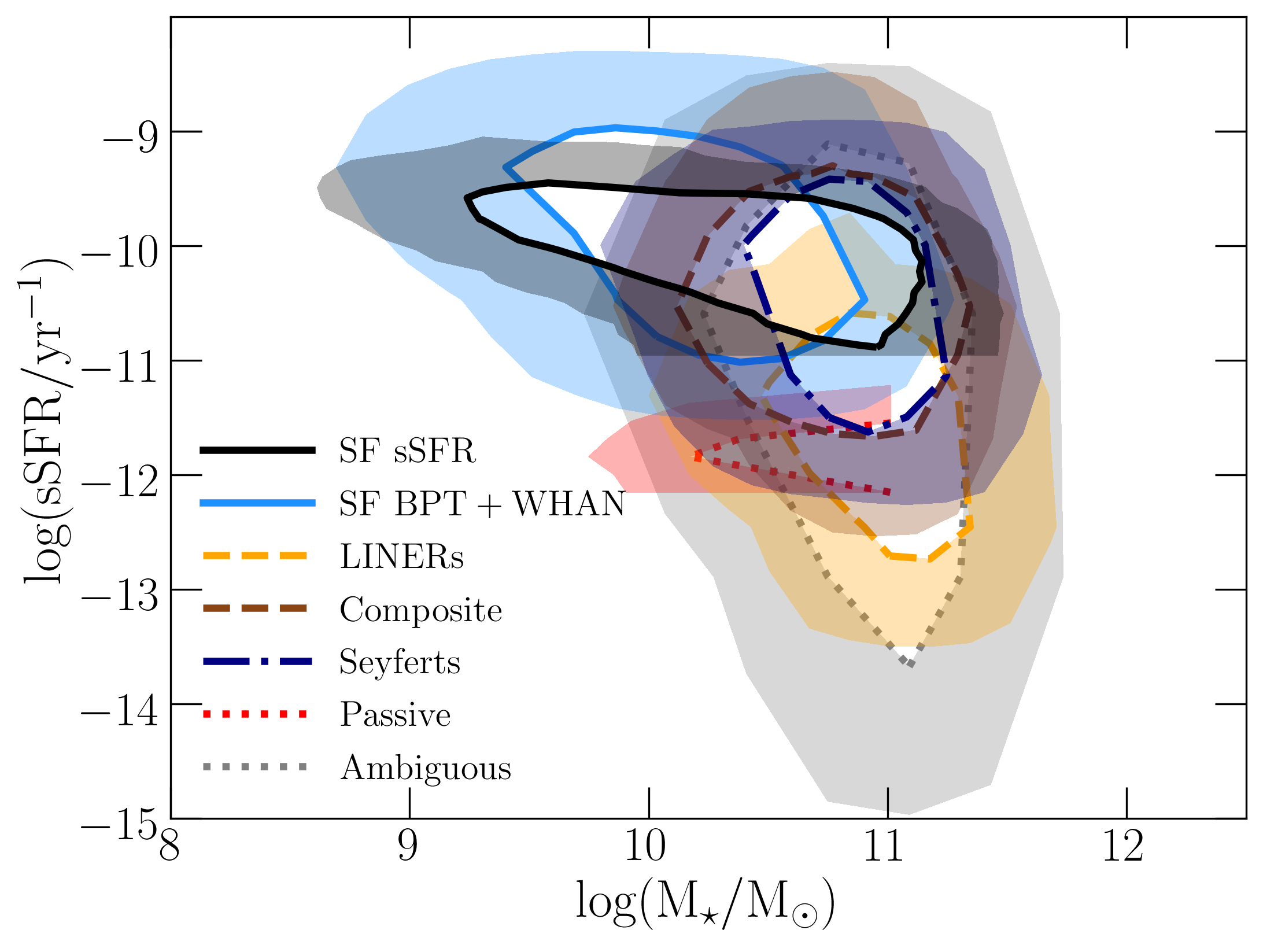}
\caption{sSFR as a function of stellar mass for all the \mainsel contributions, each one represented by contours. In each set of contours, the inner line (outer shade) represents the 68 (95) percent confidence regions. The contours of the SF population at ${\rm sSFR}>10^{-11}{\rm yr^{-1}}$ (solid black) are broken due to the sharp sSFR cut; those of the passive component (dotted red) are broken due to the very low number density of this population (0.1 percent of the total; see Table\,\ref{tab:types}).}
  \label{fig:ssfrmass}
  \end{figure}

Fig.\,\ref{fig:ssfrmass} compares, in the sSFR -- stellar mass plane, our ELG classification based on BPT+WHAN with the one based on sSFR (black contours). Both SF ELG classifications overlap well and concentrate in the upper region of the sSFR -- stellar mass plane, that is, at higher sSFR and lower mass values. In particular, while LINERs and passive ELGs mainly inhabit the lower tail of the distribution, toward lower sSFR values, composite and Seyfert galaxies populate the entire sSFR range. In terms of stellar mass, while SF ELGs span smaller values, down to $10^8\,\rm M_\odot$, the other ELG types concentrate above $10^{10}\,\rm M_\odot$.

\subsection{Old populations}\label{sec:oldpops}
\begin{figure}
\centering
    \includegraphics[width=\linewidth]{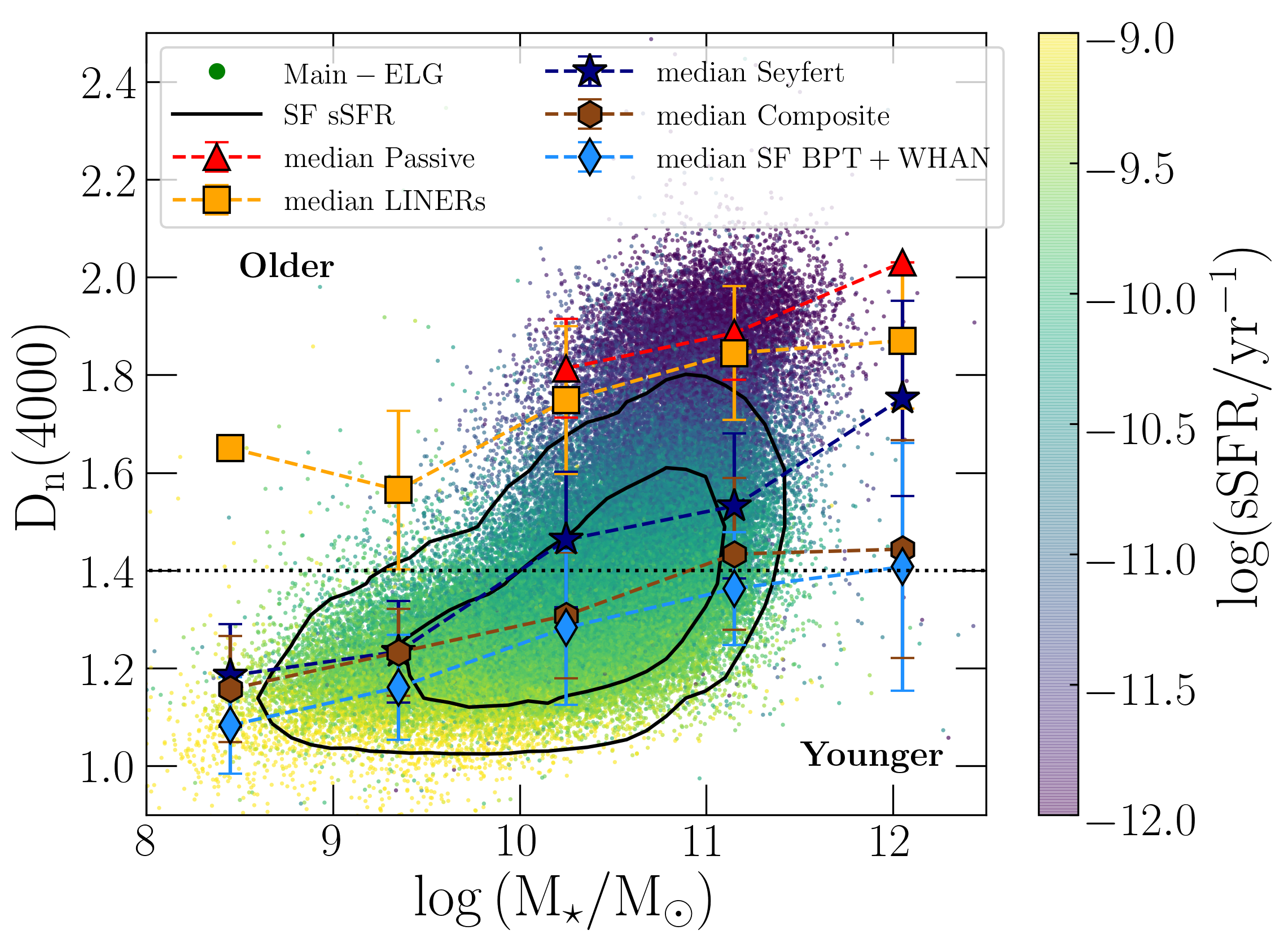}
\caption{$\rm D_n(4000)$ break index as a function of the galaxy stellar mass, color-coded by sSFR. We compare the \mainsel sample (small dots on the background) with its SF population selected from sSFR (black contours, not including weights), and with the median $\pm\sigma$ results of the SF (light blue diamonds), composite (brown hexagons), Seyfert (navy blue stars), LINERs (orange squares), and passive (red triangles) ELG contributions selected using the BPT+WHAN diagrams. The points shown here are a random subset of the \mainsel sample, 60 percent of the total, to avoid saturation. The horizontal dotted line indicates the typical separation between younger and older stellar populations. }
  \label{fig:D4000}
  \end{figure}

Galaxies that are passively evolving can present an excess of UV flux due to an old but hot stellar population, such as hot horizontal branch stars burning Helium \citep[e.g.,][]{phillipps2020}. 
We find that 16.4 (0.03) volume-corrected percent of the sample are passive according to the sSFR (BPT+WHAN) classification used in this study (see Table\,\ref{tab:types}).

To better understand the contribution of old stellar populations to the \mainsel sample, we study the 4000\,\AA\,break index, or ${\rm D_n(4000)}$, as a function of stellar mass and sSFR. The $\rm D_n(4000)$ index is reddening insensitive and traces SFRs on a time scale of 300--1000 Myrs. We employ the ${\rm D_n(4000)}$ values provided in the MPA-JHU catalog. These were estimated as the ratio of the flux in the red continuum to that in the blue continuum \citep[see e.g.,][]{Balogh1999, Angthopo2020}:
\begin{equation}
    {\rm D_n(4000)}=\frac{\langle F_{\rm c}^{ \rm r}\rangle}{\langle F_{\rm c}^{ \rm b}\rangle},\,\,\,\,{\rm where}
\end{equation}
\label{eq:D4000eq}
\begin{equation}
    \langle F_{\rm c}^{ i}\rangle=\frac{1}{(\lambda_2^i-\lambda_1^i)}\int_{\lambda_1^i}^{\lambda_2^1} F_{\rm c}(\lambda)\,d\lambda\,,
\end{equation}
and $(\lambda_1^{\rm b},\lambda_2^{\rm b},\lambda_1^{\rm r},\lambda_2^{\rm r})=(3850,3950,4000,4100)$\AA.

Fig.\,\ref{fig:D4000} compares the $\rm D_n(4000)$ index as a function of the galaxy stellar mass\footnote{We have investigated the evolution of the $\rm D_n(4000)$ -- $\log{\rm M_\star}$ relation, finding no significant variation over the redshift range $0.02<z<0.22$.}, color-coded by sSFR, for the \mainsel sample and its different components.

SF ELGs, no matter if selected from sSFR or from BPT+WHAN, are fully dominated by young stellar components. Considering the error bars, their $\rm D_n(4000)$ values range between 1 and 1.7, but most of them concentrate below 1.4.
Above ${\rm M_\star}\sim10^{10.5}\rm M_\odot$, they also show some contributions from old stellar components, which are negligible (2.5 percent) for the SF ELGs based on BPT+WHAN, but significant (25.5 percent) for SF ELGs selected from sSFR. Here we are quantifying the portion of passive ELGs falling in the younger, SF region at $\rm D_n(4000)<1.4$ in Figure\,\ref{fig:D4000}. All these numbers are volume-corrected.

On the other extreme, ELGs classified as LINERs or passive exhibit higher $\rm D_n(4000)$ values, mostly between 1.4 and 2. These ELGs are thus not only characterized by small sSFR values, as we have seen in the previous section, but they are also located outside the contour defined by the galaxies with sSFR$> 10^{-11}$~yr$^-1$ on the $\rm D_n(4000)$ -- $\log{\rm M_\star}$ plane.
According to their $\rm D_n(4000)$ values, 99.1 percent of the BPT+WHAN ELGs classified as passive are dominated by an old stellar component, with their $\rm D_n(4000)$ indices ranging between 1.6 and 2. LINERs also exhibit very high $\rm D_n(4000)$ values. About 99 percent of LINERs are dominated by older stars, with $\rm D_n(4000)$ between 1.4 and 2. The sources of the ionizing photons in LINERS are expected to be different from star forming regions. The origins could be hot low-mass evolved stars \citep[e.g.,][]{holmes}, diffuse ionized gas \citep[e.g.,][]{dig}, and X-ray busters \citep[e.g.,][]{xrb}. As the EW of LINERs are low, the origin of the emission lines is expected to be less energetic than AGN or shocks.

The situation is much more complex for Seyfert galaxies.
Note that the sSFR values and stellar masses of these objects cover the whole range $10^{-11.4}-10^{-9.5}\,{\rm yr^{-1}}$ and $10^{8.5}-10^{12}{\rm M_\odot}$, respectively.
66.6 percent of Seyfert galaxies are dominated by old stellar components, with $\rm D_n(4000)\sim1.5$, and the relation between $\rm D_n(4000)$ and stellar mass is in between the trends observed for SF and passive systems.

Composite ELGs, on the other hand, are consistent with the high-mass end of the main sequence of star formation (stellar masses above $10^{11}{\rm M_\odot}$, and sSFRs in the range $10^{-10.6}-10^{-10.2}\,{\rm yr^{-1}}$).
Only 34.9 percent of them are dominated by old stars, with $\rm D_n(4000)$ only slightly above 1.4, and they follow the same scaling relation as the SF population.

  
\section{Luminosity functions}

\label{sec:LFresults}
    \begin{figure*}
\centering\vspace{-0.2cm}
    \includegraphics[width=0.42\linewidth]{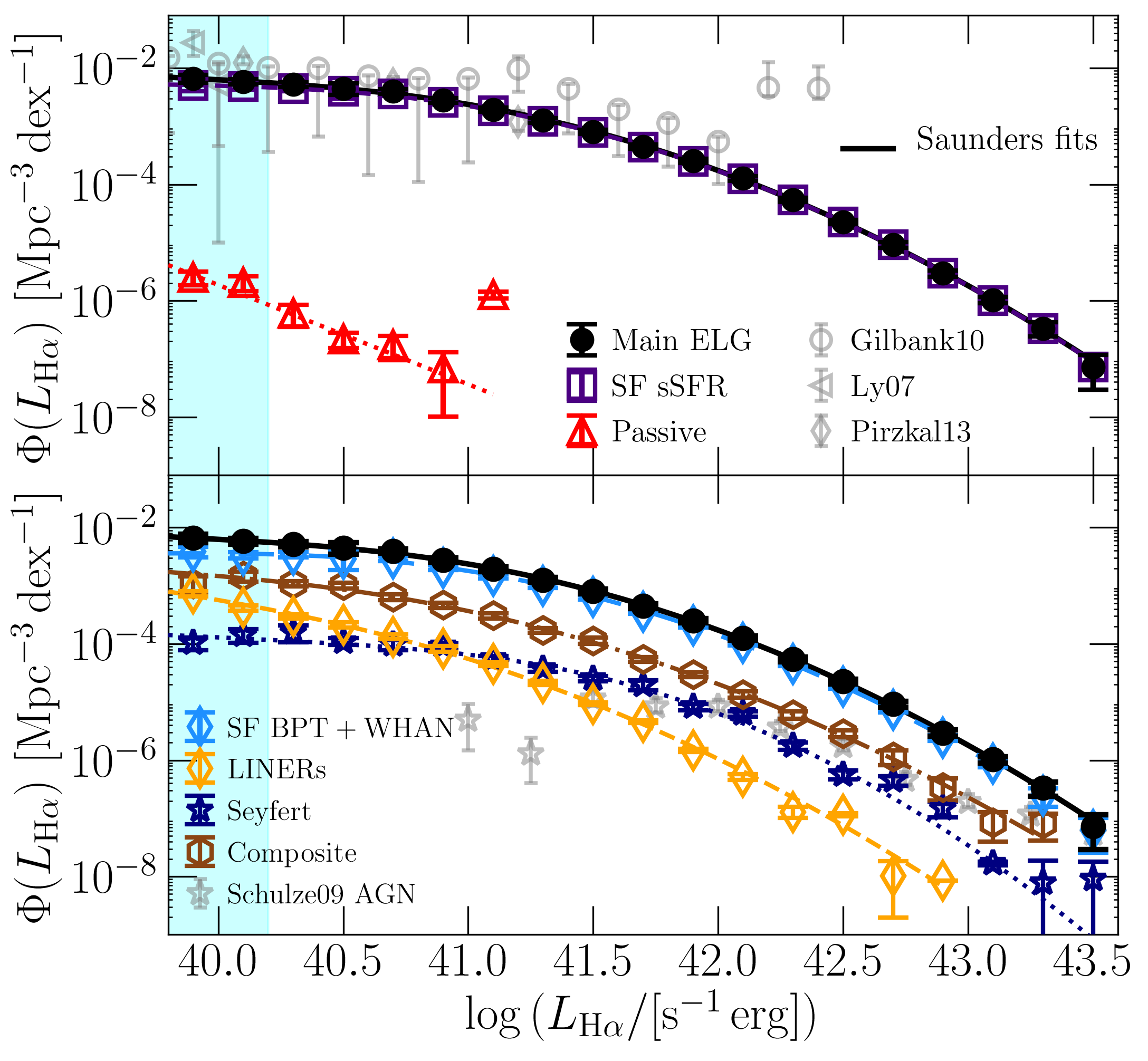}\quad
    \includegraphics[width=0.42\linewidth]{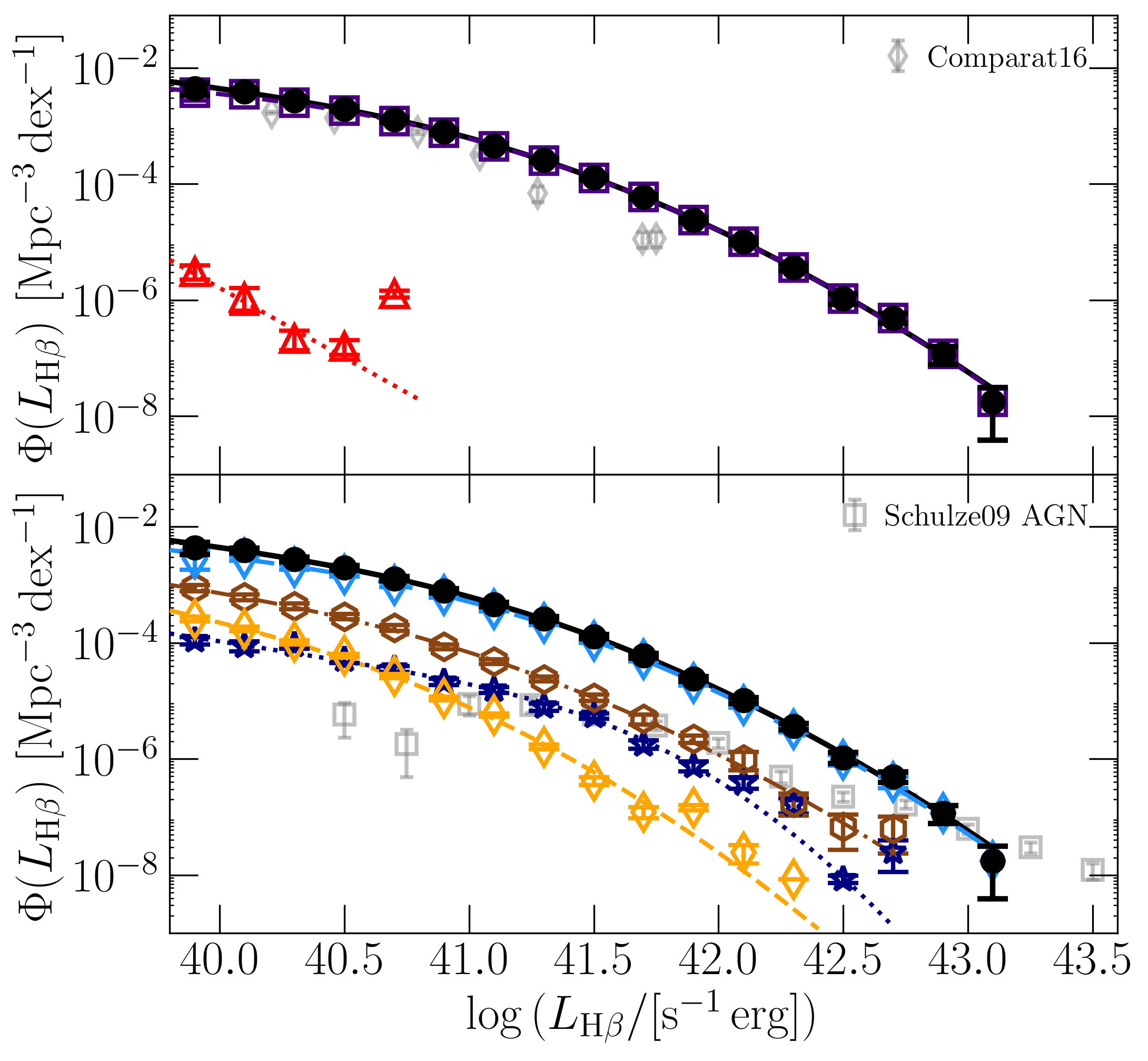}
    \includegraphics[width=0.42\linewidth]{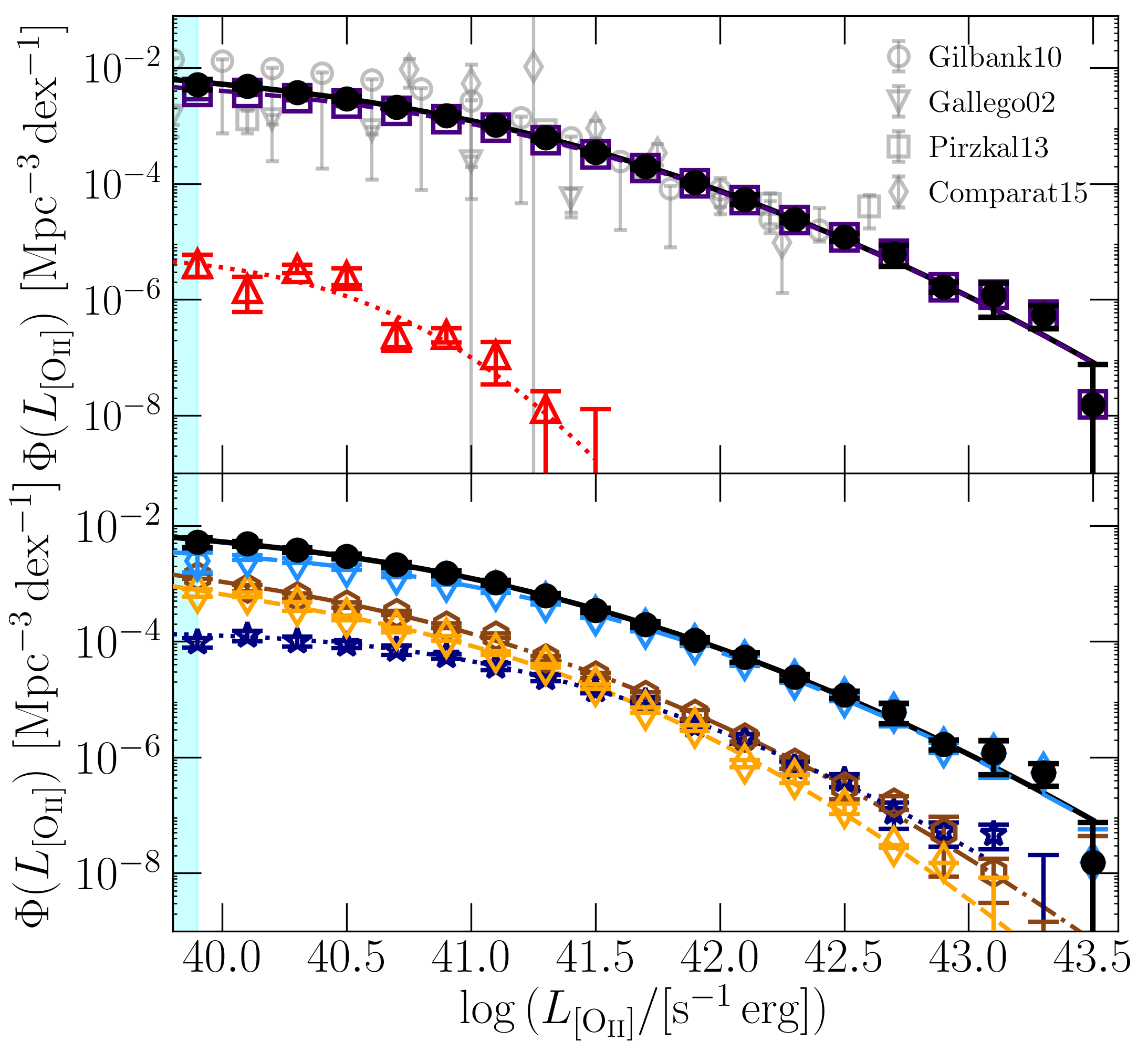}\quad
    \includegraphics[width=0.42\linewidth]{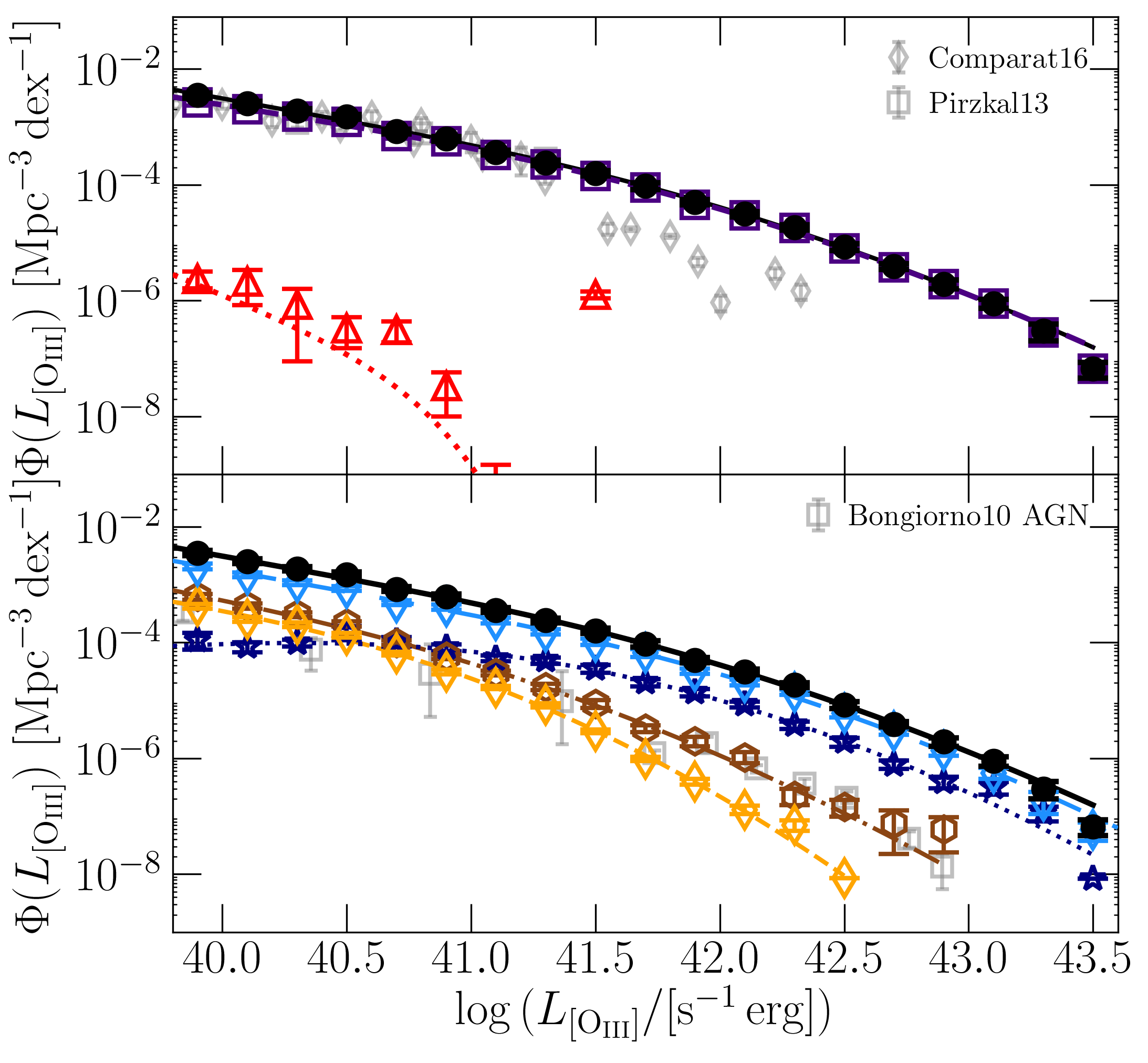}
    \includegraphics[width=0.42\linewidth]{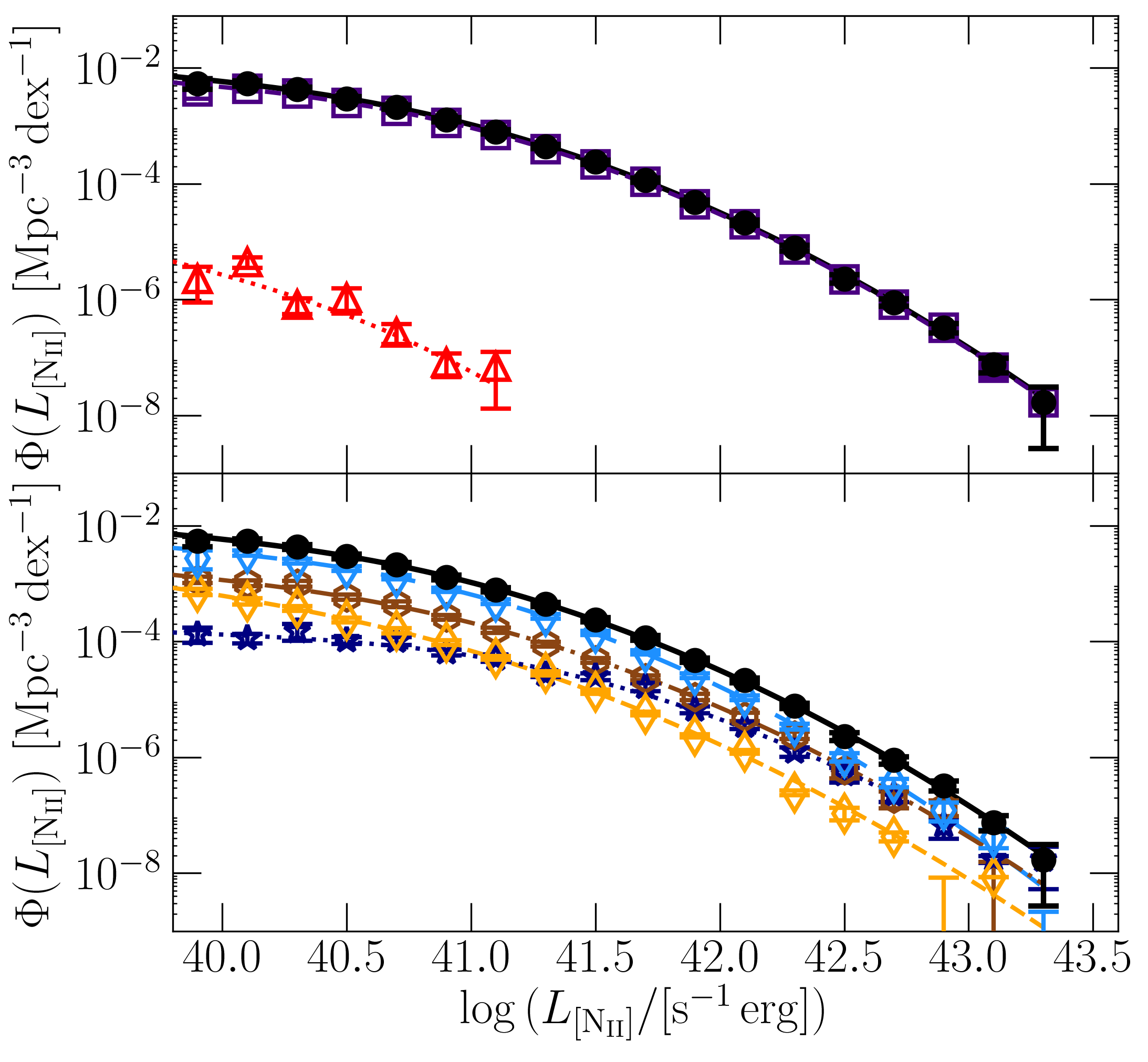}\quad
   \includegraphics[width=0.42\linewidth]{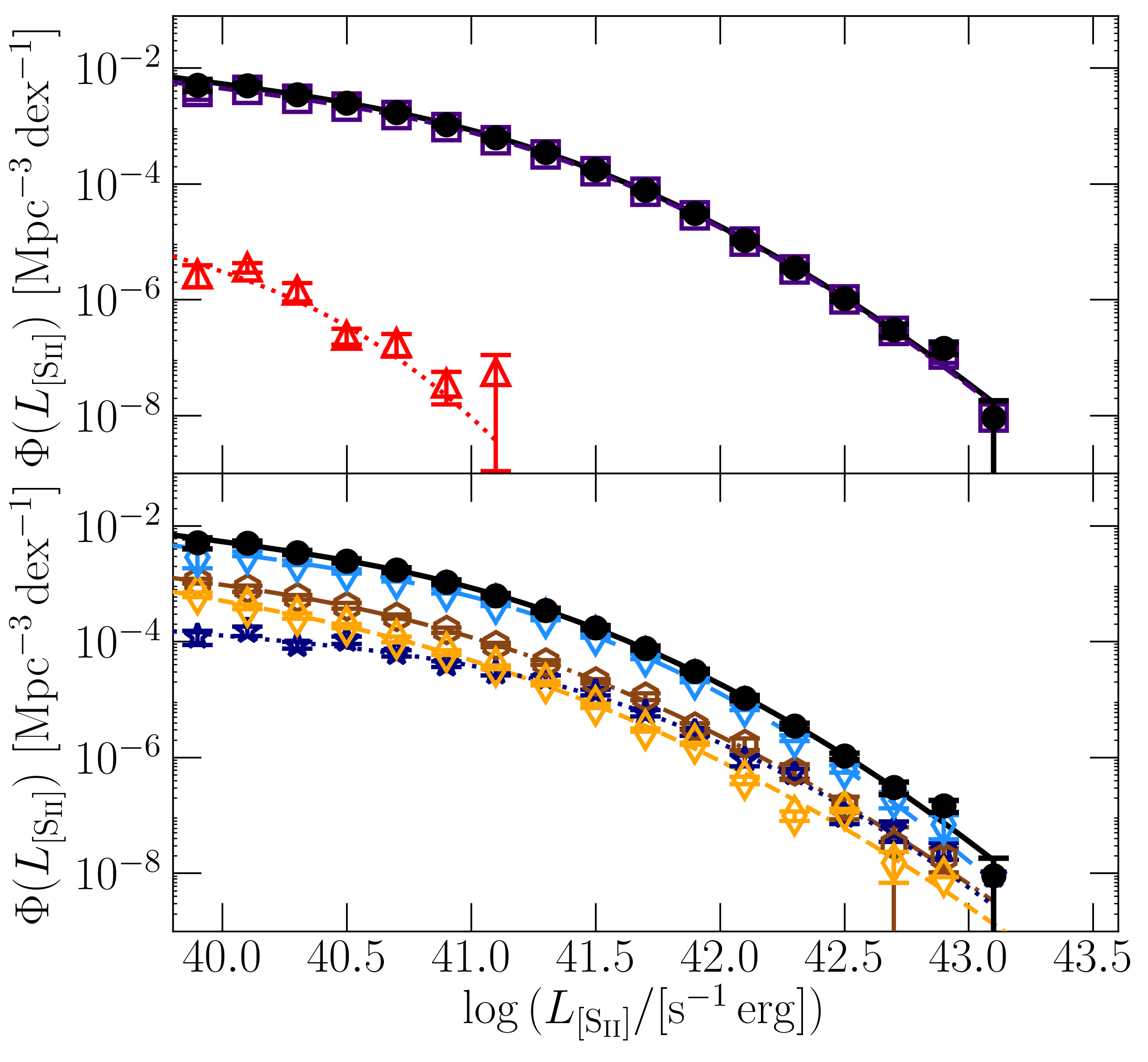}
\caption{From top to bottom and from left to right: \Ha, \Hb, \OII, \OIII, \NII, and \SII observed (i.e., dust extincted) luminosity functions of the \mainsel sample (full black dots). The contributions of ELGs classified in different ways are shown by empty colored markers, with colors as indicated in the legend. We compare our results -- both tabulated in Appendix\,\ref{sec:appendixLF} and as online material -- with several observed published measurements in the local Universe: \Ha ELGs from \citet{Ly2007} at $z=0.07-0.09$, \citet{Gilbank2010} at $0.032<z<0.2$, \citet{Pirzkal2013} at $0<z<0.5$, and \Ha AGN from \citet{Schulze2009} at $z<0.3$; \Hb ELGs from \citet{Comparat2016} at $z=0.3$, \Hb AGN from \citet{Schulze2009} at $z<0.3$; \OII ELGs from \citet{Gilbank2010} at $0.032<z<0.2$, \citet{Gallego2002} at $z\leq0.045$, \citet{Comparat2015} at $z=0.17$, and \citet{Pirzkal2013} at $0.5<z<1.5$; \OIII ELGs from \citet{Comparat2016} at $z=0.3$, from \citet{Pirzkal2013} at $0.1<z<0.9$, and \OIII AGN from \citet{Bongiorno2010} at $0.15<z<0.92$. We overplot our Saunders fits as lines; the parameters are in Table\,\ref{tab:fitparsaunders} and were obtained considering only the points above the luminosity completeness thresholds discussed in Sec.\,\ref{sec:incompleteness} and Appendix\,\ref{sec:appendix1} and represented as cyan shades in the panels (for those lines whose completeness limit falls within the $L$ range shown in the figure). The error bars are computed from 50 jackknife resamplings (see Sec.~\ref{sec:jackknife}).}
\label{fig:LFplotsaunders}
  \end{figure*}

\begin{table*}
 \setlength{\tabcolsep}{5pt}
  \centering
 \begin{tabular}{l c c c c c}
    \toprule
    &\multicolumn{5}{c}{Saunders (observed LFs)}\\
    &$\log{(\Phi_\star/[{\rm Mpc^{-3}dex^{-1}}])}$&$\log{(L_\star/[\rm erg\,s^{-1}])}$&$\alpha$&$\sigma$&$\chi^2_{\rm red}$\\
    \midrule
    &\multicolumn{5}{c}{\Ha}\\
    Full sample&-2.24$\pm$0.03&40.29$\pm$0.11&-0.19$\pm$0.06&0.73$\pm$0.01&0.2\\ 
    SF sSFR&-2.29$\pm$0.02&40.18$\pm$0.13&-0.04$\pm$0.02&0.72$\pm$0.01&0.2\\
    SF BPT+WHAN&-2.41$\pm$0.02&40.07$\pm$0.20&0.07$\pm$0.14&0.72$\pm$0.01&0.3\\
    LINERs&-3.21$\pm$0.74&40.00$\pm$0.02&-0.65$\pm$0.31&0.77$\pm$0.08&5.6\\
    Composite&-2.80$\pm$0.16&40.00$\pm$0.51&-0.27$\pm$0.24&0.80$\pm$0.03&0.5\\
    Seyfert&-4.05$\pm$0.12&40.81$\pm$0.24&-0.20$\pm$0.15&0.59$\pm$0.03&2.3\\
    Passive&-5.77$\pm$0.49&40.03$\pm$0.34&-1.71$\pm$0.79&-7.99$\pm$0.48&1.5\\
        \midrule
        &\multicolumn{5}{c}{\Hb}\\
    Full sample&-2.46$\pm$0.12&40.21$\pm$0.18&-0.56$\pm$0.11&-0.73$\pm$0.02&0.3\\
    SF sSFR&-2.41$\pm$0.08&40.00$\pm$0.16&-0.34$\pm$0.09&0.72$\pm$0.02&0.4\\
    SF BPT+WHAN&-2.56$\pm$0.11&40.15$\pm$0.20&-0.46$\pm$0.13&0.72$\pm$0.02&0.3\\
    LINERs&-3.60$\pm$1.48&40.00$\pm$0.68&-0.92$\pm$0.38&0.64$\pm$0.18&5.6\\
    Composite&-3.11$\pm$0.24&40.00$\pm$0.30&-0.64$\pm$0.14&-0.75$\pm$0.04&0.5\\
    Seyfert&-4.72$\pm$0.21&41.12$\pm$0.21&-0.67$\pm$0.09&0.42$\pm$0.05&1.9\\
    Passive&-5.81$\pm$1.37&40.01$\pm$0.38&-2.40$\pm$0.71&7.98$\pm$0.74&3.1\\
        \midrule
    &\multicolumn{5}{c}{\OII}\\
    Full sample&-2.25$\pm$0.10&40.00$\pm$0.27&-0.33$\pm$0.14&0.85$\pm$0.02&0.2\\
    SF sSFR&-2.37$\pm$0.09&40.00$\pm$0.28&-0.26$\pm$0.14&-0.83$\pm$0.02&0.2\\
    SF BPT+WHAN&-2.49$\pm$0.09&40.00$\pm$0.33&-0.21$\pm$0.18&-0.83$\pm$0.02&0.2\\
    LINERs&-3.47$\pm$0.22&40.44$\pm$0.25&-0.68$\pm$0.14&0.66$\pm$0.03&1.5\\
    Composite&-2.93$\pm$0.32&40.00$\pm$0.46&-0.54$\pm$0.28&0.78$\pm$0.05&0.9\\
    Seyfert&-3.86$\pm$0.08&40.00$\pm$0.83&-0.05$\pm$0.03&0.74$\pm$0.04&0.6\\
    Passive&-5.32$\pm$1.67&40.00$\pm$0.37&-0.18$\pm$0.09&-0.40$\pm$0.31&3.6\\
        \midrule
    &\multicolumn{5}{c}{\OIII}\\
    Full sample&-3.08$\pm$0.22&40.74$\pm$0.25&-0.77$\pm$0.05&-1.01$\pm$0.06&0.6\\
    SF sSFR&-3.02$\pm$0.23&40.60$\pm$0.30&-0.69$\pm$0.07&1.02$\pm$0.06&0.7\\
    SF BPT+WHAN&-3.47$\pm$0.24&40.96$\pm$0.27&-0.76$\pm$0.05&0.94$\pm$0.08&0.7\\
    LINERs&-3.42$\pm$0.55&40.00$\pm$0.04&-0.71$\pm$0.42&0.69$\pm$0.07&1.9\\
    Composite&-3.25$\pm$0.54&40.00$\pm$0.39&-0.83$\pm$0.19&-0.91$\pm$0.10&1.0\\
    Seyfert&-3.99$\pm$0.10&40.00$\pm$0.03&0.24$\pm$0.13&0.77$\pm$0.04&0.9\\
    Passive&-7.87$\pm$0.93&41.14$\pm$0.43&-1.73$\pm$0.64&0.10$\pm$0.05&3.4\\
        \midrule
    &\multicolumn{5}{c}{\NII}\\
    Full sample&-2.20$\pm$0.11&40.00$\pm$0.26&-0.35$\pm$0.18&-0.74$\pm$0.01&0.4\\
    SF sSFR&-2.29$\pm$0.10&40.00$\pm$0.29&-0.30$\pm$0.20&-0.73$\pm$0.02&0.4\\
    SF BPT+WHAN&-2.44$\pm$0.11&40.05$\pm$0.28&-0.31$\pm$0.21&-0.69$\pm$0.02&0.5\\
    LINERs&-3.17$\pm$0.47&40.00$\pm$0.02&-0.61$\pm$0.35&-0.79$\pm$0.07&1.3\\
    Composite&-2.90$\pm$0.19&40.00$\pm$0.42&-0.37$\pm$0.24&-0.76$\pm$0.04&0.7\\
    Seyfert&-3.92$\pm$0.15&40.38$\pm$0.48&-0.15$\pm$0.10&-0.70$\pm$0.07&1.2\\
    Passive&-5.51$\pm$1.04&40.00$\pm$0.85&-1.00$\pm$0.78&-0.56$\pm$0.48&3.5\\
        \midrule
    &\multicolumn{5}{c}{\SII}\\
    Full sample&-2.45$\pm$0.12&40.35$\pm$0.17&-0.54$\pm$0.11&-0.65$\pm$0.02&0.6\\
    SF sSFR&-2.52$\pm$0.09&40.36$\pm$0.14&-0.49$\pm$0.10&-0.64$\pm$0.02&0.4\\
    SF BPT+WHAN&-2.74$\pm$0.09&40.52$\pm$0.12&-0.56$\pm$0.08&0.60$\pm$0.02&0.4\\
    LINERs&-3.36$\pm$0.75&40.13$\pm$0.90&-0.72$\pm$0.44&0.76$\pm$0.13&3.7\\
    Composite&-2.98$\pm$0.18&40.00$\pm$0.13&-0.50$\pm$0.17&-0.72$\pm$0.03&0.4\\
    Seyfert&-4.09$\pm$0.19&40.58$\pm$0.33&-0.34$\pm$0.19&0.62$\pm$0.06&1.4\\
    Passive&-5.38$\pm$2.20&40.00$\pm$0.28&-0.98$\pm$0.51&0.38$\pm$0.16&4.0\\
\bottomrule
  \end{tabular}
   \caption{Saunders best-fit model parameters to the observed LFs shown in Fig.\,\ref{fig:LFplotsaunders}.} 
 \label{tab:fitparsaunders}
\end{table*}

\begin{table*}
 \setlength{\tabcolsep}{5pt}
  \centering
 \begin{tabular}{l c c c c c}
    \toprule
    &\multicolumn{5}{c}{Saunders (intrinsic LFs)}\\
    &$\log{(\Phi_\star/[{\rm Mpc^{-3}dex^{-1}}])}$&$\log{(L_\star/[\rm erg\,s^{-1}])}$&$\alpha$&$\sigma$&$\chi^2_{\rm red}$\\
    \midrule
    &\multicolumn{5}{c}{\Ha}\\
    Full sample&-2.24$\pm$0.09&40.00$\pm$0.05&0.23$\pm$0.08&-0.76$\pm$0.01&0.9\\ 
    SF sSFR&-2.34$\pm$0.12&40.00$\pm$0.02&0.30$\pm$0.14&0.74$\pm$0.01&0.8\\
    SF BPT+WHAN&-2.49$\pm$0.13&40.00$\pm$0.38&0.34$\pm$0.29&-0.73$\pm$0.01&0.6\\
    LINERs&-3.52$\pm$0.17&40.49$\pm$0.21&-0.61$\pm$0.10&0.67$\pm$0.04&0.7\\
    Composite&-2.95$\pm$0.11&40.00$\pm$0.63&0.18$\pm$0.12&-0.77$\pm$0.03&1.4\\
    Seyfert&-4.06$\pm$1.152&40.00$\pm$0.83&0.51$\pm$0.25&0.73$\pm$0.05&1.5\\
    Passive&-8.05$\pm$1.34&41.49$\pm$0.58&-1.53$\pm$0.94&-7.98$\pm$0.91&1.2\\
        \midrule
        &\multicolumn{5}{c}{\Hb}\\
    Full sample&-2.21$\pm$0.04&40.00$\pm$0.07&-0.12$\pm$0.10&0.75$\pm$0.01&0.6\\
    SF sSFR&-2.28$\pm$0.03&40.00$\pm$0.24&-0.05$\pm$0.01&-0.74$\pm$0.01&0.7\\
    SF BPT+WHAN&-2.43$\pm$0.03&40.00$\pm$0.28&0.03$\pm$0.01&-0.71$\pm$0.01&1.4\\
    LINERs&-3.44$\pm$0.27&40.00$\pm$0.35&-0.64$\pm$0.24&-0.68$\pm$0.04&0.5\\
    Composite&-2.89$\pm$0.11&40.00$\pm$0.42&-0.20$\pm$0.12&-0.79$\pm$0.02&0.5\\
    Seyfert&-3.88$\pm$0.25&40.00$\pm$0.39&0.11$\pm$0.07&0.72$\pm$0.36&1.2\\
    Passive&-5.79$\pm$1.52&40.00$\pm$0.71&-2.53$\pm$0.94&7.98$\pm$1.38&5.1\\
        \midrule
    &\multicolumn{5}{c}{\OII}\\
    Full sample&-2.20$\pm$0.03&40.00$\pm$0.40&0.09$\pm$0.03&0.83$\pm$0.02&0.9\\
    SF sSFR&-2.29$\pm$0.04&40.02$\pm$0.41&0.11$\pm$0.06&-0.82$\pm$0.02&0.9\\
    SF BPT+WHAN&-2.45$\pm$0.05&40.10$\pm$0.37&0.14$\pm$0.08&-0.76$\pm$0.02&1.0\\
    LINERs&-3.39$\pm$0.15&40.51$\pm$0.32&-0.34$\pm$0.12&-0.80$\pm$0.06&0.8\\
    Composite&-2.89$\pm$0.03&40.00$\pm$0.78&0.04$\pm$0.02&0.86$\pm$0.03&0.5\\
    Seyfert&-4.03$\pm$0.55&40.00$\pm$0.27&0.29$\pm$0.09&-0.90$\pm$0.10&1.8\\
    Passive&-5.58$\pm$1.23&40.00$\pm$0.38&-1.27$\pm$0.44&-7.99$\pm$0.58&2.3\\
        \midrule
    &\multicolumn{5}{c}{\OIII}\\
    Full sample&-2.27$\pm$0.27&40.00$\pm$0.54&-0.46$\pm$0.17&-1.07$\pm$0.04&1.2\\
    SF sSFR&-2.36$\pm$0.25&40.00$\pm$0.54&-0.43$\pm$0.17&1.06$\pm$0.04&1.0\\
    SF BPT+WHAN&-2.51$\pm$0.27&40.00$\pm$0.53&-0.47$\pm$0.17&1.02$\pm$0.05&1.2\\
    LINERs&-3.32$\pm$0.18&40.00$\pm$0.36&-0.39$\pm$0.23&0.70$\pm$0.04&0.7\\
    Composite&-2.91$\pm$0.47&40.00$\pm$0.11&-0.45$\pm$0.29&0.88$\pm$0.06&1.6\\
    Seyfert&-4.40$\pm$1.11&40.00$\pm$0.01&0.83$\pm$0.23&-0.72$\pm$0.04&1.7\\
    Passive&-5.60$\pm$1.48&40.00$\pm$0.89&-1.00$\pm$0.03&2.03$\pm$0.04&7.2\\
        \midrule
    &\multicolumn{5}{c}{\NII}\\
    Full sample&-2.17$\pm$0.06&40.00$\pm$0.34&-0.13$\pm$0.11&-0.79$\pm$0.02&1.0\\
    SF sSFR&-2.27$\pm$0.05&40.00$\pm$0.30&-0.08$\pm$0.03&0.78$\pm$0.02&1.1\\
    SF BPT+WHAN&-2.41$\pm$0.05&40.02$\pm$0.31&-0.05$\pm$0.02&0.72$\pm$0.02&1.4\\
    LINERs&-3.55$\pm$0.18&40.59$\pm$0.23&-0.64$\pm$0.09&0.69$\pm$0.05&0.8\\
    Composite&-2.84$\pm$0.09&40.00$\pm$0.10&-0.11$\pm$0.04&0.82$\pm$0.03&0.6\\
    Seyfert&-3.97$\pm$0.46&40.00$\pm$0.28&0.36$\pm$0.21&0.75$\pm$0.04&0.9\\
    Passive&-8.51$\pm$6.71&42.90$\pm$0.56&-0.92$\pm$0.08&-0.02$\pm$0.01&1.6\\
        \midrule
    &\multicolumn{5}{c}{\SII}\\
    Full sample&-2.23$\pm$0.07&40.18$\pm$0.18&-0.27$\pm$0.12&-0.72$\pm$0.02&0.9\\
    SF sSFR&-2.27$\pm$0.04&40.02$\pm$0.23&-0.10$\pm$0.06&0.72$\pm$0.03&0.8\\
    SF BPT+WHAN&-2.46$\pm$0.05&40.19$\pm$0.17&-0.16$\pm$0.12&0.67$\pm$0.02&1.2\\
    LINERs&-3.82$\pm$0.25&40.77$\pm$0.26&-0.75$\pm$0.11&0.58$\pm$0.06&1.7\\
    Composite&-2.92$\pm$0.17&40.14$\pm$0.39&-0.34$\pm$0.21&0.78$\pm$0.03&1.0\\
    Seyfert&-3.82$\pm$0.55&40.22$\pm$0.94&-0.13$\pm$0.06&0.81$\pm$0.19&6.8\\
    Passive&-5.60$\pm$1.31&40.00$\pm$0.48&-1.21$\pm$0.28&-0.51$\pm$0.15&2.2\\
\bottomrule
  \end{tabular}
   \caption{Same result as Table\,\ref{tab:fitparsaunders}, but for the intrinsic (i.e., dust corrected) LFs.}
 \label{tab:fitparsaundersintr}
\end{table*}

We have obtained observed and dust-corrected luminosity functions for the six emission lines of interest in 3 redshift bins. All these luminosity functions are available in Appendix\,\ref{sec:appendixLF}, and are tabulated as online material.

Fig.\,\ref{fig:LFplotsaunders} presents our \mainsel luminosity functions for \Ha, \Hb, \OII, \OIII, \NII, and \SII emission lines in the whole redshift range, $0.02<z<0.22$. Note that all these LFs are observed (i.e., dust attenuated). It is important to also highlight that we only trust our measurements at luminosities higher than the completeness thresholds established in Sec.\,\ref{sec:incompleteness} and indicated in Fig.\,\ref{fig:LFplotsaunders} by the shaded yellow regions. Those emission lines for which we do not show the shaded region have the completeness limit falling outside the luminosity range displayed in the figure. 

Our \mainsel LF measurements are in good agreement with several published results in the local Universe. However, in this work we are able to measure the \mainsel LFs beyond the limit $10^{43}{\rm erg\,s^{-1}}$ that previous studies show. This is thanks to the high statistics and large volume that the \mainsel sample offers, as well as the particular redshift selection performed.

Our \Ha LF is consistent up to $L_{\rm H\alpha}\sim10^{42}{\rm erg\,s^{-1}}$ with results from \citet{Gilbank2010} at $0.032<z<0.2$, and from \citet{Ly2007} at $z=0.07-0.09$. The latter only spans the faint tail of our distribution around $10^{40}{\rm erg\,s^{-1}}$. Our \Ha Seyfert LF shows good consistency with the AGN LF from \citet{Pirzkal2013} at $0<z<0.5$.

The \mainsel \OII LF is in good agreement with the results from \citet{Gilbank2010} at $0.032<z<0.2$ up to $\sim10^{42.3}{\rm erg\,s^{-1}}$. Below $10^{41}{\rm erg\,s^{-1}}$, our measurements are consistent with the results from \citet{Gallego2002} at $z\leq 0.045$. In the \LOII range between $10^{41}-10^{42.3}{\rm erg\,s^{-1}}$ our measurements are consistent with the results from \citet{Comparat2015} at $z=0.17$, and with the LF from \citet{Pirzkal2013} at $0.5<z<1.5$ in the range $10^{40.2}-10^{42.3}{\rm erg\,s^{-1}}$ .

The \mainsel \Hb LF agrees, up to $\sim10^{41}{\rm erg\,s^{-1}}$, with the results from \citet{Comparat2016} at slightly higher redshift, $z=0.3$. Our \Hb Seyfert LF is in reasonable agreement with the AGN LF from \citet{Schulze2009} at $0.1<z<0.9$ only in the luminosity range $10^{41.2}-10^{41.5}{\rm erg\,s^{-1}}$, while at higher luminosities we obtain up to 0.5\,dex less AGN.

Our \mainsel \OIII LF is in good agreement with the result from \citet{Comparat2016} at $z=0.3$ and \citet{Pirzkal2013} at $0.1<z<0.9$ below $10^{41.5}{\rm erg\,s^{-1}}$. Above this luminosity, we find about 1\,dex more luminous \OIII emitters than \citet{Comparat2016}. Our LF trend is smoother with no bump around $10^{41.5}{\rm erg\,s^{-1}}$.

\subsection{Fitting the emission-line LFs}

\begin{figure}
\centering
\includegraphics[width=0.9\linewidth]{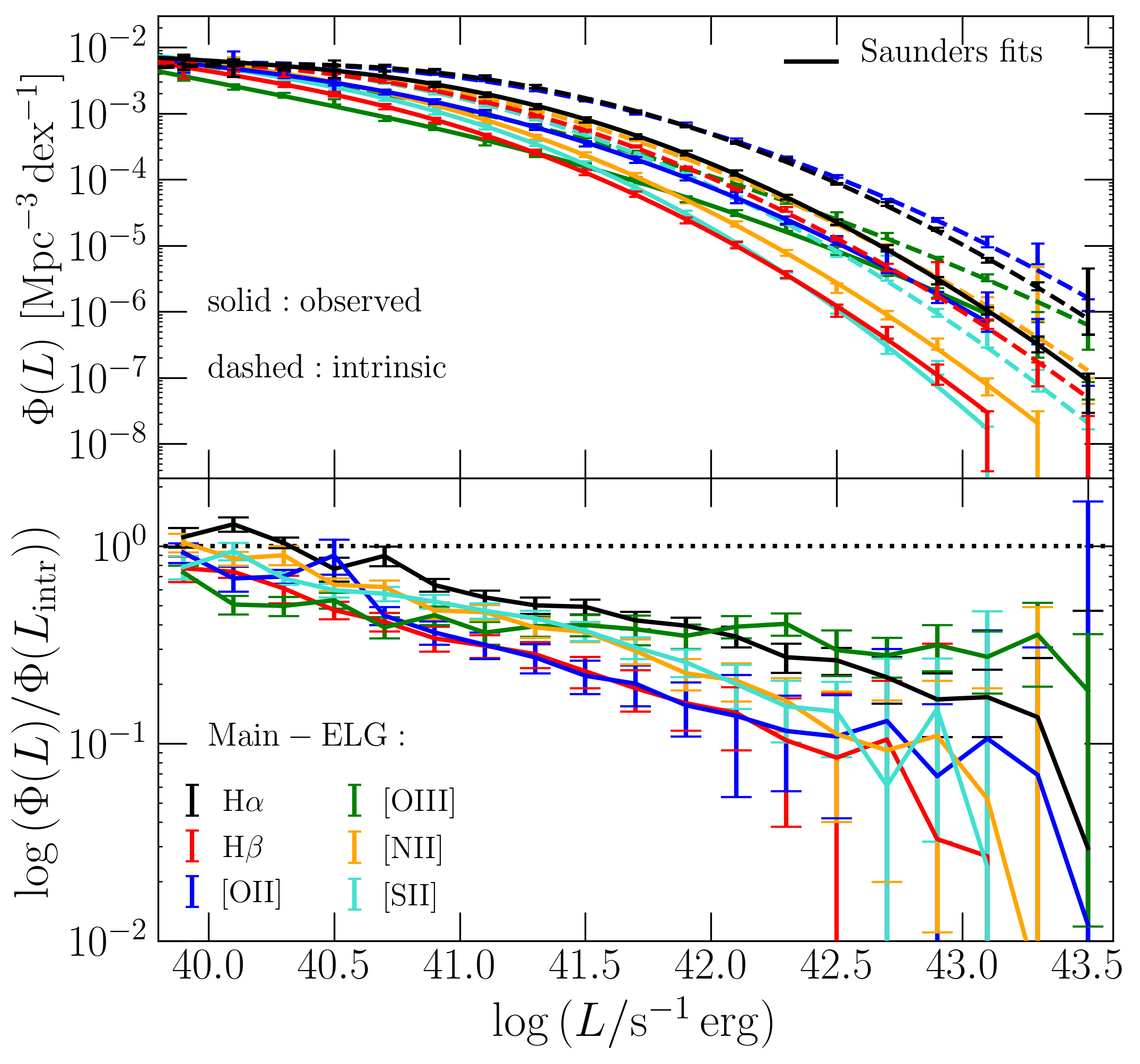}\quad
\caption{LFs compared for the six emission lines under study. {\em Top panel:} Best Saunders fit to the observed (solid lines) and dust-corrected (dashed lines) line LFs, for the six studied emission lines, as indicated in the legend. {\em Bottom panel:} Ratios between the observed and the intrinsic (dust-corrected) LF, for each emission line. In both panels the error bars are obtained from 50 jackknife realizations (\S~\ref{sec:jackknife}).}
  \label{fig:LFdustatt}
\end{figure}

For each measurement in Figs.\,\ref{fig:LFdustatt} and \ref{fig:LFplotsaundersintr}, we overplot the best fit obtained using the \citet{Saunders1990} function:
\begin{equation} 
    \Phi(L)= \Phi_\star\left(\frac{L}{L_\star}\right)^\alpha \exp\left[-\left(\frac{\log(1+L/L_\star)}{\sqrt{2} \sigma}\right)^2\right]\,.
    \label{eq:saunders}
\end{equation}
depending on four parameters.
For each emission line, we fit the quantity $\log(\Phi(L))$ considering only the points above the luminosity completeness threshold established in Sec.\,\ref{sec:incompleteness}.
The optimal parameters for each line LF are reported in Table\,\ref{tab:fitparsaunders} and they are overall consistent within the error bars with those provided by \citet{Comparat2016} as a function of redshift.
Our reduced $\chi^2$ values indicate that the Saunders model statistically provides a very good fit both to the \mainsel LFs and their different contributions.

In Fig.\,\ref{fig:LFdustatt} we compare the best Saunders models for the \mainsel LFs of the six studied emission lines. We do not find a clear trend with metallicity, however the $[\rm OIII]$ LF is flatter than the rest. 

Beyond Saunders, we also fit the ELG LFs using a single Schechter function \citep{Schechter1976}:
\begin{equation} 
    \Phi(L)= \Phi_\star \left(\frac{L}{L_\star}\right)^{\alpha}\exp{\left(-\frac{L}{L_\star}\right)}\,,
    \label{eq:singleschechter}
\end{equation}
a double Schechter one \citep[e. g.][]{Blanton2005lf}:
\begin{equation} 
    \Phi(L)= \left[\Phi_1^\star \left(\frac{L}{L_\star}\right)^{\alpha_1}+\Phi_2^\star \left(\frac{L}{L_\star}\right)^{\alpha_2}\right]\exp{\left(-\frac{L}{L_\star}\right)}\, ,
    \label{eq:doubleschechter}
\end{equation}
and a double power law.
Their best-fit parameters and results are tabulated as online material. 

The reduced $\chi^2$ values in Table~\ref{tab:fitparrestschechter} indicate that a single Schechter function provides a poor fit to the observational data.
The measured line LFs do show an excess in the very bright end, as already observed by \citet{Blanton2007} and \citet{MonteroDorta2009}, who justified this excess by the presence of AGN and QSOs.

The double Schechter model statistically provides a good fit to the \mainsel LF, as shown in Table~\ref{tab:fitparrest2schechter}, but it produces a bump in the bright end that seems to suggest overfitting rather than a physical feature of the LF.
Moreover, in Fig.~\ref{fig:LFplotsaunders}, when splitting the \mainsel LF in its different components, we see no evidence that the LF can be explained as the combination of two or more Schechter functions representing distinct galaxy populations. 
On the contrary, we argue that the bright end of both the individual and the combined LFs decrease more slowly than the exponential decay assumed by the Schechter parametric form.

The exact asymptotic behavior of very luminous galaxies is fundamental in order to make extrapolations at higher redshift, and it has profound implications on the expected duration of reionization and the type of galaxies contributing to it \citep[see e.g.,][]{Mason2015, Sharma2018}.
Therefore, we further test a double power law model \citep[e.g.,][]{Pei1995} with five parameters, that is, slightly more flexible than the Saunders function:
\begin{equation}
    \Phi(L)= \Phi_0 \left(\frac{L}{L_0}\right)^{-\alpha_0}\left[1+\left(\frac{L}{L_0}\right)^\beta\right]^{(\alpha_0-\alpha_1)/\beta}\,.
\end{equation}
As shown in Appendix\,\ref{sec:otherLFfits}, our power-law fit reaches the same level of agreement with the observations as the Saunders model, both for the \mainsel population as well as its different components. Therefore, in our analysis we choose to adopt a Saunders functional form for the fit, as it performs significantly better than any Schechter model and at a similar level than a model with more free parameters.

\subsection{LF uncertainties}
\label{sec:jackknife}

The uncertainties in the LFs are computed from 50 jackknife resamplings using the method presented in \citet[][]{Favolejackk}. We split the SDSS footprint into a grid of $5 \times 10 = 50$~cells, with 5 RA and 10 DEC bins. Each cell spans $\sim146\,{\rm deg}^2$ and  contains about 3500 \mainsel galaxies. We then estimate 50 times the LF of the \mainsel sample removing a different cell each time. From these estimates we compute the jackknife covariance matrix as \citep[e.g.,][]{Favole2016cameron}:
\begin{equation}
C_{ij}=\frac{(\rm N_{res}-1)}{\rm N_{res}}\sum_{a=1}^{\rm N_{res}}(\Phi_i^a-\bar{\Phi}_i)(\Phi_j^a-\bar{\Phi}_j)\,,
\end{equation}
where the indices $i$ and $j$ run over the bins in luminosity, and $a$ runs over the number of resamplings, $\rm N_{res}=50$. The $\bar{\Phi}$ term represents the mean of the $\rm N_{res}$ LFs, and the multiplicative factor outside the sum takes into account that, in each jackknife configuration, $(\rm N_{res}-2)$ copies are not independent from each other \citep[see][]{Norberg2011}. The 1\,$\sigma$ jackknife uncertainties are obtained as the square root of the diagonal elements of the covariance matrix.

\subsection{Contributions to the luminosity functions}
\label{sec:LFcomponents}

We find that the \mainsel LFs at $z\sim0.1$ are dominated by star-forming galaxies, independently from the emission line considered.
This is true for the two classifications we have made, based on sSFR and the BPT+WHAN diagrams. 
For most spectral lines, the second contributing population is that classified as ``composite", which could actually be mostly massive SF galaxies with weaker emission lines.
The shape of the composite component of each emission line is similar to the full and SF results, but its amplitude is about one order of magnitude lower. 

Our measurements of the LFs for the Seyfert and LINER components are in reasonable agreement with results in the literature \citep[see e.g.,][]{Bongiorno2010,Ermash2013}. In particular, the Seyfert contributions to the \Ha and \Hb \, \mainsel LFs are consistent with the AGN LFs at $z<0.3$ measured by \citet{Schulze2009}. The Seyfert contribution to the \OIII line is in agreement, up to $\sim10^{41.8}\,{\rm erg\,s^{-1}}$, with the \OIII AGN LF at $0.15<z<0.92$ from \citet{Bongiorno2010}, but it drops by about 1\,dex at $10^{42.5}\,{\rm erg\,s^{-1}}$.

In general, Seyfert galaxies contribute significantly to the \mainsel LFs only in the bright end, while passive galaxies and LINERs are nonnegligible only in the faint end.
One may notice in Fig. \ref{fig:LFplotsaunders} that the Seyfert contribution to \OIII at $10^{42}\,{\rm erg\,s^{-1}}$ is higher than that from composite galaxies by $\sim 1$\,dex.
For the other lines (e.g., \NII and \SII), the contribution from Seyfert ELGs is either subdominant or similar to that of composite galaxies.
This is somewhat expected to happen by construction, as in the BPT diagram we are requiring that these emission lines are strong for a galaxy to be assigned to the Seyfert class.

\subsection{LF evolution}
\label{sec:evolution}
\begin{figure}
\centering
\includegraphics[width=\linewidth]{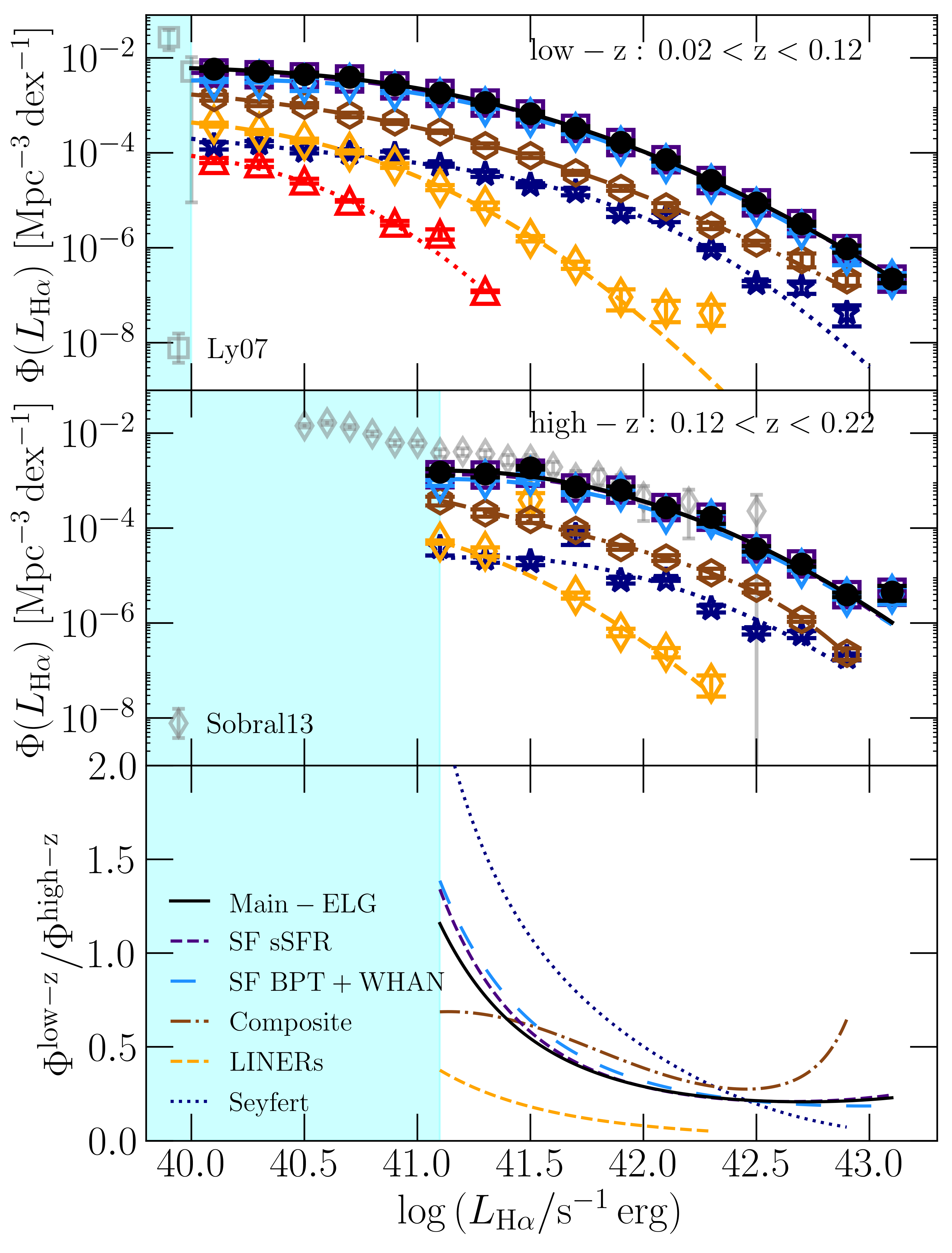}
\caption{\mainsel observed LF in the low (top) and high (middle) redshift bins, together with their Saunders fits. The markers, lines and colors are the same as in Fig.\,\ref{fig:LFplotsaunders}. The cyan shades indicate where the incompleteness starts to dominate and our LF measurements cannot be trusted. The lower completeness limits are set to $L_{\rm H\alpha}=10^{40},\,10^{41.1}\,{\rm s^{-1}\,erg}$ for the lower-$z$ and higher-$z$, respectively (see Appendix\,\ref{sec:appendixevol}). We compare them to the LF results at slightly higher redshift from \citet{Ly2007} ($z=0.08$) and \citet{Sobral2013} ($z=0.4$). The bottom panel shows the ratios of the low- to high-$z$ LF Saunders fits. We compare these ratios to better understand the change in the line luminosity functions. At $z>0.12$, there are no passive galaxies above the completeness threshold considered: \LHa$=10^{41.1}\,{\rm s^{-1}erg}$.}
  \label{fig:evolution}
\end{figure}

We further explore the evolution of the observed \mainsel LFs by separating the sample into two redshift bins: low-$z$, $0.02<z\leq0.12$, and high-$z$, $0.12<z<0.22$. Fig.\,\ref{fig:evolution} shows that our \mainsel results are consistent with observations from \citet{Ly2007} at $z=0.08$ and \citet{Sobral2013} at $z=0.4$. Other lines are presented in Appendix\,\ref{sec:appendixevol}. Similar consistency is found for the other lines compared to observations.

We fit our LFs in the two redshift bins using a Saunders model and compare them. 
The \Ha best-fit Saunders parameters in both $z$ bins are reported in Table\,\ref{tab:fitparsaunderszbins}; those for the rest of lines are in Tables\,\ref{tab:fitparsaunderszbinsall} and\,\ref{tab:fitparsaunderszbinsN2S2}.

The global increase with redshift of the number of the \mainsel is clear from Figs.\,\ref{fig:evolution},  \ref{fig:LFevolotherlines} and \ref{fig:LFevolotherlines2}. The differences are larger for the brightest objects, except when low number statistics appear to affect the observations. Such a trend is expected, as the \mainsel are predominantly star-forming galaxies and the star formation density increases with redshift (i.e., decreases with cosmic time since the Big Bang) within the range considered.

In terms of ELG contributions, SF, LINERs and Composite ELGs follow similar trends to those reported for the \mainsel, with some differences mostly happening at the brightest end. 
There are no passive ELGs brighter than $L\sim10^{41.5}\,{\rm erg\,s^{-1}}$ in the low-$z$ bin. In the high-$z$ bin, we do not have enough statistics to measure the passive contributions to the \Ha, \Hb and \SII LF.

From the low- to the high-$z$ bin, the luminosities of the full ELG sample increase by $0.2-0.3\,{\rm dex}$ (a factor of $\sim 0.5$). Part of the decrease in numbers is due to the effect of dust attenuation. However, there is also an expected decline in the star formation rates at lower redshifts, consistent with that reported for star-forming main sequence \citep{Speagle2014}. A similar behaviour is found for the different types of ELGs, although number statistics start to become a problem for Seyfert galaxies at low 
luminosities. The evolution of Seyfert ELGs is not trivial and will be worth examining in more detail in the future.

\begin{table*}
 \setlength{\tabcolsep}{5pt}
  \centering
 \begin{tabular}{l c c c c c}
    \toprule
    &\multicolumn{5}{c}{Saunders \Ha (observed LF)}\\ 
 &$\log{(\Phi_\star/[{\rm Mpc^{-3}dex^{-1}}])}$&$\log{(L_\star/[\rm s^{-1}\,erg])}$&$\alpha$&$\sigma$&$\chi^2_{\rm red}$\\
    \midrule  
    $0.02<z<0.12$& & & & & \\
    Full sample&-2.21$\pm$0.02&40.20$\pm$0.16&-0.07$\pm$0.13&0.65$\pm$0.01&0.2\\ 
        SF sSFR&-2.27$\pm$0.03&40.10$\pm$0.10&0.08$\pm$0.19&0.65$\pm$0.01&0.2\\
    SF BPT+WHAN&-2.42$\pm$0.04&40.00$\pm$0.01&0.25$\pm$0.19&0.64$\pm$0.01&0.3\\
   LINERs&-3.27$\pm$0.21&40.00$\pm$0.79&-0.10$\pm$0.92&-0.47$\pm$0.05&1.3\\
    Composite&-2.73$\pm$0.13&40.00$\pm$0.39&-0.32$\pm$0.24&0.75$\pm$0.03&0.8\\
    Seyfert&-4.31$\pm$0.12&40.72$\pm$0.31&-0.17$\pm$0.25&0.53$\pm$0.05&2.7\\
    Passive&-3.92$\pm$0.53&41.29$\pm$0.90&-0.47$\pm$0.25&0.43$\pm$0.12&3.4\\
        \midrule
    $0.12<z<0.22$& & & & & \\
    Full sample&-3.61$\pm$0.42&40.00$\pm$1.66&2.12$\pm$0.16&-0.52$\pm$0.07&4.8\\ 
    SF sSFR&-3.29$\pm$0.73&40.34$\pm$0.56&1.30$\pm$0.45&0.51$\pm$0.04&2.8\\
    SF BPT+WHAN&-3.85$\pm$0.31&40.00$\pm$1.50&1.71$\pm$1.08&0.53$\pm$0.09&4.0\\
    LINERs&-3.84$\pm$0.51&40.00$\pm$0.23&0.81$\pm$0.72&0.45$\pm$0.18&4.5\\
    Composite&-5.12$\pm$0.37&42.59$\pm$0.30&-1.14$\pm$0.06&0.20$\pm$0.08&0.6\\
    Seyfert&-4.92$\pm$0.90&40.55$\pm$0.76&1.19$\pm$0.48&0.49$\pm$0.23&10.6\\
    Passive&--&--&--&--&--\\
\bottomrule
  \end{tabular}
   \caption{Best-fit Saunders parameters of the observed \Ha LF fits in two redshift bins, $0.02<z<0.12$ (top) and $0.12<z<0.22$ (bottom), to better understand their evolution. Note that there are no passive galaxies above the completeness threshold \LHa$=10^{41.1}\,{\rm s^{-1}erg}$, in the high-$z$ bin.} 
 \label{tab:fitparsaunderszbins}
\end{table*}

\subsection{Dust effect in the luminosity functions}
\label{sec:dust}
The analysis carried out so far shows observed (i.e., dust attenuated) emission-line luminosities. In this Section we study the effect that dust extinction has on the LFs. We correct the line fluxes from dust attenuation using the Balmer decrement as implemented in \citet{CorchoCaballero20} and assuming a \citet{Calzetti2000} extinction curve. The intrinsic Balmer decrement remains roughly constant for typical gas conditions in star-forming galaxies \citep{Osterbrock1989}. Therefore, we assume the standard intrinsic value of (\Ha\Hb)$_{\rm int}=2.86$, commonly used in the literature for star forming galaxies.\footnote{This corresponds to a gas temperature of $T=10^4$\,K and an electron density of $n_e=10^2$\,cm$^{-3}$ for Case B recombination \citep{Osterbrock1989}.} For the small fraction of galaxies, 5.3 percent, with an observed ratio \Ha/\Hb below the theoretical value of $2.86$, no correction is applied.

The intrinsic (i.e., dust extinction corrected) \mainsel luminosity functions for the six lines of interest are presented in Figure\,\ref{fig:LFplotsaundersintr} and tabulated in Tables\,\ref{tab:LFtable1intr}-\ref{tab:LFtable3intr} in the Appendix. Their best-fit Saunders parameters are shown in Table\,\ref{tab:fitparsaundersintr} to facilitate the comparison with the observed LF parameters in Table\,\ref{tab:fitparsaunders}. 

Our intrinsic LFs are consistent with several published results in the literature, with different levels of agreement. In particular, beyond $L_{\rm H\alpha}10^{41}\,{\rm erg\,s^{-1}}$, our LFs are in good agreement with \citet{Guna2013} and \citet{James2008} results at $z<0.1$, while at fainter luminosities they measure up to 3 times more \Ha ELGs than us. Our LFs agree with the results from \cite{Sullivan2000} at $z<0.4$ above $L_{\rm H\alpha}10^{42}\,$s$^{-1}\,$erg. Below this value, we measure 2 times less \Ha emitters. The result by \citet{Ly2007} at $z=0.07-0.09$ only spans the very faint end of the \Ha LF, at $L_{\rm H\alpha}10^{40}\,$s$^{-1}\,$erg, where it is consistent with our findings. Our \Ha LFs are consistent with those from \citet{Fujita2003} at $z=0.24$ around $10^{42}\,{\rm erg\,s^{-1}}$ \Ha, but at fainter luminosities our LFs are lower by about 0.8\,dex. With \cite{1995ApJ...455L...1G} \Ha LF at $z\leq 0.045$ we agree around $10^{42}\,{\rm erg\,s^{-1}}$, while at lower (higher) \LHa our LF is higher (lower) by about 2\,dex. 

In the \OII line, our \mainsel LF is in between those from \citet{Sullivan2000} and \citet{Gallego2002}.

Fig.\,\ref{fig:LFdustatt} compares the observed and intrinsic \mainsel LFs for the six studied emission lines. We find that the effect of dust increases with luminosity. As shown by \citet{Duarte2017}, this is motivated by the fact that the actual amount of dust increases with stellar mass and SFR, which correlate strongly with line luminosity. Similar results were found also by \citet{Gilbank2010}, \citet{Lumbreras2019} and \citet{Vilella2021}.

For the six lines, the number of galaxies is affected by less than a factor of 10 up to $L\gtrsim 10^{42}\,{\rm erg\,s^{-1}}$. For brighter galaxies there is a clear decline in numbers beyond a factor of 10 for \Hb, \NII and \SII.
Since the impact of the extinction corrections on the LFs is significant only at $L\gtrsim 10^{42}\,{\rm erg\,s^{-1}}$, for the intrinsic LFs we maintain the same luminosity completeness thresholds of the observed ones (see Sec.\,\ref{sec:incompleteness}).

Dust attenuation changes the slope of the Saunders fits to the line LFs. Observed LFs are systematically steeper (i.e., smaller $\alpha$ values) than the intrinsic ones. However, most of the best fit $\alpha$ values are compatible with zero both with or without dust attenuation. This indicates a small variation.

\section{Summary and conclusions}
\label{sec:disc}
We have studied the properties of emission-line galaxies (ELGs) selected from the SDSS DR7 main galaxy sample \citep[]{Strauss2002} at $0.02<z<0.22$ (i.e., 2.4\,Gyrs). We have obtained the spectral properties of these galaxies from the MPA-JHU catalog\footref{mpanote}. Here we only study galaxies with a line flux of $F>2\times 10^{-16}\,{\rm erg\,s^{-1}\,cm^{-2}}$ and error $\sigma_{\rm F}<10^{-12}\,{\rm erg\,s^{-1}\,cm^{-2}}$, a signal-to-noise S/N$>2$, and an equivalent width EW$\geq 0$\,\AA\,in the six lines of interest: \Ha, \Hb, \OII, \OIII, \NII, and \SII. The resulting \mainsel is composed of 174572 ELGs (see Sec.~\ref{sec:selections}). The performed cuts guarantee the line luminosity function (LF) to be complete up to certain luminosity threshold.

We have measured the \mainsel luminosity function (LF) -- both observed and corrected from dust extinction (i.e., intrinsic) -- of the \Ha, \Hb, \OII, \OIII and, for the first time, of the \NII, and \SII emission lines. To this purpose, we have developed a generalized $1/V_{\rm max}$ weighting scheme to account for the different incompleteness effects in the LF due to the sample selection: the one due to the SDSS $r$-band magnitude limit, the spectroscopic selection, and those related to the thresholds imposed to each studied spectral line in our \mainsel sample. However, we have not taken into account the effect that the correlations between the different sources of incompleteness  might have. In fact, when selecting galaxies based on emission-line flux, we are implicitly removing a fraction of objects fainter than a given $M_r$ (see Fig.\,\ref{fig:LHaMr}). Neither the standard $1/V_{\rm max}$ estimator nor our modified method are capable of correcting from this source of incompleteness.

We have fit the \Ha, \Hb, \OII, \OIII, \NII, and \SII LFs using several functional forms (Sec.~\ref{sec:LFresults}): Saunders \citep{Saunders1990}, Schechter \citep{Schechter1976}, double Schechter \cite[e.g.,][]{Blanton2005lf}, and a double power law \citep[e. g.][]{Pei1995}. Globally, the smallest reduced $\chi^2$ are achieved using double power laws, however this function has five free parameters. Comparable values of reduced $\chi^2$ are obtained using Saunders models, with four free parameters. We therefore conclude that Saunders functions are the most appropriate ones to describe the emission-line LFs.

We have investigated the contributions of different ELG types to the emission-line LFs, both observed and intrinsic, and we also explored their redshift evolution. Our \mainsel sample has been classified both according to the specific star formation rate, ${\rm sSFR}>10^{-11}{\rm yr^{-1}}$ for star-forming (SF) galaxies, and using the line ratios (Sec.~\ref{sec:classification}). In particular, we have measured the \NII and \SII~BPT diagrams, as well as the WHAN one. Using these three diagrams, we have separated the \mainsel sample into star-forming (SF), passive, LINER, Seyfert and composite galaxies. We have also used the $\rm D_n(4000)$ break index to quantify the contribution of older stellar components to the \mainsel sample.

Our main findings on the ELG types and their contributions to the line LFs are summarized below:
\begin{itemize}
    \item The \mainsel sample is dominated by star-forming galaxies, independently from how they are selected and from the specific emission line considered. Including the volume correction, we find that 84 (63.3) percent of the sample are SF when selected from sSFR (BPT+WHAN).
    \item ELGs selected using a combination of line flux and signal-to-noise cuts are not equivalent to ELGs selected using a sharp cut in sSFR. In order to minimize the incompleteness in the faint end of their luminosity function, it is preferable to select ELGs based on line flux and S/N.  
    \item Besides the SF population, composite galaxies and LINERs are the ones that contribute the most to the ELG production below $10^{41}\,{\rm erg\,s^{-1}}$.
    \item The Seyfert contribution is nonnegligible only in the bright end of the line LF for the \OIII and \NII lines, $L_{[\rm NII]}>10^{42}\,{\rm erg\,s^{-1}}$, $L_{[\rm OIII]}>10^{43}\,{\rm erg\,s^{-1}}$. 
    \item The effect of dust in the LFs becomes significant only at $L\gtrsim10^{42}\,{\rm erg\,s^{-1}}$, independently from the emission line chosen. Correcting from dust extinction does not change the LF shape, and both observed and intrinsic LFs are best fitted using Saunders functions.
    \item The number of ELGs decline with redshift, with the exception of passive ELGs and Seyfert ELGs. Most of the passive ELGs are detected at $z<0.12$. The evolution of Seyfert ELGs is not trivial and needs a more detailed study.
\end{itemize}

The \mainsel sample can be considered as a low-redshift laboratory to test the robustness of our ELG selection methods and our ability to correct for survey incompleteness.
The ongoing DESI \citep{Schlegel2015, desi} and near future Euclid \citep{euclid, Sartoris2016}, 4MOST \citep{dejong2012} or Rubin \citep{Abell2009,2012arXiv1211.0310L} surveys will target millions of galaxies out to $z\sim2$ with strong emission spectral lines. These will be used as tracers of the dark matter field, in an attempt to build the most detailed 3D maps of the Universe to date. 
The methods used in cosmological surveys for validating different inference pipelines are based on model catalogs of galaxies, and the results of this study, together with the \Ha\,\mainsel clustering and bias results from Favole et al. in prep., can be used as guidelines to prepare these and other future science cases at higher redshifts.
A detailed comparison of the results presented here with those from a range of semi-analytic galaxy models will be instrumental in order to constrain their parameters and make realistic predictions of the statistics of the galaxy population at earlier cosmic epochs.

The observational samples were selected from the SDSS NYU--VAGC (\url{http://cosmo.nyu.edu/blanton/vagc/}) and spectroscopically matched to the MPA-JHU DR7 spectral relase (\url{http://www.mpa-garching.mpg.de/SDSS/DR7/}) to obtain the emission-line properties. 

\section*{Acknowledgements}
The \mainsel selections and all the results of our analysis are publicly available as {\em A\&A} online material and at \url{http://research.iac.es/proyecto/cosmolss/pages/en/dataresults.php}. 

The observational samples were selected from the SDSS NYU--VAGC (\url{http://cosmo.nyu.edu/blanton/vagc/}) and spectroscopically matched to the MPA-JHU DR7 spectral relase (\url{http://www.mpa-garching.mpg.de/SDSS/DR7/}) to obtain the emission-line properties. 

GF is supported by a {\em Juan de la Cierva Incorporación} grant n.\,IJC2020-044343-I. GF acknowledges the MICINN ``Big Data of the Cosmic Web" research grant (P.I. F.-S. Kitaura) for additional support, as well as the SNF 175751 ``Cosmology with 3D Maps of the Universe" research grant and the LASTRO group at the Observatoire de Sauverny for hosting and supporting the first stage of this project. She further thanks Andr\'es Balaguera for insightful discussion on the computational aspects of this work.

VGP is supported by the Atracci\'{o}n de Talento Contract no. 2019-T1/TIC-12702 granted by the Comunidad de Madrid in Spain. VGP and AK are also supported by the Ministerio de Ciencia e Innovaci\'{o}n (MICINN) under research grant PID2021-122603NB-C21.
YA and PC acknowledge financial support from grant PID2019-107408GB-C42 of the Spanish State Research Agency (AEI/10.13039/501100011033).
AK and further thanks Dan Lacksman for the flamenco moog.
SAC acknowledges funding from {\em Consejo Nacional de Investigaciones Cient\'{\i}ficas y T\'ecnicas} (CONICET, PIP-2876), {\em  
Agencia Nacional de Promoci\'on de la Investigaci\'on, el Desarrollo Tecnol\'ogico y la Innovaci\'on} (Agencia I+D+i, PICT-2018-3743), and {\em Universidad Nacional de La Plata} (G11-150), Argentina. ADMD thanks Fondecyt for financial support through the Fondecyt Regular 2021 grant 1210612. GF and coauthors are thankful to the anonymous referee for comments that have improved the quality and scope of the paper.

Funding for the SDSS and SDSS-II has been provided by the Alfred P. Sloan Foundation, the Participating Institutions, the National Science Foundation, the U.S. Department of Energy, the National Aeronautics and Space Administration, the Japanese Monbukagakusho, the Max Planck Society, and the Higher Education Funding Council for England. The SDSS Web Site is \url{http://www.sdss.org/}. 
The SDSS is managed by the Astrophysical Research Consortium for the Participating Institutions. The Participating Institutions are the American Museum of Natural History, Astrophysical Institute Potsdam, University of Basel, University of Cambridge, Case Western Reserve University, University of Chicago, Drexel University, Fermilab, the Institute for Advanced Study, the Japan Participation Group, Johns Hopkins University, the Joint Institute for Nuclear Astrophysics, the Kavli Institute for Particle Astrophysics and Cosmology, the Korean Scientist Group, the Chinese Academy of Sciences (LAMOST), Los Alamos National Laboratory, the Max-Planck-Institute for Astronomy (MPIA), the Max-Planck-Institute for Astrophysics (MPA), New Mexico State University, Ohio State University, University of Pittsburgh, University of Portsmouth, Princeton University, the United States Naval Observatory, and the University of Washington.

\bibliographystyle{aa}
\bibliography{biblio}

\begin{appendix} 
\section{Selection effects and ELG properties for all the six lines of interest}
\label{sec:appendix1}
\begin{figure*}
\centering \hspace{-0.8cm}
     \includegraphics[width=0.31\linewidth]{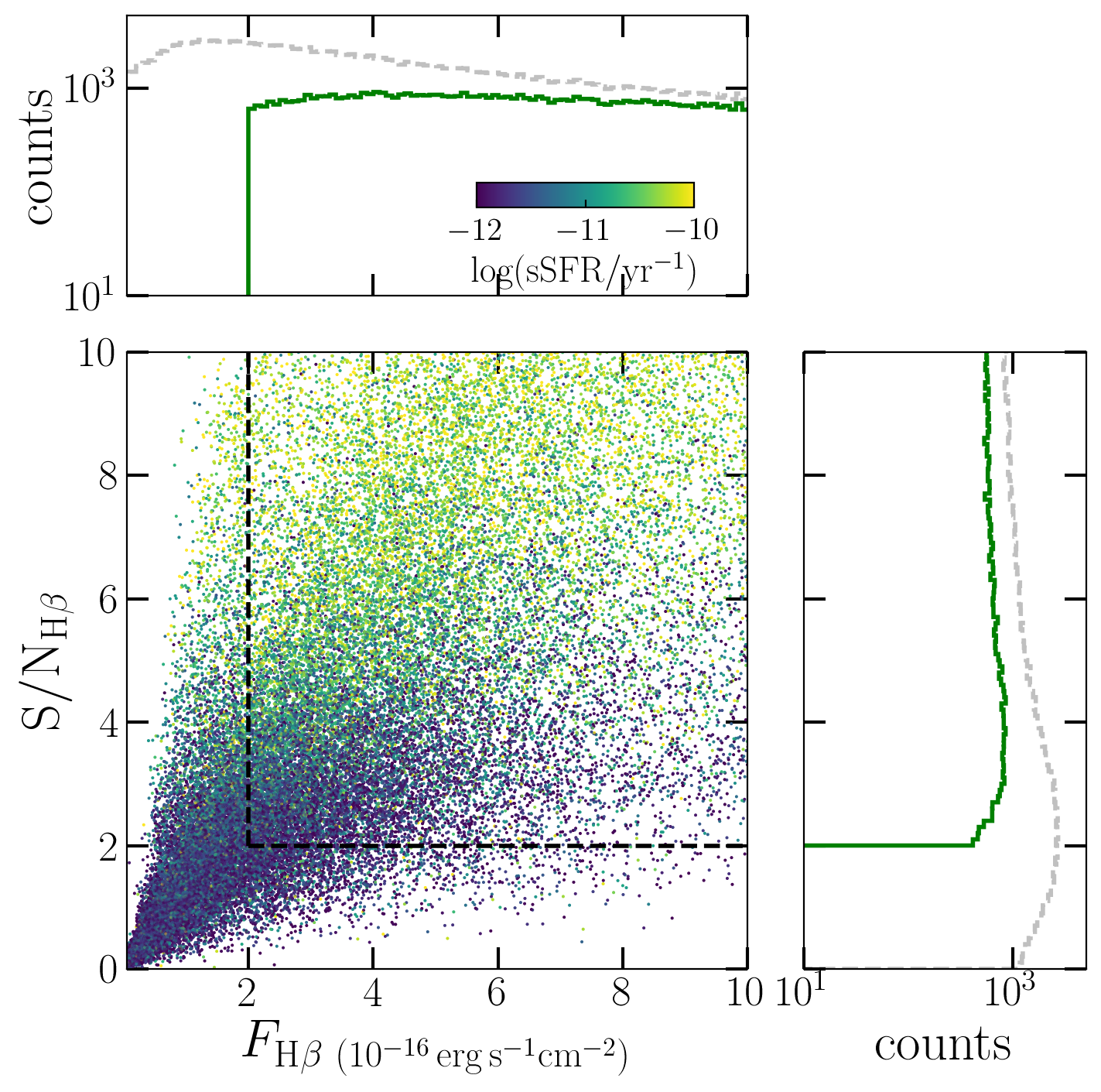}\quad
   \includegraphics[width=0.31\linewidth]{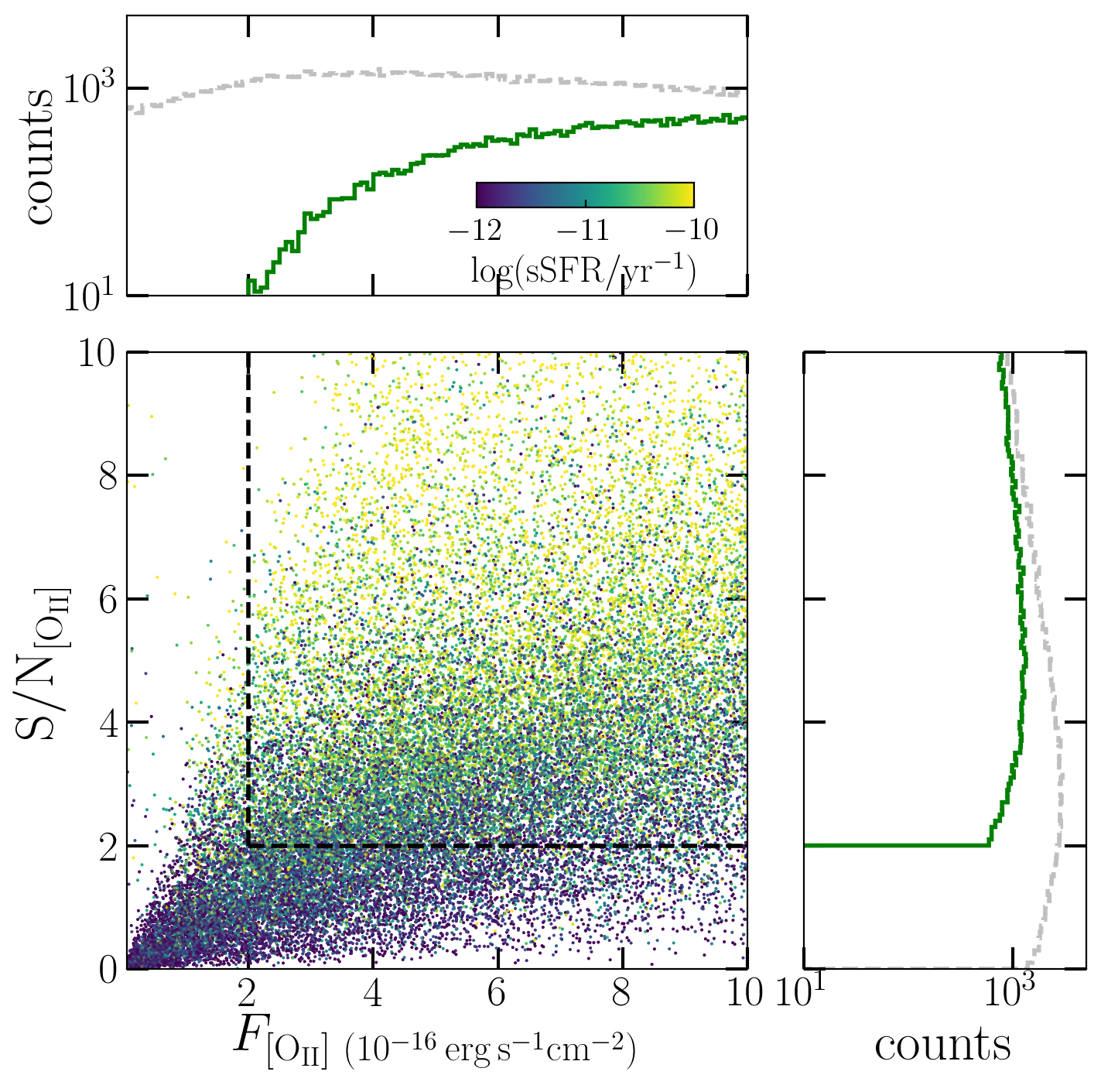}\quad
  \includegraphics[width=0.31\linewidth]{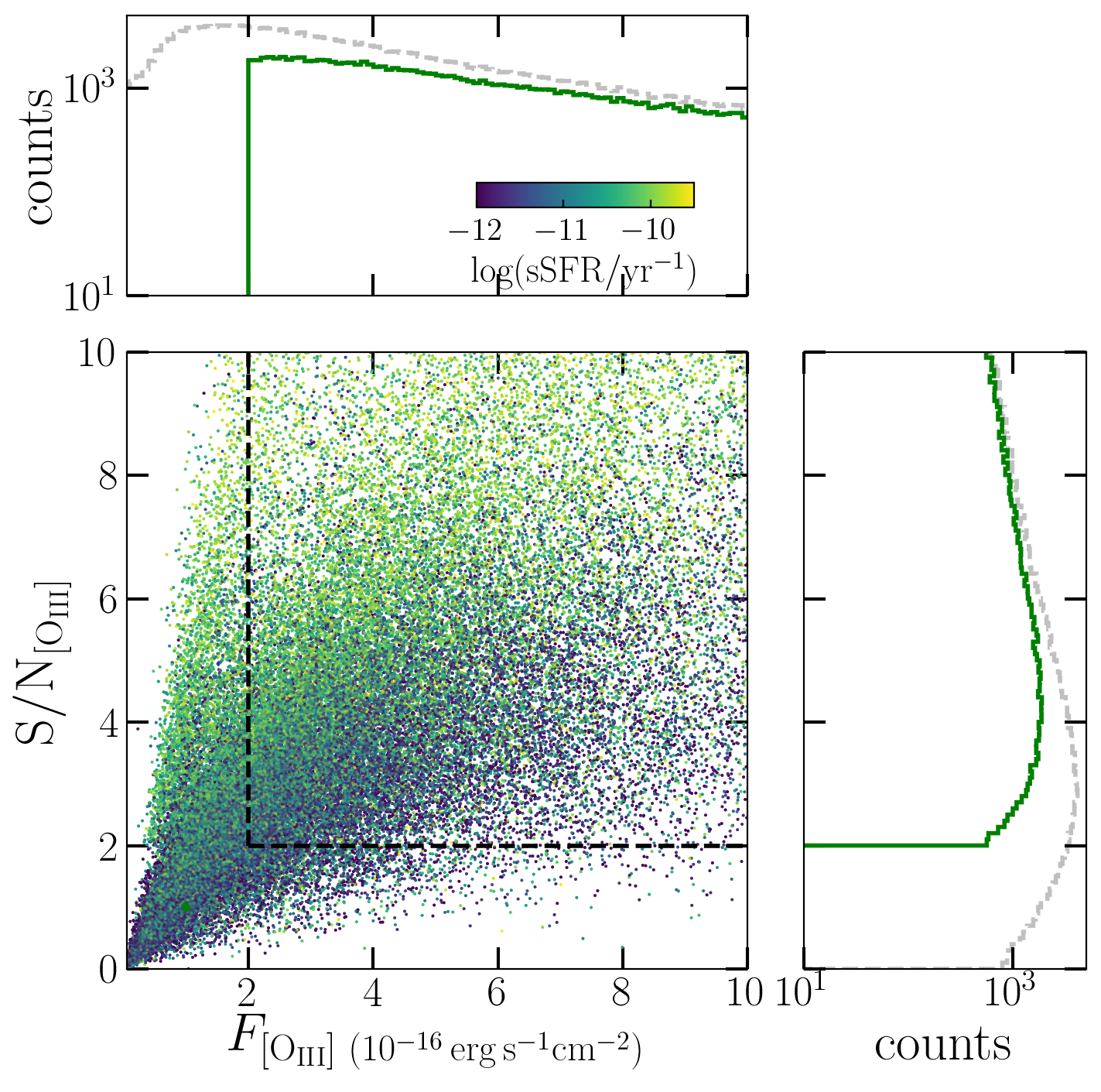}\hspace{-0.8cm}
  \includegraphics[width=0.31\linewidth]{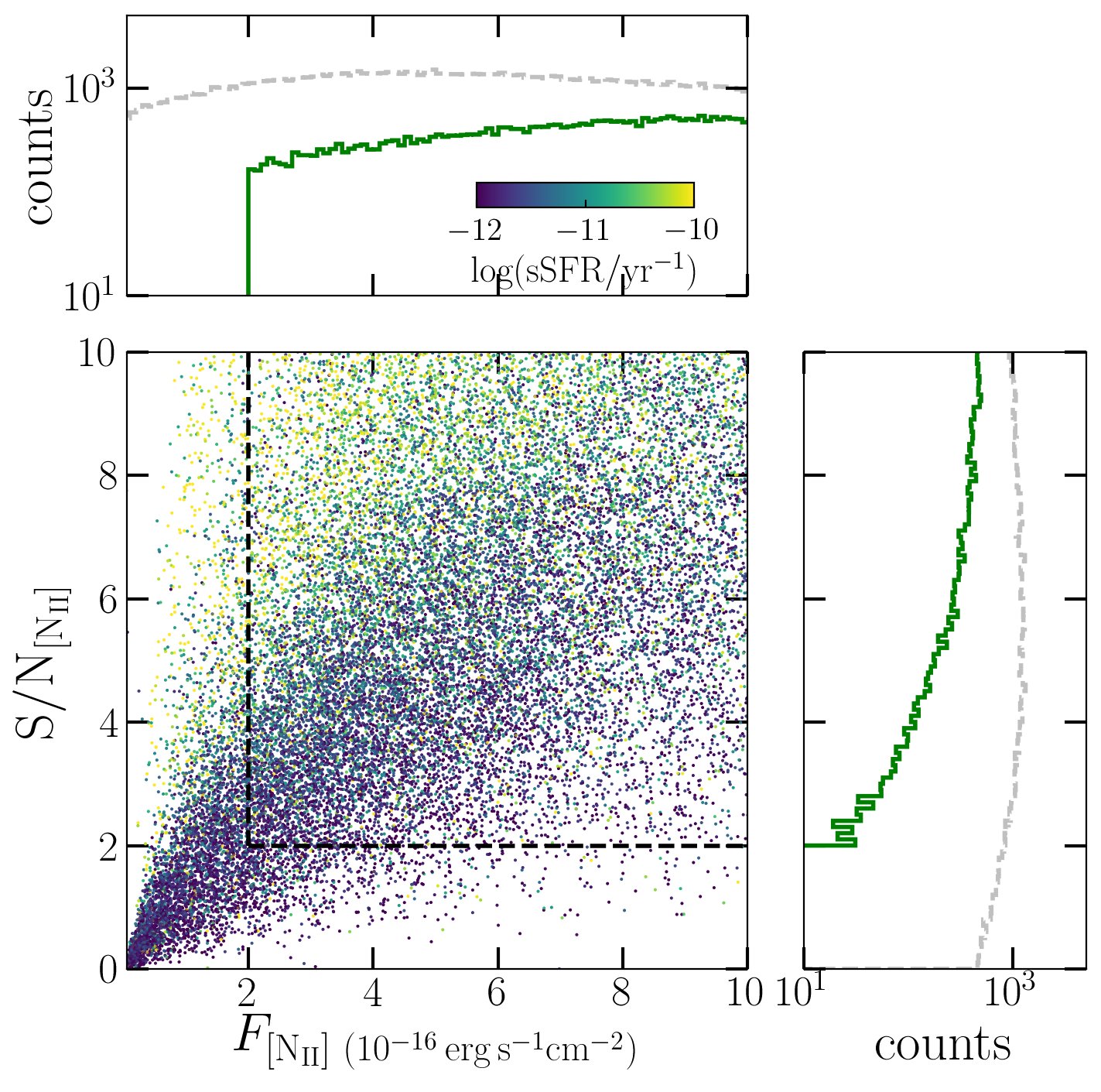}\quad
 \includegraphics[width=0.31\linewidth]{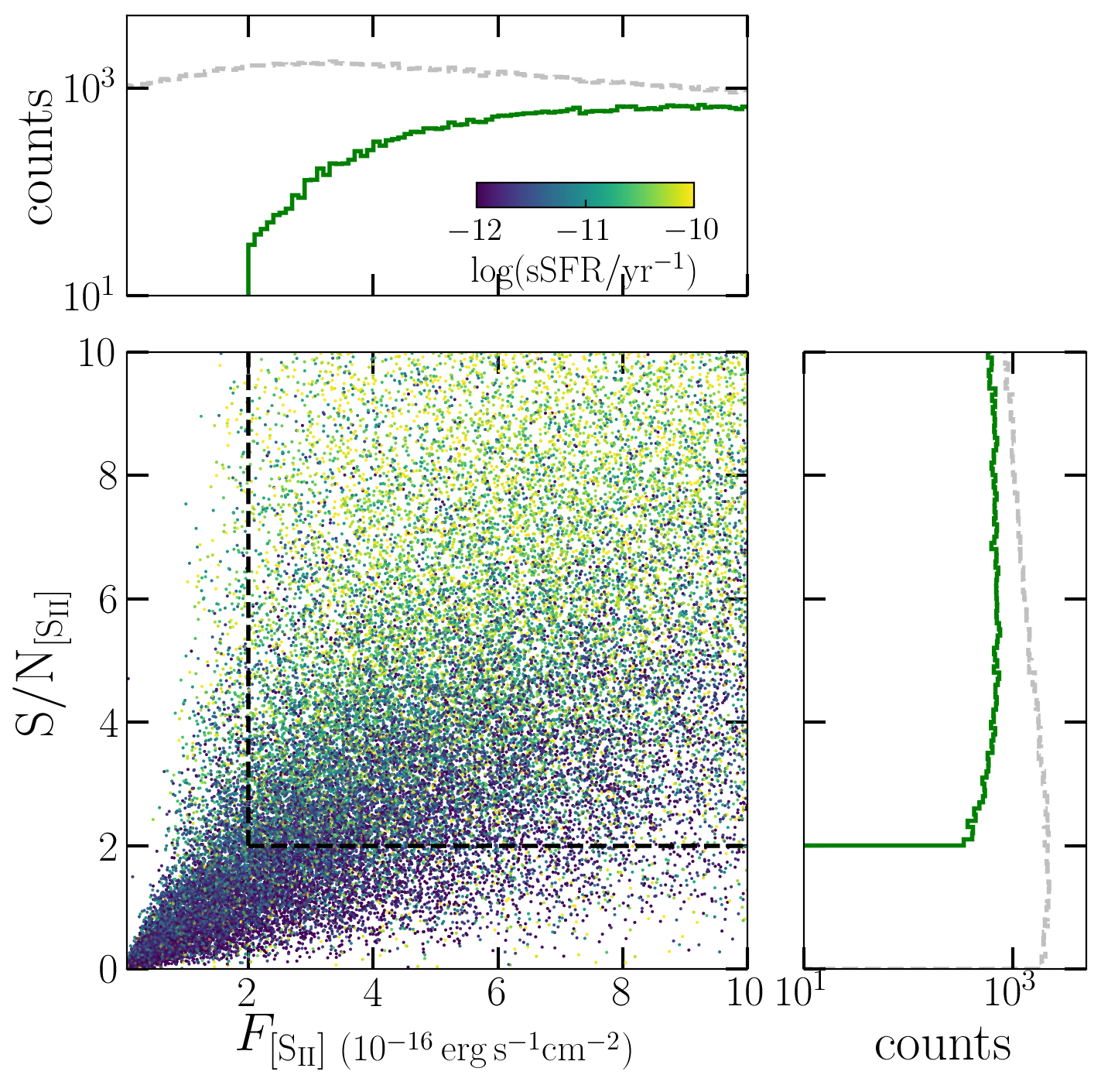}\quad
\caption{Same result as Fig.\,\ref{fig:megacorte} but for the rest of the lines of interest, color-coded by sSFR. From top to bottom and from left to right we show the \Hb, \OII, \OIII, \NII and \SII lines.}
  \label{fig:megacortealllines}
  \end{figure*}

In Fig.\,\ref{fig:megacortealllines} below we show the impact of the line flux and SN selection cuts in all six lines of interest, color-coded by sSFR (upper 6 panels) and EW (lower 6 panels). The results of the different lines are overall consistent, with \OII spanning larger EW values compared to the rest of the lines.
\begin{figure*}
\centering \hspace{-0.8cm}
    \includegraphics[width=0.31\linewidth]{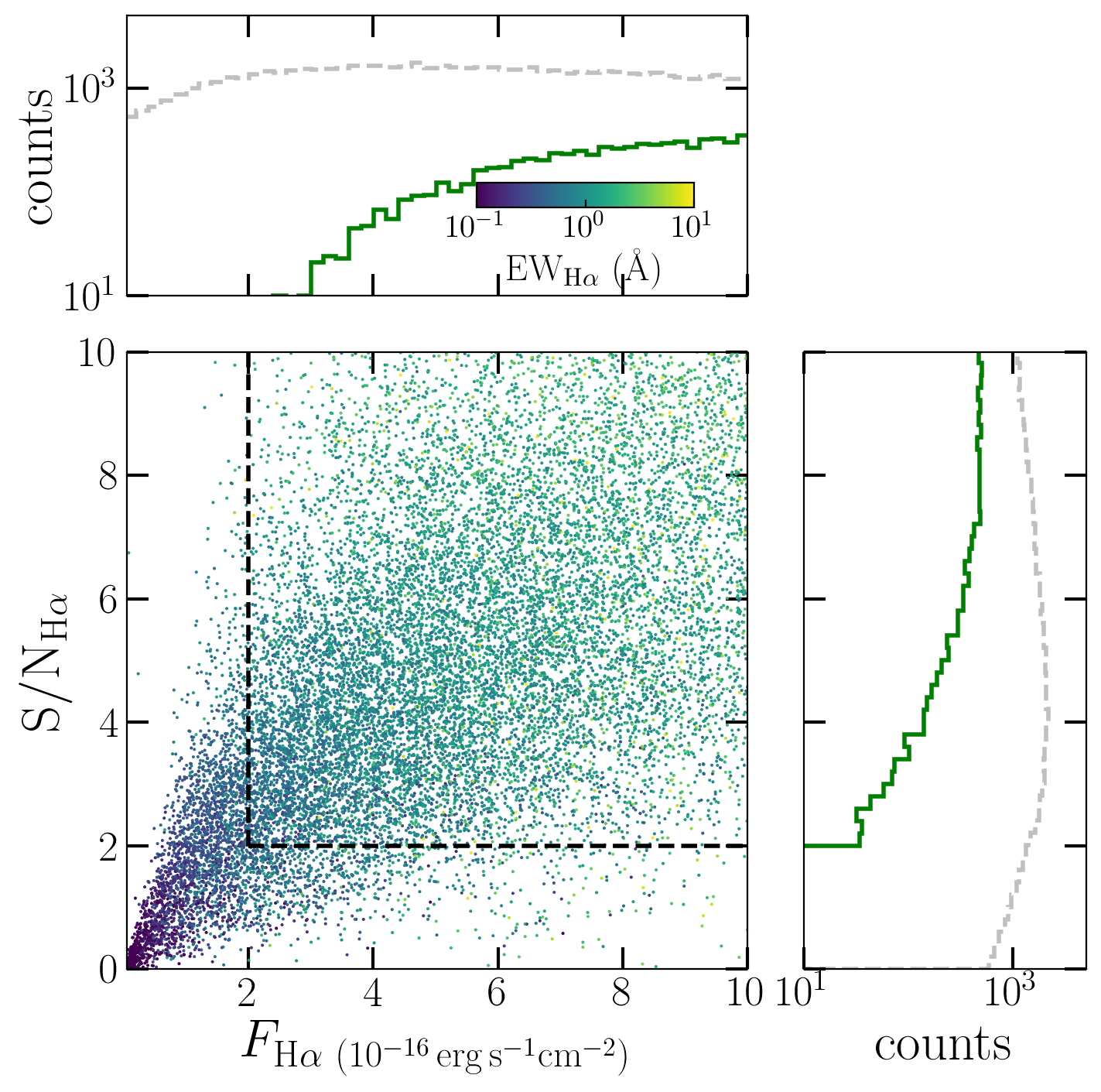}\quad
    \includegraphics[width=0.31\linewidth]{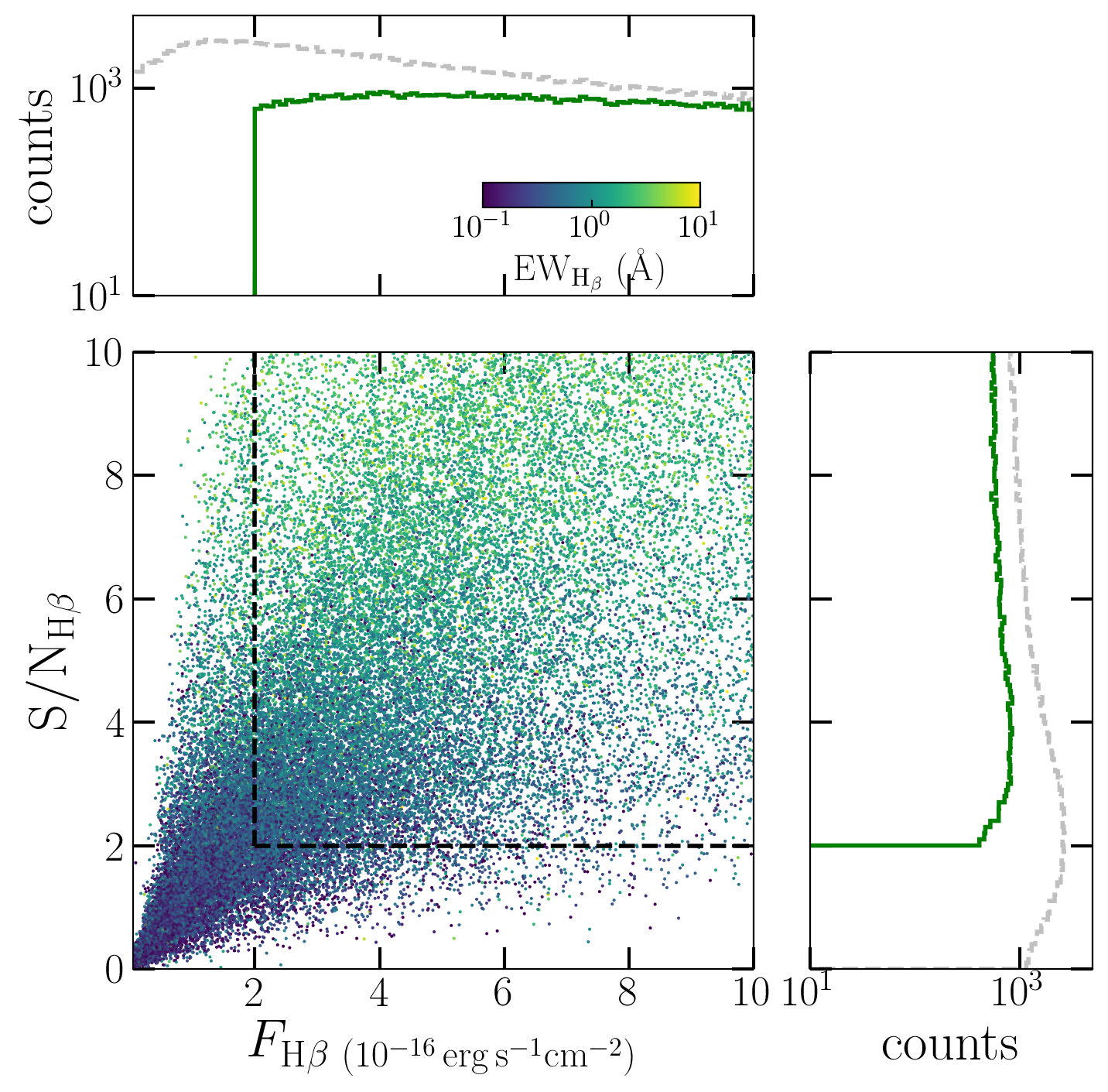}\quad
  \includegraphics[width=0.31\linewidth]{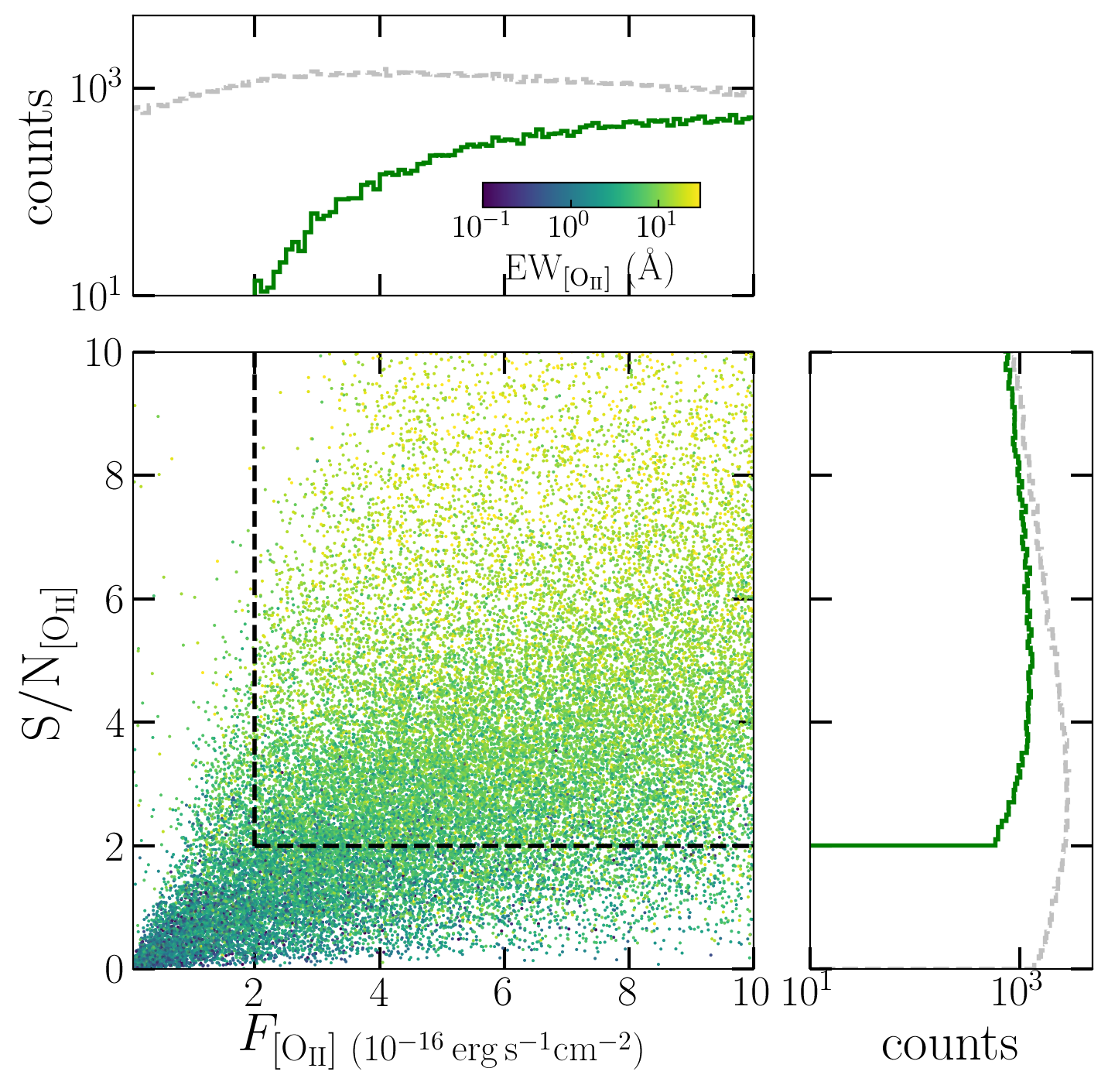}
  \includegraphics[width=0.31\linewidth]{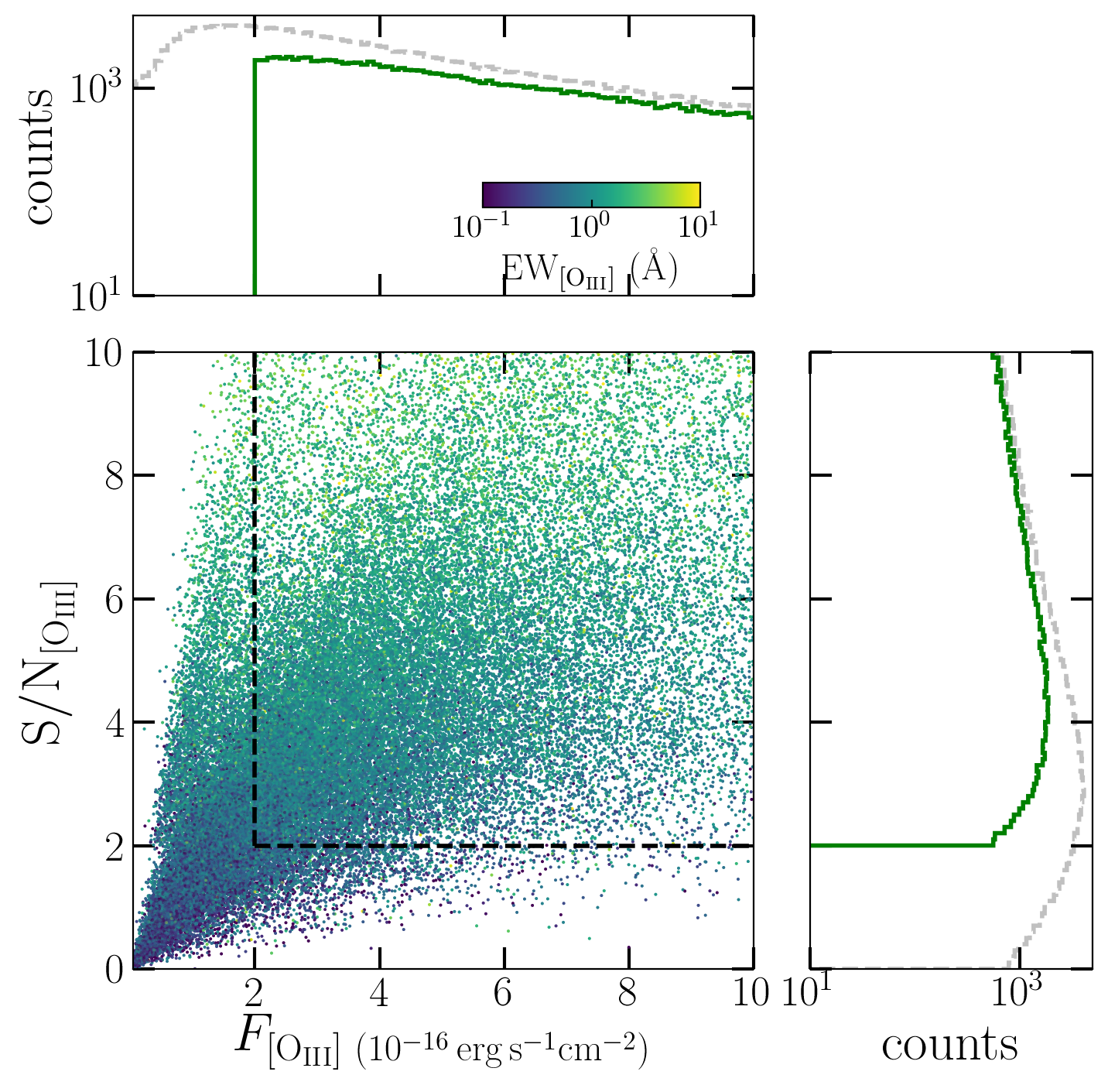}\quad
  \includegraphics[width=0.31\linewidth]{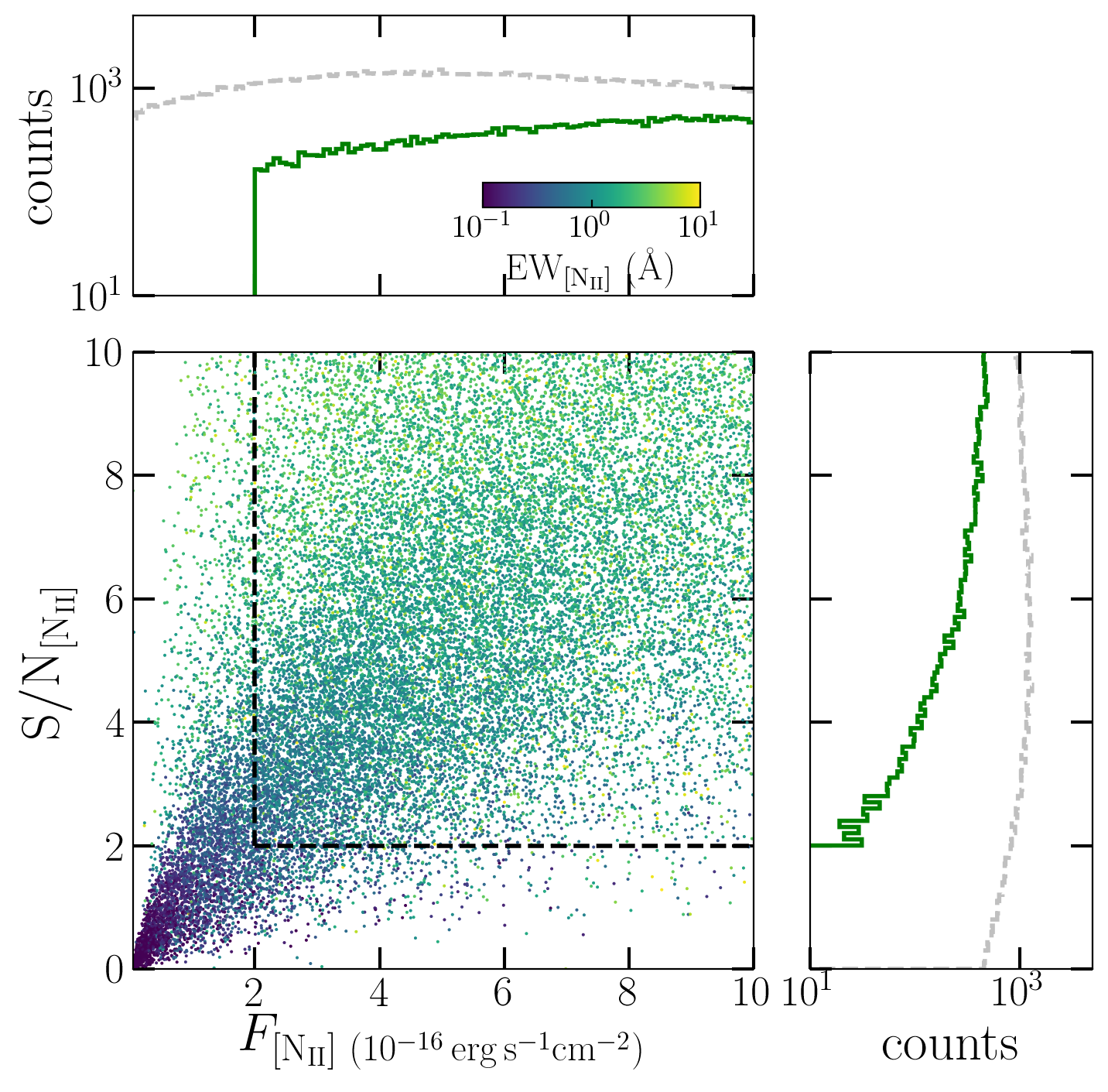}\quad
 \includegraphics[width=0.31\linewidth]{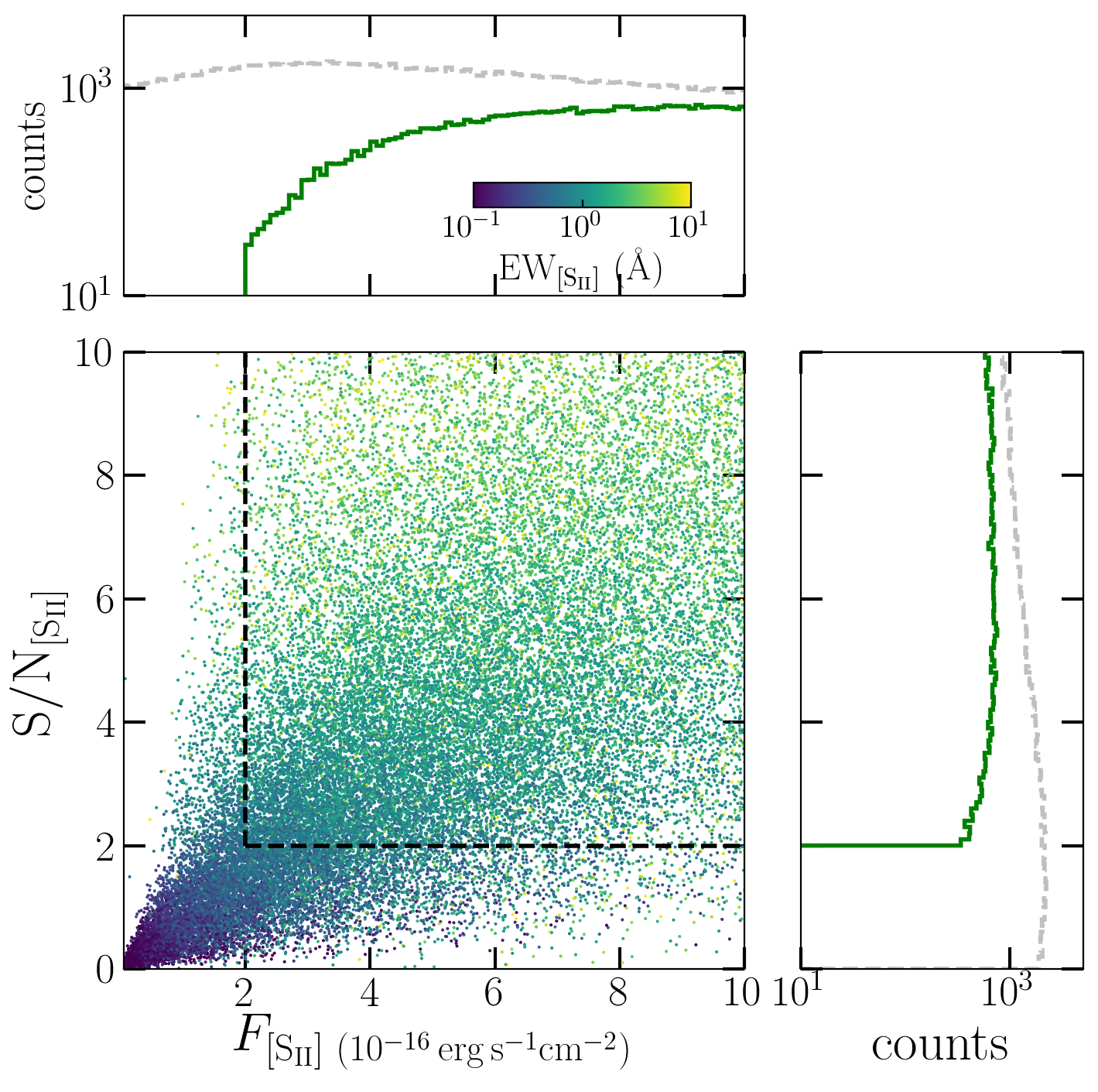}\quad
\caption{Same result as Fig.\,\ref{fig:megacortealllines} but color-coded by EW.}
  \label{fig:megacortealllinesEW}
  \end{figure*}
  
\begin{figure*}
\centering \hspace{-0.8cm}
     \includegraphics[width=0.42\linewidth]{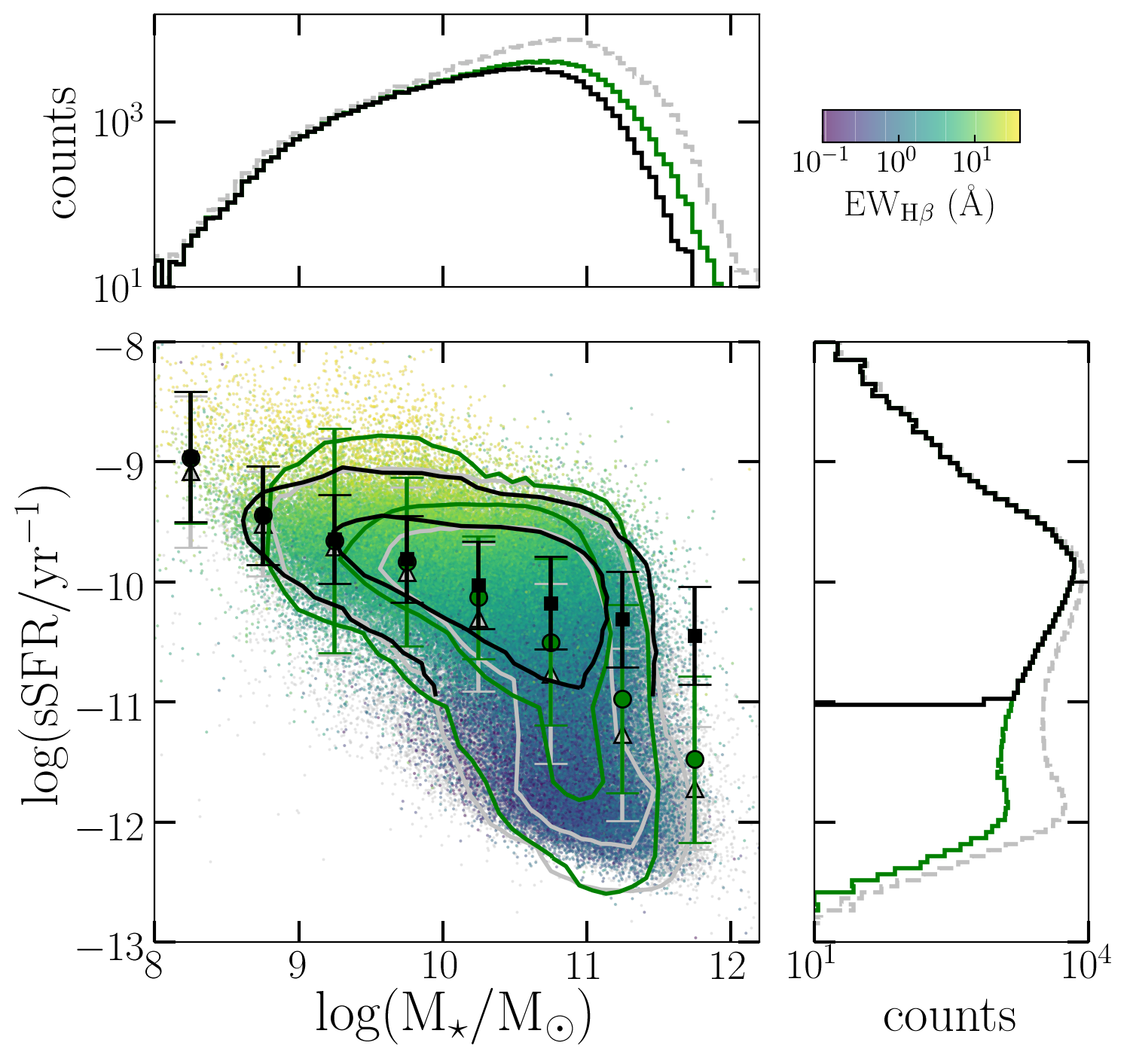}\hfill\vspace{0.5cm}
   \includegraphics[width=0.42\linewidth]{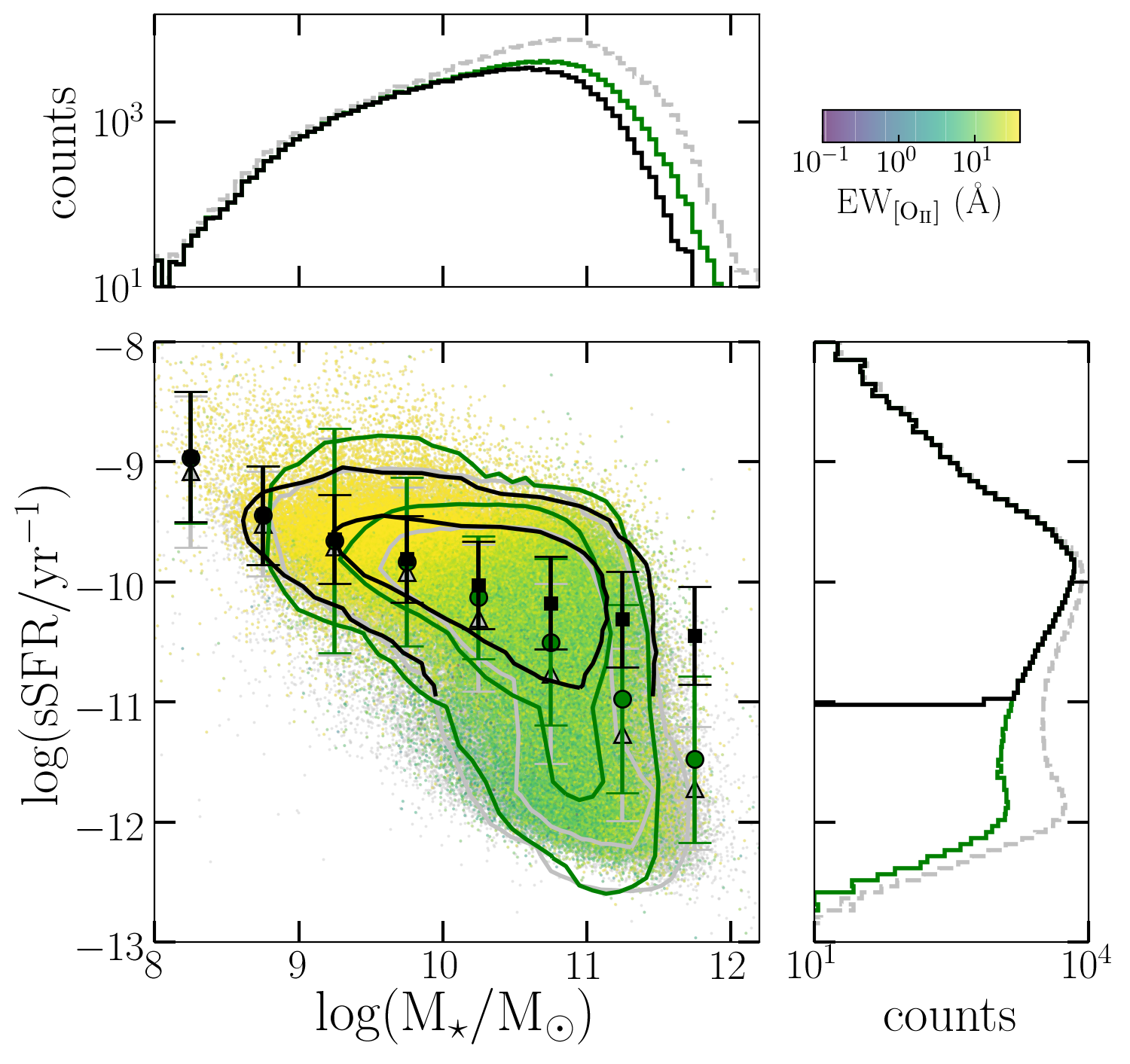}\hfill\quad
  \includegraphics[width=0.42\linewidth]{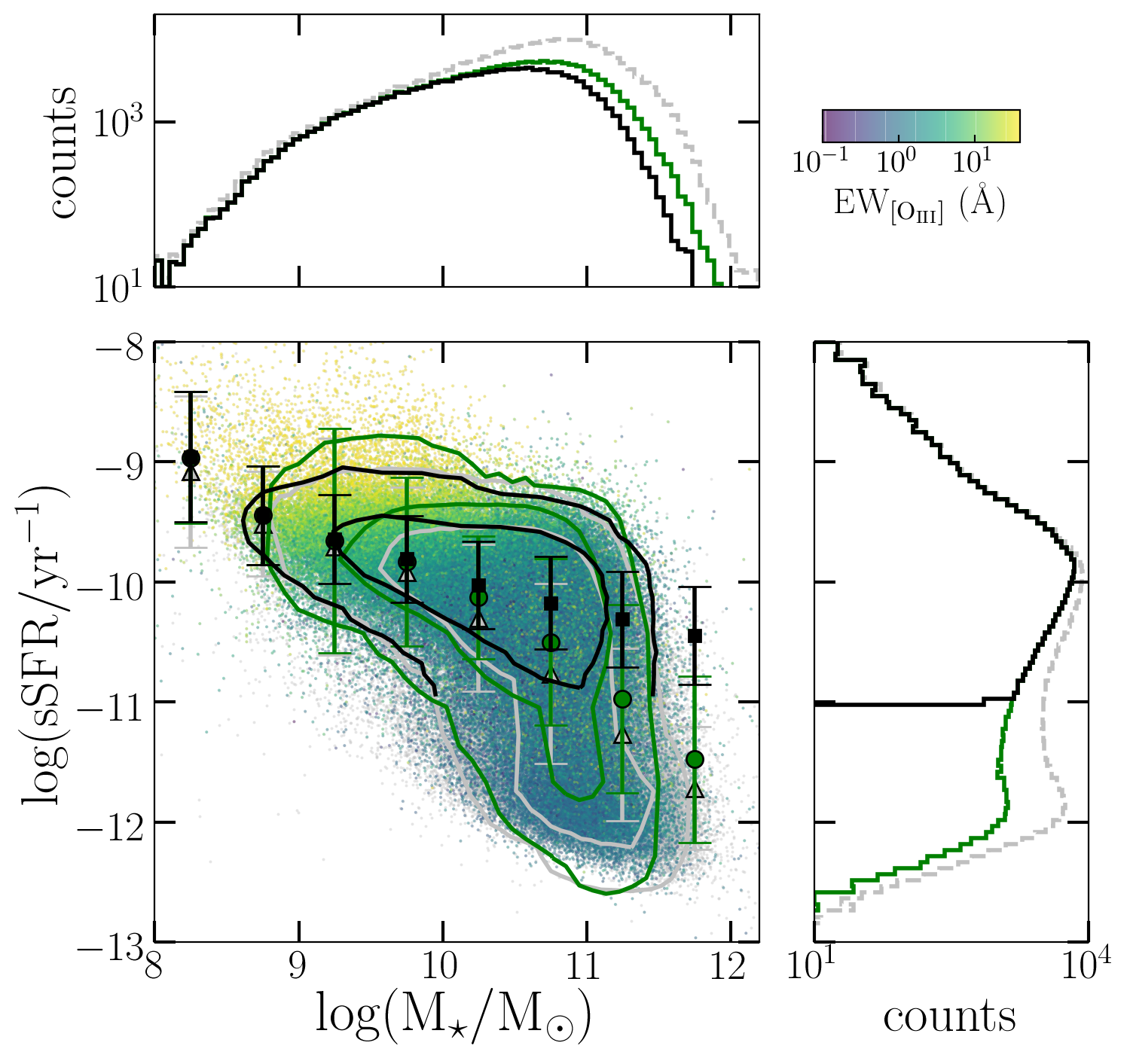}\hfill\vspace{0.5cm}
   \includegraphics[width=0.42\linewidth]{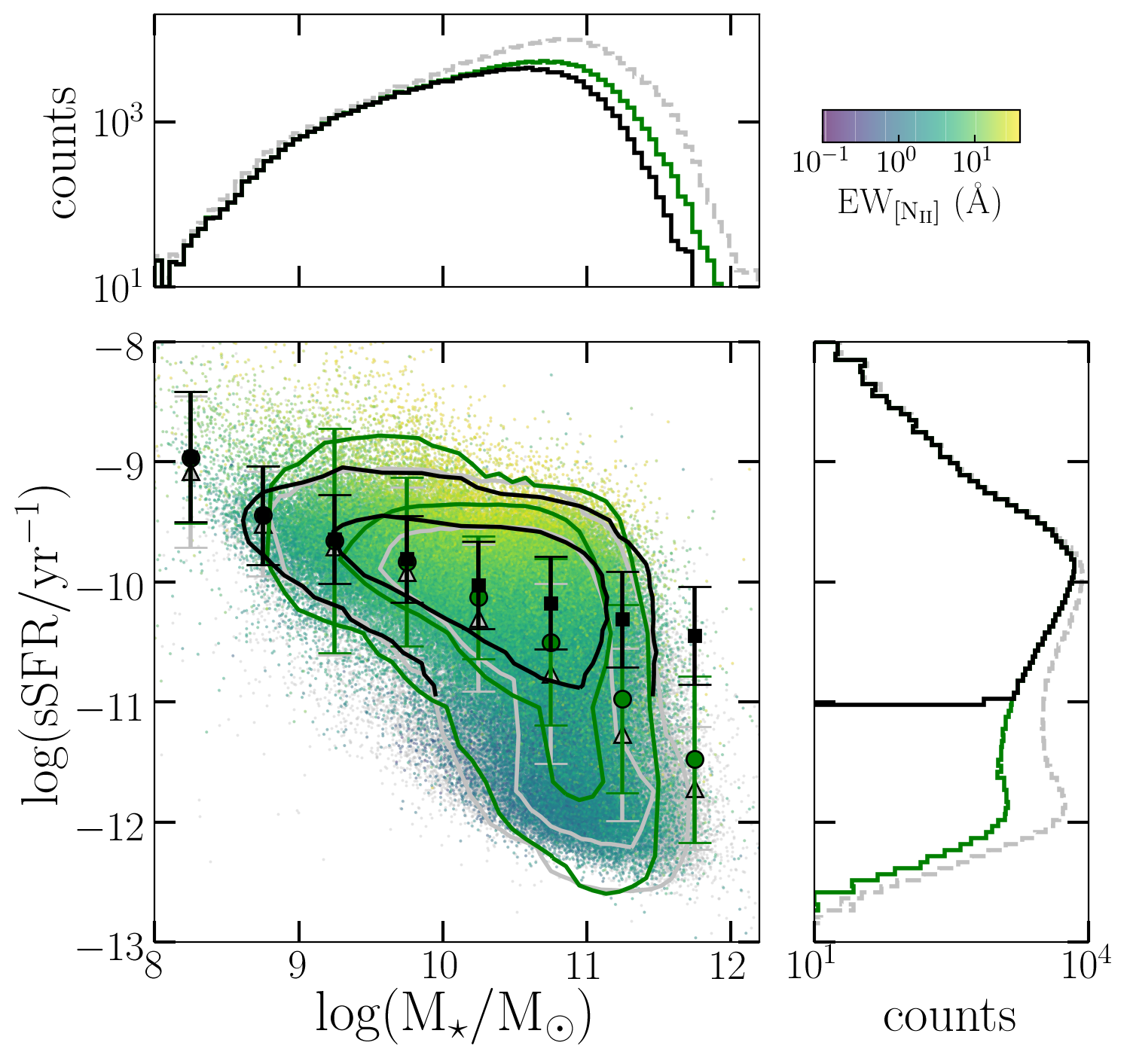}\hfill\quad
   \includegraphics[width=0.42\linewidth]{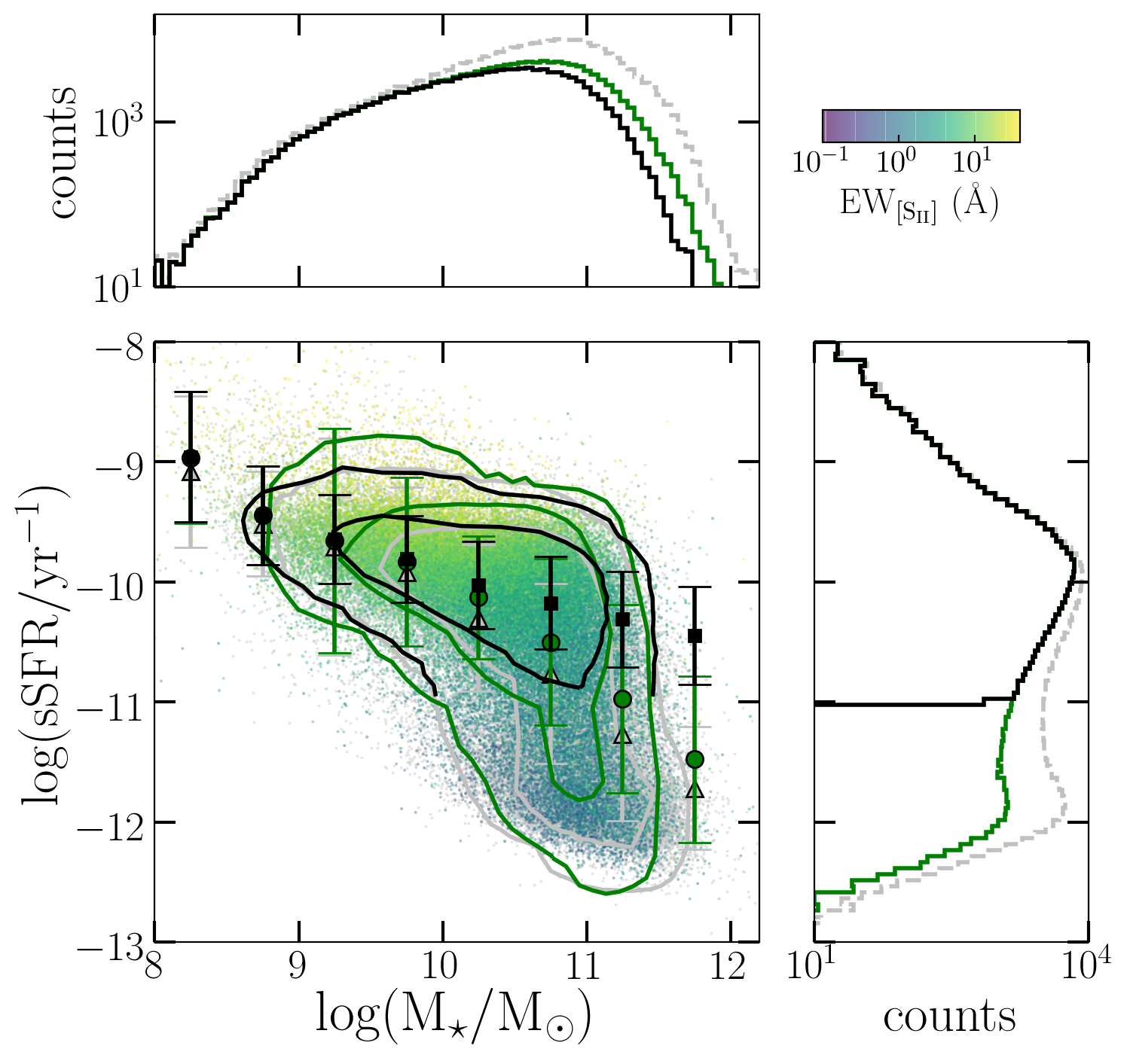}\vspace{0.5cm}
  \caption{Same result as shown in the left panel of Fig.\,\ref{fig:charplot} for the other lines of interest. From top to bottom and from left to right we show the \Hb, \OII, \OIII, \NII and \SII lines.}
  \label{fig:charplotallines}
  \end{figure*}

  Fig.\,\ref{fig:charplotallines} displays the \mainsel sSFR as a function of stellar mass color-coded by EW for the six lines of interest. The contours change in each panel as they are weighted by the Ew of each line. Overall the results are all consistent. The \OII line is the one showing higher EW values, while the \OIII and \Hb EW are more concentrated toward smaller values.

  \begin{figure*}
\centering 
   \includegraphics[width=0.4\linewidth]{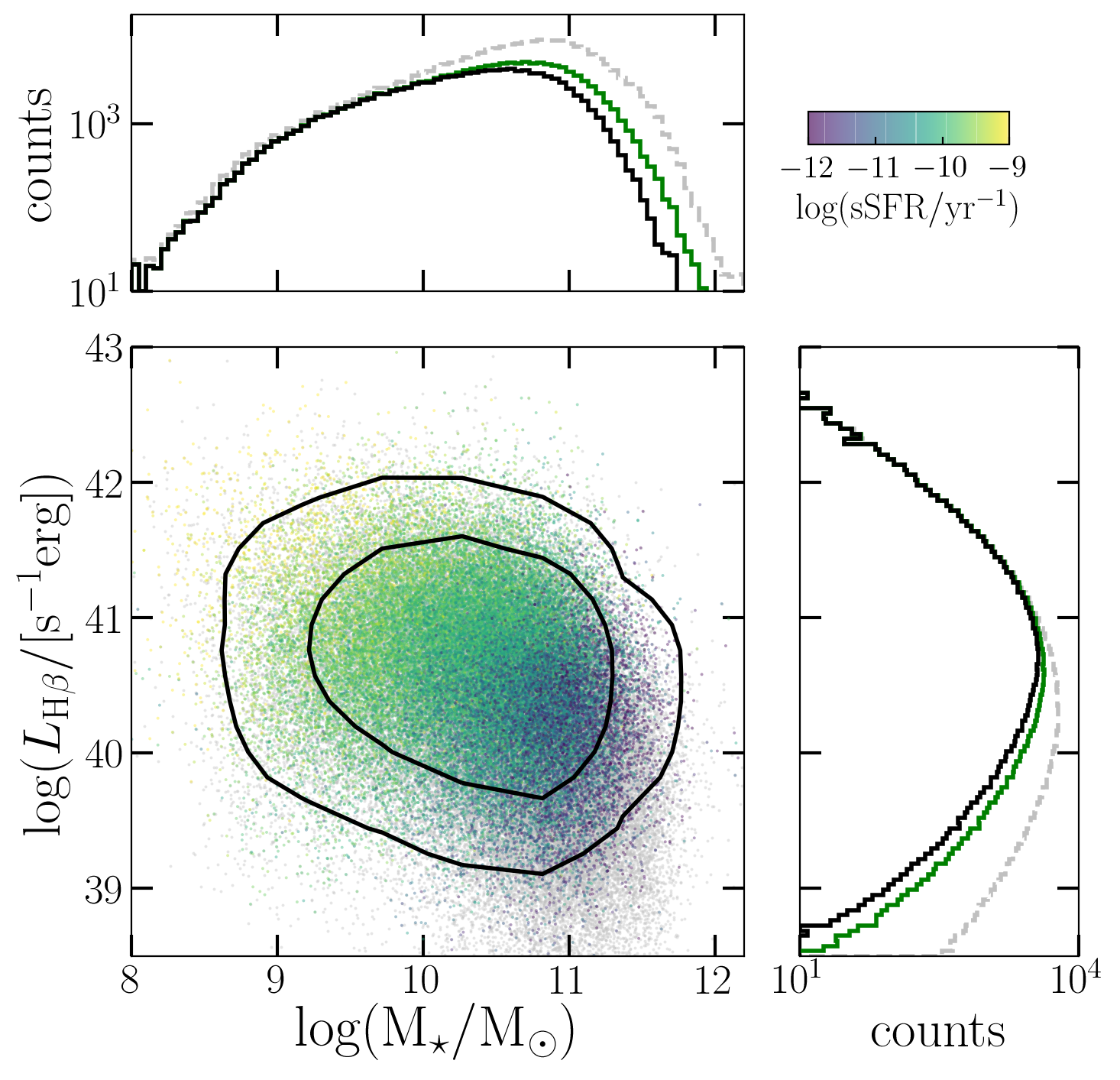}\hfill\vspace{0.5cm}
  \includegraphics[width=0.4\linewidth]{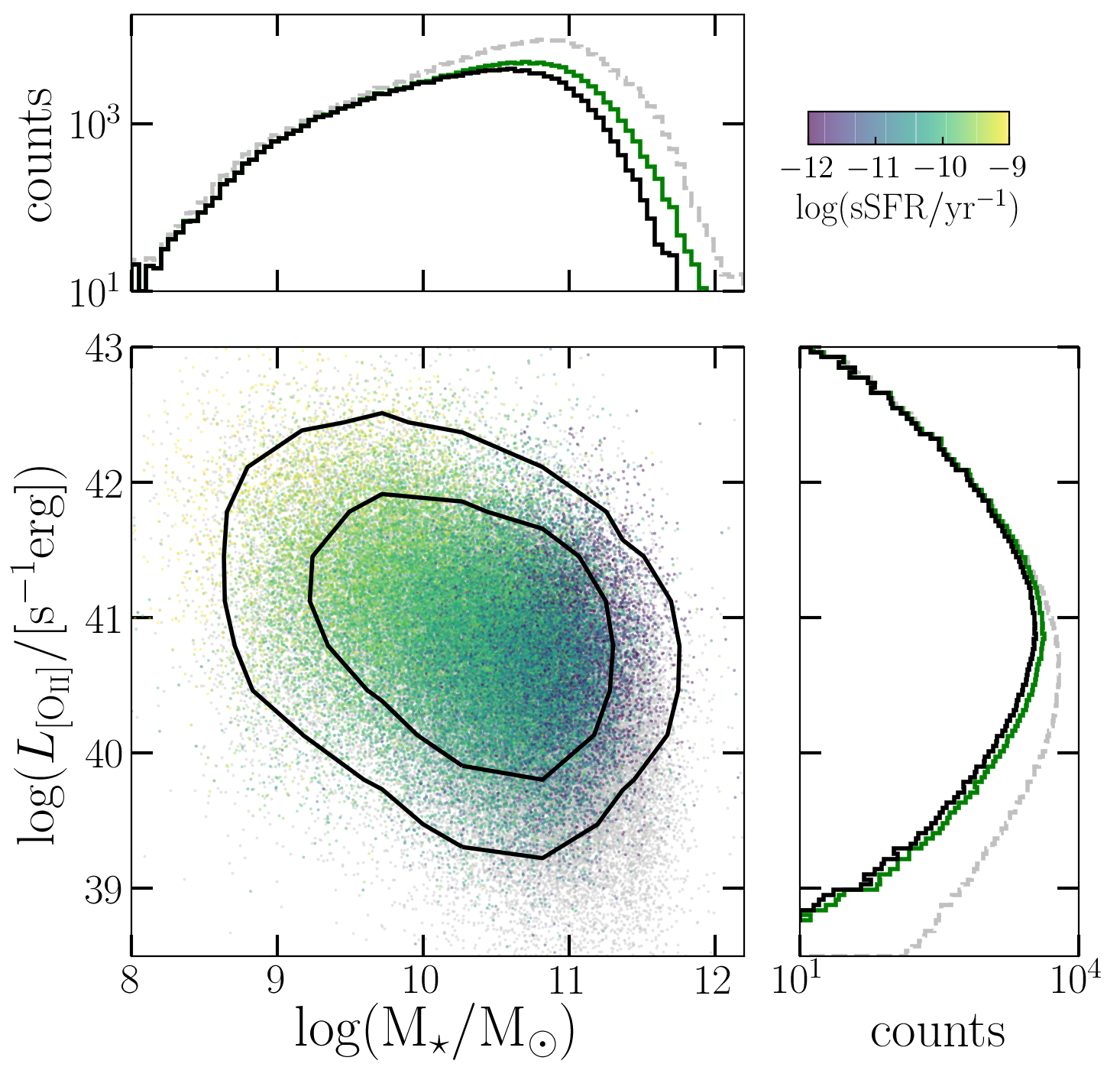}\hfill\quad
   \includegraphics[width=0.4\linewidth]{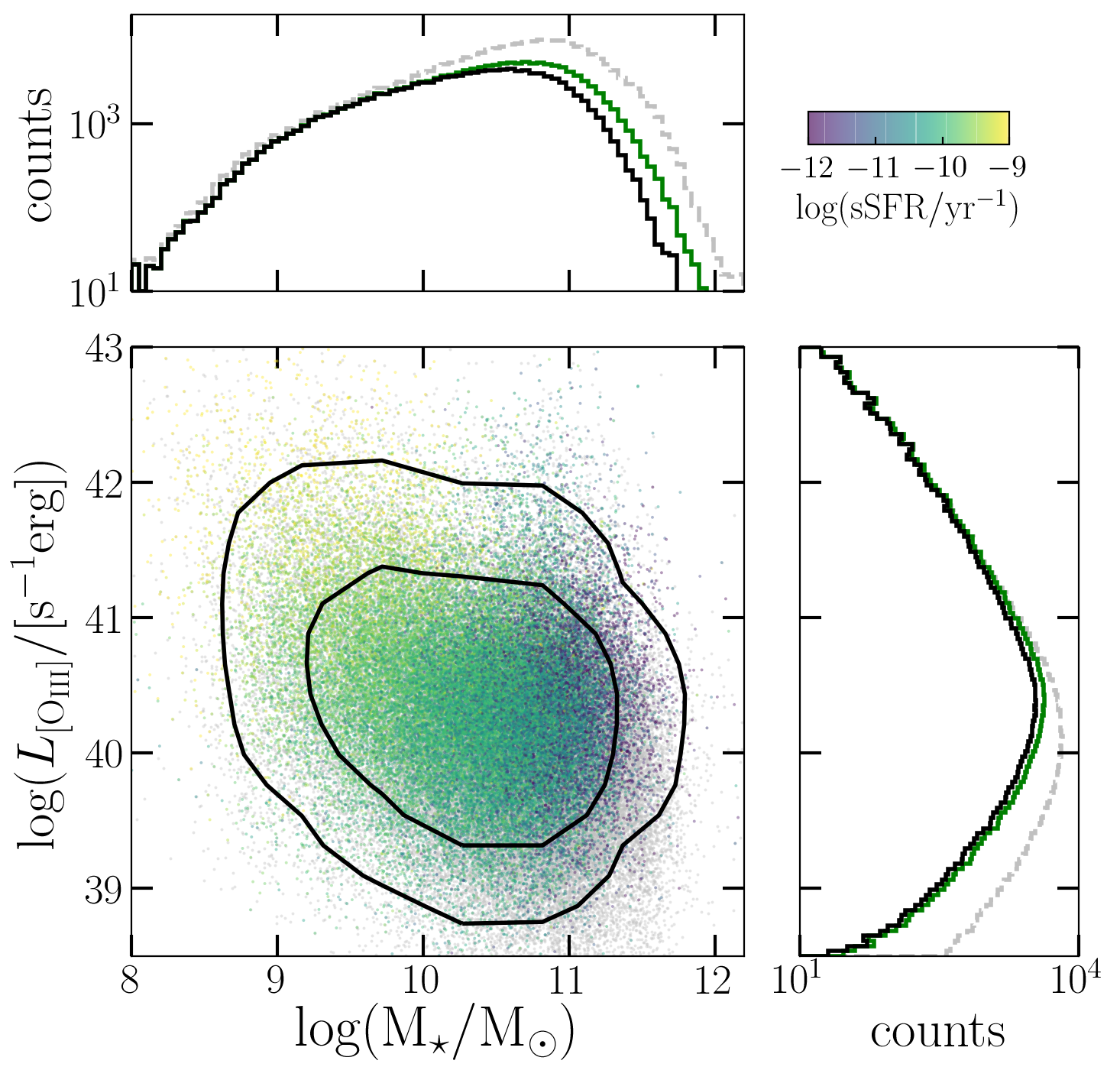}\hfill\vspace{0.5cm}
   \includegraphics[width=0.4\linewidth]{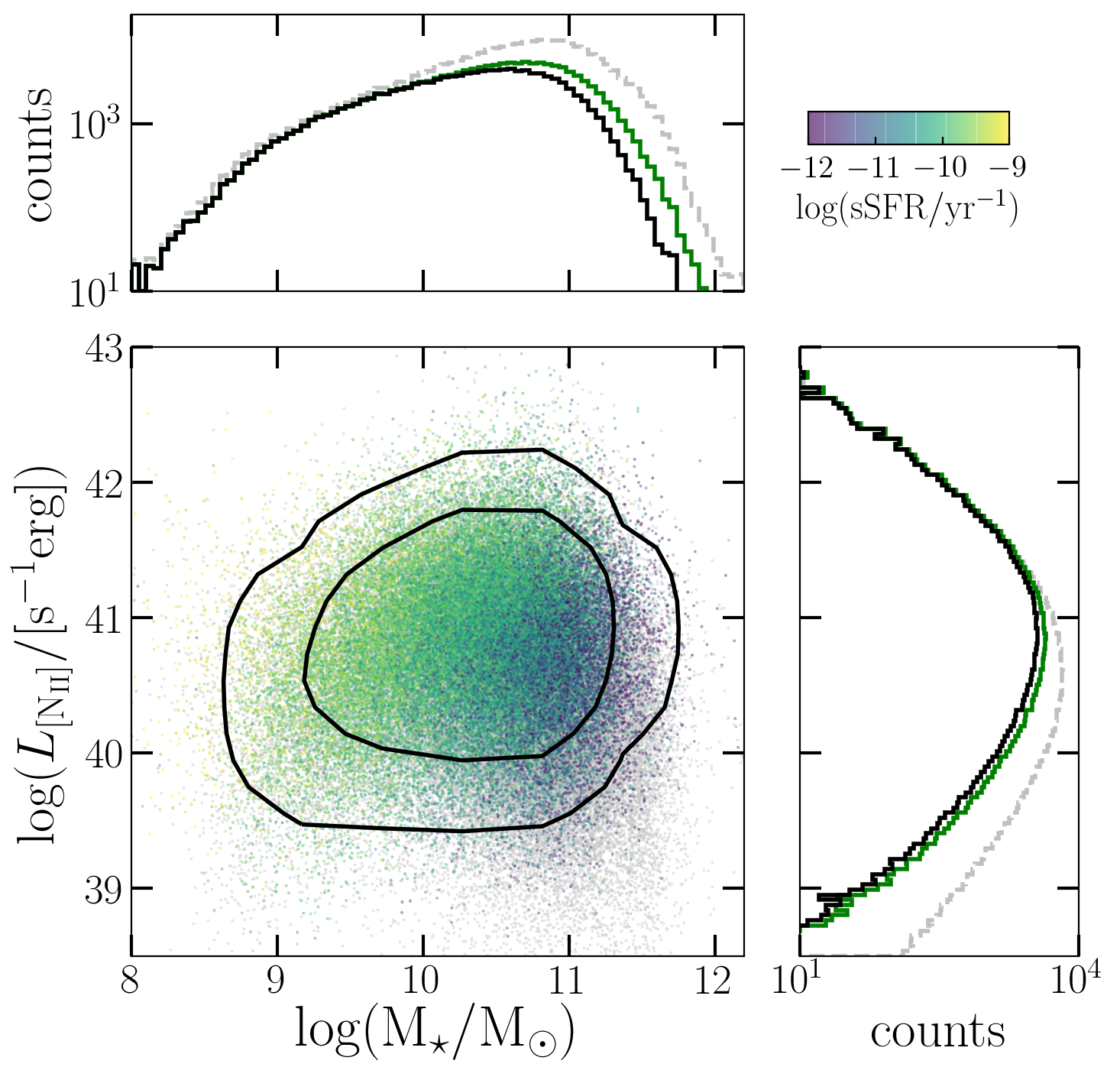}\hfill\quad
   \includegraphics[width=0.4\linewidth]{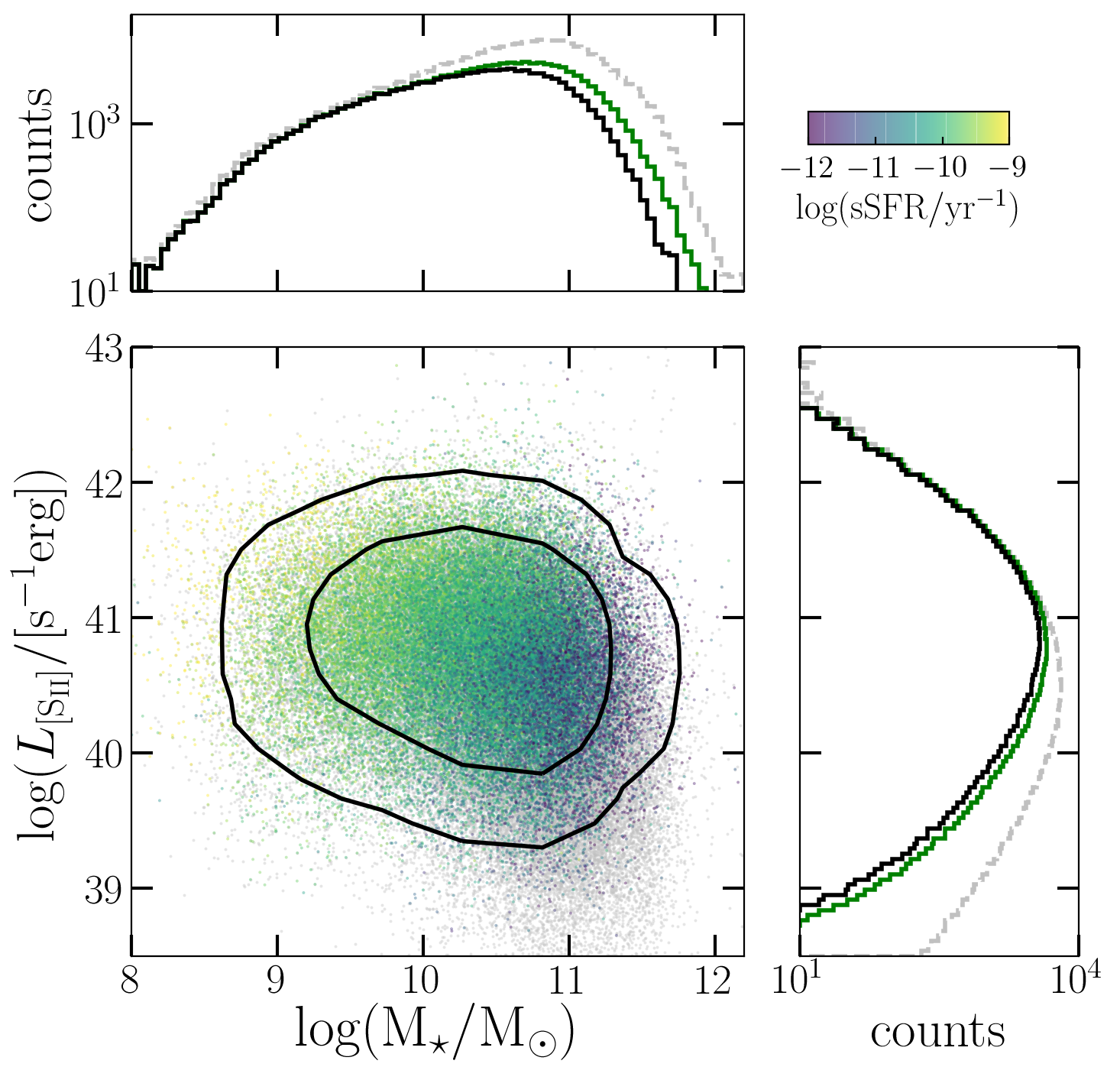}\vspace{0.5cm}
  \caption{Same result as shown in the right panel of Fig.\,\ref{fig:charplot} for the rest of the lines. From top to bottom and from left to right we show the \Hb, \OII, \OIII, \NII and \SII lines.}
  \label{fig:lummassallines}
  \end{figure*}
  
  \begin{figure*}
\centering 
  \includegraphics[width=1\linewidth]{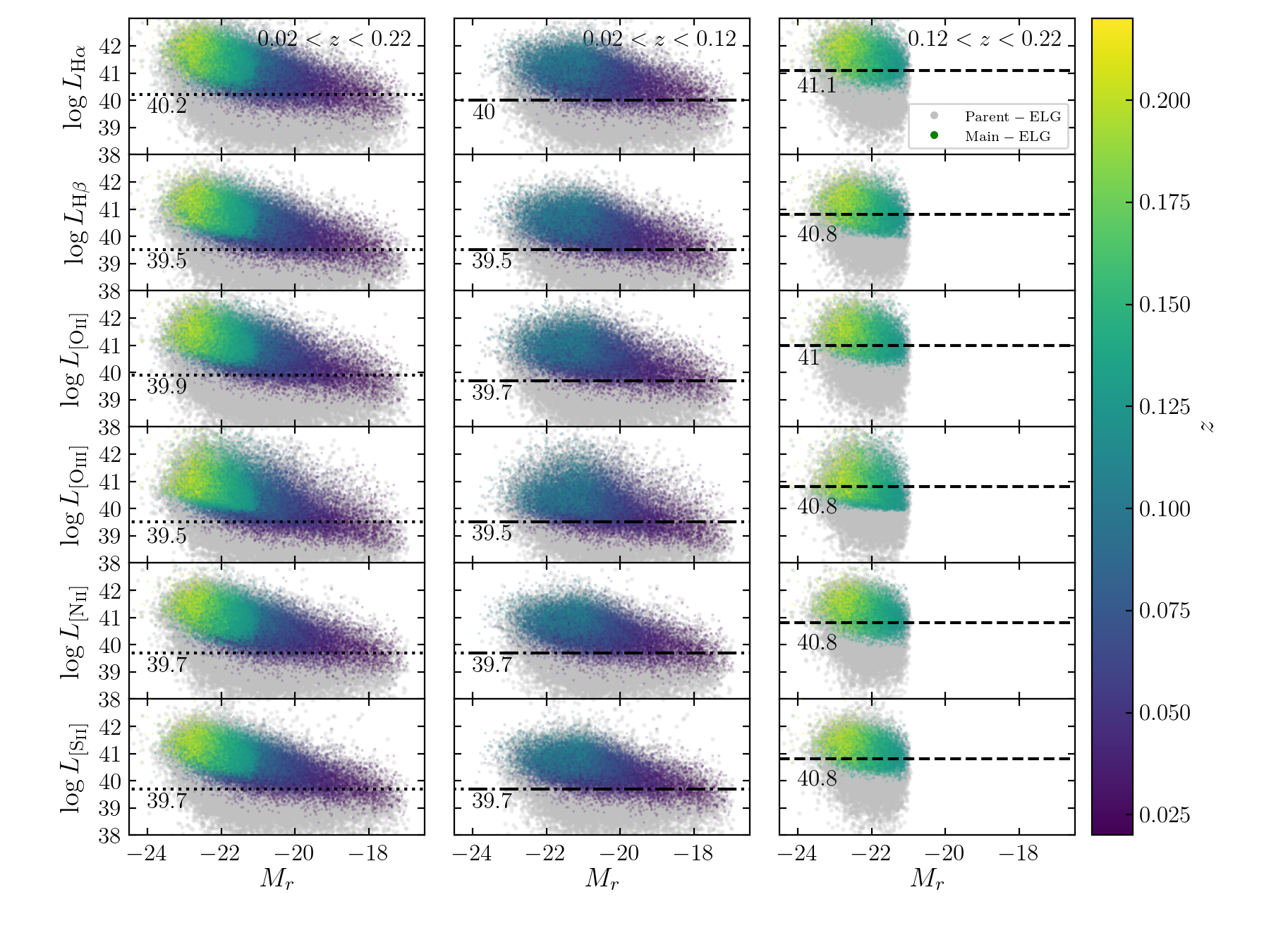}
  \caption{Same result as Fig.\,\ref{fig:LHaMr} for all the six lines of interest. The horizontal lines indicate the luminosity completeness threshold we establish by eye as the $L$ value where the distribution starts to degrade (see Sec.\,\ref{sec:incompleteness}).}
  \label{fig:LMralllines}
  \end{figure*}
  
In Fig.\,\ref{fig:LMralllines} we show the emission line luminosity, in the six lines of interest, as a function of the $r-$band absolute magnitude, color-coded by redshift. From left to right we show our result in three redshift bins to better analyze their evolution: the full sample at $0.02<z<0.22$, the lower $0.02<z<0.12$, and the upper $0.12<z<0.22$ bins. We overplot as horizontal lines the completeness limits chosen by eye as the luminosity below which the galaxy number density falls significantly. This threshold changes for each one of the six emission lines as a function of redshift. These thresholds for the full, low-$z$ and high-$z$ samples are summarized in Table\,\ref{tab:complimitsall}.

\begin{table}
    \centering
    \setlength{\tabcolsep}{3pt}
    \begin{tabular}{|c|c|c|c|}
    \hline
     &&  &  \\ 
    line & $0.02<z<0.22$&$0.02<z<0.12$ & $0.12<z<0.22$\\
    \hline
    \Ha& $10^{40.2}$ & $10^{40}$&$10^{41}$ \\
    \Hb& $10^{39.5}$ & $10^{39.5}$&$10^{40.8}$ \\
    \OII& $10^{39.9}$ & $10^{39.7}$&$10^{41}$ \\
    \OIII& $10^{39.5}$ & $10^{39.5}$&$10^{40.8}$ \\
    \NII& $10^{39.7}$ & $10^{39.7}$&$10^{40.8}$\\
    \SII& $10^{39.7}$ & $10^{39.7}$&$10^{40.8}$ \\
    \hline
    \end{tabular}
    \caption{Emission-line luminosity completeness limits, in units of $\rm{s^{-1}\,erg}$, for the full (left column), low-$z$ (center) and high-$z$ (right) samples.}
    \label{tab:complimitsall}
\end{table}
This result tells us that, when selecting ELGs by cutting in line flux (i.e., in luminosity), we are implicitly removing a fraction of the sample fainter than a given $r-$band magnitude, meaning that we are making our sample incomplete in $M_r$. Our $1/V_{\rm max}$ LF estimator is not able to correct from this incompleteness effect.


\section{{\em Main-ELG} luminosity function values}
\label{sec:appendixLF}
The numerical values of the \Ha, \Hb, \OII, \OIII, \NII, and \SII\,\mainsel luminosity functions and its different components are provided in Tables\,\ref{tab:LFtable1} - \ref{tab:LFtable3}.

 \begin{table*}
  \setlength{\tabcolsep}{2pt}
  \centering
 \begin{tabular}{c c c c c c c c}
    \toprule
    &\multicolumn{7}{c}{observed $\log(\Phi(L_{\rm H\alpha]}/\rm{Mpc^{-3}\,dex^{-1}})$)}\\
$\log{L_{\rm H\alpha}}$&Full sample&SF sSFR&SF BPT+WHAN&LINERs&Composite&Seyferts&Passive\\ 
\midrule
39.9&$-2.180\pm0.071$&$-2.300\pm0.085$&$-2.409\pm0.090$&$-3.133\pm0.043$&$-2.968\pm0.152$&$-3.985\pm0.107$&$-5.598\pm0.114$\\
40.1&$-2.233\pm0.049$&$-2.315\pm0.053$&$-2.475\pm0.052$&$-3.377\pm0.046$&$-2.833\pm0.088$&$-3.845\pm0.118$&$-5.689\pm0.124$\\
40.3&$-2.283\pm0.041$&$-2.358\pm0.042$&$-2.463\pm0.042$&$-3.526\pm0.040$&$-2.965\pm0.047$&$-3.801\pm0.138$&$-6.221\pm0.181$\\
40.5&$-2.350\pm0.097$&$-2.401\pm0.108$&$-2.542\pm0.149$&$-3.673\pm0.043$&$-2.995\pm0.052$&$-3.959\pm0.058$&$-6.665\pm0.127$\\
40.7&$-2.402\pm0.039$&$-2.441\pm0.040$&$-2.533\pm0.040$&$-3.878\pm0.039$&$-3.195\pm0.046$&$-4.010\pm0.087$&$-6.764\pm0.197$\\
40.9&$-2.557\pm0.038$&$-2.588\pm0.038$&$-2.690\pm0.039$&$-4.083\pm0.045$&$-3.332\pm0.043$&$-4.035\pm0.049$&$-7.159\pm0.371$\\
41.1&$-2.715\pm0.038$&$-2.738\pm0.038$&$-2.833\pm0.038$&$-4.348\pm0.030$&$-3.514\pm0.046$&$-4.254\pm0.051$&$-5.898\pm0.060$\\
41.3&$-2.898\pm0.039$&$-2.917\pm0.039$&$-2.999\pm0.040$&$-4.669\pm0.032$&$-3.766\pm0.039$&$-4.427\pm0.048$&$-$\\
41.5&$-3.095\pm0.038$&$-3.108\pm0.039$&$-3.192\pm0.039$&$-5.029\pm0.026$&$-3.945\pm0.053$&$-4.598\pm0.042$&$-$\\
41.7&$-3.350\pm0.036$&$-3.359\pm0.036$&$-3.445\pm0.036$&$-5.337\pm0.024$&$-4.265\pm0.039$&$-4.702\pm0.085$&$-$\\
41.9&$-3.598\pm0.036$&$-3.603\pm0.036$&$-3.679\pm0.037$&$-5.830\pm0.027$&$-4.526\pm0.038$&$-5.083\pm0.056$&$-$\\
42.1&$-3.899\pm0.038$&$-3.904\pm0.038$&$-3.984\pm0.038$&$-6.283\pm0.050$&$-4.866\pm0.051$&$-5.199\pm0.045$&$-$\\
42.3&$-4.265\pm0.039$&$-4.268\pm0.039$&$-4.340\pm0.039$&$-6.888\pm0.092$&$-5.216\pm0.063$&$-5.753\pm0.062$&$-$\\
42.5&$-4.646\pm0.038$&$-4.648\pm0.038$&$-4.729\pm0.038$&$-6.937\pm0.028$&$-5.542\pm0.056$&$-6.249\pm0.076$&$-$\\
42.7&$-5.037\pm0.048$&$-5.037\pm0.048$&$-5.132\pm0.046$&$-7.989\pm0.351$&$-5.938\pm0.124$&$-6.350\pm0.077$&$-$\\
42.9&$-5.530\pm0.053$&$-5.532\pm0.053$&$-5.614\pm0.050$&$-8.072\pm0.376$&$-6.464\pm0.178$&$-6.825\pm0.131$&$-$\\
43.1&$-5.988\pm0.054$&$-5.998\pm0.055$&$-6.059\pm0.055$&$-$&$-7.073\pm0.228$&$-7.772\pm0.030$&$-$\\
43.3&$-6.479\pm0.120$&$-6.479\pm0.120$&$-6.615\pm0.151$&$-$&$-7.088\pm0.211$&$-8.111\pm0.620$&$-$\\
43.5&$-7.138\pm0.259$&$-7.138\pm0.259$&$-7.196\pm0.258$&$-$&$-$&$-8.039\pm0.425$&$-$\\
  \end{tabular}
 \begin{tabular}{c c c c c c c c}
    \midrule
        &\multicolumn{7}{c}{observed $\log(\Phi(L_{\rm H\beta]}/\rm{Mpc^{-3}\,dex^{-1}})$)}\\
{$\log{L_{\rm H\beta}}$}&Full sample&SF sSFR&SF BPT+WHAN&LINERs&Composite&Seyfert&Passive\\ 
\midrule
39.9&$-2.360\pm0.102$&$-2.426\pm0.116$&$-2.549\pm0.152$&$-3.590\pm0.041$&$-3.053\pm0.056$&$-3.959\pm0.065$&$-5.510\pm0.117$\\
40.1&$-2.406\pm0.038$&$-2.453\pm0.038$&$-2.541\pm0.040$&$-3.762\pm0.042$&$-3.215\pm0.045$&$-4.053\pm0.081$&$-5.959\pm0.204$\\
40.3&$-2.562\pm0.039$&$-2.601\pm0.039$&$-2.696\pm0.039$&$-3.988\pm0.036$&$-3.369\pm0.046$&$-4.041\pm0.053$&$-6.667\pm0.159$\\
40.5&$-2.707\pm0.038$&$-2.738\pm0.039$&$-2.822\pm0.039$&$-4.223\pm0.031$&$-3.547\pm0.048$&$-4.291\pm0.054$&$-6.803\pm0.123$\\
40.7&$-2.894\pm0.038$&$-2.921\pm0.038$&$-3.000\pm0.038$&$-4.575\pm0.035$&$-3.743\pm0.049$&$-4.447\pm0.049$&$-5.894\pm0.060$\\
40.9&$-3.101\pm0.038$&$-3.117\pm0.038$&$-3.181\pm0.039$&$-4.957\pm0.034$&$-4.067\pm0.038$&$-4.673\pm0.054$&$-$\\
41.1&$-3.341\pm0.037$&$-3.352\pm0.037$&$-3.421\pm0.038$&$-5.246\pm0.028$&$-4.321\pm0.041$&$-4.809\pm0.049$&$-$\\
41.3&$-3.588\pm0.037$&$-3.595\pm0.037$&$-3.656\pm0.037$&$-5.793\pm0.038$&$-4.625\pm0.040$&$-5.101\pm0.062$&$-$\\
41.5&$-3.892\pm0.038$&$-3.898\pm0.038$&$-3.962\pm0.038$&$-6.376\pm0.071$&$-4.941\pm0.054$&$-5.259\pm0.048$&$-$\\
41.7&$-4.221\pm0.040$&$-4.225\pm0.040$&$-4.281\pm0.039$&$-6.916\pm0.093$&$-5.310\pm0.085$&$-5.748\pm0.070$&$-$\\
41.9&$-4.619\pm0.041$&$-4.624\pm0.041$&$-4.689\pm0.041$&$-6.836\pm0.044$&$-5.653\pm0.063$&$-6.122\pm0.082$&$-$\\
42.1&$-4.993\pm0.048$&$-4.994\pm0.048$&$-5.068\pm0.047$&$-7.615\pm0.149$&$-6.006\pm0.152$&$-6.405\pm0.099$&$-$\\
42.3&$-5.437\pm0.054$&$-5.441\pm0.054$&$-5.489\pm0.054$&$-$&$-6.748\pm0.176$&$-6.785\pm0.127$&$-$\\
42.5&$-5.975\pm0.091$&$-5.975\pm0.091$&$-6.017\pm0.089$&$-$&$-7.165\pm0.262$&$-8.069\pm0.061$&$-$\\
42.7&$-6.307\pm0.087$&$-6.307\pm0.087$&$-6.392\pm0.098$&$-$&$-7.203\pm0.270$&$-7.598\pm0.241$&$-$\\
42.9&$-6.929\pm0.147$&$-6.929\pm0.147$&$-6.929\pm0.147$&$-$&$-$&$-$&$-$\\
43.1&$-7.756\pm0.338$&$-7.756\pm0.338$&$-7.756\pm0.338$&$-$&$-$&$-$&$-$\\
43.3&$-$&$-$&$-$&$-$&$-$&$-$&$-$\\
43.5&$-$&$-$&$-$&$-$&$-$&$-$&$-$\\
  \end{tabular}
    \caption{\Ha and \Hb luminosity functions of the \mainsel sample and their different components. } 
 \label{tab:LFtable1}
\end{table*}

\begin{table*}
\setlength{\tabcolsep}{2pt}
  \centering
 \begin{tabular}{c c c c c c c c}
    \toprule
       &\multicolumn{7}{c}{observed $\log(\Phi(L_{\rm [O_{II}]]}/\rm{Mpc^{-3}\,dex^{-1}})$)}\\
$\log{L_{\rm [O_{II}]}}$&Full sample&SF sSFR&SF BPT+WHAN&LINERs&Composite&Seyferts&Passive\\ 
\hline
39.9&$-2.286\pm0.084$&$-2.418\pm0.108$&$-2.605\pm0.163$&$-3.160\pm0.062$&$-2.870\pm0.047$&$-4.015\pm0.078$&$-5.375\pm0.185$\\
40.1&$-2.313\pm0.039$&$-2.427\pm0.039$&$-2.549\pm0.040$&$-3.190\pm0.043$&$-3.078\pm0.054$&$-3.898\pm0.085$&$-5.813\pm0.261$\\
40.3&$-2.419\pm0.041$&$-2.518\pm0.041$&$-2.600\pm0.041$&$-3.384\pm0.073$&$-3.212\pm0.046$&$-3.993\pm0.087$&$-5.465\pm0.072$\\
40.5&$-2.531\pm0.041$&$-2.612\pm0.043$&$-2.716\pm0.044$&$-3.590\pm0.044$&$-3.365\pm0.045$&$-4.049\pm0.061$&$-5.584\pm0.138$\\
40.7&$-2.675\pm0.038$&$-2.740\pm0.038$&$-2.844\pm0.039$&$-3.797\pm0.040$&$-3.506\pm0.039$&$-4.143\pm0.081$&$-6.596\pm0.209$\\
40.9&$-2.821\pm0.038$&$-2.881\pm0.039$&$-2.972\pm0.040$&$-3.945\pm0.039$&$-3.725\pm0.042$&$-4.251\pm0.041$&$-6.601\pm0.118$\\
41.1&$-2.983\pm0.038$&$-3.034\pm0.038$&$-3.121\pm0.039$&$-4.199\pm0.034$&$-3.898\pm0.038$&$-4.431\pm0.061$&$-6.959\pm0.297$\\ 
41.3&$-3.205\pm0.040$&$-3.250\pm0.040$&$-3.319\pm0.041$&$-4.417\pm0.031$&$-4.285\pm0.040$&$-4.630\pm0.054$&$-7.879\pm0.420$\\ 
41.5&$-3.457\pm0.038$&$-3.490\pm0.037$&$-3.550\pm0.037$&$-4.779\pm0.040$&$-4.586\pm0.043$&$-4.837\pm0.054$&$-$\\ 
41.7&$-3.702\pm0.044$&$-3.725\pm0.044$&$-3.779\pm0.045$&$-5.185\pm0.045$&$-4.918\pm0.048$&$-5.028\pm0.047$&$-$\\
41.9&$-3.975\pm0.040$&$-3.994\pm0.040$&$-4.038\pm0.039$&$-5.491\pm0.048$&$-5.251\pm0.077$&$-5.463\pm0.070$&$-$\\ 
42.1&$-4.268\pm0.081$&$-4.281\pm0.083$&$-4.322\pm0.088$&$-6.105\pm0.064$&$-5.655\pm0.069$&$-5.673\pm0.080$&$-$\\
42.3&$-4.614\pm0.053$&$-4.623\pm0.053$&$-4.658\pm0.054$&$-6.372\pm0.068$&$-6.090\pm0.098$&$-6.171\pm0.108$&$-$\\
42.5&$-4.919\pm0.063$&$-4.927\pm0.063$&$-4.963\pm0.064$&$-6.883\pm0.089$&$-6.502\pm0.174$&$-6.388\pm0.113$&$-$\\
42.7&$-5.207\pm0.169$&$-5.210\pm0.170$&$-5.233\pm0.178$&$-7.535\pm0.025$&$-6.777\pm0.113$&$-6.950\pm0.209$&$-$\\
42.9&$-5.777\pm0.085$&$-5.787\pm0.087$&$-5.827\pm0.088$&$-$&$-7.286\pm0.149$&$-7.283\pm0.193$&$-$\\
43.1&$-5.914\pm0.255$&$-5.914\pm0.255$&$-5.935\pm0.268$&$-$&$-7.985\pm0.361$&$-7.325\pm0.196$&$-$\\
43.3&$-6.261\pm0.183$&$-6.261\pm0.183$&$-6.261\pm0.183$&$-$&$-$&$-$&$-$\\
43.5&$-7.809\pm1.668$&$-7.809\pm1.668$&$-7.809\pm1.182$&$-$&$-$&$-$&$-$\\
\end{tabular}
 \begin{tabular}{c c c c c c c c}
    \toprule
    &\multicolumn{7}{c}{observed $\log(\Phi(L_{\rm [O_{III}]]}/\rm{Mpc^{-3}\,dex^{-1}})$)}\\
$\log{L_{\rm [O_{III}]}}$&Full sample&SF sSFR&SF BPT+WHAN&LINERs&Composite&Seyferts&Passive\\ 
\midrule
39.9&$-2.453\pm0.045$&$-2.573\pm0.050$&$-2.683\pm0.053$&$-3.371\pm0.042$&$-3.210\pm0.046$&$-3.956\pm0.138$&$-5.622\pm0.139$\\
40.1&$-2.597\pm0.040$&$-2.696\pm0.041$&$-2.829\pm0.044$&$-3.599\pm0.041$&$-3.371\pm0.045$&$-4.072\pm0.086$&$-5.674\pm0.261$\\
40.3&$-2.723\pm0.038$&$-2.818\pm0.039$&$-2.966\pm0.041$&$-3.690\pm0.042$&$-3.519\pm0.038$&$-4.004\pm0.058$&$-6.070\pm0.388$\\
40.5&$-2.824\pm0.041$&$-2.913\pm0.041$&$-3.064\pm0.042$&$-3.870\pm0.041$&$-3.673\pm0.045$&$-3.819\pm0.118$&$-6.477\pm0.238$\\
40.7&$-3.076\pm0.039$&$-3.158\pm0.038$&$-3.305\pm0.040$&$-4.189\pm0.032$&$-4.002\pm0.041$&$-3.977\pm0.081$&$-6.503\pm0.168$\\
40.9&$-3.204\pm0.041$&$-3.252\pm0.041$&$-3.384\pm0.046$&$-4.506\pm0.040$&$-4.242\pm0.042$&$-4.087\pm0.060$&$-7.472\pm0.306$\\
41.1&$-3.438\pm0.042$&$-3.481\pm0.043$&$-3.619\pm0.047$&$-4.775\pm0.028$&$-4.526\pm0.041$&$-4.273\pm0.048$&$-$\\
41.3&$-3.616\pm0.042$&$-3.650\pm0.043$&$-3.812\pm0.046$&$-5.079\pm0.036$&$-4.765\pm0.052$&$-4.307\pm0.054$&$-$\\
41.5&$-3.797\pm0.044$&$-3.844\pm0.045$&$-3.985\pm0.051$&$-5.533\pm0.031$&$-5.056\pm0.063$&$-4.448\pm0.065$&$-5.898\pm0.061$\\
41.7&$-4.020\pm0.054$&$-4.044\pm0.055$&$-4.173\pm0.067$&$-6.005\pm0.036$&$-5.480\pm0.058$&$-4.686\pm0.063$&$-$\\
41.9&$-4.300\pm0.043$&$-4.384\pm0.044$&$-4.491\pm0.047$&$-6.397\pm0.050$&$-5.705\pm0.078$&$-4.860\pm0.061$&$-$\\
42.1&$-4.504\pm0.045$&$-4.526\pm0.046$&$-4.688\pm0.049$&$-6.866\pm0.070$&$-5.979\pm0.094$&$-5.067\pm0.047$&$-$\\
42.3&$-4.732\pm0.042$&$-4.748\pm0.043$&$-4.856\pm0.046$&$-7.150\pm0.092$&$-6.643\pm0.138$&$-5.424\pm0.062$&$-$\\
42.5&$-5.074\pm0.059$&$-5.097\pm0.061$&$-5.216\pm0.071$&$-$&$-6.837\pm0.139$&$-5.714\pm0.075$&$-$\\
42.7&$-5.407\pm0.048$&$-5.421\pm0.048$&$-5.532\pm0.052$&$-$&$-7.128\pm0.302$&$-6.105\pm0.072$&$-$\\
42.9&$-5.712\pm0.073$&$-5.718\pm0.072$&$-5.868\pm0.081$&$-$&$-7.221\pm0.263$&$-6.404\pm0.095$&$-$\\
43.1&$-6.046\pm0.079$&$-6.046\pm0.079$&$-6.250\pm0.094$&$-$&$-$&$-6.502\pm0.110$&$-$\\
43.3&$-6.523\pm0.143$&$-6.549\pm0.150$&$-6.732\pm0.176$&$-$&$-$&$-6.941\pm0.122$&$-$\\
43.5&$-7.178\pm0.131$&$-7.178\pm0.130$&$-7.237\pm0.149$&$-$&$-$&$-$&$-$\\
\bottomrule
  \end{tabular}
   \caption{\OII and \OIII luminosity functions of the \mainsel sample and their different components. } 
 \label{tab:LFtable2}
\end{table*}

  \begin{table*}
  \setlength{\tabcolsep}{2pt}
  \centering
 \begin{tabular}{c c c c c c c c}
    \toprule
    &\multicolumn{7}{c}{observed $\log(\Phi(L_{\rm [N_{II}]]}/\rm{Mpc^{-3}\,dex^{-1}})$)}\\
{$\log{L_{\rm [N_{II}]}}$}&Full sample&SF sSFR&SF BPT+WHAN&LINERs&Composite&Seyfert&Passive\\ 
\hline
39.9&$-2.264\pm0.088$&$-2.401\pm0.114$&$-2.561\pm0.155$&$-3.141\pm0.049$&$-2.932\pm0.052$&$-3.875\pm0.128$&$-5.640\pm0.267$\\
40.1&$-2.267\pm0.042$&$-2.359\pm0.042$&$-2.473\pm0.044$&$-3.305\pm0.054$&$-2.993\pm0.048$&$-3.939\pm0.080$&$-5.351\pm0.091$\\
40.3&$-2.367\pm0.040$&$-2.444\pm0.041$&$-2.610\pm0.040$&$-3.428\pm0.039$&$-2.999\pm0.051$&$-3.820\pm0.137$&$-6.088\pm0.131$\\
40.5&$-2.531\pm0.038$&$-2.603\pm0.038$&$-2.740\pm0.037$&$-3.629\pm0.039$&$-3.230\pm0.046$&$-3.976\pm0.058$&$-5.969\pm0.201$\\
40.7&$-2.676\pm0.039$&$-2.736\pm0.040$&$-2.888\pm0.040$&$-3.812\pm0.043$&$-3.356\pm0.048$&$-3.989\pm0.067$&$-6.563\pm0.164$\\
40.9&$-2.894\pm0.037$&$-2.952\pm0.038$&$-3.106\pm0.039$&$-4.016\pm0.040$&$-3.578\pm0.040$&$-4.189\pm0.052$&$-7.085\pm0.190$\\
41.1&$-3.103\pm0.037$&$-3.154\pm0.037$&$-3.311\pm0.037$&$-4.282\pm0.032$&$-3.799\pm0.045$&$-4.299\pm0.045$&$-7.159\pm0.352$\\
41.3&$-3.356\pm0.036$&$-3.401\pm0.036$&$-3.563\pm0.036$&$-4.544\pm0.036$&$-4.045\pm0.039$&$-4.568\pm0.044$&$-$\\
41.5&$-3.624\pm0.036$&$-3.668\pm0.036$&$-3.850\pm0.036$&$-4.864\pm0.034$&$-4.327\pm0.037$&$-4.604\pm0.057$&$-$\\
41.7&$-3.935\pm0.037$&$-3.966\pm0.038$&$-4.183\pm0.037$&$-5.243\pm0.029$&$-4.627\pm0.038$&$-4.787\pm0.065$&$-$\\
41.9&$-4.322\pm0.038$&$-4.351\pm0.038$&$-4.590\pm0.039$&$-5.631\pm0.027$&$-4.955\pm0.052$&$-5.177\pm0.057$&$-$\\
42.1&$-4.669\pm0.041$&$-4.697\pm0.042$&$-4.970\pm0.042$&$-5.902\pm0.031$&$-5.295\pm0.069$&$-5.452\pm0.053$&$-$\\
42.3&$-5.109\pm0.045$&$-5.137\pm0.047$&$-5.466\pm0.046$&$-6.606\pm0.040$&$-5.615\pm0.066$&$-5.903\pm0.078$&$-$\\
42.5&$-5.638\pm0.069$&$-5.651\pm0.068$&$-5.996\pm0.072$&$-6.967\pm0.103$&$-6.215\pm0.127$&$-6.293\pm0.123$&$-$\\
42.7&$-6.049\pm0.063$&$-6.082\pm0.062$&$-6.440\pm0.070$&$-7.368\pm0.078$&$-6.710\pm0.137$&$-6.649\pm0.108$&$-$\\
42.9&$-6.487\pm0.083$&$-6.487\pm0.082$&$-6.914\pm0.158$&$-$&$-6.860\pm0.131$&$-7.179\pm0.176$&$-$\\
43.1&$-7.120\pm0.123$&$-7.172\pm0.139$&$-7.380\pm0.155$&$-$&$-$&$-7.761\pm0.061$&$-$\\
43.3&$-7.773\pm0.365$&$-7.773\pm0.365$&$-$&$-$&$-$&$-7.773\pm0.298$&$-$\\
43.5&$-$&$-$&$-$&$-$&$-$&$-$&$-$\\
  \end{tabular}
 \begin{tabular}{c c c c c c c c}
    \midrule
    &\multicolumn{7}{c}{observed $\log(\Phi(L_{\rm [S_{II}]]}/\rm{Mpc^{-3}\,dex^{-1}})$)}\\
{$\log{L_{\rm [S_{II}]}}$}&Full sample&SF sSFR&SF BPT+WHAN&LINERs&Composite&Seyfert&Passive\\ 
\midrule
39.9&$-2.294\pm0.095$&$-2.412\pm0.120$&$-2.542\pm0.157$&$-3.192\pm0.043$&$-2.958\pm0.041$&$-3.917\pm0.119$&$-5.557\pm0.178$\\
40.1&$-2.303\pm0.039$&$-2.376\pm0.041$&$-2.486\pm0.041$&$-3.403\pm0.040$&$-3.084\pm0.056$&$-3.813\pm0.080$&$-5.442\pm0.078$\\
40.3&$-2.462\pm0.038$&$-2.531\pm0.038$&$-2.638\pm0.039$&$-3.560\pm0.042$&$-3.238\pm0.042$&$-4.060\pm0.056$&$-5.844\pm0.152$\\
40.5&$-2.606\pm0.038$&$-2.665\pm0.039$&$-2.788\pm0.038$&$-3.731\pm0.037$&$-3.385\pm0.049$&$-3.973\pm0.063$&$-6.620\pm0.133$\\
40.7&$-2.765\pm0.038$&$-2.814\pm0.039$&$-2.923\pm0.039$&$-3.963\pm0.046$&$-3.567\pm0.039$&$-4.199\pm0.056$&$-6.749\pm0.181$\\
40.9&$-2.975\pm0.037$&$-3.018\pm0.038$&$-3.127\pm0.038$&$-4.187\pm0.039$&$-3.799\pm0.045$&$-4.391\pm0.049$&$-7.444\pm0.246$\\
41.1&$-3.205\pm0.036$&$-3.244\pm0.036$&$-3.352\pm0.037$&$-4.444\pm0.031$&$-4.058\pm0.038$&$-4.526\pm0.052$&$-7.251\pm0.426$\\
41.3&$-3.457\pm0.036$&$-3.494\pm0.036$&$-3.599\pm0.036$&$-4.722\pm0.034$&$-4.360\pm0.039$&$-4.639\pm0.061$&$-$\\
41.5&$-3.758\pm0.036$&$-3.786\pm0.037$&$-3.894\pm0.037$&$-5.112\pm0.035$&$-4.685\pm0.038$&$-4.879\pm0.051$&$-$\\
41.7&$-4.113\pm0.037$&$-4.133\pm0.037$&$-4.262\pm0.037$&$-5.526\pm0.026$&$-4.949\pm0.057$&$-5.231\pm0.054$&$-$\\
41.9&$-4.510\pm0.038$&$-4.544\pm0.037$&$-4.660\pm0.038$&$-5.763\pm0.021$&$-5.465\pm0.055$&$-5.562\pm0.058$&$-$\\
42.1&$-4.974\pm0.046$&$-4.996\pm0.047$&$-5.133\pm0.047$&$-6.394\pm0.062$&$-5.764\pm0.096$&$-6.045\pm0.096$&$-$\\
42.3&$-5.454\pm0.048$&$-5.479\pm0.049$&$-5.667\pm0.051$&$-7.006\pm0.078$&$-6.234\pm0.131$&$-6.276\pm0.084$&$-$\\
42.5&$-5.977\pm0.053$&$-5.995\pm0.055$&$-6.199\pm0.062$&$-6.913\pm0.027$&$-6.830\pm0.182$&$-7.007\pm0.122$&$-$\\
42.7&$-6.520\pm0.106$&$-6.542\pm0.108$&$-6.718\pm0.132$&$-7.821\pm0.239$&$-7.510\pm0.442$&$-7.247\pm0.167$&$-$\\
42.9&$-6.837\pm0.101$&$-6.947\pm0.126$&$-7.161\pm0.192$&$-$&$-7.727\pm0.193$&$-7.601\pm0.138$&$-$\\
43.1&$-8.039\pm0.425$&$-8.039\pm0.425$&$-$&$-$&$-$&$-8.039\pm0.425$&$-$\\
43.3&$-$&$-$&$-$&$-$&$-$&$-$&$-$\\
43.5&$-$&$-$&$-$&$-$&$-$&$-$&$-$\\
\bottomrule
  \end{tabular}
   \caption{Observed \NII and \SII luminosity functions of the \mainsel sample and their different components. }
 \label{tab:LFtable3}
\end{table*}

\section{{\em Main-ELG} intrinsic LFs}
\label{sec:appendixLFintr}
\begin{figure*}
\centering
    \includegraphics[width=0.42\linewidth]{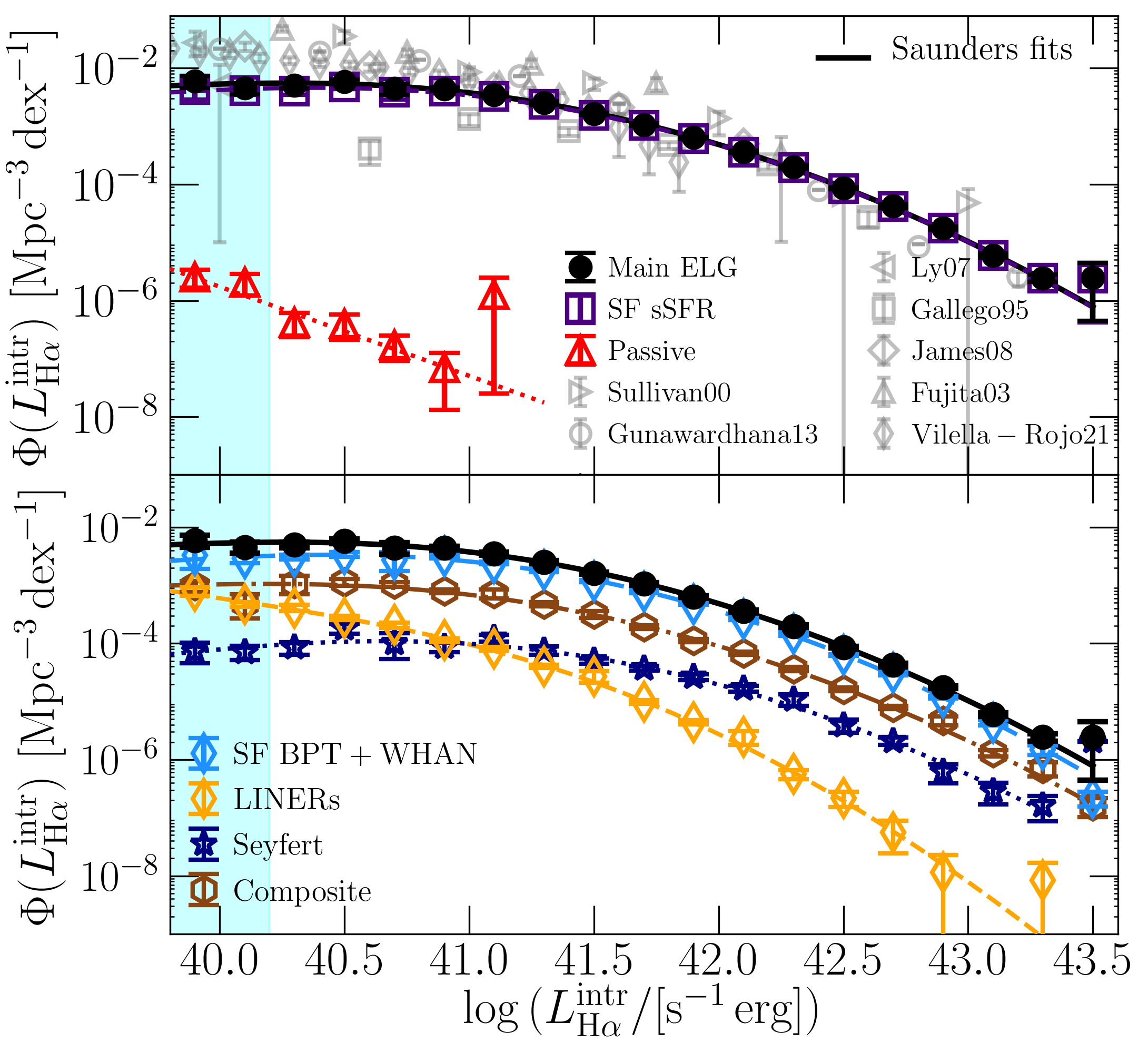}\quad\hfill
    \includegraphics[width=0.42\linewidth]{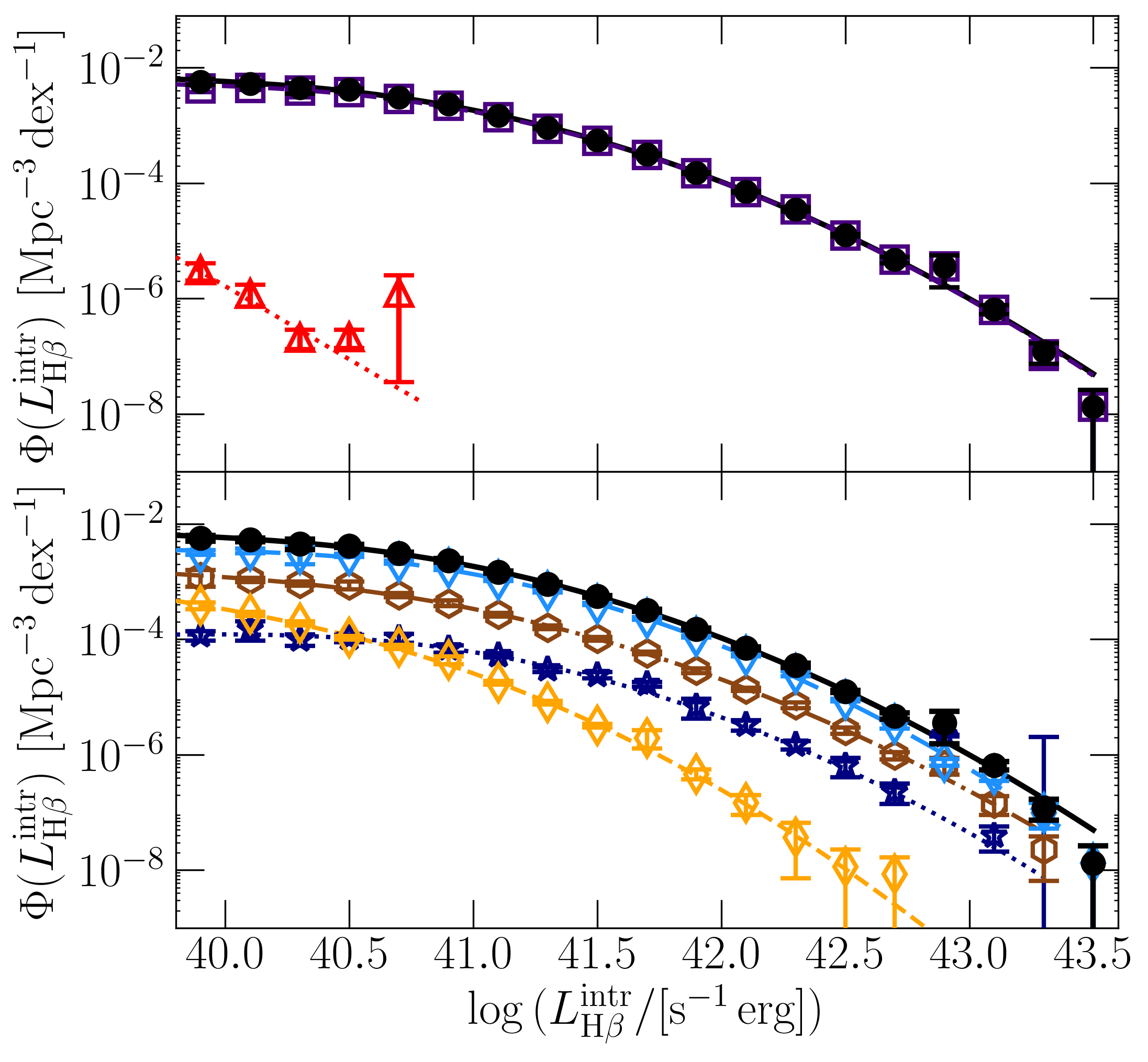}
    \includegraphics[width=0.42\linewidth]{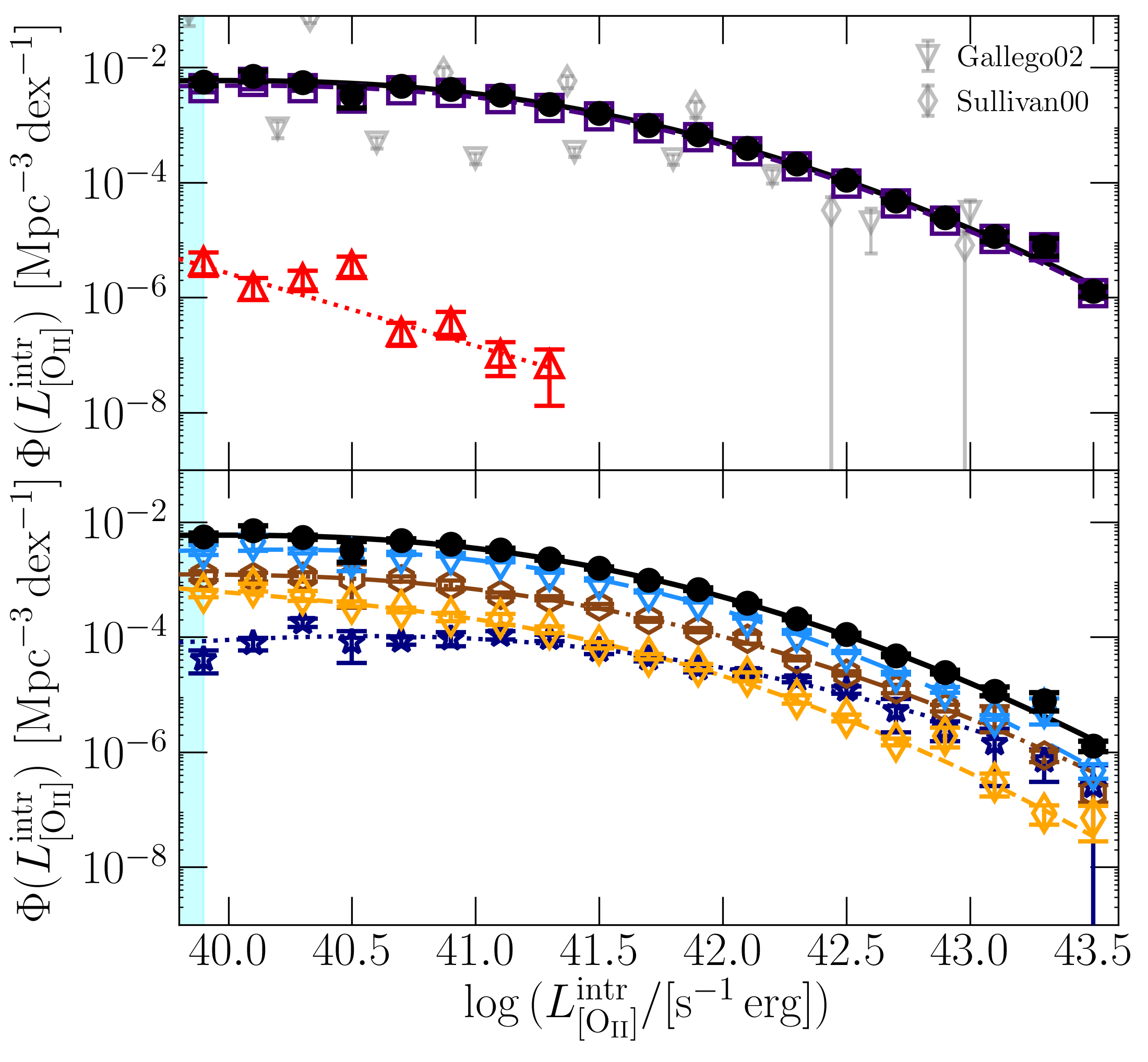}\quad\hfill
    \includegraphics[width=0.42\linewidth]{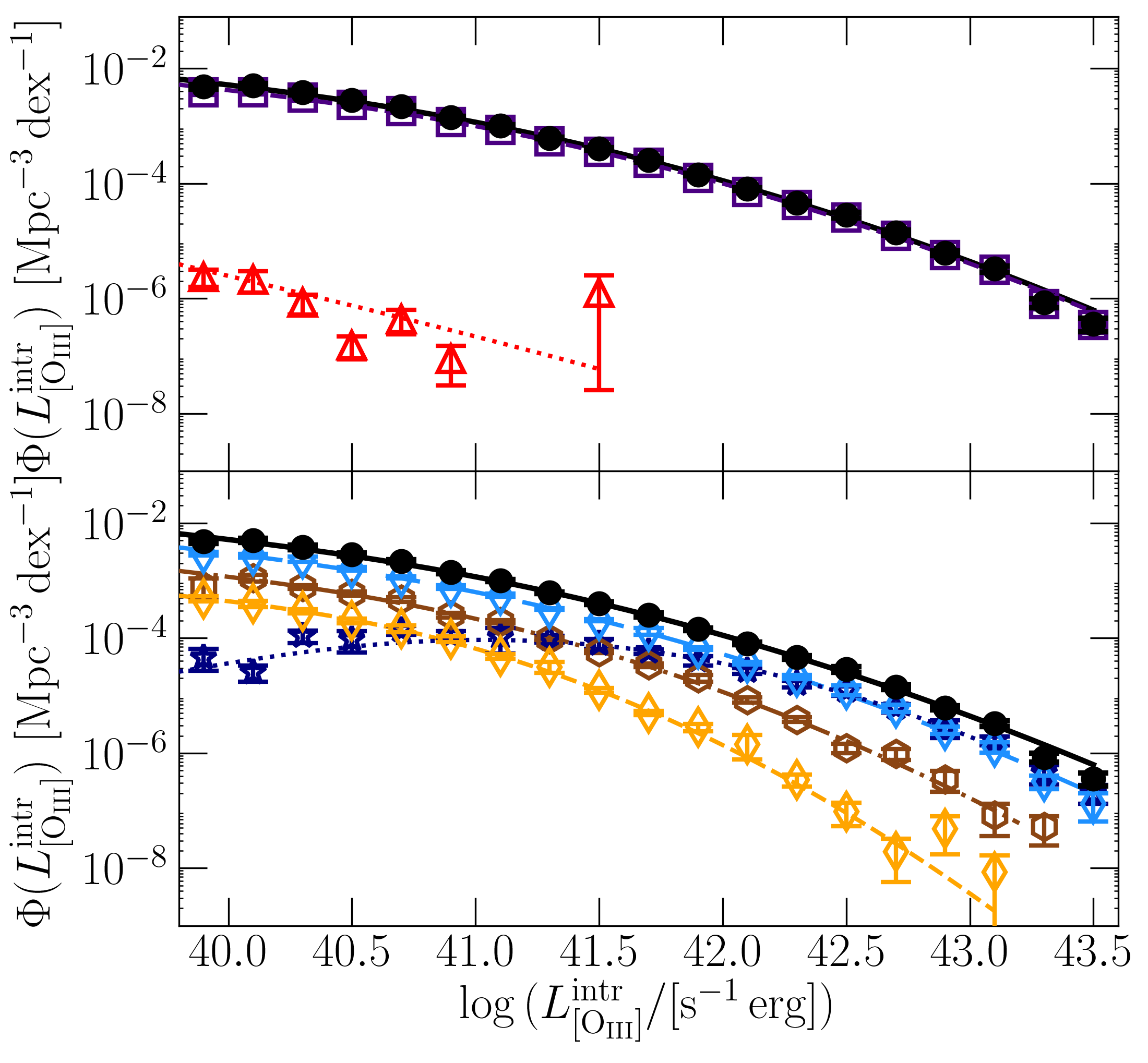}
    \includegraphics[width=0.42\linewidth]{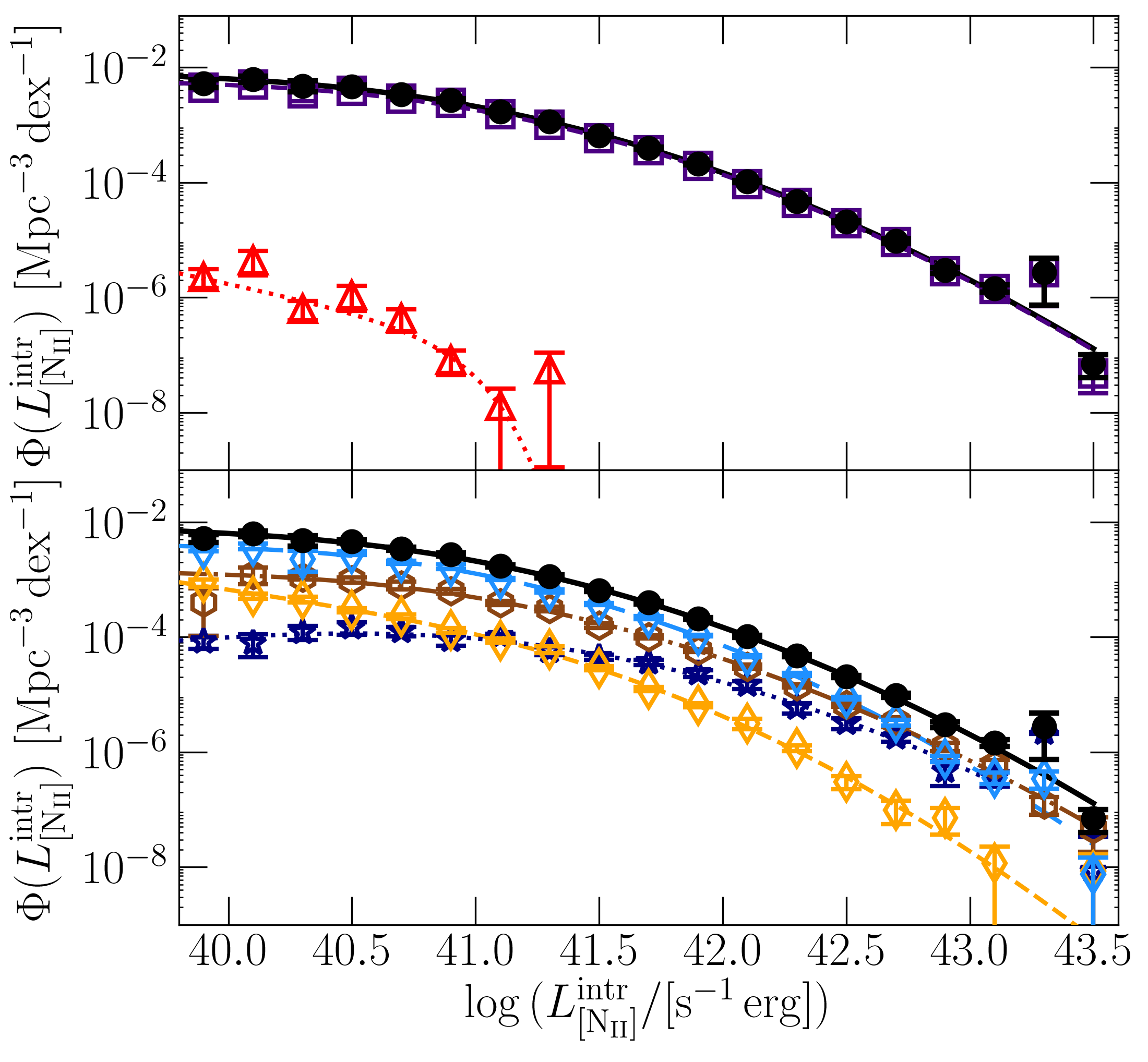}\quad\hfill
   \includegraphics[width=0.42\linewidth]{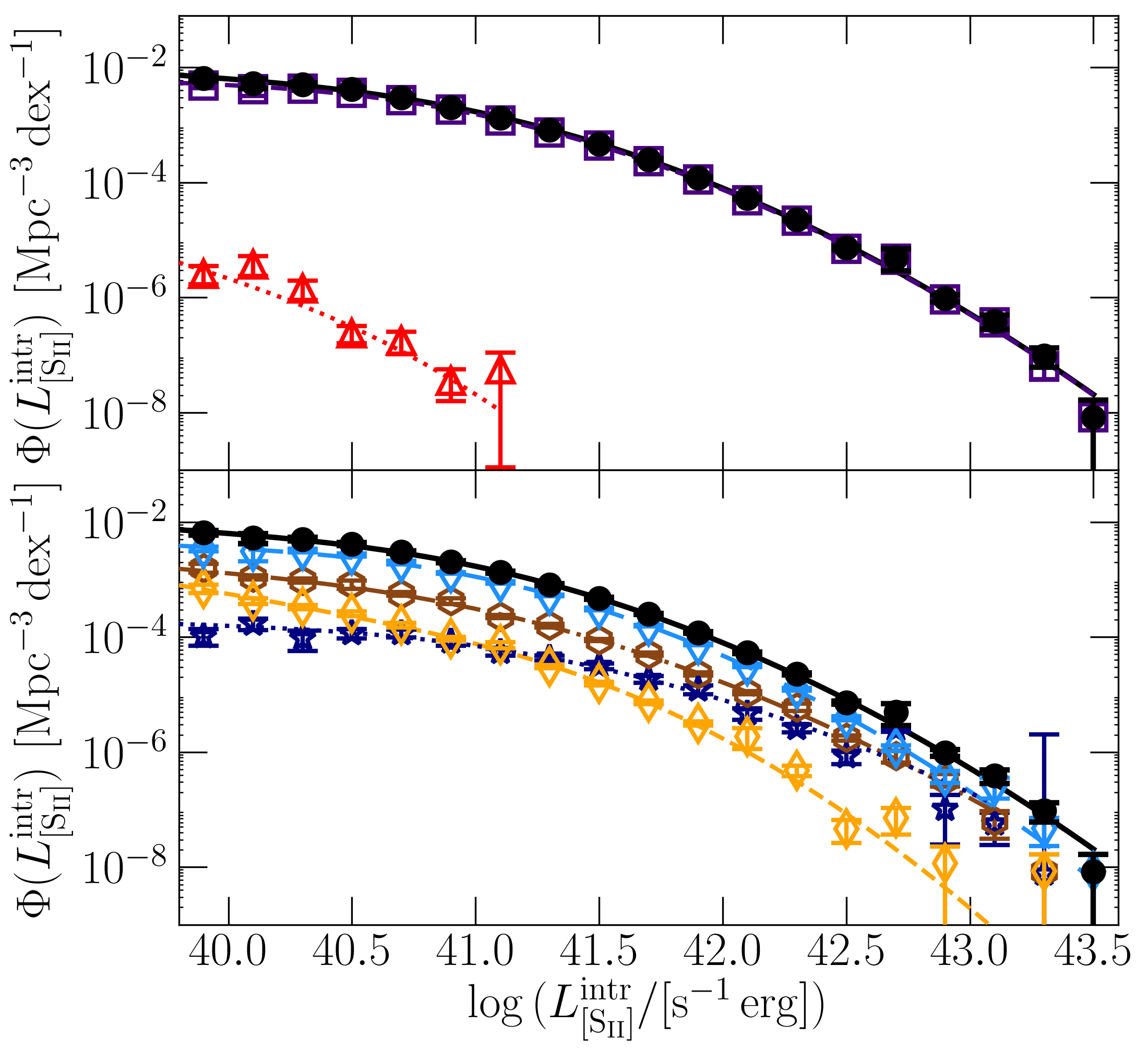}
    \vspace{-0.2cm}
\caption{Intrinsic emission-line LFs. From top to bottom and from left to right: \Ha, \Hb, \OII, \OIII, \NII, and \SII intrinsic (i.e., dust corrected) luminosity functions of the \mainsel sample (full black dots). The contribution of ELGs classified in different ways are shown by empty colored markers, with colors as indicated in the legend. We overplot the Saunders fits as thick black lines; the parameters are tabulated in Table\,\ref{tab:fitparsaundersintr} and were obtained considering only the points above the luminosity completeness threshold established in Sec. \ref{sec:incompleteness}. For those lines having the lower completeness limit in the $L$ range of the figure, we highlight with a yellow shade the region of incompleteness, where our LFs cannot be trusted. The error bars are computed from 70 jackknife resamplings.}
\label{fig:LFplotsaundersintr}
  \end{figure*}

The intrinsic (i.e., dust corrected) \Ha, \Hb, \OII, \OIII, \NII, and \SII \mainsel luminosity functions are shown in Figure\,\ref{fig:LFplotsaundersintr}. The numerical values are provided in Tables\,\ref{tab:LFtable1intr} - \ref{tab:LFtable3intr} and the best-fit Saunders parameters are given in Table\,\ref{tab:fitparsaundersintr}.

 \begin{table*}
  \setlength{\tabcolsep}{2pt}
  \centering
 \begin{tabular}{c c c c c c c c}
    \toprule
    &\multicolumn{7}{c}{intrinsic $\log(\Phi(L_{\rm H\alpha}/\rm{Mpc^{-3}\,dex^{-1}})$)}\\
$\log{L_{\rm H\alpha}}$&Full sample&SF sSFR&SF BPT+WHAN&LINERs&Composite&Seyferts&Passive\\ 
\midrule
39.9&$-2.224\pm0.103$&$-2.372\pm0.140$&$-2.483\pm0.179$&$-3.099\pm0.070$&$-3.009\pm0.093$&$-4.128\pm0.168$&$-5.598\pm0.145$\\
40.1&$-2.343\pm0.094$&$-2.389\pm0.066$&$-2.531\pm0.080$&$-3.325\pm0.043$&$-3.314\pm0.193$&$-4.149\pm0.118$&$-5.689\pm0.176$\\
40.3&$-2.299\pm0.049$&$-2.397\pm0.058$&$-2.511\pm0.041$&$-3.385\pm0.056$&$-2.968\pm0.151$&$-4.070\pm0.107$&$-6.376\pm0.192$\\
40.5&$-2.235\pm0.041$&$-2.324\pm0.037$&$-2.467\pm0.038$&$-3.559\pm0.038$&$-2.942\pm0.061$&$-3.738\pm0.083$&$-6.402\pm0.202$\\
40.7&$-2.352\pm0.095$&$-2.417\pm0.110$&$-2.558\pm0.159$&$-3.682\pm0.043$&$-2.982\pm0.043$&$-3.993\pm0.202$&$-6.764\pm0.202$\\
40.9&$-2.359\pm0.029$&$-2.396\pm0.030$&$-2.500\pm0.036$&$-3.942\pm0.031$&$-3.102\pm0.032$&$-4.061\pm0.080$&$-7.159\pm0.352$\\
41.1&$-2.454\pm0.027$&$-2.485\pm0.029$&$-2.613\pm0.024$&$-4.089\pm0.035$&$-3.142\pm0.078$&$-3.907\pm0.061$&$-5.898\pm0.426$\\
41.3&$-2.598\pm0.028$&$-2.621\pm0.029$&$-2.734\pm0.036$&$-4.393\pm0.031$&$-3.326\pm0.036$&$-4.119\pm0.071$&$-$\\
41.5&$-2.787\pm0.019$&$-2.809\pm0.019$&$-2.927\pm0.017$&$-4.569\pm0.098$&$-3.497\pm0.055$&$-4.299\pm0.047$&$-$\\
41.7&$-2.973\pm0.021$&$-2.987\pm0.022$&$-3.103\pm0.026$&$-5.005\pm0.035$&$-3.717\pm0.031$&$-4.428\pm0.041$&$-$\\
41.9&$-3.194\pm0.014$&$-3.208\pm0.015$&$-3.320\pm0.016$&$-5.352\pm0.036$&$-3.950\pm0.024$&$-4.578\pm0.052$&$-$\\
42.1&$-3.440\pm0.016$&$-3.451\pm0.017$&$-3.576\pm0.016$&$-5.610\pm0.117$&$-4.169\pm0.028$&$-4.777\pm0.050$&$-$\\
42.3&$-3.701\pm0.024$&$-3.709\pm0.024$&$-3.834\pm0.030$&$-6.249\pm0.076$&$-4.438\pm0.034$&$-4.966\pm0.099$&$-$\\
42.5&$-4.066\pm0.014$&$-4.072\pm0.014$&$-4.199\pm0.016$&$-6.665\pm0.127$&$-4.792\pm0.032$&$-5.410\pm0.111$&$-$\\
42.7&$-4.371\pm0.031$&$-4.378\pm0.032$&$-4.506\pm0.040$&$-7.250\pm0.248$&$-5.100\pm0.030$&$-5.682\pm0.063$&$-$\\
42.9&$-4.753\pm0.025$&$-4.755\pm0.026$&$-4.926\pm0.023$&$-7.938\pm0.426$ &$-5.327\pm0.075$&$-6.212\pm0.157$&$-$\\
43.1&$-5.222\pm0.033$&$-5.222\pm0.033$&$-5.362\pm0.044$&$-$              &$-5.886\pm0.052$&$-6.546\pm0.174$&$-$\\
43.3&$-5.612\pm0.061$&$-5.613\pm0.062$&$-5.829\pm0.082$&$-$              &$-6.154\pm0.119$&$-6.792\pm0.199$&$-$\\
43.5&$-5.608\pm0.036$&$-5.612\pm0.359$&$-6.664\pm0.122$&$-$              &$-6.794\pm0.157$&$-5.686\pm0.006$&$-$\\
  \end{tabular}
 \begin{tabular}{c c c c c c c c}
    \midrule
        &\multicolumn{7}{c}{intrinsic $\log(\Phi(L_{\rm H\beta}/\rm{Mpc^{-3}\,dex^{-1}})$)}\\
{$\log{L_{\rm H\beta}}$}&Full sample&SF sSFR&SF BPT+WHAN&LINERs&Composite&Seyfert&Passive\\ 
\midrule
39.9&$-2.247\pm0.051$&$-2.361\pm0.056$&$-2.496\pm0.041$&$-3.415\pm0.057$&$-2.921\pm0.140$&$-3.937\pm0.080$&$-5.511\pm0.144$\\
40.1&$-2.276\pm0.033$&$-2.350\pm0.036$&$-2.468\pm0.040$&$-3.562\pm0.037$&$-2.967\pm0.037$&$-3.821\pm0.160$&$-5.912\pm0.179$\\
40.3&$-2.347\pm0.091$&$-2.398\pm0.103$&$-2.527\pm0.014$&$-3.720\pm0.037$&$-3.035\pm0.039$&$-4.005\pm0.098$&$-6.677\pm0.160$\\
40.5&$-2.382\pm0.027$&$-2.421\pm0.028$&$-2.546\pm0.029$&$-3.972\pm0.032$&$-3.063\pm0.065$&$-3.982\pm0.060$&$-6.670\pm0.154$\\
40.7&$-2.510\pm0.023$&$-2.543\pm0.024$&$-2.661\pm0.027$&$-4.133\pm0.037$&$-3.232\pm0.034$&$-3.975\pm0.077$&$-5.894\pm0.422$\\
40.9&$-2.633\pm0.030$&$-2.658\pm0.032$&$-2.765\pm0.037$&$-4.358\pm0.068$&$-3.376\pm0.049$&$-4.163\pm0.066$&$-$\\
41.1&$-2.836\pm0.017$&$-2.856\pm0.017$&$-2.969\pm0.018$&$-4.753\pm0.031$&$-3.575\pm0.033$&$-4.274\pm0.035$&$-$\\
41.3&$-3.040\pm0.020$&$-3.055\pm0.021$&$-3.166\pm0.023$&$-5.098\pm0.032$&$-3.794\pm0.031$&$-4.525\pm0.044$&$-$\\
41.5&$-3.257\pm0.016$&$-3.270\pm0.016$&$-3.389\pm0.016$&$-5.496\pm0.034$&$-3.990\pm0.029$&$-4.620\pm0.053$&$-$\\
41.7&$-3.500\pm0.019$&$-3.511\pm0.019$&$-3.638\pm0.023$&$-5.709\pm0.152$&$-4.238\pm0.024$&$-4.777\pm0.041$&$-$\\
41.9&$-3.823\pm0.015$&$-3.829\pm0.015$&$-3.956\pm0.015$&$-6.334\pm0.088$&$-4.544\pm0.040$&$-5.169\pm0.165$&$-$\\
42.1&$-4.147\pm0.014$&$-4.152\pm0.014$&$-4.282\pm0.015$&$-6.837\pm0.166$&$-4.867\pm0.028$&$-5.481\pm0.089$&$-$\\
42.3&$-4.452\pm0.037$&$-4.457\pm0.038$&$-4.591\pm0.046$&$-7.434\pm0.349$&$-5.143\pm0.050$&$-5.832\pm0.071$&$-$\\
42.5&$-4.902\pm0.020$&$-4.904\pm0.020$&$-5.052\pm0.020$&$-7.938\pm0.425$&$-5.582\pm0.053$&$-6.190\pm0.158$&$-$\\
42.7&$-5.327\pm0.054$&$-5.328\pm0.055$&$-5.470\pm0.070$&$-8.072\pm0.426$&$-6.013\pm0.061$&$-6.641\pm0.168$&$-$\\
42.9&$-5.444\pm0.245$&$-5.444\pm0.245$&$-6.127\pm0.058$&$-$&$-6.175\pm0.139$&$-5.667\pm0.13$&$-$\\
43.1&$-6.184\pm0.074$&$-6.200\pm0.072$&$-6.348\pm0.094$&$-$&$-6.847\pm0.155$&$-7.406\pm0.202$&$-$\\
43.3&$-6.916\pm0.171$&$-6.970\pm0.188$&$-7.005\pm0.202$&$-$&$-7.646\pm0.308$&$-$&$-$\\
43.5&$-7.876\pm0.426$&$-7.876\pm0.426$&$-7.876\pm0.425$&$-$&$-$&$-$&$-$\\
  \end{tabular}
    \caption{\Ha and \Hb intrinsic luminosity functions of the \mainsel sample and their different components.} 
 \label{tab:LFtable1intr}
\end{table*}

\begin{table*}
\setlength{\tabcolsep}{2pt}
  \centering
 \begin{tabular}{c c c c c c c c}
    \toprule
       &\multicolumn{7}{c}{intrinsic $\log(\Phi(L_{\rm [O_{II}]]}/\rm{Mpc^{-3}\,dex^{-1}})$)}\\
$\log{L_{\rm [O_{II}]}}$&Full sample&SF sSFR&SF BPT+WHAN&LINERs&Composite&Seyferts&Passive\\ 
\hline
39.9&$-2.252\pm0.059$&$-2.359\pm0.071$&$-2.478\pm0.088$&$-3.239\pm0.063$&$-2.924\pm0.080$&$-4.392\pm0.186$&$-5.375\pm0.192$\\
40.1&$-2.149\pm0.092$&$-2.248\pm0.111$&$-2.341\pm0.134$&$-3.135\pm0.071$&$-2.954\pm0.077$&$-4.110\pm0.107$&$-5.813\pm0.177$\\
40.3&$-2.264\pm0.038$&$-2.363\pm0.041$&$-2.504\pm0.039$&$-3.274\pm0.087$&$-2.946\pm0.081$&$-3.746\pm0.070$&$-5.666\pm0.155$\\
40.5&$-2.483\pm0.174$&$-2.542\pm0.151$&$-2.632\pm0.177$&$-3.423\pm0.044$&$-2.971\pm0.302$&$-4.093\pm0.243$&$-5.433\pm0.164$\\
40.7&$-2.321\pm0.027$&$-2.393\pm0.029$&$-2.545\pm0.027$&$-3.527\pm0.044$&$-2.989\pm0.046$&$-4.064\pm0.065$&$-6.597\pm0.186$\\
40.9&$-2.382\pm0.031$&$-2.443\pm0.031$&$-2.568\pm0.035$&$-3.649\pm0.071$&$-3.089\pm0.075$&$-4.023\pm0.120$&$-6.421\pm0.212$\\
41.1&$-2.482\pm0.030$&$-2.548\pm0.032$&$-2.665\pm0.028$&$-3.683\pm0.093$&$-3.266\pm0.037$&$-3.963\pm0.075$&$-6.980\pm0.257$\\ 
41.3&$-2.641\pm0.022$&$-2.702\pm0.023$&$-2.860\pm0.021$&$-3.872\pm0.061$&$-3.329\pm0.037$&$-4.015\pm0.054$&$-7.159\pm0.352$\\ 
41.5&$-2.799\pm0.020$&$-2.847\pm0.020$&$-3.002\pm0.020$&$-4.142\pm0.054$&$-3.465\pm0.043$&$-4.222\pm0.071$&$-$\\ 
41.7&$-3.005\pm0.015$&$-3.058\pm0.015$&$-3.209\pm0.016$&$-4.343\pm0.046$&$-3.703\pm0.033$&$-4.372\pm0.053$&$-$\\
41.9&$-3.168\pm0.026$&$-3.213\pm0.028$&$-3.361\pm0.034$&$-4.519\pm0.053$&$-3.879\pm0.038$&$-4.546\pm0.064$&$-$\\ 
42.1&$-3.407\pm0.025$&$-3.457\pm0.026$&$-3.661\pm0.016$&$-4.687\pm0.078$&$-4.040\pm0.089$&$-4.613\pm0.054$&$-$\\
42.3&$-3.676\pm0.023$&$-3.717\pm0.022$&$-3.912\pm0.030$&$-5.081\pm0.073$&$-4.375\pm0.031$&$-4.740\pm0.048$&$-$\\
42.5&$-3.954\pm0.022$&$-4.019\pm0.019$&$-4.258\pm0.018$&$-5.402\pm0.058$&$-4.634\pm0.035$&$-4.918\pm0.064$&$-$\\
42.7&$-4.320\pm0.025$&$-4.362\pm0.026$&$-4.626\pm0.020$&$-5.820\pm0.057$&$-4.901\pm0.070$&$-5.276\pm0.254$&$-$\\
42.9&$-4.610\pm0.029$&$-4.663\pm0.030$&$-4.915\pm0.047$&$-5.717\pm0.161$&$-5.232\pm0.060$&$-5.606\pm0.163$&$-$\\
43.1&$-4.938\pm0.078$&$-4.983\pm0.086$&$-5.393\pm0.042$&$-6.530\pm0.184$&$-5.375\pm0.208$&$-5.852\pm0.355$&$-$\\
43.3&$-5.101\pm0.150$&$-5.124\pm0.158$&$-5.240\pm0.020$&$-7.061\pm0.161$&$-6.061\pm0.100$&$-6.151\pm0.247$&$-$\\
43.5&$-5.892\pm0.087$&$-5.922\pm0.092$&$-6.320\pm0.124$&$-7.142\pm0.266$&$-6.700\pm0.150$&$-6.591\pm0.571$&$-$\\
\end{tabular}
 \begin{tabular}{c c c c c c c c}
    \toprule
    &\multicolumn{7}{c}{intrinsic $\log(\Phi(L_{\rm [O_{III}]]}/\rm{Mpc^{-3}\,dex^{-1}})$)}\\
$\log{L_{\rm [O_{III}]}}$&Full sample&SF sSFR&SF BPT+WHAN&LINERs&Composite&Seyferts&Passive\\ 
\midrule
39.9&$-2.319\pm0.038$&$-2.415\pm0.044$&$-2.529\pm0.030$&$-3.317\pm0.045$&$-3.101\pm0.161$&$-4.340\pm0.176$&$-5.622\pm0.145$\\
40.1&$-2.300\pm0.037$&$-2.417\pm0.030$&$-2.567\pm0.030$&$-3.404\pm0.054$&$-2.954\pm0.062$&$-4.599\pm0.134$&$-5.674\pm0.174$\\
40.3&$-2.420\pm0.039$&$-2.510\pm0.045$&$-2.623\pm0.046$&$-3.533\pm0.035$&$-3.112\pm0.031$&$-3.963\pm0.107$&$-6.071\pm0.165$\\
40.5&$-2.549\pm0.037$&$-2.636\pm0.027$&$-2.792\pm0.033$&$-3.702\pm0.039$&$-3.226\pm0.035$&$-4.027\pm0.176$&$-6.814\pm0.187$\\
40.7&$-2.664\pm0.036$&$-2.737\pm0.025$&$-2.944\pm0.021$&$-3.834\pm0.061$&$-3.328\pm0.041$&$-3.860\pm0.077$&$-6.359\pm0.196$\\
40.9&$-2.853\pm0.034$&$-2.929\pm0.027$&$-3.115\pm0.031$&$-4.016\pm0.051$&$-3.601\pm0.037$&$-3.954\pm0.076$&$-7.046\pm0.286$\\
41.1&$-2.999\pm0.022$&$-3.065\pm0.021$&$-3.252\pm0.025$&$-4.318\pm0.042$&$-3.718\pm0.029$&$-3.991\pm0.205$&$-$\\
41.3&$-3.210\pm0.014$&$-3.268\pm0.016$&$-3.494\pm0.017$&$-4.503\pm0.092$&$-4.012\pm0.020$&$-4.014\pm0.058$&$-$\\
41.5&$-3.396\pm0.020$&$-3.453\pm0.022$&$-3.689\pm0.022$&$-4.906\pm0.040$&$-4.241\pm0.022$&$-4.093\pm0.085$&$-5.898\pm0.426$\\
41.7&$-3.598\pm0.034$&$-3.642\pm0.039$&$-3.878\pm0.059$&$-5.303\pm0.034$&$-4.471\pm0.030$&$-4.236\pm0.038$&$-$\\
41.9&$-3.844\pm0.022$&$-3.906\pm0.023$&$-4.184\pm0.025$&$-5.550\pm0.062$&$-4.694\pm0.059$&$-4.390\pm0.078$&$-$\\
42.1&$-4.095\pm0.021$&$-4.148\pm0.022$&$-4.437\pm0.027$&$-5.846\pm0.197$&$-5.076\pm0.047$&$-4.573\pm0.051$&$-$\\
42.3&$-4.338\pm0.029$&$-4.373\pm0.031$&$-4.679\pm0.034$&$-6.464\pm0.104$&$-5.419\pm0.041$&$-4.797\pm0.055$&$-$\\
42.5&$-4.546\pm0.048$&$-4.588\pm0.053$&$-4.897\pm0.085$&$-7.017\pm0.191$&$-5.911\pm0.078$&$-4.915\pm0.079$&$-$\\
42.7&$-4.852\pm0.041$&$-4.908\pm0.037$&$-5.213\pm0.063$&$-7.018\pm0.304$&$-6.014\pm0.094$&$-5.215\pm0.070$&$-$\\
42.9&$-5.210\pm0.039$&$-5.245\pm0.041$&$-5.598\pm0.067$&$-7.313\pm0.281$&$-6.454\pm0.170$&$-5.559\pm0.146$&$-$\\
43.1&$-5.484\pm0.052$&$-5.501\pm0.055$&$-5.896\pm0.089$&$-8.072\pm0.426$&$-7.081\pm0.247$&$-5.783\pm0.078$&$-$\\
43.3&$-6.074\pm0.075$&$-6.100\pm0.081$&$-6.494\pm0.114$&$-$&$-7.280\pm0.228$&$-6.348\pm0.160$&$-$\\
43.5&$-6.445\pm0.113$&$-6.459\pm0.117$&$-6.881\pm0.219$&$-$&$-$&$-6.687\pm0.155$&$-$\\
\bottomrule
  \end{tabular}
   \caption{\OII and \OIII intrinsic luminosity functions of the \mainsel sample and their different components.} 
 \label{tab:LFtable2intr}
\end{table*}

  \begin{table*}
  \setlength{\tabcolsep}{2pt}
  \centering
 \begin{tabular}{c c c c c c c c}
    \toprule
    &\multicolumn{7}{c}{intrinsic $\log(\Phi(L_{\rm [N_{II}]]}/\rm{Mpc^{-3}\,dex^{-1}})$)}\\
{$\log{L_{\rm [N_{II}]}}$}&Full sample&SF sSFR&SF BPT+WHAN&LINERs&Composite&Seyfert&Passive\\ 
\hline
39.9&$-2.282\pm0.067$&$-2.351\pm0.044$&$-2.461\pm0.049$&$-3.056\pm0.062$&$-3.398\pm0.321$&$-4.080\pm0.111$&$-5.640\pm0.154$\\
40.1&$-2.202\pm0.052$&$-2.294\pm0.058$&$-2.417\pm0.049$&$-3.287\pm0.045$&$-2.914\pm0.137$&$-4.109\pm0.185$&$-5.351\pm0.198$\\
40.3&$-2.320\pm0.092$&$-2.455\pm0.115$&$-2.645\pm0.173$&$-3.347\pm0.051$&$-2.959\pm0.039$&$-3.912\pm0.122$&$-6.196\pm0.162$\\
40.5&$-2.335\pm0.028$&$-2.410\pm0.029$&$-2.544\pm0.030$&$-3.533\pm0.036$&$-3.005\pm0.043$&$-3.838\pm0.109$&$-5.964\pm0.201$\\
40.7&$-2.467\pm0.026$&$-2.538\pm0.029$&$-2.701\pm0.024$&$-3.648\pm0.038$&$-3.096\pm0.068$&$-3.896\pm0.079$&$-6.356\pm0.183$\\
40.9&$-2.567\pm0.028$&$-2.619\pm0.030$&$-2.776\pm0.039$&$-3.884\pm0.038$&$-3.202\pm0.037$&$-4.054\pm0.088$&$-7.085\pm0.195$\\
41.1&$-2.767\pm0.028$&$-2.818\pm0.019$&$-2.993\pm0.019$&$-4.059\pm0.038$&$-3.409\pm0.034$&$-3.990\pm0.053$&$-7.879\pm0.425$\\
41.3&$-2.943\pm0.024$&$-2.994\pm0.026$&$-3.199\pm0.024$&$-4.216\pm0.056$&$-3.511\pm0.043$&$-4.241\pm0.065$&$-7.251\pm0.425$\\
41.5&$-3.188\pm0.017$&$-3.232\pm0.018$&$-3.429\pm0.017$&$-4.539\pm0.038$&$-3.800\pm0.039$&$-4.350\pm0.044$&$-$\\
41.7&$-3.403\pm0.020$&$-3.445\pm0.021$&$-3.661\pm0.024$&$-4.900\pm0.035$&$-3.997\pm0.026$&$-4.440\pm0.042$&$-$\\
41.9&$-3.677\pm0.014$&$-3.713\pm0.015$&$-3.974\pm0.017$&$-5.183\pm0.036$&$-4.236\pm0.022$&$-4.632\pm0.061$&$-$\\
42.1&$-3.988\pm0.018$&$-4.026\pm0.018$&$-4.329\pm0.017$&$-5.503\pm0.091$&$-4.511\pm0.038$&$-4.832\pm0.066$&$-$\\
42.3&$-4.325\pm0.019$&$-4.352\pm0.020$&$-4.651\pm0.029$&$-5.932\pm0.050$&$-4.849\pm0.029$&$-5.227\pm0.090$&$-$\\
42.5&$-4.684\pm0.018$&$-4.707\pm0.018$&$-5.061\pm0.024$&$-6.516\pm0.114$&$-5.173\pm0.037$&$-5.496\pm0.093$&$-$\\
42.7&$-5.014\pm0.035$&$-5.038\pm0.037$&$-5.487\pm0.046$&$-7.004\pm0.189$&$-5.470\pm0.085$&$-5.745\pm0.069$&$-$\\
42.9&$-5.525\pm0.051$&$-5.540\pm0.053$&$-6.115\pm0.053$&$-7.144\pm0.211$&$-5.903\pm0.074$&$-6.326\pm0.196$&$-$\\
43.1&$-5.839\pm0.060$&$-5.850\pm0.060$&$-6.442\pm0.097$&$-7.936\pm0.425$&$-6.227\pm0.102$&$-6.449\pm0.125$&$-$\\
43.3&$-5.560\pm0.318$&$-5.564\pm0.321$&$-6.465\pm0.143$&$-$&$-6.906\pm0.149$&$-5.669\pm0.016$&$-$\\
43.5&$-7.153\pm0.185$&$-7.322\pm0.238$&$-8.128\pm0.425$&$-8.072\pm0.425$&$-7.338\pm0.260$&$-8.069\pm1.294$&$-$\\
  \end{tabular}
 \begin{tabular}{c c c c c c c c}
    \midrule
    &\multicolumn{7}{c}{intrinsic $\log(\Phi(L_{\rm [S_{II}]]}/\rm{Mpc^{-3}\,dex^{-1}})$)}\\
{$\log{L_{\rm [S_{II}]}}$}&Full sample&SF sSFR&SF BPT+WHAN&LINERs&Composite&Seyfert&Passive\\ 
\midrule
39.9&$-2.187\pm0.042$&$-2.312\pm0.044$&$-2.469\pm0.039$&$-3.154\pm0.073$&$-2.797\pm0.073$&$-3.984\pm0.140$&$-5.586\pm0.149$\\
40.1&$-2.276\pm0.086$&$-2.376\pm0.105$&$-2.514\pm0.140$&$-3.371\pm0.048$&$-2.963\pm0.037$&$-3.771\pm0.096$&$-5.421\pm0.161$\\
40.3&$-2.298\pm0.030$&$-2.372\pm0.032$&$-2.494\pm0.033$&$-3.479\pm0.039$&$-3.016\pm0.038$&$-4.026\pm0.167$&$-5.844\pm0.158$\\
40.5&$-2.381\pm0.027$&$-2.444\pm0.028$&$-2.572\pm0.026$&$-3.591\pm0.041$&$-3.084\pm0.073$&$-3.938\pm0.081$&$-6.620\pm0.141$\\
40.7&$-2.523\pm0.026$&$-2.573\pm0.028$&$-2.705\pm0.034$&$-3.791\pm0.042$&$-3.251\pm0.030$&$-3.934\pm0.064$&$-6.749\pm0.184$\\
40.9&$-2.693\pm0.019$&$-2.736\pm0.019$&$-2.886\pm0.015$&$-4.027\pm0.032$&$-3.370\pm0.048$&$-4.079\pm0.065$&$-7.444\pm0.242$\\
41.1&$-2.876\pm0.018$&$-2.925\pm0.019$&$-3.048\pm0.021$&$-4.136\pm0.053$&$-3.643\pm0.024$&$-4.263\pm0.057$&$-7.251\pm0.425$\\
41.3&$-3.090\pm0.015$&$-3.127\pm0.015$&$-3.275\pm0.015$&$-4.513\pm0.025$&$-3.801\pm0.034$&$-4.364\pm0.041$&$-$\\
41.5&$-3.328\pm0.017$&$-3.360\pm0.018$&$-3.511\pm0.021$&$-4.807\pm0.029$&$-4.053\pm0.020$&$-4.505\pm0.048$&$-$\\
41.7&$-3.599\pm0.012$&$-3.631\pm0.012$&$-3.803\pm0.012$&$-5.134\pm0.032$&$-4.295\pm0.026$&$-4.732\pm0.060$&$-$\\
41.9&$-3.921\pm0.016$&$-3.951\pm0.016$&$-4.129\pm0.016$&$-5.502\pm0.037$&$-4.640\pm0.027$&$-4.910\pm0.078$&$-$\\
42.1&$-4.274\pm0.017$&$-4.308\pm0.018$&$-4.496\pm0.023$&$-5.732\pm0.170$&$-4.964\pm0.027$&$-5.328\pm0.101$&$-$\\
42.3&$-4.641\pm0.020$&$-4.663\pm0.021$&$-4.906\pm0.022$&$-6.317\pm0.092$&$-5.218\pm0.062$&$-5.584\pm0.067$&$-$\\
42.5&$-5.139\pm0.027$&$-5.152\pm0.028$&$-5.408\pm0.035$&$-7.334\pm0.186$&$-5.752\pm0.045$&$-6.077\pm0.114$&$-$\\
42.7&$-5.306\pm0.178$&$-5.324\pm0.186$&$-5.931\pm0.047$&$-7.145\pm0.211$&$-6.068\pm0.098$&$-5.616\pm0.028$&$-$\\
42.9&$-6.012\pm0.060$&$-6.032\pm0.062$&$-6.418\pm0.101$&$-7.936\pm0.425$&$-6.478\pm0.108$&$-6.987\pm0.329$&$-$\\
43.1&$-6.412\pm0.116$&$-6.428\pm0.120$&$-6.589\pm0.172$&$-$&$-7.200\pm0.219$&$-7.247\pm0.248$&$-$\\
43.3&$-7.016\pm0.158$&$-7.196\pm0.179$&$-7.324\pm0.223$&$-8.072\pm0.425$&$-8.075\pm0.425$&$-8.111\pm0.112$&$-$\\
43.5&$-8.078\pm0.425$&$-8.078\pm0.425$&$-8.078\pm0.425$&$-$&$-$&$-$&$-$\\
\bottomrule
  \end{tabular}
   \caption{\NII and \SII intrinsic luminosity functions of the \mainsel sample and their different components. }
 \label{tab:LFtable3intr}
\end{table*}

\section{Evolution of the LFs in all the lines of interest}
\label{sec:appendixevol}
  \begin{figure*}
    \includegraphics[width=0.48\linewidth]{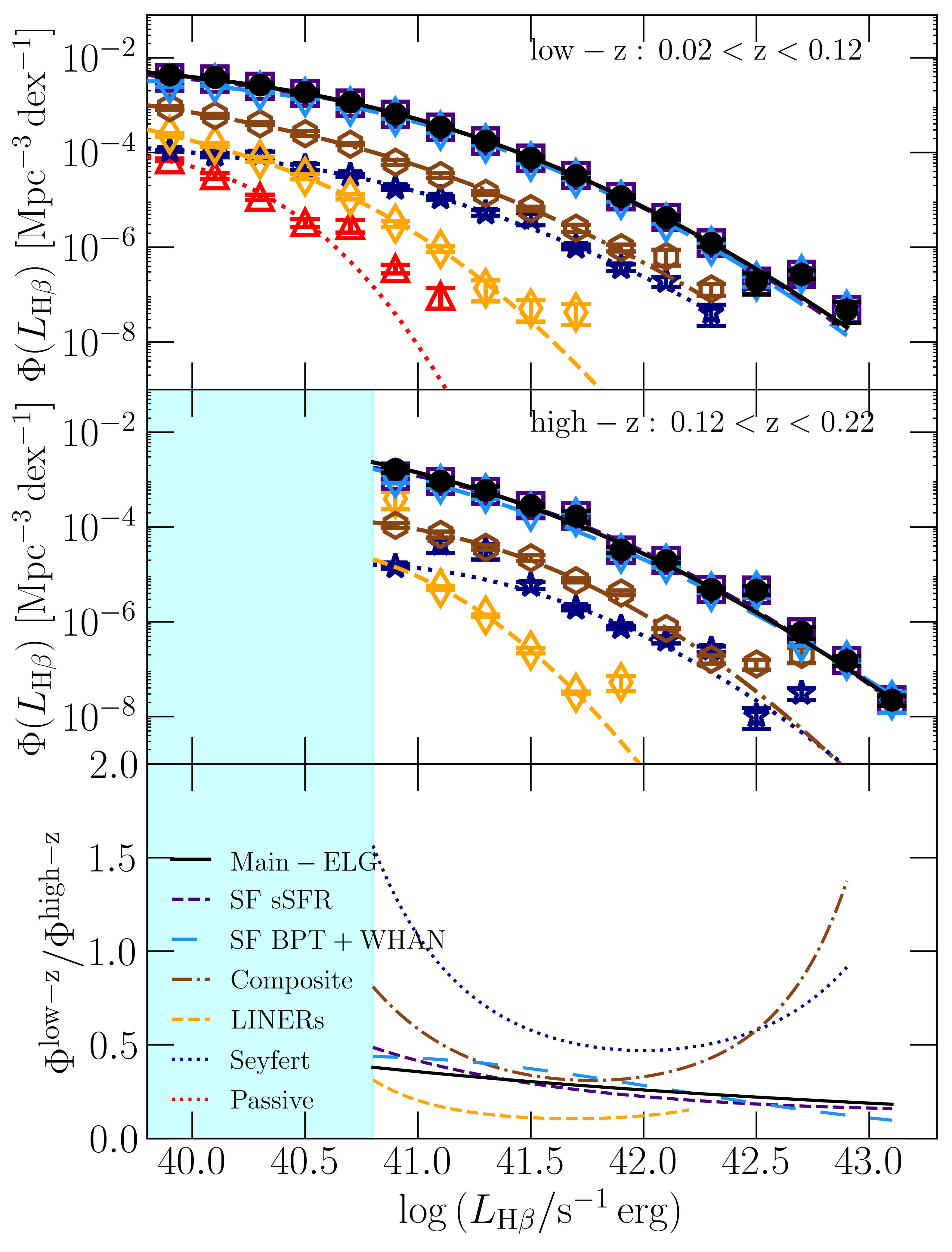}\quad
    \includegraphics[width=0.48\linewidth]{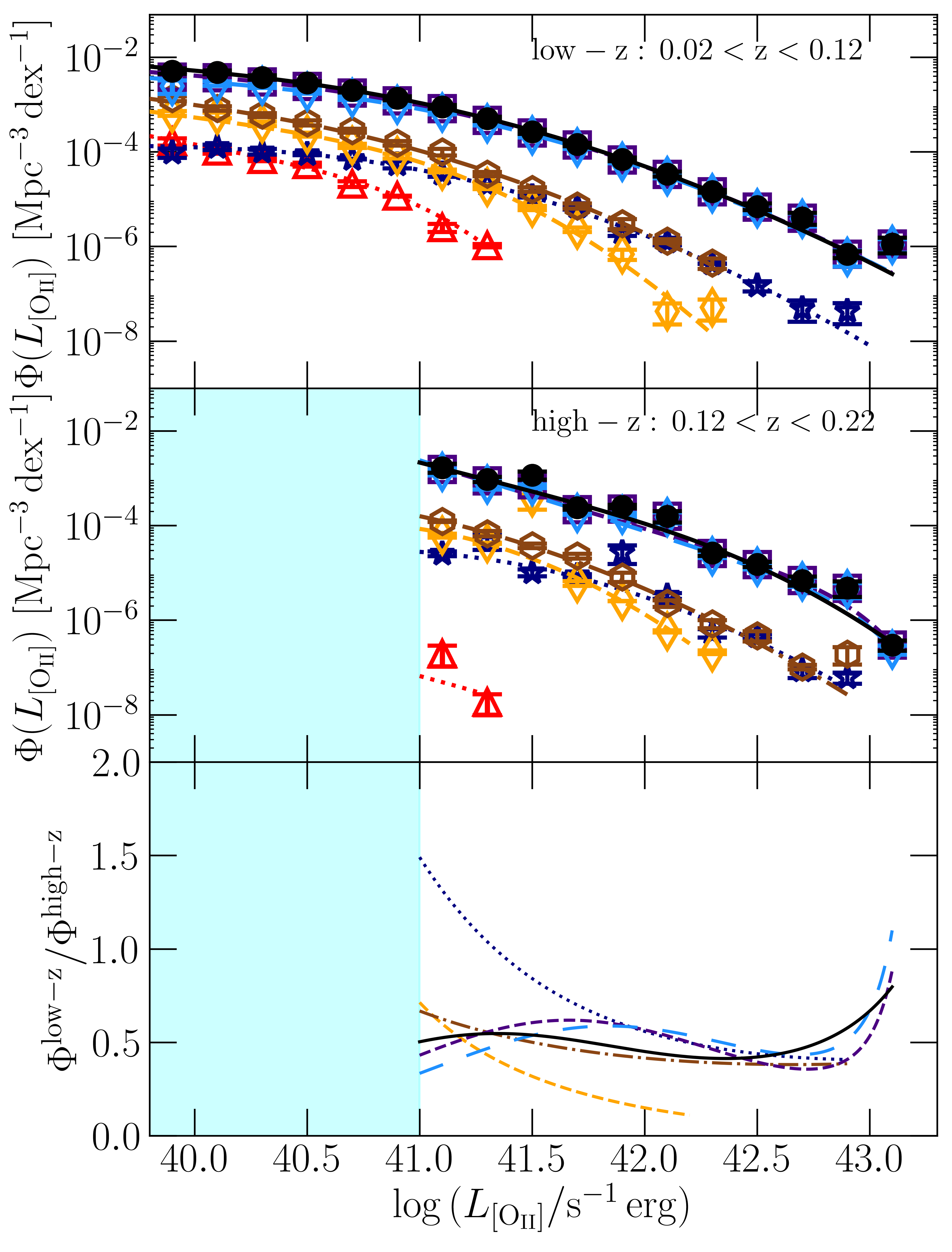}\\\vspace{0.4cm}
    \includegraphics[width=0.48\linewidth]{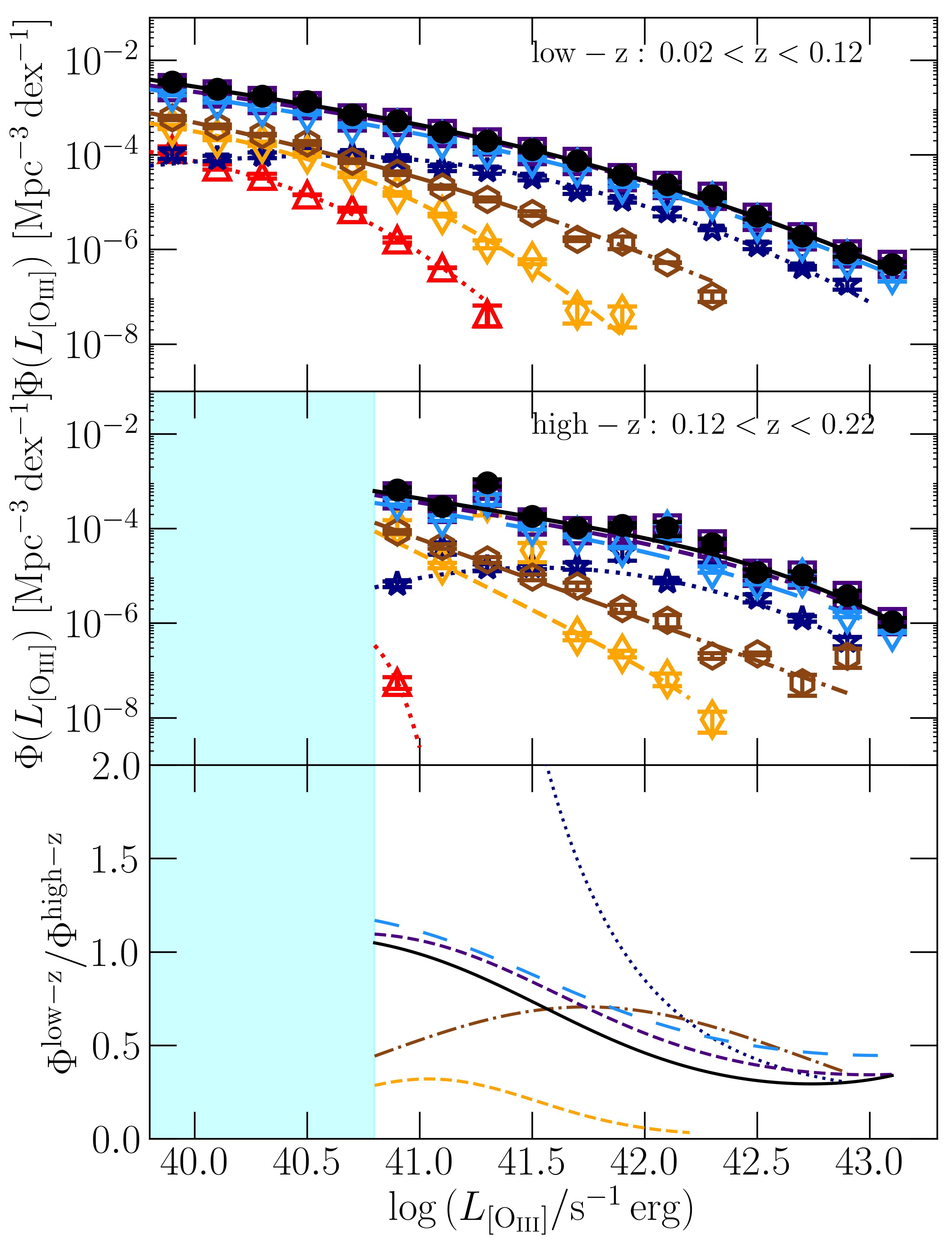}\quad
    \includegraphics[width=0.48\linewidth]{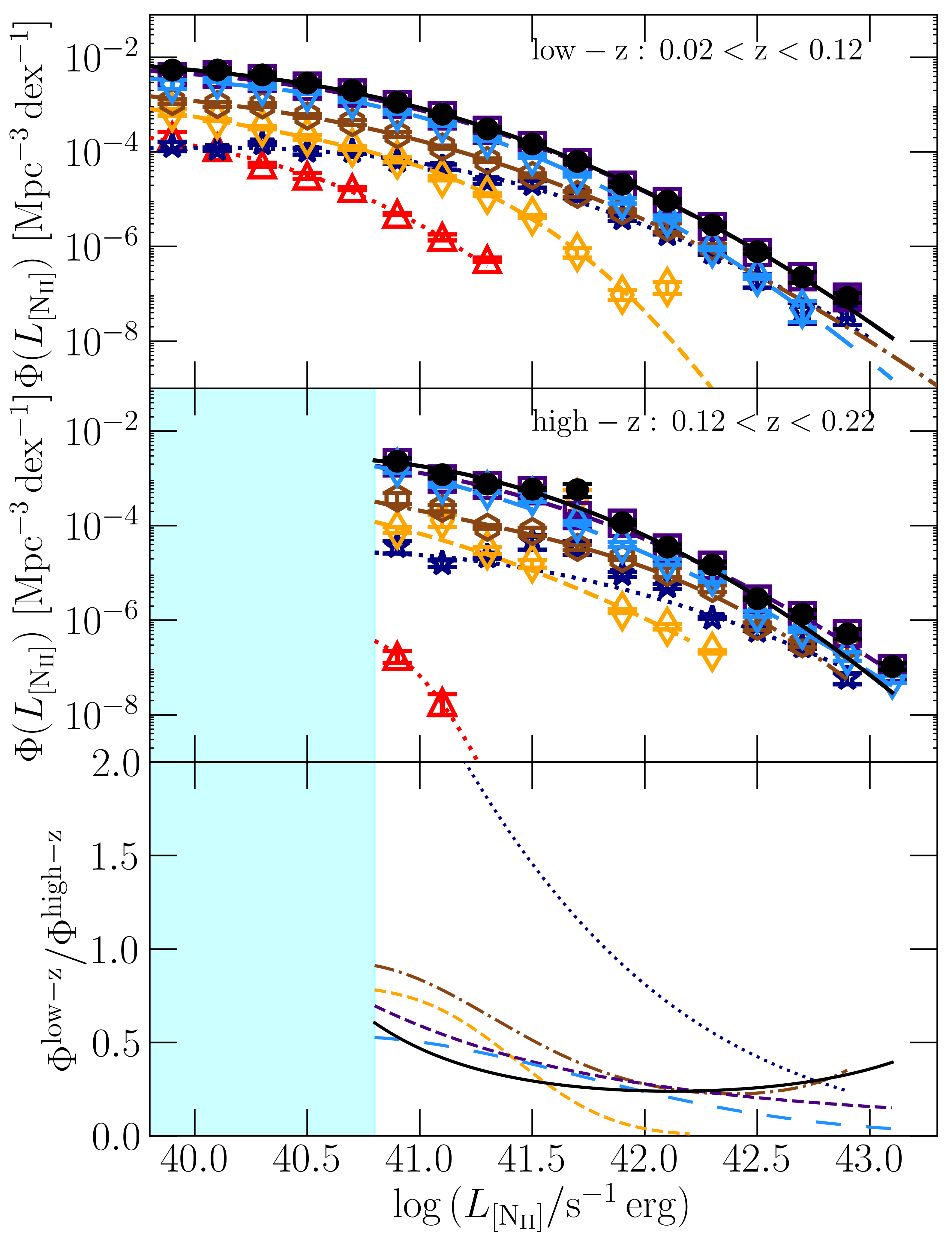}\vspace{-0.5cm}
    \caption{From top left to bottom right we show the observed \Hb, \OII, \OIII and \NII LFs and their different ELG components in the low-$z$ (top panels) and high-$z$ (middle panels) bins, together with their Saunders fits. The bottom panels in each figure displays the ratios between the high- and low-$z$ LF fits. The shaded cyan regions indicate where the incompleteness starts to dominate and our LF results cannot be trusted. The completeness limit value for each line is provided in Table\,\ref{tab:complimitsall}. For those lines for which the shade is not visible, the completeness limit is at lower $L$, outside the figure range. }
\label{fig:LFevolotherlines}
  \end{figure*}
\begin{figure}
    \includegraphics[width=\linewidth]{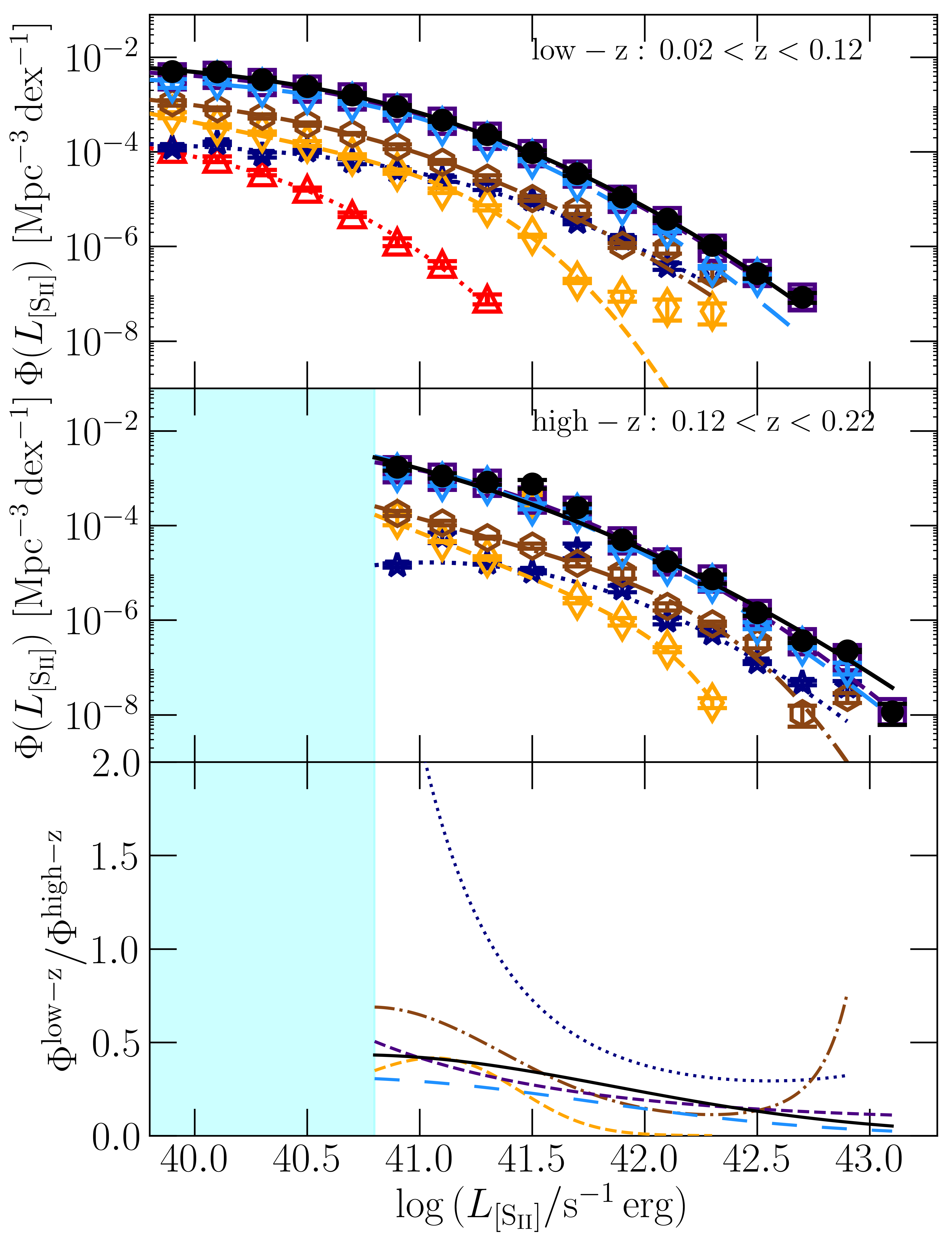}\quad
   \caption{Same as Fig.\,\ref{fig:LFevolotherlines}, but for the \SII line. }
\label{fig:LFevolotherlines2}
  \end{figure}
  
\begin{table*}
 \setlength{\tabcolsep}{5pt}
  \centering
 \begin{tabular}{l c c c c c}
    \toprule
    &\multicolumn{5}{c}{Saunders \Hb LF (observed)}\\    
    &$\log{(\Phi_\star/[{\rm Mpc^{-3}dex^{-1}}])}$&$\log{(L_\star/[\rm erg\,s^{-1}])}$&$\alpha$&$\sigma$&$\chi^2_{\rm red}$\\
    \midrule
    $0.02<z<0.12$ & & & & &  \\
    Full sample&-2.35$\pm$0.12&40.00$\pm$0.01&-0.40$\pm$0.11&-0.66$\pm$0.02&1.0\\ 
    SF sSFR&-2.40$\pm$0.17&40.00$\pm$0.16&-0.34$\pm$0.21&0.64$\pm$0.04&3.5\\
    SF BPT+WHAN&-2.52$\pm$0.08&40.04$\pm$0.17&-0.24$\pm$0.13&0.61$\pm$0.02&1.3\\
    LINERs&-3.63$\pm$0.29&40.00$\pm$0.26&-0.77$\pm$0.23&-0.42$\pm$0.05&1.3\\
    Composite&-3.12$\pm$0.32&40.00$\pm$0.41&-0.62$\pm$0.17&-0.67$\pm$0.08&2.8\\
    Seyfert&-3.93$\pm$0.18&40.00$\pm$0.34&-0.36$\pm$0.17&-0.66$\pm$0.05&1.4\\
    Passive&-4.05$\pm$1.22&40.00$\pm$1.35&-0.44$\pm$1.14&-0.26$\pm$0.32&8.3\\
        \midrule
    $0.12<z<0.22$ & & & & &  \\
    Full sample&-4.58$\pm$1.13&42.03$\pm$0.62&-1.60$\pm$0.13&0.45$\pm$0.21&3.1\\ 
    SF sSFR&-2.19$\pm$1.13&40.00$\pm$0.30&-0.27$\pm$0.17&0.68$\pm$0.21&3.3\\
    SF BPT+WHAN&-2.36$\pm$0.56&40.00$\pm$0.21&-0.12$\pm$0.49&0.65$\pm$0.18&3.4\\
    LINERs&-7.24$\pm$2.72&41.69$\pm$1.89&-3.33$\pm$2.10&-8.00$\pm$2.38&8.6\\
    Composite&-3.74$\pm$0.80&40.00$\pm$0.89&0.58$\pm$0.61&0.52$\pm$0.09&3.2\\
    Seyfert&-5.18$\pm$0.33&40.00$\pm$0.60&-1.37$\pm$0.25&0.47$\pm$0.22&5.9\\
    Passive&--&--&--&--&--\\
\midrule
    &\multicolumn{5}{c}{\OII }\\     
    \midrule
    $0.02<z<0.12$ & & & & &  \\
    Full sample&-2.28$\pm$0.11&40.08$\pm$0.20&-0.37$\pm$0.08&0.78$\pm$0.02&1.3\\ 
    SF sSFR&-2.35$\pm$0.14&40.00$\pm$0.06&-0.28$\pm$0.20&0.79$\pm$0.03&1.3\\
    SF BPT+WHAN&-2.55$\pm$0.12&40.18$\pm$0.27&-0.27$\pm$0.15&0.74$\pm$0.03&1.6\\
    LINERs&-3.55$\pm$0.32&40.54$\pm$0.32&-0.76$\pm$0.16&-0.49$\pm$0.08&5.4\\
    Composite&-2.97$\pm$0.26&40.00$\pm$1.17&-0.52$\pm$0.21&0.71$\pm$0.06&1.2\\
    Seyfert&-3.85$\pm$0.09&40.00$\pm$0.89&-0.03$\pm$0.01&-0.69$\pm$0.06&0.7\\
    Passive&-4.53$\pm$0.54&40.82$\pm$0.45&-0.86$\pm$0.18&-0.27$\pm$0.12&3.7\\
        \midrule
    $0.12<z<0.22$ & & & & &  \\
    Full sample&-3.27$\pm$1.85&41.56$\pm$1.44&-1.10$\pm$0.83&0.59$\pm$0.28&8.5\\ 
    SF sSFR&-6.86$\pm$2.31&43.98$\pm$1.46&-1.40$\pm$0.13&0.03$\pm$0.47&3.7\\
    SF BPT+WHAN&-6.91$\pm$1.53&43.99$\pm$0.22&-1.39$\pm$0.12&0.02$\pm$0.54&4.6\\
    LINERs&-3.35$\pm$0.52&40.00$\pm$0.93&-0.02$\pm$0.77&0.59$\pm$0.84&3.6\\
    Composite&-3.06$\pm$0.82&40.00$\pm$0.23&-0.24$\pm$0.51&-0.69$\pm$0.45&2.4\\
    Seyfert&-3.85$\pm$0.07&40.00$\pm$0.36&-0.04$\pm$0.21&0.70$\pm$0.003&1.0\\
    Passive&-5.71$\pm$1.42&40.60$\pm$0.56&4.15$\pm$2.50&0.16$\pm$0.06&0.2\\
\midrule
    &\multicolumn{5}{c}{\OIII }\\     
    \midrule
    $0.02<z<0.12$ & & & & &  \\
    Full sample&-2.86$\pm$0.24&40.44$\pm$0.29&-0.71$\pm$0.06&0.97$\pm$0.06&1.6\\ 
    SF sSFR&-3.17$\pm$0.23&40.71$\pm$0.27&-0.70$\pm$0.05&0.87$\pm$0.07&1.5\\
    SF BPT+WHAN&-3.03$\pm$0.25&40.37$\pm$0.31&-0.69$\pm$0.07&0.98$\pm$0.07&1.5\\
    LINERs&-3.42$\pm$0.31&40.04$\pm$0.30&-0.71$\pm$0.21&-0.48$\pm$0.06&1.8\\
    Composite&-3.28$\pm$0.51&40.00$\pm$0.79&-0.76$\pm$0.19&0.77$\pm$0.12&3.7\\
    Seyfert&-4.09$\pm$0.11&40.00$\pm$0.15&0.46$\pm$0.26&-0.67$\pm$0.04&3.1\\
    Passive&0.30$\pm$1.21&41.65$\pm$1.84&5.19$\pm$1.73&-0.02$\pm$0.06&0.6\\
        \midrule
    $0.12<z<0.22$ & & & & &  \\
    Full sample&-3.94$\pm$1.40&41.73$\pm$1.31&-0.91$\pm$0.27&0.74$\pm$0.45&6.9\\ 
    SF sSFR&-6.28$\pm$3.57&43.99$\pm$0.53&-0.96$\pm$0.08&-0.03$\pm$0.90&5.0\\
    SF BPT+WHAN&-5.26$\pm$1.67&42.81$\pm$1.59&-0.96$\pm$0.10&-0.24$\pm$0.57&6.4\\
    LINERs&-10.98$\pm$3.02&43.77$\pm$2.07&-2.33$\pm$0.32&-0.01$\pm$0.18&4.0\\
    Composite&-4.78$\pm$2.80&41.43$\pm$1.84&-1.35$\pm$0.65&-0.52$\pm$0.60&2.9\\
    Seyfert&-6.68$\pm$2.58&40.00$\pm$0.91&2.57$\pm$1.48&0.50$\pm$0.04&2.4\\
    Passive&-4.26$\pm$0.35&40.28$\pm$0.31&-0.74$\pm$0.19&0.32$\pm$0.07&1.7\\
\bottomrule
  \end{tabular}
   \caption{Best-fit Saunders parameters of the observed \Hb, \OII and \OIII LF fits shown in Fig.\,\ref{fig:LFevolotherlines} in the two redshift bins, $0.02<z<0.12$ and $0.12<z<0.22$.}
 \label{tab:fitparsaunderszbinsall}
\end{table*}

\begin{table*}
 \setlength{\tabcolsep}{5pt}
  \centering
 \begin{tabular}{l c c c c c}
    \toprule
    &\multicolumn{5}{c}{Saunders \NII LF (observed)}\\    &$\log{(\Phi_\star/[\rm erg\,s^{-1}])}$&$\log{(L_\star/[\rm s^{-1}\,erg])}$&$\alpha$&$\sigma$&$\chi^2_{\rm red}$\\
    \midrule
    $0.02<z<0.12$ & & & & &  \\
    Full sample&-2.20$\pm$0.12&40.00$\pm$0.34&-0.35$\pm$0.16&0.67$\pm$0.02&1.2\\ 
    SF sSFR&-2.32$\pm$0.10&40.00$\pm$0.24&-0.23$\pm$0.17&0.65$\pm$0.02&1.2\\
    SF BPT+WHAN&-2.46$\pm$0.09&40.00$\pm$0.25&-0.13$\pm$0.21&0.59$\pm$0.02&2.6\\
    LINERs&-3.19$\pm$1.03&40.00$\pm$0.72&-0.64$\pm$0.87&-0.63$\pm$0.12&7.4\\
    Composite&-3.17$\pm$0.56&40.00$\pm$0.16&-0.47$\pm$0.58&0.56$\pm$0.10&1.0\\
    Seyfert&-3.88$\pm$0.06&40.00$\pm$0.54&0.21$\pm$0.45&0.64$\pm$0.04&2.5\\
    Passive&-3.82$\pm$0.79&40.00$\pm$0.81&-0.69$\pm$0.57&0.48$\pm$0.17&2.7\\
        \midrule
    $0.12<z<0.22$ & & & & &  \\
    Full sample&-2.33$\pm$0.52&40.31$\pm$1.25&-0.16$\pm$0.22&0.62$\pm$0.06&4.3\\ 
    SF sSFR&-2.38$\pm$1.77&40.39$\pm$1.58&-0.35$\pm$0.07&0.64$\pm$0.07&3.7\\
    SF BPT+WHAN&-2.42$\pm$1.09&40.61$\pm$1.18&-0.91$\pm$0.65&-0.70$\pm$0.11&4.7\\
    LINERs&-3.84$\pm$1.03&40.00$\pm$0.76&0.59$\pm$0.37&-0.51$\pm$0.09&4.4\\
    Composite&-4.39$\pm$1.54&41.75$\pm$1.19&-1.10$\pm$0.38&0.42$\pm$0.40&5.7\\
    Seyfert&-5.92$\pm$0.64&40.01$\pm$0.96&2.15$\pm$0.32&-0.48$\pm$0.25&5.6\\
    Passive&-7.55$\pm$0.93&40.03$\pm$1.11&7.99$\pm$1.93&-0.17$\pm$0.004&0.1\\
\midrule
    &\multicolumn{5}{c}{\SII}\\
    \midrule
    $0.02<z<0.12$ & & & & &  \\
    Full sample&-2.24$\pm$0.06&40.00$\pm$0.13&-0.29$\pm$0.11&0.61$\pm$0.01&0.4\\ 
    SF sSFR&-2.35$\pm$0.09&40.00$\pm$0.17&-0.20$\pm$0.19&0.59$\pm$0.02&1.3\\
    SF BPT+WHAN&-2.54$\pm$0.11&40.21$\pm$0.21&-0.23$\pm$0.17&-0.53$\pm$0.02&2.1\\
    LINERs&-3.53$\pm$0.24&40.30$\pm$0.26&-0.65$\pm$0.16&-0.46$\pm$0.05&2.3\\
    Composite&-2.98$\pm$0.28&40.03$\pm$0.43&-0.45$\pm$0.31&0.59$\pm$0.05&2.2\\
    Seyfert&-4.36$\pm$0.18&41.01$\pm$0.25&-0.46$\pm$0.08&0.42$\pm$0.07&3.4\\
    Passive&-3.90$\pm$0.45&40.00$\pm$0.06&-0.52$\pm$0.50&0.36$\pm$0.09&1.2\\
        \midrule
    $0.12<z<0.22$ & & & & &  \\
    Full sample&-2.30$\pm$0.69&40.00$\pm$0.25&0.35$\pm$0.16&0.55$\pm$0.11&5.6\\ 
    SF sSFR&-2.42$\pm$0.72&40.00$\pm$0.84&0.24$\pm$0.75&-0.58$\pm$0.13&2.7\\
    SF BPT+WHAN&-2.41$\pm$0.24&40.00$\pm$1.33&0.16$\pm$0.31&-0.58$\pm$0.15&5.1\\
    LINERs&-9.92$\pm$0.62&43.97$\pm$0.80&-1.93$\pm$0.26&-0.08$\pm$0.60&2.0\\
    Composite&-4.24$\pm$0.73&41.42$\pm$0.55&-0.99$\pm$0.47&0.40$\pm$0.11&5.1\\
    Seyfert&-5.66$\pm$0.76&40.00$\pm$0.78&1.94$\pm$0.38&0.47$\pm$0.06&5.4\\
    Passive&--&--&--&--&--\\
\bottomrule
  \end{tabular}
   \caption{Same as previous Table, but for \NII and \SII.} 
 \label{tab:fitparsaunderszbinsN2S2}
\end{table*}

\section{Other functional forms for the LF fits}
\label{sec:otherLFfits}
\begin{figure*}
\centering\vspace{-0.3cm}
    \includegraphics[width=0.45\linewidth]{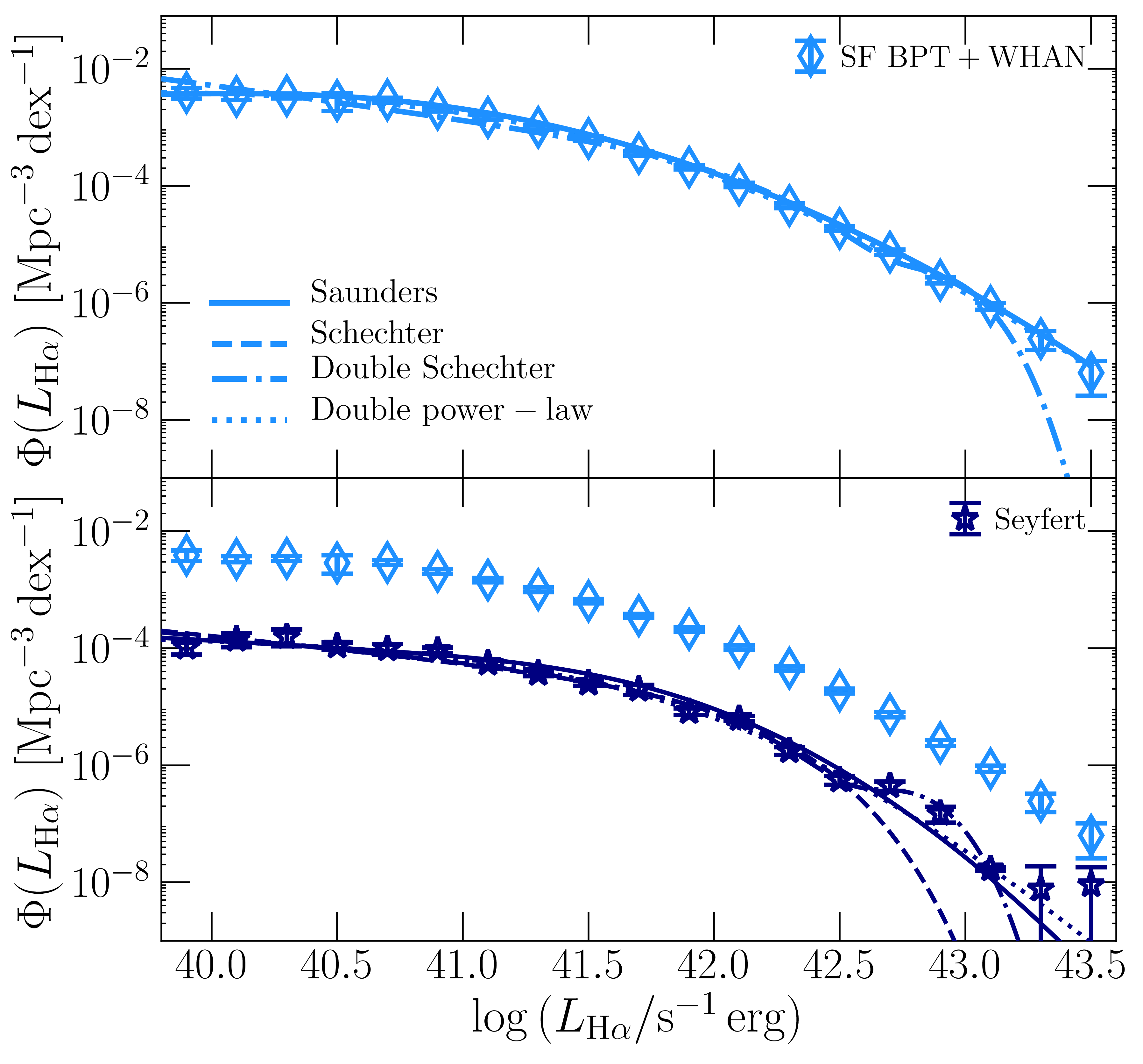}\hfill\quad
    \includegraphics[width=0.45\linewidth]{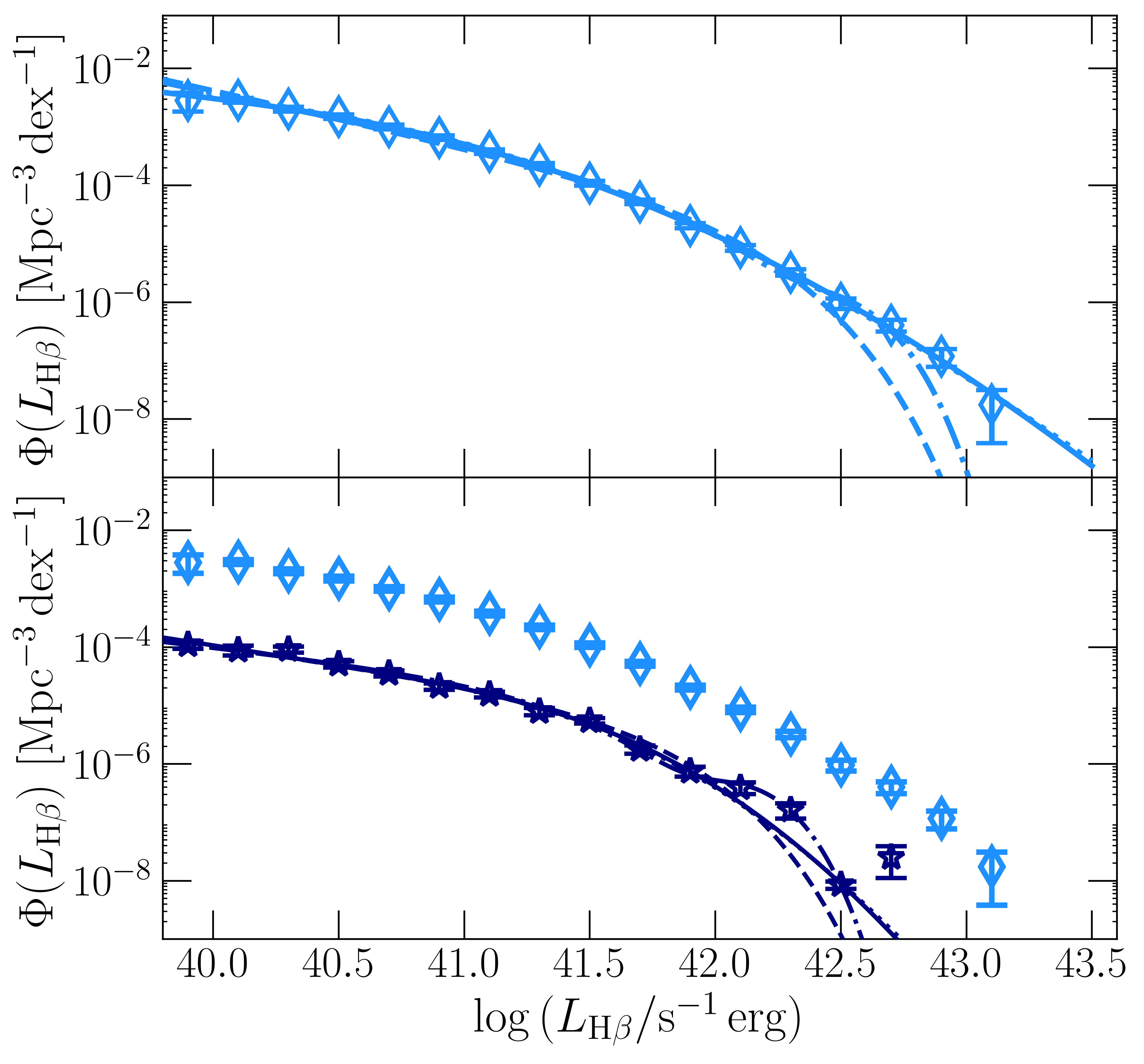}
    \includegraphics[width=0.45\linewidth]{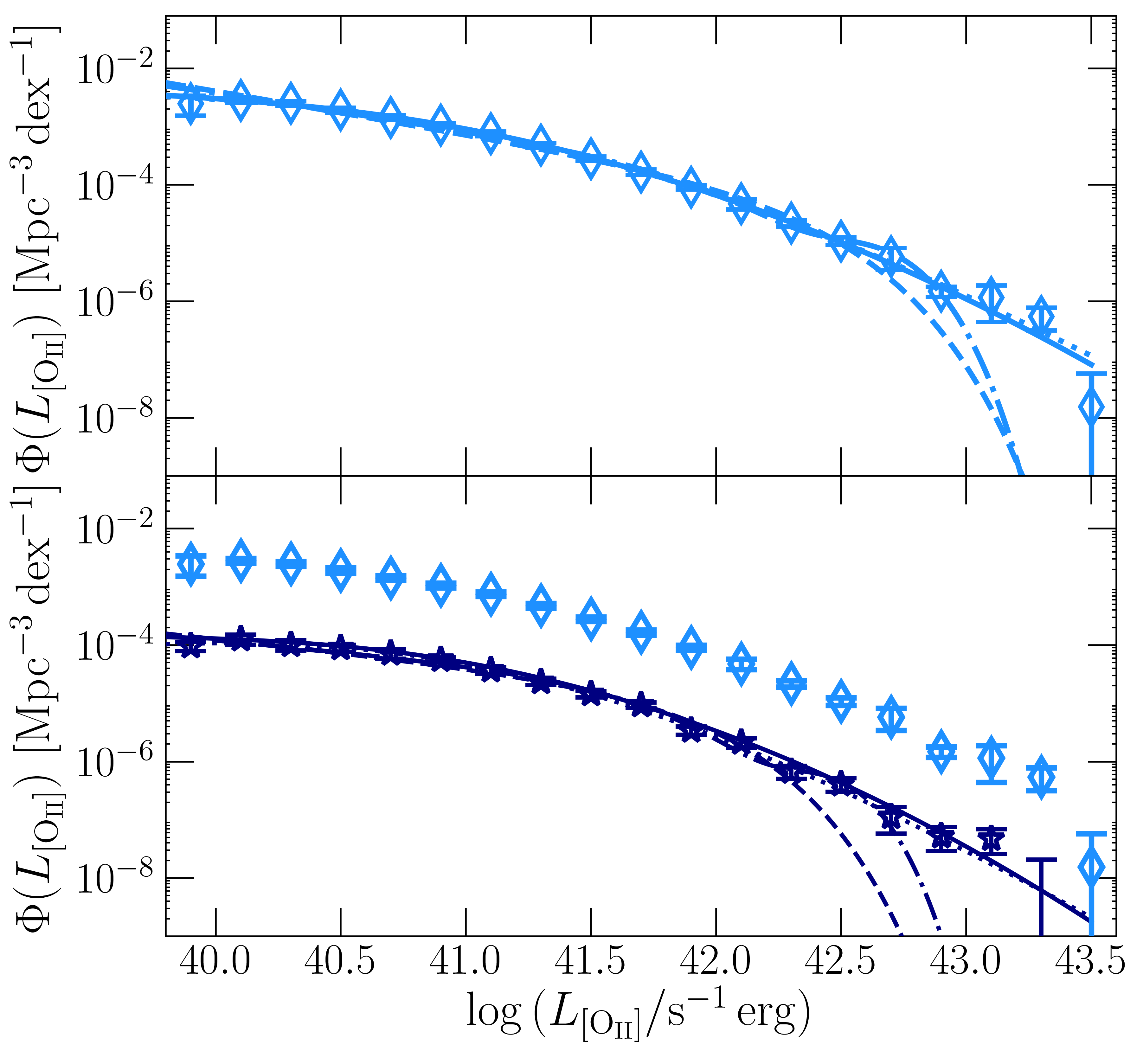}\hfill\quad
    \includegraphics[width=0.45\linewidth]{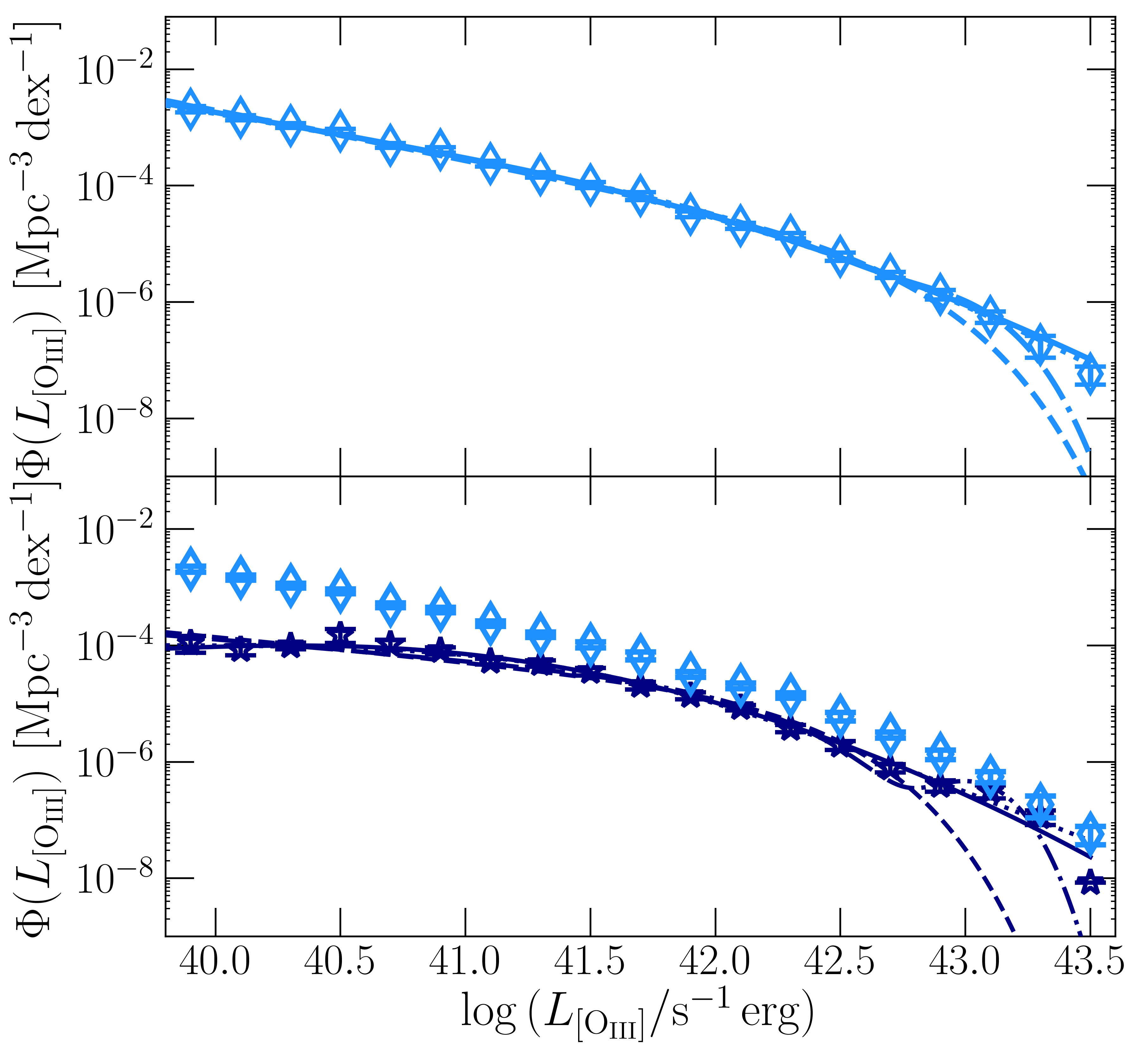}
    \includegraphics[width=0.45\linewidth]{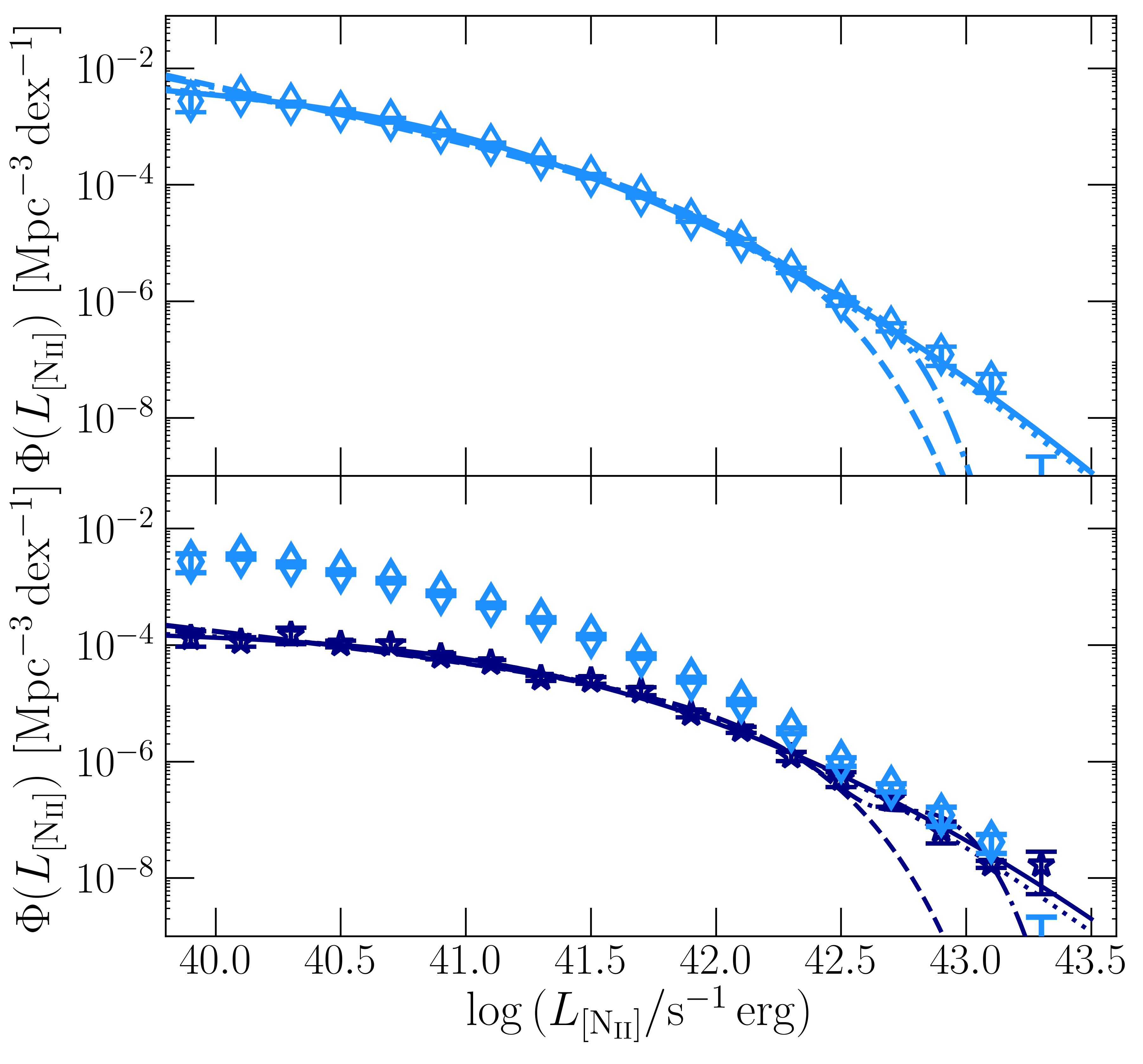}\hfill\quad
    \includegraphics[width=0.45\linewidth]{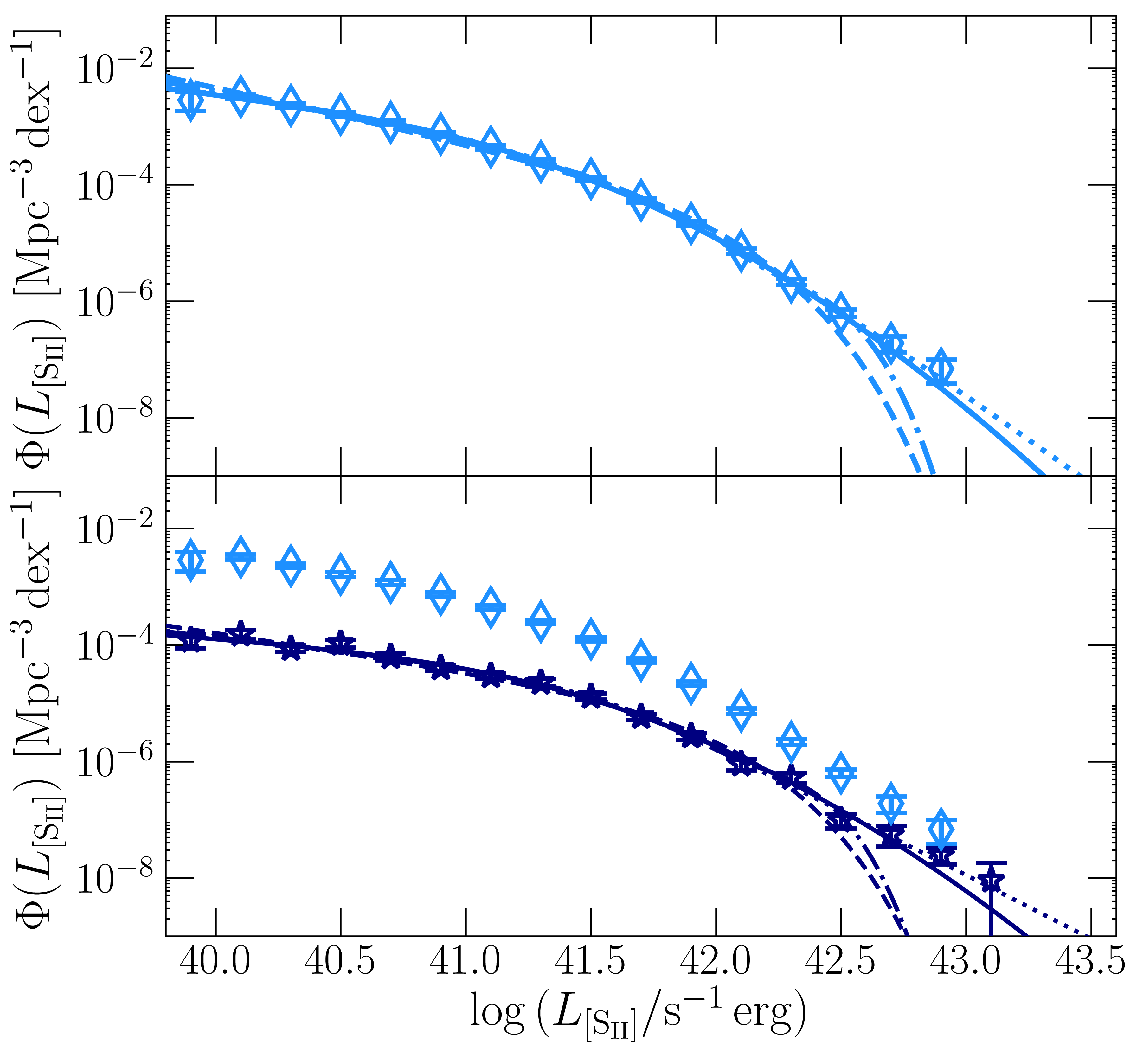}
    \vspace{-0.2cm}
    \caption{SF BPT+WHAN and Seyfert contributions to the six ELG luminosity functions of interest. We compare the performance of the Saunders fits already shown in Fig.\,\ref{fig:LFplotsaunders} (solid curves), with the Schechter (dashed), double Schechter (dot-dashed), and double power-law (dotted) functions.}
\label{fig:LFplotallfits}
  \end{figure*}

In Tables\,\ref{tab:fitparrestschechter}--\ref{tab:fitparrest2pl} we present the best-fit parameters of the LF fits using the models beyond Saunders, as described in Sec.~\ref{sec:LFcalculation}. The corresponding results are shown in Fig.\,\ref{fig:LFplotallfits}.

\begin{table*}
 \setlength{\tabcolsep}{3pt}
  \centering
 \begin{tabular}{l c c c c  }
    \toprule
    \multicolumn{5}{c}{Schechter (observed LF)}\\
    &$\log{(\Phi_\star/[{\rm Mpc^{-3}dex^{-1}}])}$&$\log{(L_\star/[\rm erg\,s^{-1}])}$&$\alpha$&$\chi^2_{\rm red}$\\
    \midrule
    &\multicolumn{4}{c}{\Ha}\\
    Full sample&-3.67$\pm$0.14&42.27$\pm$0.07& -0.71$\pm$0.06&8.7\\ 
    SF$_{\rm sSFR}$&-3.64$\pm$0.14&42.25$\pm$0.07&-0.67$\pm$0.06&9.5\\
    SF$_{\rm BPT+WHAN}$&-3.55$\pm$0.12&42.15$\pm$0.06&-0.59$\pm$0.06&7.3\\
    LINERs&-5.00$\pm$0.17&41.75$\pm$0.13&-1.03$\pm$0.12&19.7\\
    Composite&-4.49$\pm$0.15&42.19$\pm$0.09&-0.83$\pm$0.06&4.1\\
    Seyfert&-4.67$\pm$0.25&42.00$\pm$0.14&-0.44$\pm$0.16&18.9\\
    Passive&-3.00$\pm$1.01&40.00$\pm$1.85&-2.03$\pm$1.14&38.4\\
        \midrule
        &\multicolumn{4}{c}{\Hb}\\
    Full sample&-4.24$\pm$0.15&41.98$\pm$0.08&-1.01$\pm$0.06&4.8\\
    SF$_{\rm sSFR}$&-4.21$\pm$0.15&41.96$\pm$0.08&-0.98$\pm$0.06&4.9\\
    SF$_{\rm BPT+WHAN}$&-4.25$\pm$0.15&41.95$\pm$0.08&-0.96$\pm$0.06&4.7\\
    LINERs&-5.00$\pm$0.01&41.17$\pm$0.15&-1.18$\pm$0.18&12.8\\
    Composite&-5.00$\pm$0.14&41.80$\pm$0.08&-1.07$\pm$0.05&1.5\\
    Seyfert&-5.00$\pm$0.03&41.61$\pm$0.10&-0.64$\pm$0.11&5.9\\
    Passive&-4.00$\pm$0.01&40.00$\pm$1.43&-1.00$\pm$0.56&32.7\\
        \midrule
    &\multicolumn{4}{c}{\OII}\\
    Full sample&-4.00$\pm$0.15&42.24$\pm$0.09&-0.81$\pm$0.05&4.3\\
    SF$_{\rm sSFR}$&-3.68$\pm$0.12&42.02$\pm$0.08&-0.71$\pm$0.05&16.8\\
    SF$_{\rm BPT+WHAN}$&-4.02$\pm$0.14&42.23$\pm$0.09&-0.73$\pm$0.05&3.9\\
    LINERs&-5.00$\pm$0.24&41.92$\pm$0.12&-0.97$\pm$0.12&24.2\\
    Composite&-4.69$\pm$0.18&41.86$\pm$0.10&-0.96$\pm$0.06&3.9\\
    Seyfert&-4.71$\pm$0.14&41.80$\pm$0.09&-0.46$\pm$0.08&3.3\\
    Passive&-5.00$\pm$0.22&40.06$\pm$0.30&0.84$\pm$0.65&3.6\\
        \midrule
    &\multicolumn{4}{c}{\OIII}\\
    Full sample&-5.00$\pm$0.01&42.76$\pm$0.07&-0.92$\pm$0.03&2.7\\
    SF$_{\rm sSFR}$&-5.00$\pm$0.01&42.76$\pm$0.08&-0.89$\pm$0.03&3.0\\
    SF$_{\rm BPT+WHAN}$&-4.91$\pm$0.11&42.59$\pm$0.07&-0.85$\pm$0.03&1.3\\
    LINERs&-5.00$\pm$0.21&41.48$\pm$0.09&-1.06$\pm$0.10&7.2\\
    Composite&-5.00$\pm$0.05&41.72$\pm$0.13&-1.01$\pm$0.08&3.3\\
    Seyfert&-4.80$\pm$0.14&42.26$\pm$0.09&-0.42$\pm$0.07&4.9\\
    Passive&-4.00$\pm$0.85&40.01$\pm$1.23&-1.02$\pm$0.76&26.3\\
        \midrule
    &\multicolumn{4}{c}{\NII}\\
    Full sample&-4.06$\pm$0.15&42.05$\pm$0.08&-0.96$\pm$0.06&6.7\\
    SF$_{\rm sSFR}$&-4.09$\pm$0.15&42.05$\pm$0.08&-0.93$\pm$0.06&6.8\\
    SF$_{\rm BPT+WHAN}$&-4.16$\pm$0.16&41.95$\pm$0.07&-0.94$\pm$0.07&6.8\\
    LINERs&-5.00$\pm$0.18&41.84$\pm$0.07&-1.00$\pm$0.07&7.1\\
    Composite&-4.63$\pm$0.16&42.01$\pm$0.09&-0.87$\pm$0.06&4.5\\
    Seyfert&-4.78$\pm$0.18&41.97$\pm$0.11&-0.51$\pm$0.11&6.2\\
    Passive&-4.00$\pm$1.42&40.00$\pm$1.68&-1.00$\pm$0.78&23.2\\
        \midrule
    &\multicolumn{4}{c}{\SII}\\
    Full sample&-4.05$\pm$0.15&41.92$\pm$0.07&-0.99$\pm$0.06&6.6\\
    SF$_{\rm sSFR}$&-4.05$\pm$0.14&41.92$\pm$0.06&-0.96$\pm$0.06&6.3\\
    SF$_{\rm BPT+WHAN}$&-4.05$\pm$0.12&41.85$\pm$0.06&-0.93$\pm$0.06&4.8\\
    LINERs&-5.00$\pm$0.19&41.72$\pm$0.15&-1.01$\pm$0.14&25.2\\
    Composite&-4.51$\pm$0.15&41.69$\pm$0.08&-0.91$\pm$0.06&2.9\\
    Seyfert&-4.98$\pm$0.14&41.86$\pm$0.08&-0.64$\pm$0.07&1.5\\
    Passive&-6.40$\pm$1.94&40.66$\pm$0.80&-1.46$\pm$0.99&3.2\\
\bottomrule
  \end{tabular}
   \caption{Best-fit Schechter parameters to the measured luminosity functions shown in Tables\,\ref{tab:LFtable1} and \ref{tab:LFtable2}. } 
 \label{tab:fitparrestschechter}
\end{table*}

\begin{table*}
 \setlength{\tabcolsep}{3pt}
  \centering
 \begin{tabular}{l c c c c c c }
    \toprule
    \multicolumn{7}{c}{Double Schechter (observed LF)}\\
    &$\log{(\Phi_1^\star/[{\rm Mpc^{-3}dex^{-1}}])}$&$\log{(\Phi_2^\star/[\rm erg\,s^{-1}])}$&$\log{(L_\star/[\rm s^{-1}\,erg])}$&$\alpha_1$&$\alpha_2$&$\chi^2_{\rm red}$\\
    \midrule
    &\multicolumn{6}{c}{\Ha}\\
    Full sample&-3.78$\pm$0.10&-6.94$\pm$0.65&42.11$\pm$0.05&-1.63$\pm$0.05&3.78$\pm$0.87&4.8\\ 
    SF$_{\rm sSFR}$&-3.76$\pm$0.10&-7.02$\pm$0.70&42.10$\pm$0.06&-1.59$\pm$0.05&3.92$\pm$0.94&5.3\\
    SF$_{\rm BPT+WHAN}$&-3.84$\pm$0.10&-7.33$\pm$0.74&42.10$\pm$0.06&-1.57$\pm$0.06&4.19$\pm$0.97&3.6\\
    LINERs&-4.98$\pm$0.34&-7.27$\pm$0.56&41.57$\pm$0.36&-1.86$\pm$0.43&-2.78$\pm$0.87&17.8\\
Composite&-4.39$\pm$0.12&-6.45$\pm$0.48&41.86$\pm$0.07&-1.69$\pm$0.06&2.74$\pm$0.80&2.0\\
    Seyfert&-4.94$\pm$0.08&-9.10$\pm$0.74&41.93$\pm$0.04&-1.40$\pm$0.06&5.27$\pm$0.92&2.5\\
    Passive&-5.61$\pm$0.67&-8.89$\pm$3.86&40.00$\pm$0.48&-1.31$\pm$0.77&4.42$\pm$0.53&1.2\\
        \midrule
        &\multicolumn{6}{c}{\Hb}\\
    Full sample&-4.38$\pm$0.11&-7.29$\pm$0.59&41.85$\pm$0.59&-1.95$\pm$0.04&3.37$\pm$0.82&2.2\\
    SF$_{\rm sSFR}$&-4.35$\pm$0.11&-7.34$\pm$0.61&41.84$\pm$0.06&-1.91$\pm$0.05&3.47$\pm$0.83&2.3\\
    SF$_{\rm BPT+WHAN}$&-4.32$\pm$0.12&-6.94$\pm$0.54&41.78$\pm$0.06&-1.87$\pm$0.05&3.02$\pm$0.79&2.4\\
    LINERs&-5.00$\pm$0.12&-9.86$\pm$0.78&41.00$\pm$0.05&-2.01$\pm$0.08&5.76$\pm$0.94&2.0\\
    Composite&-4.73$\pm$0.14&-6.16$\pm$0.31&41.40$\pm$0.08&-1.91$\pm$0.06&1.75$\pm$0.71&0.9\\
    Seyfert&-5.00$\pm$0.09&-8.18$\pm$0.72&41.34$\pm$0.05&-1.48$\pm$0.08&4.18$\pm$0.99&1.8\\
    Passive&-5.61$\pm$3.24&-11.00$\pm$6.01&41.58$\pm$1.75&0.33$\pm$0.19&-4.02$\pm$2.19&31.4\\
        \midrule
    &\multicolumn{6}{c}{\OII}\\
    Full sample&-3.93$\pm$0.10&-6.02$\pm$0.40&41.93$\pm$0.06&-1.70$\pm$0.04&2.99$\pm$0.71&1.8\\
    SF$_{\rm sSFR}$&-3.88$\pm$0.08&-5.97$\pm$0.38&41.88$\pm$0.05&-1.66$\pm$0.04&3.15$\pm$0.66&12.7\\
    SF$_{\rm BPT+WHAN}$&-3.98$\pm$0.09&-6.10$\pm$0.39&41.93$\pm$0.06&-1.62$\pm$0.04&3.04$\pm$0.70&1.7\\
    LINERs&-4.96$\pm$0.09&-8.18$\pm$0.52&41.67$\pm$0.04&-1.88$\pm$0.05&3.81$\pm$0.72&2.0\\
    Composite&-4.86$\pm$0.13&-7.69$\pm$0.69&41.75$\pm$0.07&-1.91$\pm$0.05&3.57$\pm$0.99&2.0\\
    Seyfert&-4.88$\pm$0.10&-8.14$\pm$0.87&41.64$\pm$0.07&-1.37$\pm$0.07&4.31$\pm$1.17&2.0\\
    Passive&-6.68$\pm$4.25&-10.99$\pm$1.63&41.00$\pm$3.07&-1.48$\pm$0.45&7.24$\pm$4.65&13.1\\
        \midrule
    &\multicolumn{6}{c}{\OIII}\\
    Full sample&-5.00$\pm$0.08&-6.60$\pm$0.28&42.50$\pm$0.05&-1.86$\pm$0.02&2.17$\pm$0.51&1.2\\
    SF$_{\rm sSFR}$&-5.00$\pm$0.09&-6.62$\pm$0.32&42.49$\pm$0.05&-1.83$\pm$0.02&2.21$\pm$0.56&1.5\\
    SF$_{\rm BPT+WHAN}$&-5.00$\pm$0.02&-6.19$\pm$0.28&42.38$\pm$0.09&-1.81$\pm$0.03&1.59$\pm$0.62&1.7\\
    LINERs&-5.00$\pm$0.15&-7.42$\pm$0.48&41.28$\pm$0.07&-1.93$\pm$0.08&2.75$\pm$0.74&3.2\\
    Composite&-5.00$\pm$0.19&-6.23$\pm$0.34&41.46$\pm$0.11&-1.95$\pm$0.07&1.73$\pm$0.80&1.9\\
    Seyfert&-5.00$\pm$0.09&-9.97$\pm$1.07&42.13$\pm$0.06&-1.35$\pm$0.06&6.46$\pm$1.24&2.6\\
    Passive&-5.00$\pm$2.30&-6.72$\pm$3.42&40.00$\pm$0.58& -6.13$\pm$4.32&1.70$\pm$1.12&1.1\\
        \midrule
    &\multicolumn{6}{c}{\NII}\\
    Full sample&-4.24$\pm$0.12&-7.89$\pm$0.80&41.95$\pm$0.06&-1.91$\pm$0.05&4.22$\pm$1.04&3.8\\
    SF$_{\rm sSFR}$&-4.26$\pm$0.11&-8.05$\pm$0.83&41.94$\pm$0.06&-1.87$\pm$0.05&4.39$\pm$1.07&3.8\\
    SF$_{\rm BPT+WHAN}$&-4.20$\pm$0.12&-7.10$\pm$0.56&41.78$\pm$0.06&-1.85$\pm$0.06&3.23$\pm$0.78&3.0\\
    LINERs&-4.66$\pm$0.15&-6.39$\pm$0.35&41.42$\pm$0.08&-1.78$\pm$0.09&2.12$\pm$0.78&4.5\\
    Composite&-4.37$\pm$0.14&-6.23$\pm$0.49&41.59$\pm$0.08&-1.69$\pm$0.07&2.47$\pm$0.85&2.8\\
    Seyfert&-5.12$\pm$0.10&-9.91$\pm$1.21&41.97$\pm$0.06&-1.51$\pm$0.06&5.76$\pm$1.39&2.2\\
    Passive&-6.51$\pm$3.60&-11.00$\pm$4.56&40.68$\pm$1.93&-1.96$\pm$0.98&9.71$\pm$3.55&5.0\\
        \midrule
    &\multicolumn{6}{c}{\SII}\\
    Full sample&-3.98$\pm$0.13&-6.68$\pm$0.57&41.69$\pm$0.07&-1.84$\pm$0.06&2.91$\pm$0.79&3.1\\
    SF$_{\rm sSFR}$&-3.97$\pm$0.11&-6.69$\pm$0.51&41.67$\pm$0.06&-1.80$\pm$0.06&2.98$\pm$0.72&2.6\\
    SF$_{\rm BPT+WHAN}$&-4.04$\pm$0.10&-6.78$\pm$0.46&41.64$\pm$0.05&-1.79$\pm$0.05&2.91$\pm$0.64&1.9\\
    LINERs&-5.36$\pm$0.16&-8.93$\pm$0.85&41.70$\pm$0.08&-2.02$\pm$0.07&4.42$\pm$1.14&6.7\\
    Composite&-4.56$\pm$0.12&-6.66$\pm$0.39&41.49$\pm$0.07&-1.82$\pm$0.05&2.60$\pm$0.66&1.5\\
    Seyfert&-4.92$\pm$0.13&-7.31$\pm$0.64&41.57$\pm$0.09&-1.45$\pm$0.09&2.98$\pm$0.99&2.2\\
    Passive&-5.49$\pm$1.13&-11.00$\pm$6.54&40.06$\pm$0.75&-1.38$\pm$0.78&6.47$\pm$1.65&4.9\\
\bottomrule
  \end{tabular}
   \caption{Best-fit double Schechter parameters to the measured luminosity functions in Tables\,\ref{tab:LFtable1} and \ref{tab:LFtable2}. } 
 \label{tab:fitparrest2schechter}
\end{table*}

\begin{table*}
 \setlength{\tabcolsep}{3pt}
  \centering
 \begin{tabular}{l c c c c c c}
    \toprule
    &\multicolumn{6}{c}{Double power law (observed LF)}\\
    &$\log{(\Phi_0/[{\rm Mpc^{-3}dex^{-1}}])}$&$\log{(L_0/[\rm erg\,s^{-1}])}$&$\alpha_0$&$\alpha_1$&$\beta$&$\chi^2_{\rm red}$\\
    \midrule
    &\multicolumn{6}{c}{\Ha}\\
    Full sample&-2.12$\pm$0.52&41.94$\pm$0.15&-0.05$\pm$0.03&6.00$\pm$0.08&0.52$\pm$0.12&0.2\\ 
    SF$_{\rm sSFR}$&-1.99$\pm$0.62&41.87$\pm$0.14&-0.17$\pm$0.26&6.00$\pm$2.60&0.52$\pm$0.13&0.3\\
    SF$_{\rm BPT+WHAN}$&-1.98$\pm$0.74&41.82$\pm$0.14&-0.24$\pm$0.31&6.00$\pm$0.29&0.52$\pm$0.15&0.3\\
    LINERs&-3.93$\pm$0.27&41.16$\pm$0.15&0.86$\pm$0.22&1.46$\pm$0.46&2.06$\pm$1.68&4.5\\
    Composite&-2.23$\pm$1.17&41.34$\pm$0.44&-0.27$\pm$0.14&6.00$\pm$0.31&0.46$\pm$0.43&0.6\\
    Seyfert&-4.03$\pm$1.07&42.25$\pm$0.83&0.02$\pm$0.0.1&6.00$\pm$4.09&0.64$\pm$0.54&2.4\\
    Passive&-4.86$\pm$2.01&39.01$\pm$1.43&1.68$\pm$0.34&5.02$\pm$2.12&0.02$\pm$0.01&3.0\\
        \midrule
        &\multicolumn{6}{c}{\Hb}\\
    Full sample&-2.85$\pm$0.89&41.74$\pm$0.30&0.29$\pm$0.18&6.00$\pm$0.47&0.53$\pm$0.25&0.3\\
    SF$_{\rm sSFR}$&-2.78$\pm$0.94&41.71$\pm$0.28&0.23$\pm$0.16&6.00$\pm$1.63&0.53$\pm$0.25&0.3\\
    SF$_{\rm BPT+WHAN}$&-2.68$\pm$1.03&41.63$\pm$0.23&0.14$\pm$0.06&6.00$\pm$1.38&0.53$\pm$0.24&0.3\\
    LINERs&-3.88$\pm$0.56&40.61$\pm$0.40&0.88$\pm$0.30&1.55$\pm$1.07&2.09$\pm$1.19&0.7\\
    Composite&-3.51$\pm$1.07&41.47$\pm$1.23&0.34$\pm$0.21&6.00$\pm$2.23&0.49$\pm$0.24&0.5\\
    Seyfert&-5.00$\pm$1.48&42.07$\pm$1.48&0.47$\pm$0.44&6.00$\pm$2.52&0.86$\pm$0.74&2.3\\
    Passive&-4.90$\pm$2.67&40.17$\pm$1.22&-1.40$\pm$0.94&1.50$\pm$0.82&2.01$\pm$1.37&23.2\\
        \midrule
    &\multicolumn{6}{c}{\OII}\\
    Full sample&-1.67$\pm$1.64&41.27$\pm$0.36&-0.23$\pm$0.13&6.00$\pm$0.73&0.42$\pm$0.33&0.3\\
    SF$_{\rm sSFR}$&-1.86$\pm$0.81&41.41$\pm$0.22&-0.22$\pm$0.17&6.00$\pm$2.07&0.43$\pm$0.26&28.5\\
    SF$_{\rm BPT+WHAN}$&-1.99$\pm$1.66&41.47$\pm$0.33&-0.23$\pm$0.12&6.00$\pm$0.44&0.43$\pm$0.38&0.4\\
    LINERs&-4.20$\pm$0.17&41.50$\pm$0.13&0.74$\pm$0.17&2.35$\pm$0.93&1.21$\pm$0.55&1.5\\
    Composite&-3.61$\pm$0.14&41.16$\pm$0.10&0.66$\pm$0.17&1.72$\pm$0.50&1.32$\pm$0.55&0.7\\
    Seyfert&-2.66$\pm$1.02&41.08$\pm$0.58&-0.79$\pm$0.51&6.00$\pm$1.48&0.49$\pm$0.33&0.6\\
    Passive&-4.56$\pm$0.60&40.32$\pm$0.40&-1.03$\pm$0.18&-0.43$\pm$0.26&6.00$\pm$2.19&3.0\\
        \midrule
    &\multicolumn{6}{c}{\OIII}\\
    Full sample&-4.57$\pm$1.12&42.95$\pm$0.92&0.61$\pm$0.31&6.00$\pm$1.37&0.42$\pm$0.36&0.5\\
    SF$_{\rm sSFR}$&-4.28$\pm$0.67&42.79$\pm$0.88&0.52$\pm$0.42&6.00$\pm$1.02&0.41$\pm$0.30&0.6\\
    SF$_{\rm BPT+WHAN}$&-5.00$\pm$1.75&43.12$\pm$1.29&0.64$\pm$0.25&6.00$\pm$0.97&0.48$\pm$0.32&0.7\\
    LINERs&-3.93$\pm$0.49&40.90$\pm$0.24&0.67$\pm$0.57&2.05$\pm$1.34&1.19$\pm$0.99&1.6\\
    Composite&-2.26$\pm$1.43&40.01$\pm$0.42&-0.43$\pm$0.13&3.82$\pm$1.04&0.55$\pm$0.31&0.9\\
    Seyfert&-3.95$\pm$0.32&41.52$\pm$0.17&-0.03$\pm$0.01&1.82$\pm$1.42&0.95$\pm$0.58&1.0\\
    Passive&-5.00$\pm$0.64&40.72$\pm$0.93&-0.36$\pm$0.21&6.00$\pm$2.23&1.33$\pm$0.40&2.3\\
        \midrule
    &\multicolumn{6}{c}{\NII}\\
    Full sample&-2.29$\pm$1.02&41.60$\pm$0.18&0.08$\pm$0.05&6.00$\pm$0.50&0.51$\pm$0.21&0.4\\
    SF$_{\rm sSFR}$&-2.31$\pm$1.14&41.59$\pm$0.19&0.04$\pm$0.02&6.00$\pm$3.35&0.52$\pm$0.23&0.4\\
    SF$_{\rm BPT+WHAN}$&-2.96$\pm$0.87&41.84$\pm$0.33&0.26$\pm$0.21&6.00$\pm$1.07&0.59$\pm$0.28&0.9\\
    LINERs&-4.17$\pm$0.64&41.55$\pm$0.48&0.59$\pm$0.54&3.62$\pm$2.54&0.76$\pm$0.70&2.5\\
    Composite&-2.67$\pm$1.18&41.41$\pm$0.46&-0.06$\pm$0.04&6.00$\pm$2.50&0.48$\pm$0.36&1.0\\
    Seyfert&-4.06$\pm$1.16&42.18$\pm$0.81&0.05$\pm$0.01&6.00$\pm$0.55&0.59$\pm$0.48&1.1\\
    Passive&-4.57$\pm$1.11&40.00$\pm$0.21&-1.48$\pm$0.70&-0.92$\pm$0.69&6.00$\pm$1.81&4.4\\
        \midrule
    &\multicolumn{6}{c}{\SII}\\
    Full sample&-2.84$\pm$0.60&41.71$\pm$0.24&0.31$\pm$0.22&5.29$\pm$3.31&0.63$\pm$0.25&0.6\\
    SF$_{\rm sSFR}$&-2.74$\pm$0.64&41.75$\pm$0.24&0.22$\pm$0.17&5.95$\pm$3.56&0.59$\pm$0.21&0.4\\
    SF$_{\rm BPT+WHAN}$&-3.28$\pm$0.13&41.60$\pm$0.10&0.53$\pm$0.13&3.25$\pm$0.96&0.94$\pm$0.22&0.3\\
    LINERs&-4.02$\pm$0.43&41.13$\pm$0.27&0.89$\pm$0.37&1.51$\pm$0.80&1.86$\pm$2.51&10.1\\
    Composite&-2.98$\pm$1.78&41.34$\pm$0.31&0.10$\pm$0.06&5.83$\pm$1.53&0.51$\pm$0.37&0.4\\
    Seyfert&-4.34$\pm$0.21&41.50$\pm$0.10&0.53$\pm$0.11&1.33$\pm$0.50&2.08$\pm$1.17&1.2\\
    Passive&-4.29$\pm$1.67&40.01$\pm$1.53&-3.69$\pm$0.88&-2.60$\pm$0.57&6.00$\pm$1.71&2.0\\
\bottomrule
  \end{tabular}
   \caption{Best-fit parameters of our double power law fits to the measured luminosity functions in Tables\,\ref{tab:LFtable1} and \ref{tab:LFtable2}. } 
 \label{tab:fitparrest2pl}
\end{table*}

\end{appendix}
\end{document}